\documentclass[prb,aps,amssymb,twocolumn,superscriptaddress,10pt]{revtex4-2}
\usepackage{amsmath}
\usepackage{amssymb}
\usepackage{amsthm}
\usepackage{amsfonts}
\usepackage{listings}
\usepackage{longtable}
\usepackage{enumerate}
\usepackage{latexsym}
\usepackage{color}
\usepackage{setspace} 
\usepackage{blindtext}
\usepackage{dsfont}
\usepackage{mathrsfs}

\usepackage{array,etoolbox}
\preto\tabular{\setcounter{magicrownumbers}{0}}
\newcounter{magicrownumbers}

\usepackage{bm}
\usepackage{hyperref}
\hypersetup{
pdfnewwindow=true, colorlinks=true,
linkcolor=blue, anchorcolor=blue,
citecolor=blue, filecolor=blue,
menucolor=blue, urlcolor=blue}

\usepackage{graphicx}
\usepackage{subfigure}

\begin{document}

\newcommand{\reftosuppmat}[1]{(See~\ref{#1})} 
\newcommand{\extref}[1]{(See Ref. \ref{#1})}

\tolerance 10000

\newcommand{\vk}{\mathbf{k}}
\newcommand{\vp}{\mathbf{p}}
\newcommand{\vq}{\mathbf{q}}

\title{Twisted bilayer graphene.VI. An Exact Diagonalization Study of Twisted Bilayer Graphene at Non-Zero Integer Fillings}

\author{Fang Xie}
\affiliation{Department of Physics, Princeton University, Princeton, New Jersey 08544, USA}
\author{Aditya Cowsik}
\affiliation{Department of Physics, Princeton University, Princeton, New Jersey 08544, USA}
\author{Zhi-Da Song}
\affiliation{Department of Physics, Princeton University, Princeton, New Jersey 08544, USA}
\author{Biao Lian}
\affiliation{Department of Physics, Princeton University, Princeton, New Jersey 08544, USA}
\author{B. Andrei Bernevig}
\affiliation{Department of Physics, Princeton University, Princeton, New Jersey 08544, USA}
\author{Nicolas Regnault}
\affiliation{Department of Physics, Princeton University, Princeton, New Jersey 08544, USA}
\affiliation{Laboratoire de Physique de l'Ecole normale superieure, ENS, Universit\'e PSL, CNRS,
Sorbonne Universit\'e, Universit\'e Paris-Diderot, Sorbonne Paris Cit\'e, Paris, France}

\date{\today}

\begin{abstract}
Using exact diagonalization, we study the projected Hamiltonian with Coulomb interaction in the 8 flat bands of first magic angle twisted bilayer graphene.
Employing the U(4) (U(4)$\times$U(4)) symmetries in the nonchiral (chiral) flat band limit, we reduced the Hilbert space to an extent which allows for study around $\nu=\pm 3,\pm2,\pm1$ fillings. In the first chiral limit $w_0/w_1=0$ where $w_0$ ($w_1$) is the $AA$ ($AB$) stacking hopping, we find that the ground states at these fillings are extremely well-described by Slater determinants in a so-called Chern basis, and the exactly solvable charge $\pm1$ excitations found in Bernevig {\it et al}. [Phys. Rev. B 103, 205415 (2021)] are the lowest charge excitations up to system sizes $8\times8$ (for restricted Hilbert space) in the chiral-flat limit.
We also find that the Flat Metric Condition (FMC) used in Bernevig {\it et al}. [Phys. Rev. B 103, 205411 (2021)], Song {\it et al}. [Phys. Rev. B 103, 205412 (2021)], Bernevig {\it et al}. [Phys. Rev. B 103, 205413 (2021)], Lian {\it et al}. [Phys. Rev. B 103, 205414 (2021)], and Bernevig {\it et al}. [Phys. Rev. B 103, 205415 (2021)] for obtaining a series of exact ground states and excitations holds in a large parameter space. For $\nu=-3$, the ground state is the spin and valley polarized Chern insulator with $\nu_C=\pm1$ at  $w_0/w_1\lesssim0.9$ (0.3) with (without) FMC. At $\nu=-2$, we can only numerically access the valley polarized sector, and we find a spin ferromagnetic phase when $w_0/w_1\gtrsim0.5t$ where $t\in[0,1]$ is the factor of rescaling of the actual TBG bandwidth, and  a spin singlet phase otherwise, confirming the perturbative calculation [Lian. {\it et al}., Phys. Rev. B 103, 205414 (2021), Bultinck {\it et al}., Phys. Rev. X 10, 031034 (2020)]. The analytic FMC ground state is, however, predicted in the intervalley coherent sector which we cannot access [Lian {\it et al}., Phys. Rev. B 103, 205414 (2021), Bultinck et al., Phys. Rev. X 10, 031034 (2020)]. For $\nu=-3$ with/without FMC, when $w_0/w_1$ is large, the finite-size gap $\Delta$ to the neutral excitations vanishes, leading to phase transitions. Further analysis of the ground state momentum sectors at $\nu=-3$ suggests a competition among (nematic) metal, momentum $M_M$ ($\pi$) stripe and $K_M$-CDW orders at large $w_0/w_1$.
\end{abstract}

\maketitle

\section{Introduction}\label{sec:introduction}

The physics of the insulating states in twisted bilayer graphene (TBG) at integer electron number per unit cell has attracted considerable experimental and theoretical interest \cite{bistritzer_moire_2011,cao_correlated_2018,cao_unconventional_2018, lu2019superconductors, yankowitz2019tuning, sharpe_emergent_2019, saito_independent_2020, stepanov_interplay_2020, liu2020tuning, arora_2020, serlin_QAH_2019, cao_strange_2020, polshyn_linear_2019,  xie2019spectroscopic, choi_imaging_2019, kerelsky_2019_stm, jiang_charge_2019,  wong_cascade_2020, zondiner_cascade_2020,  nuckolls_chern_2020, choi2020tracing, saito2020,das2020symmetry, wu_chern_2020,park2020flavour, saito2020isospin,rozen2020entropic, lu2020fingerprints, burg_correlated_2019,shen_correlated_2020, cao_tunable_2020, liu_spin-polarized_2019, chen_evidence_2019, chen_signatures_2019, chen_tunable_2020, burg2020evidence, tarnopolsky_origin_2019, zou2018, fu2018magicangle, liu2019pseudo, Efimkin2018TBG, kang_symmetry_2018, song_all_2019,po_faithful_2019,ahn_failure_2019,Slager2019WL, hejazi_multiple_2019, lian2020, hejazi_landau_2019, padhi2020transport, xu2018topological,  koshino_maximally_2018, ochi_possible_2018, xux2018, guinea2018, venderbos2018, you2019,  wu_collective_2020, Lian2019TBG,Wu2018TBG-BCS, isobe2018unconventional,liu2018chiral, bultinck2020, zhang2019nearly, liu2019quantum,  wux2018b, thomson2018triangular,  dodaro2018phases, gonzalez2019kohn, yuan2018model,kang_strong_2019,bultinck_ground_2020,seo_ferro_2019, hejazi2020hybrid, khalaf_charged_2020,po_origin_2018,xie_superfluid_2020,julku_superfluid_2020, hu2019_superfluid, kang_nonabelian_2020, soejima2020efficient, pixley2019, knig2020spin, christos2020superconductivity,lewandowski2020pairing, xie_HF_2020,liu2020theories, cea_band_2020,zhang_HF_2020,liu2020nematic, daliao_VBO_2019,daliao2020correlation, classen2019competing, kennes2018strong, eugenio2020dmrg, huang2020deconstructing, huang2019antiferromagnetically,guo2018pairing, ledwith2020, repellin_EDDMRG_2020,abouelkomsan2020,repellin_FCI_2020, vafek2020hidden, fernandes_nematic_2020, Wilson2020TBG, wang2020chiral, ourpaper1, ourpaper2,ourpaper3,ourpaper4,ourpaper5}. Both  scanning tunneling microscope \cite{xie2019spectroscopic, jiang_charge_2019, choi_imaging_2019, zondiner_cascade_2020,wong_cascade_2020,nuckolls_chern_2020,choi2020tracing} and transport \cite{cao_correlated_2018,cao_unconventional_2018, lu2019superconductors, yankowitz2019tuning, sharpe_emergent_2019, saito_independent_2020, stepanov_interplay_2020, arora_2020, serlin_QAH_2019, cao_strange_2020, polshyn_linear_2019, saito2020,das2020symmetry,saito2020isospin, wu_chern_2020,park2020flavour} experiments 
show correlated insulators at integer fillings. Correlated Chern insulators originating at integer filling are also observed in either zero or finite  magnetic field \cite{serlin_QAH_2019,sharpe_emergent_2019}, but most importantly even without hBN substrate alignment \cite{nuckolls_chern_2020,choi2020tracing,saito2020,das2020symmetry,wu_chern_2020,park2020flavour}.
In the latter case, the single-particle picture predicts a gapless state at electron number $\pm3,\pm2,\pm1$ and hence the insulating states have to follow from many-body interactions. 

The initial observations of the insulating states were followed by the experimental discovery that these states might exhibit Chern numbers. So far, a rather intriguing picture of insulating states of Chern numbers $\pm (4-|\nu|) $, with or without the presence of a magnetic field, at integer filling $\nu \in (-4,4)$ has been discovered in spectroscopic \cite{nuckolls_chern_2020,choi2020tracing,saito2020,das2020symmetry,wu_chern_2020,park2020flavour}  
experiements. Superconductivity also appears in TBG samples, mostly at finite doping away from integer fillings \cite{cao_unconventional_2018,lu2019superconductors,yankowitz2019tuning,stepanov_interplay_2020,arora_2020,liu2020tuning} but also at or extremely close to integer fillings \cite{saito_independent_2020,stepanov_interplay_2020}, with or without enhanced screening by another graphene layer \cite{saito_independent_2020,stepanov_interplay_2020,liu2020tuning}. 

Theoretically, the initial important insight in the physics behind the many-body insulating states was the strong-coupling projected Coulomb interaction in the two flat bands of TBG obtained by Kang and Vafek \cite{kang_strong_2019}. By projecting into a set of Wannier orbitals, they found a positive semidefinite Hamiltonian (PSDH), of an enhanced approximate U(4) symmetry \cite{bultinck_ground_2020,kang_strong_2019,seo_ferro_2019,ourpaper3}. They then proceeded to show that some of the insulating groundstates (in their case the $\nu=\pm 2$ filling from charge neutrality) of this model can be obtained exactly. They also found one extended excitation of the model. These represent exact results. 
The large unit cell, large number of orbitals per moir\'e unit cell, strong interactions and topological obstruction \cite{ahn_failure_2019,po_faithful_2019,song_all_2019,ourpaper5, kang_symmetry_2018} of maximally symmetric Wannier orbitals make the numerical simulation of the TBG many-body physics unusually difficult. For magic angle TBG without hBN substrate alignment (where the Hamiltonian respects a $C_{2z}T$ symmetry), the theoretical efforts so far have focused on the Hartree-Fock (HF) studies employing momentum/hybrid basis of the Bistritzer-Macdonald (BM) continuum model \cite{xie_HF_2020,bultinck_ground_2020,liu2020theories,hejazi2020hybrid,cea_band_2020,zhang_HF_2020,liu2020nematic}, quantum Monte Carlo (QMC) simulation \cite{xux2018,daliao_VBO_2019,daliao2020correlation}, functional RG \cite{classen2019competing, kennes2018strong} and ED \cite{ochi_possible_2018, dodaro2018phases} with non-maximally-symmetric Wannier orbitals, and density matrix renormalization group (DMRG) simulation with hybrid Wannier wavefunctions \cite{soejima2020efficient,kang_nonabelian_2020} or simplified models \cite{eugenio2020dmrg, huang2020deconstructing}. The HF numerical calculations predicted various phases at integer fillings, including spin-valley polarized (Chern) insulators \cite{liu2020theories,hejazi2020hybrid}, intervalley coherent states \cite{zhang_HF_2020, bultinck_ground_2020} and nematic semimetals \cite{liu2020nematic}. The QMC studies predicted valley Hall insulator, intervalley coherent states or Kekul\'e valence bond orders at charge neutrality \cite{xux2018,daliao_VBO_2019,daliao2020correlation}, and unconventional superconductivity at non-integer fillings \cite{huang2019antiferromagnetically,guo2018pairing,kennes2018strong}. The recent DMRG studies using hybrid Wannier basis \cite{soejima2020efficient,kang_nonabelian_2020} predicted Chern number $\pm1$ insulator, $C_{2z}T$ symmetric ($C_{3z}$-breaking) nematic semimetal and $C_{2z}T$ symmetric stripe insulator at momentum $\pi$ (in either direction) as candidate ground states at $\nu=-3$. At $\nu=-3$, these studies find that the ground state in the chiral limit is quantum anomalous Hall (i.e., Chern insulator), and that in the nonchiral  limit the  nematic or stripe order takes over around $w_0/w_1\gtrsim0.8$ of the Bistritzer-MacDonald parameters. Besides, for TBG with hBN alignment which breaks $C_{2z}$, the single-particle bands form valley Chern bands, and exact diagonalizations (ED) or DMRG have been performed \emph{only} within single valley-spin polarized Chern band \cite{repellin_EDDMRG_2020,abouelkomsan2020,repellin_FCI_2020}, where fractional Chern insulators are proposed. The particularization to \emph{only} within single valley-spin polarized Chern bands renders their Hilbert space manageable, but potentially biases the system as the time-reversal symmetry breaking is introduced by hand.

Over the five previous parts \cite{ourpaper1, ourpaper2,ourpaper3, ourpaper4, ourpaper5}   of our series of six works on TBG, we have paved the way for employing  the  momentum-space  projected TBG Hamiltonian derived in Ref.~\cite{ourpaper3} which is of the PSDH  Kang-Vafek  type \cite{kang_strong_2019}. We showed that \emph{all} projected Coulomb Hamiltonians can be written in this PSDH  Kang-Vafek form \cite{ourpaper3}, and that, due to a particle-hole (PH) symmetry discovered in Ref.~\cite{song_all_2019}, they generically exhibit an enlarged symmetry group U(4) for \emph{any} number of projected bands, for any parameter regime. For projection into the lowest 8 active bands (2 per spin per valley), this U(4) was previously discovered in Ref.~\cite{bultinck_ground_2020}; in Ref.~\cite{ourpaper3} we also related our U(4) to the inital one discovered by Kang and Vafek \cite{kang_strong_2019}. In Ref.~\cite{ourpaper2}, we showed the Bistritzer-MacDonald model with the PH symmetry is always  anomalous \cite{ourpaper2} - meaning it is incompatible with the lattice, proving stable (\emph{not fragile}) topology of this model. We further discovered \emph{two} chiral limits \cite{ourpaper2}, in both of which the symmetry is enhanced to a U(4)$\times$U(4) symmetry (again of any number of bands) in the exactly flat band (projected Coulomb) model \cite{ourpaper3}. The U(4)$\times$U(4) of the first chiral-flat band limit in the lowest 8 bands was first shown in Ref.~\cite{bultinck_ground_2020}. When kinetic energy is added to the chiral limit, the symmetry is  lowered to U(4).

In papers Refs.~\cite{ourpaper4, ourpaper5} we have found a series of exact eigenstates of the PSDH Hamiltonians. Using a condition called the Flat Metric Condition (FMC) \cite{ourpaper1} Eq.~(\ref{ed:eq:FMC}), we have proved that some of these states form the \emph{exact  ground states} at all integer fillings in the (first) chiral limit \cite{ourpaper4}, and at even fillings away from the chiral limit. Our results,  presented in the Chern basis defined in Ref.~\cite{ourpaper3} (see also definition in Refs.~\cite{bultinck_ground_2020,liu2019pseudo}) are: in the (first) chiral-flat limit with relexation parameter $w_0/w_1=0$ (with U(4)$\times$U(4) symmetry), with the FMC Eq.~(\ref{ed:eq:FMC}),
the exact ground states at each integer filling $\nu$ ($|\nu|\le 4$) relative to the charge neutral point (CNP) are obtained by fully occupying any $\nu+4$ Chern bands (of either Chern number $\pm1$). This leads to exactly degenerate Chern insulator ground states with total Chern number $\nu_C=4-|\nu|,2-|\nu|,\cdots,|\nu|-4$.   When tuned to the nonchiral-flat limit (with U(4) symmetry \cite{bultinck_ground_2020, ourpaper3}),  we found \cite{ourpaper4} that the lowest possible   Chern number is favored: all the even fillings $\nu=0,\pm2$ have Chern number $0$ insulator exact U(4) ferromagnetic (FM) ground states, while all the odd fillings $\nu=\pm1,\pm3$ have Chern number $\pm1$ insulator U(4) ferromagnetic (FM) perturbative ground states. Perturbing in another direction, we obtain the (first) chiral-nonflat limit with a nonzero kinetic energy (with another U(4) symmetry \cite{bultinck_ground_2020, ourpaper3}) where we find all the different Chern number states at a fixed integer filling $\nu$ to be degenerate up to second order in kinetic energy. Upon further reducing to the nonchiral-nonflat case (with U(2)$\times$U(2) symmetry \cite{bultinck_ground_2020, ourpaper3}), we showed \cite{ourpaper4} that in second order perturbation, the U(4) ground states at all integer fillings $\nu$ favor intervalley coherent states if the Chern number $|\nu_C|<4-|\nu|$, and favor valley polarized states if $|\nu_C|=4-|\nu|$. At even fillings, this agrees with the K-IVC state proposed in Ref.~\cite{bultinck_ground_2020}. We note, however, that the possibility of other ground states in various limits are not ruled out in Ref.~\cite{ourpaper4}.

In paper \cite{ourpaper5}  we showed that exact expressions of the charge $\pm 1$, $\pm 2$ and neutral  excitation can be obtained for the exact ground states we found in Ref.~\cite{ourpaper5}. We predicted gaps, Goldstone stiffness, and the representations of each of the excitations. The neutral excitation has an exact zero mode, which we identify with the FM U(4)-spin wave. While these excitations are above the ground state for FMC, we could not prove that they are the \emph{lowest} charge excitations. The Goldstone branch, far away from $\mathbf{k}$, cannot be analytically proved to be the \emph{lowest} extitation, either.

In this paper, we present some of the first full Hilbert-space unbiased exact diagonalization (ED)  numerical calculations on the TBG problem. Our purpose is three-fold. First, we address the question of the robustness of the FMC model in the (first) chiral-flat limit for which exact (Chern) insulator ground states and excitations at integer fillings \cite{ourpaper4,ourpaper5, kang_strong_2019, vafek2020hidden} can be obtained. Away from the FMC (except for $\nu=0$), one cannot prove analytically that the exact states found in Refs.~\cite{ourpaper4, kang_strong_2019} are the (only) ground states, although they are still exact eigenstates. Hence we use ED to show that they are still ground states and are unique in the chiral-flat limit (for $\nu=-1$ this is verified only within nearly valley polarized Hilbert space due to computational capability). Moreover, for the exact excited states (we here focus on the charges $\pm 1$ and neutral excitations), even with the FMC, we cannot prove that they are the \emph{lowest} excitations. We hence use ED to show that (up to potential finite size effects) at all fillings $\nu=-3,-2,-1$ these exact excited states are the lowest, except for charge $+1$ excitations at $\nu=-1$ without FMC. We thus confirm the validity of the FMC for more realistic parameters, which allows finite kinetic energy ($t$), finite $w_0$ and breaking ($\lambda$) of the FMC.   

Second, we then check the validity of our analytic approximations for the  full $w_0/w_1$ and kinetic energy $t$ range, to obtain a phase diagram (with or without quantum number constraints) for both ground states and excited states at $\nu= -3,-2,-1$ (note that $\nu$ and $-\nu$ are PH symmetric \cite{ourpaper3}). In the process, we confirm the theoretical calculations that the kinetic energy has a minimal effect on the phase diagram, as was also pointed out in Refs. \cite{soejima2020efficient,kang_nonabelian_2020}.
At $\nu=-3$, we find the projected Coulomb Hamiltonian stabilizes the spin-valley polarized Chern number $\pm1$ insulator in a large range of $t\in [0,1]$ for $w_0/w_1\lesssim 0.9$ ($w_0/w_1\lesssim 0.3$) when the FMC is assumed (not assumed). At small $w_0/w_1$, this agrees with our conclusions in \cite{ourpaper4} from the perturbation theory (where $w_0$ is treated perturbatively).  At $\nu=-2$, our computational power is restricted in the \emph{fully valley polarized} sector or the \emph{fully spin polarized} sector. In the \emph{fully valley polarized} sector, we find that the ground state with FMC is the U(4) FM state with zero Chern number when $w_0/w_1\gtrsim0.5t$, and the valley-polarized spin-singlet state with zero Chern number when $w_0/w_1\lesssim0.5t$. Without FMC, the U(4) FM state is further restricted within $w_0/w_1\lesssim 0.6$. These findings are in agreement with our exact/perturbation analytis in Ref.~\cite{ourpaper4} for the nonchiral-flat and chiral-nonflat limits (see a similar analysis in Ref.~\cite{bultinck_ground_2020}).  In the \emph{fully spin polarized} sector, we find the ground state in the range $w_0/w_1\lesssim 1$ when the FMC is assumed (or $w_0/w_1\lesssim 0.6$ when the FMC is not assumed) agrees well with the intervalley coherent states predicted at $\nu=-2$ \cite{ourpaper4,bultinck_ground_2020}. The ground state energy in the fully spin polarized sector is lower than that in the fully valley polarized sector. 
We also identify that the lowest charge neutral excitations, for for both the $\nu=-3$ and $\nu=-2$ states, as the Goldstone branches predicted in Ref. \cite{ourpaper5}.

Third, toward the isotropic limit, i.e., with $w_0/w_1$ being increased above the phase boundaries, we observe phase transitions to different ground states. In particular, at $\nu=-3$, the phase transition at $w_0/w_1\sim 0.9$ with FMC goes into a new state with zero momentum (relative to the Chern insulator ground state at small $w_0/w_1$), while the phase transition at $w_0/w_1\sim 0.3$ without FMC is at nonzero momentum close to $\Gamma_M$, $M_M$ or $K_M$ points of the moir\'e Brillouin zone, depending on system sizes, $t$ and $w_0/w_1$. We thus conjecture the possible competing orders include nematic, momentum $\pi$ ($M_M$) CDW (stripe), or momentum $K_M$ CDW in this parameter range. The nematic and $M_M$ CDW (stripe) orders was recently predicted by DMRG to arise when $w_0/w_1\gtrsim 0.8$ in Refs. \cite{kang_nonabelian_2020,soejima2020efficient}; while we conjecture $K_M$ CDW is another possibility, which was not mentioned in previous works. The phase transition is due to softening of collective modes at finite or zero momenta and hence may break the translation symmetry. The momentum of the translation breaking phase may depend on detailed model parameters. 

This article is organized as follows. In Sec.~\ref{sec:interactinghamTBG}, we give a short review of the TBG single-particle Hamiltonian, the projected interacting Hamiltonian in the active bands and the symmetries in the different limits. Sec.~\ref{sec:nu3} is devoted to the study of the integer filling factor $\nu=-3$, in the chiral-flat limit for the projected Hamiltonian with and without the FMC, including the ground states, the charge and neutral excitations. We also provide the phase diagrams in the nonchiral-nonflat limit, with and without the FMC. In Sec.~\ref{sec:nu2}, we perform a similar analysis for the filling factor $\nu=-2$, discussing in details the phase diagrams and the dominance of the trivial insulating phase and its magnetic properties. Finally in Sec.~\ref{sec:nu1}, we briefly consider the filling factor $\nu=-1$ in the chiral-flat limit, focusing mostly on the charge excitations.

\section{Interacting Hamiltonian for TBG}\label{sec:interactinghamTBG}

In this section, for completeness, we give a brief overview of the TBG Hamiltonian with Coulomb interaction projected into the flat bands. The full details can be found in Refs.~\cite{ourpaper3, ourpaper2,ourpaper4, ourpaper5}.

\subsection{TBG Model}\label{sec:model}

We start with the single-body Hamiltonian of TBG whose  low energy physics is mostly dominated by states around the two Dirac points $K$ and $K'$. By focusing on one valley $K$, we further define vectors $\vq_{j} = C_{3z}^{j-1}(\mathbf{K}_- - \mathbf{K}_+)$, where $\mathbf{K}_l$ is the momentum of the Dirac point $K$ in layer $l$, and $|\mathbf{K}_l| = 1.703\,\textrm{\AA}^{-1}$. The reciprocal vectors of the triangular  moir\'e lattice, denoted by $\mathcal{Q}_0$, are spanned by basis vectors ${\mathbf{b}}_{M1} = \vq_{3} - \vq_{1}$ and ${\mathbf{b}}_{M2} = \vq_{3} - \vq_{2}$. 
Momenta lattices $\mathcal{Q}_\pm = \mathcal{Q}_0 \pm \vq_1$ form a hexagonal lattice in momentum space, and they stand for Dirac points of the top and bottom layers, respectively. The single-particle Bistritzer-MacDonald (BM) model \cite{bistritzer_moire_2011} of the TBG Hamiltonian is
{\small
\begin{equation}
    \hat{H}_0 = \sum_{\vk\in{\rm MBZ}}\sum_{\mathbf{Q},\mathbf{Q}'\in{\mathcal{Q}_{\pm}}}\sum_{\eta,s, \alpha,\beta}\Big[h^{(\eta)}_{\mathbf{Q} \mathbf{Q}'}(\vk)\Big]_{\alpha\beta}c^\dagger_{\vk,\mathbf{Q},\eta, \alpha, s}c_{\vk,\mathbf{Q}', \eta, \beta, s}
\end{equation}}
where MBZ stands for the moir\'e Brillouin zone, and the operator $c^\dagger_{\vk, \mathbf{Q}, \eta, \alpha, s}$ creates an electron at valley $\eta$, on sublattice $\alpha$, in layer $l = \ell \cdot \eta$ with spin $s$ and momentum $\vp = \eta\mathbf{K}_{\eta\cdot \ell} + \vk - \mathbf{Q}$ if $\mathbf{Q}\in\mathcal{Q}_\ell$.
The kinetic Hamiltonian at valley $\eta=+$ is given by
\begin{equation} \label{BMHamilt1}
    h^{(+)}_{\mathbf{Q} \mathbf{Q}'}(\vk) = v_F \bm{\sigma}\cdot(\vk - \mathbf{Q})\delta_{\mathbf{Q}, \mathbf{Q}'} + \sum_{j=1,2,3}T_j \delta_{\mathbf{Q} - \mathbf{Q}',\pm\vq_j}\,,
\end{equation}
where $v_F = 6104.5\,\mathrm{meV}\cdot \textrm{\AA}$ is the Fermi velocity, and $T_j$ are the interlayer hopping matrices:
{\small
\begin{equation}
    T_j = w_0 \sigma_0 + w_1 \left[\cos\frac{2\pi (j - 1)}{3}\sigma_x + \sin\frac{2\pi (j - 1)}{3}\sigma_y\right]\,.
\end{equation}}
The parameters $w_0$ and $w_1$ stand for the interlayer hopping strength at AA and AB stacking centers, respectively. In this paper we set $w_1=110\,\rm{meV}$ while $w_0$ will be used (and varied) as a parameter. The Hamiltonian at valley $\eta=-$ can be obtained by performing a $C_{2z}$ transformation in Eq.~(\ref{BMHamilt1}).

\subsection{Interaction and Projected Hamiltonian}\label{sec:interaction}

The repulsive interaction between electrons is accurately captured by the Coulomb interaction screened by the top and bottom gates. The Fourier transformation of this interaction reads:

\begin{equation}\label{eq:ed:screencoulomb}
V(\vq) = \pi \xi^2 U_\xi \frac{\tanh(\xi q/2)}{\xi q/2}
\end{equation}
where $\xi \approx 10\,\rm nm$ is the distance between the top and bottom gates in typical TBG experiments, and $U_\xi = e^2/\epsilon\xi\approx 24~\rm meV$ is the interaction strength for a dielectric constant $\epsilon\sim6$ \cite{kang_symmetry_2018,cao_correlated_2018,cao_unconventional_2018}. The second quantized interacting Hamiltonian is \cite{ourpaper3, ourpaper4}
\begin{equation}\label{ed:eqn:four-fermion-ham}
    \hat{H}_{I} = \frac{1}{2\Omega_{\rm tot}}\sum_{\vq\in{\rm MBZ}}\sum_{\mathbf{G}}V(\vq + \mathbf{G})\delta\rho_{\vq + \mathbf{G}}\delta\rho_{-\vq - \mathbf{G}}\,,
\end{equation}
where 
\begin{equation}\label{ed:eqn:shifted-density}
\delta\rho_{\vq + \mathbf{G}} =  \sum_{\vk, \eta, \alpha, s, \mathbf{Q}}(c^\dagger_{\vk + \vq, \mathbf{Q} - \mathbf{G}, \alpha, \eta, s}c_{\vk, \mathbf{Q}, \alpha, \eta, s} - \frac{1}{2}\delta_{\vq, 0}\delta_{\mathbf{G}, 0})
\end{equation}
is the density at momentum $\vq + \mathbf{G}$ relative to the charge neutral point. We neglect the  electron-phonon interaction in this study, although it should be considered in a complete study as it is conjectured to be important \cite{Lian2019TBG, Wu2018TBG-BCS, wu_phonon_linearT2019, polshyn_linear_2019, ourpaper5} for superconductivity.

The exponential complexity of the quantum many-body simulations prevents a direct numerical treatment of the full interacting Hamiltonian. Fortunately close to the (first) magic angle, the bandwitdh of the two  flat bands around charge neutral point (one valence band and one conduction band) is smaller than the Coulomb interaction. We can then greatly simplify the calculation by projecting the Hamiltonian onto these two bands. By diagonalizing the Hamiltonian $h^{(\eta)}(\vk)$, we obtain the dispersion relation $\epsilon_{\vk, m, \eta}$ and single-body wavefunctions $u_{\mathbf{Q}\alpha,  m\eta}(\vk)$ of the flat bands. Here $m =\pm 1$ is the band index. The projected kinetic energy term is:
\begin{equation}
    {H}_{0} = \sum_{\vk\in{\rm MBZ}}\sum_{\eta,s}\sum_{m=\pm 1}\epsilon_{\vk,m,\eta}c^\dagger_{\vk, m, \eta, s} c_{\vk, m, \eta, s}\,,
\end{equation}
where $c^\dagger_{\vk, m, \eta, s} = \sum_{\mathbf{Q}\alpha}u_{\mathbf{Q}\alpha, m\eta}(\vk)c^\dagger_{\vk, \mathbf{Q},\eta,\alpha,s}$ is the electron creation operator in band basis. Note that we have dropped the "hat" notation for the projected quantities such as the kinetic Hamiltonian.

Similarly, we can write the projection of the interaction term $\hat{H}_{\rm I}$ onto the flat bands
\begin{equation}
H_{ I} = \frac{1}{2\Omega_{\rm tot}}\sum_{\vk\in{\rm MBZ}}\sum_{\mathbf{G}}V(\vq + \mathbf{G})\overline{\delta \rho}_{\vq + \mathbf{G}}\overline{\delta \rho}_{-\vq - \mathbf{G}}\,\label{ed:eq:HIprojected}
\end{equation}
with $\overline{\delta\rho}_{\vq + \mathbf{G}}$ the density operator projected onto the flat bands defined as
\begin{align}
   &\overline{\delta\rho}_{\vq + \mathbf{G}} = \sum_{\vk,\eta, s}\sum_{m, n = \pm 1}M^{(\eta)}_{mn}(\vk, \vq + \mathbf{G})\nonumber\\
   & \qquad\times \left(c^\dagger_{\vk + \vq, m, \eta, s}c_{\vk, n, \eta, s} - \frac{1}{2}\delta_{\vq, 0}\delta_{m, n}\right),\\
   &M^{(\eta)}_{mn}(\vk, \vq + \mathbf{G}) =  \sum_{\mathbf{Q}\alpha}u^*_{\mathbf{Q}-\mathbf{G},\alpha, m\eta}(\vk + \vq)u_{\mathbf{Q}, \alpha, n\eta }(\vk)\,.
\end{align}

The form factors (overlaps) $M^{(\eta)}_{mn}(\vk, \vq + \mathbf{G})$ depend on the gauge choice of single-body wavefunctions. By choosing the gauge properly (see App.~\ref{ed:appsubsec:real}), the form factors can all be made real. We notice that Eq.~\ref{ed:eqn:four-fermion-ham} can be written as the summation of a normal-ordered two-body term and a quadratic term. It can be shown that the quadratic term matches with the ``Hartree-Fock'' contribution from the filled bands below the flat bands~\cite{ourpaper3} and is required in order  to recover the many-body charge-conjugation symmetry around  the charge neutral point (CNP) for the projected Hamiltonian. The effects of the normal ordering and this ``Hartree-Fock'' contribution are discussed in App.~\ref{ed:appsec:phs}.

We can also define another basis, the Chern band basis, by:
\begin{equation}\label{eq:chernbasis}
    d^\dagger_{\vk, e_Y, \eta, s} = \frac{c^\dagger_{\vk, 1, \eta, s} + ie_Yc^\dagger_{\vk, -1, \eta, s}}{\sqrt2}\,,~e_Y = \pm 1\,.
\end{equation}
It was shown in Refs.~\cite{ourpaper2, ourpaper3} that, with a consistent gauge choice, the band formed by the states $d^\dagger_{\vk, e_Y, \eta, s}$ of all $\vk$ with fixed $e_Y$, $\eta$ and $s$ carries a Chern number $e_Y = \pm 1$. (See App. \ref{ed:appsubsec:chern} for a short review.) In this basis, the form factors cannot been made all real. While this is generally more computational and memory intensive, the chiral basis greatly simplifies the identification of some strongly correlated phases as discussed in Ref.~\cite{ourpaper4}  and summarized in the following sections.

Since both the band structure and the single-body wavefunctions depend on $w_0$, the projected interacting Hamiltonian also depends on $w_0$. To probe the competition between the kinetic energy and the interaction, we introduce a dimensionless parameter $t$ to control the amplitude of the kinetic term. By adding them together, the (tunable) Hamiltonian is:
\begin{equation}
H(t, w_0) = t H_{0}(w_0) + H_{I}(w_0)\,.\label{eq:tunableham}
\end{equation}

If we assume the form factors satisfy the \emph{flat metric condtion} (FMC) \cite{ourpaper1, ourpaper4}, namely:
\begin{equation}
    M^{(\eta)}_{mn}(\vk, \mathbf{G}) = \xi(\mathbf{G})\delta_{mn},\label{ed:eq:FMC}
\end{equation}
we will obtain a simplified Hamiltonian:
\begin{equation}
    H_{\rm FMC}(t,w_0)= t H_{0}(w_0) + H_{I,FMC}(w_0)\,.\label{ed:eq:FMCham}
\end{equation}
where $H_{0}$ is identical to that in Eq.~\ref{eq:tunableham} but the interaction term $H_{I,FMC}$ is obtained from Eq.~\ref{ed:eq:HIprojected}, discarding $\vq = 0$ in the sum.
This condition is identical to the flat metric condition, which is proved in App \ref{ed:appsubsec:fock_ham}, thus the name $H_{\rm FMC}$. In Ref.~\cite{ourpaper1} it was proved to hold, with exponential accuracy, for all $G$ with $|G| \ne |\mathbf{b}_M|$, and is hence a "weak approximation".  With this flat metric condition, for $t=0$ and $w_0=0$, as well as away from the chiral flat limit, the ground state and some low energy excitations (both neutral and charged) can be derived analytically \cite{ourpaper4,ourpaper5}. In order to study the connection between this  partially solvable model and the full fledged model without the FMC, we can use a linear interpolation and the following Hamiltonian with three parameters:
\begin{equation}
    H(t, w_0, \lambda) = \lambda \cdot H(t, w_0) + (1 - \lambda) \cdot H_{\rm FMC}(t, w_0)\,.\label{ed:eq:fullinterpolatingham}
\end{equation}
where $\lambda\in[0,1]$ is the dimensionless interpolating parameter (denoted from now on the FMC parameter). Thus $H(t, w_0, 0)$ is just the FMC Hamiltonian while  $H(t, w_0, 1)$ is the full interacting TBG model, with kinetic energy multiplied by a factor $t$ and no approximations such as the flat metric condition.

\subsection{Symmetries}\label{sec:symmetries}

The Hamiltonian Eq.~\ref{ed:eq:fullinterpolatingham} has several symmetries depending on the parameters values of $t$ and $w_0$ (irrespective of the FMC parameter $\lambda$). These symmetries have been derived and discussed in details in Ref.~\cite{ourpaper3}. We here provide a further examination of their numerical implementations in App.~\ref{ed:appsec:symhamiltonian}.

For generic values of $w_0$ and $t$, the Hamiltonian Eq.~(\ref{ed:eq:fullinterpolatingham}) has the spinless crystalline symmetries $C_{2z}$, $C_{3z}$, $C_{2x}$, a spinless time reversal symmetry $T$, and a unitary particle-hole (PH) transformation $P$ which satisfies $\{P,H_0\}=[P,H_I]=0$. The combined symmetry $C_{2z}TP$ gives rise to a many-body charge-conjugation symmetry $\mathcal{P}_c$ that satisfies $\mathcal{P}_c(H_0+H_I)\mathcal{P}_c^{-1}=H_0+H_I+\text{const}$.
(See Refs.~\cite{ourpaper2,ourpaper3} for details, see also App. \ref{app:singleparticle} for a review). Furthermore, the Hamiltonian has a U(2)$\times$U(2) symmetry, which corresponds to the spin and charge rotation symmetries in each valley.

We will review and work on two limits (along with their combination) where this U(2)$\times$U(2) symmetry is extended into higher symmetries:
\begin{itemize}
\item The \emph{nonchiral-flat limit}: In the absence of kinetic term, i.e. $t=0$ in Eq.~(\ref{eq:tunableham}), the Hamiltonian is solely given by interaction $H_I$, and exhibits a U(4) symmetry in the spin/valley space due to the $C_{2z}P$ symmetry \cite{bultinck_ground_2020,ourpaper3}.

\item  The \emph{(first) chiral-nonflat limit}: when $w_0=0$, there is a first unitary chiral transformation $C$ satisfying $\{C,H_0\}=[C,H_I]=0$. the Hamiltonian also exhibits a U(4) symmetry in the spin/valley space (however, different from the U(4) in the nonchiral-flat limit) due to the $CC_{2z}P$ symmetry, as was shown in Ref.~\cite{bultinck_ground_2020,ourpaper3}. Note that in this paper, we will only focus on the first chiral symmetry, but not the second chiral limit (where $w_1=0$, and which also exhibits an extended symmetry) introduced in Ref.~\cite{ourpaper3}. 
\item The \emph{(first) chiral-flat limit}: when both of the above limits are reached, i.e., $t=0$ and $w_0=0$, the symmetry of the Hamiltonian is further enhanced into a U(4)$\times$U(4) symmetry in the band/spin/valley space \cite{bultinck_ground_2020, ourpaper3}. However, we note that the U(4) in the nonchiral-flat (or the first chiral-nonflat) limit is not one of the two tensor-producted U(4)'s in the chiral-flat limit.

\end{itemize}

Due to these symmetries, in either the flat band limit or the first chiral limit, the eigenstates of $H(t, w_0, \lambda)$ can be labeled by irreducible representations (irreps) of the corresponding U(4) symmetry group encoded in Young tableaux. For these U(4) irreps, we use the notation $[l_1,l_2,l_3]_4$ where the positive (or zero) integers $l_1 \ge l_2 \ge l_3 \ge 0$ correspond to the number of boxes in the first, second and third lines respectively (see Ref.~\cite{ourpaper4} for a review of Young tableaux notations). The integers are omitted when they are equal to zero. In particular, the fundamental U(4) irrep is $[1]_4$ (a Young tableau with one box), and the identity U(4) irrep is $[0]_4$ (an empty Young tableaux). As shown in Ref.~\cite{ourpaper3}, in either of these two limits, each electron occupies an irrep $[1]_4$ of the corresponding U(4). In many-body wavefunctions, two electrons antisymmetric (symmetric) in spin-valley indices will be in the same column (row) of a Young tableau of a U(4) irrep.

In the \emph{(first) chiral-flat limit} where $w_0=0$ and $t=0$, the eigenstates of $H(t=0, w_0=0, \lambda)$ will fall into irreps of the U(4)$\times$U(4) group, which are given by the tensor product of irreps of the two U(4)'s. We denote the U(4)$\times$U(4) irreps by $([l_1,l_2,l_3]_4,[l_1',l_2',l_3']_4)$, where $[l_1,l_2,l_3]_4$ and $[l_1',l_2',l_3']_4$ are the irreps of the two U(4)'s, respectively. In particular, the Chern basis in Eq.~(\ref{eq:chernbasis}), which are eigenbasis of the chiral symmetry $C$ with eigenvalue $e_Y$, occupy the single-electron U(4)$\times$U(4) irrep $([1]_4,[0]_4)$ if $e_Y=+1$, and irrep $([0]_4,[1]_4)$ if $e_Y=-1$. Besides, the $C_{2z}T$ symmetry of the Hamiltonian exchanges the two U(4)'s of the U(4)$\times$U(4) group. As a result, any energy level with an irrep $([l_1,l_2,l_3]_4,[l_1',l_2',l_3']_4)$ will imply another energy level with an irrep $([l_1',l_2',l_3']_4,[l_1,l_2,l_3]_4)$ at the same energy (related by $C_{2z}T$).
Thus we will show only one of these two $C_{2z}T$ related irreps in the various plots.

In U(2)$\times$U(2), chiral-nonflat U(4) and flat nonchiral U(4) cases, the electron numbers in each spin valley sectors $N_{\eta, s}$ are conserved. We use the eigenstates of these operators to perform the ED calculation, due to the fact that the interacting Hamiltonian will be block diagonal in this basis. We can also recombine these good quantum numbers into a more convenient form: $N = N_{+,\uparrow} + N_{-,\uparrow} + N_{+,\downarrow} + N_{-,\downarrow}$, $N_{v} = N_{+,\uparrow} +  N_{+,\downarrow} - N_{-,\uparrow} - N_{-,\downarrow}$, $2S_{z,\eta=+} = N_{+,\uparrow} -  N_{+,\downarrow}$ and $2S_{z,\eta=-} = N_{-,\uparrow} -  N_{-,\downarrow}$. We also define $S_{\eta=\pm}$ as the total spin in each valley $\eta$ (the $z$-component of which are $S_{z,\eta=\pm}$). Moreover, at chiral-flat limit, there are eight Cartan subalgebra operators for U(4)$\times$U(4) group. The total electron numbers in each spin, valley and Chern band $e_Y=\pm1$ are conserved separately. A detailed discussion about the symmetry sectors can be found in App. \ref{ed:appsec:symhamiltonian}.

An important discrete symmetry is the translation symmetry of the moir\'e lattice, which corresponds to the conserved total momentum. In order to perform numerical ED, we use a discrete momentum lattice in the MBZ. By imposing periodic boundary condition, we choose the momentum lattice given by 
\begin{equation}
\vk = \frac{k_1}{N_1}{\mathbf{b}}_{M1} + \frac{k_2}{N_2}{\mathbf{b}}_{M2},
\end{equation}
where $k_1 = 0, 1,\cdots, N_1-1$, and $k_2 = 0, 1, \cdots N_2 - 1$. 
Thus there are $N_M = N_1 N_2$ moir\'e unit cells in total. The conserved total momentum components are defined as 
\begin{eqnarray}
K_1 &=& \left(\sum_{i=1}^Nk_{1i}\right)~{\rm mod}~N_1,\label{ed:eq:manybodymomentumK1}\\
K_2 &=& \left(\sum_{i=1}^Nk_{2i}\right)~{\rm mod}~N_2.\label{ed:eq:manybodymomentumK2}
\end{eqnarray}
in which $k_{1i}$ and $k_{2i}$ are the momentum components of $i$-th electron along ${\mathbf{b}}_{M1}$ and ${\mathbf{b}}_{M2}$ respectively, and $N$ is the total electron number. Some momentum sectors are related by discrete symmetries. For example, $C_{2z}$ symmetry can transform a sector with $K_1$ and $K_2$ into another sector with momentum $-K_1~{\rm mod}~N_1$ and $-K_2~{\rm mod}~N_2$.

\section{Numerical results at filling factor \texorpdfstring{$\nu = -3$}{nu=-3}}\label{sec:nu3}

In this paper we define the filling factor $\nu$  as the number of electrons per moir\'e unit cell relative to the filling of the charge neutral point (CNP). Within the active bands (the lowest 2 flat bands per spin per valley), we have $-4\le \nu\le 4$. When the total electron number within the active bands is $N$, the filling factor is defined by 
\begin{equation}
\nu = \frac{N}{N_M} - 4= \frac{N}{N_1N_2} - 4.\label{ed:eq:fillingfactor}
\end{equation}
Therefore, the filling factor $\nu$ is equal to $0$ at CNP, and it is an integer if there are integer numbers of electrons in each moir\'e unit cell. The charge-conjugation symmetry $\mathcal{P}_c$ \cite{ourpaper3} around the CNP of our Hamiltonian implies that the energy spectra at $\nu$ and $-\nu$ are identical (up to a chemical potential shift). We can thus focus solely on $\nu \le 0$. Each moir\'e unit cell can host at most 8 electrons (two bands, two valley and two spin degrees of freedom). The inherent exponential complexity of the quantum many-body simulations restrain the system sizes that can be reached. As compared to other systems exhibiting the same low energy physics, the fractional quantum Hall effect, or its lattice cousin the fractional Chern insulator, the 8-fold degree of freedom per unit cell puts an even more severe cap on the maximal sizes: the closer $\nu$ is to 0, the greater the limitation. In the rest of this section, we will focus on simplest case, namely $\nu=-3$. We note that our simulations are \emph{unbiased}: we work with the full Hilbert space of the  projected $2$ orbitals (per spin per valley) of the flat active bands at the first magic angle, connected by Dirac points. We do not project further to smaller single-particle orbital spaces. 

\subsection{(First) chiral-flat limit}\label{sec:chiralflatbandnu3}

We start with the (first) chiral-flat limit. As derived in Ref.~\cite{ourpaper4}, the FMC Hamiltonian $H(0, 0, 0)$ ground state at $\nu=-3$ is exactly solvable and is built from the two following Fock states of one filled band, carrying a Chern number $\nu_C = \pm 1$ respectively, 
\begin{align}
    |\Psi^{1,0}_{\nu=-3}\rangle &= \prod_{\vk}d^\dagger_{\vk, +1, +, \uparrow}|0\rangle\,,~~\nu_C=1\label{ed:eq:nu3chern1}\ ,\\
    |\Psi^{0,1}_{\nu=-3}\rangle &= \prod_{\vk}d^\dagger_{\vk, -1, +, \uparrow}|0\rangle\,,~~\nu_C=-1\,.\label{ed:eq:nu3chernm1}
\end{align}
They are the fully band polarized states of the multiplets associated to the U(4)$\times$U(4) irreps $([N_M]_4, [0]_4)$ and $([0]_4, [N_M]_4)$  respectively. Other states of these multiplets are generated by successive application of the U(4)$\times$U(4) generators onto these two Chern insulator states. These two irreps are the ones with the most columns in their Young tableau at this filling factor (having a multiplicity of $d_{([N_M]_4, [0]_4)}=d_{([0]_4, [N_M]_4)}=N_M(N_M+1)(N_M+2)/6$ states per irrep) and form ferromagnetic multiplets for U(4)$\times$U(4) (analogous to the SU(2) spin ferromagnet). In fact, there is only one $([N_M]_4, [0]_4)$ irrep and one $([0]_4, [N_M]_4)$ irrep that can be built from the Hilbert space of $N=N_M$ electrons and $N_M$ moir\'e unit cell at this filling factor, each being given in Eq.~\ref{ed:eq:nu3chern1} and Eq.~\ref{ed:eq:nu3chernm1} respectively. As such, these two irreps, including the two Chern states $|\Psi^{1,0}_{\nu=-3}\rangle $ and $|\Psi^{0,1}_{\nu=-3}\rangle$, are \emph{always exact eigenstates} of the Hamiltonian as long as the U(4)$\times$U(4) is preserved, in particular for any value of $\lambda$ along the interpolation Eq.~\ref{ed:eq:fullinterpolatingham}. This does not imply  that these states are the ground states, with the exception of $\lambda=0$ when the nature of the ground state is known analytically; nor does it imply that, if they are the ground states, they are unique ground states, even at $\lambda=0$. We now test these issues.

\begin{figure}
    \centering
    \includegraphics[width=\linewidth]{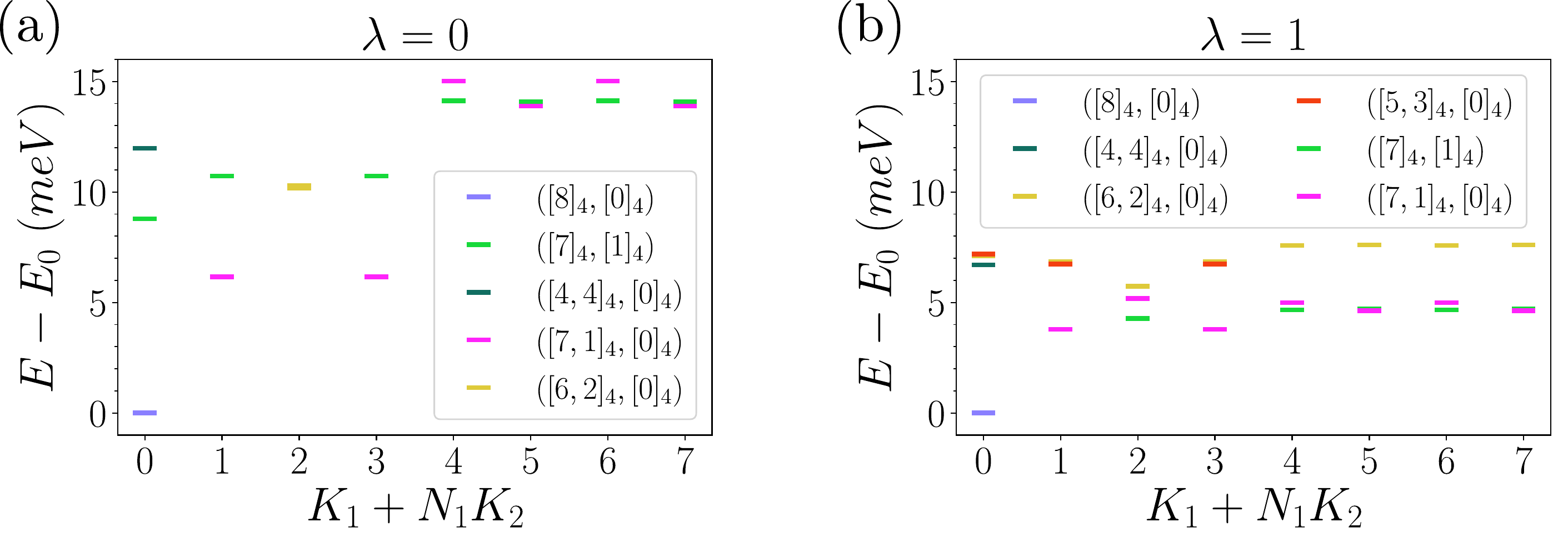}
    \caption{The spectrum of the ground state and some low energy neutral excitations at $\nu=-3$ for the FMC model (a) and the full TBG model (b) at the (first) chiral-flat limit, with the corresponding U(4)$\times$U(4) irreps. The spectrum is computed on a $N_1=4$ and $N_2=2$ lattice with a twisting angle of $\theta=1.07^\circ$. In both cases, the ground state has a total momentum $K_1 = 0$, $K_2 = 0$, with irrep $([8]_4, [0]_4)$.}
    \label{ed:fig:irrep_spectrum_4x2}
\end{figure}

Without any further assumptions, we use ED to study the spectrum of the FMC model $H(0, 0, \lambda=0)$ and full TBG model $H(0, 0, \lambda=1)$ at chiral-flat limit. The results are shown in Fig.~\ref{ed:fig:irrep_spectrum_4x2} for $N=8$ on a $N_1=4$, $N_2=2$ system. As explained in Sec.~\ref{sec:symmetries}, we only show one of the two irrep sectors related by the $C_{2z}T$ symmetry. 
In both cases, we find that the irreps of the ground states are $([8]_4, [0]_4)$ for both $\lambda=0$ (as expected) and also for $\lambda=1$. Since there is only one such representation formed by $N=N_M$ electrons this means that Eq.~(\ref{ed:eq:nu3chernm1}) are the \emph{exact} wavefunctions at this filling $\nu=-3$ in the chiral flat limit, and they are Slater determinants.

We also show in Fig.~\ref{ed:fig:irrep_spectrum_4x2} the charge neutral excitation with the corresponding irreps for each momentum sector. By comparing the spectrum of these two Hamiltonians, we find that the energy gap between ground states and the first excited state (at the system size we calculate, which may not be a gap in the thermodynamic limit) at $\lambda=1$ is noticeably smaller than the gap of the FMC model. We also notice that the irreps of the lowest states in most (but not all)  momentum sectors are identical between $\lambda=0$ and $\lambda=1$. There are  level crossings among the low \emph{but barely lowest} energy excited states when we change $\lambda$ from 0 to 1 (see  App.~\ref{ed:app:phasediagramlambdanu3} and Fig.~\ref{ed:fig:lambda_sweep_spectrum} therein). 

The ED results hint that the irreps of most of the low energy states are close to the ``fully Chern band polarized'' irreps. By close, we mean that the Young tableaux of these irreps can be built by only moving a few boxes from the Young tableaux of the ground state (including moving boxes between Chern bands). This is also something that we observe for a smaller size such as $N_M=3\times2$ or a slightly bigger one $N_M=3\times3$ (albeit for $N_M=3\times3$ we can only access very few states per quantum number sector). Physically, this means most of the low energy excited state wavefunctions differ from the ground state wavefunctions by only a few electron-hole pairs (recall that each box correspond to an electron).

For example, the \emph{lowest} irreps of the charge neutral excitations in each finite momentum of Fig.~\ref{ed:fig:irrep_spectrum_4x2} are $([7, 1]_4, [0]_4)$, $([7]_4, [1]_4)$ or $([6,2]_4, [0]_4)$,  which differ from the ground state irrep $([8]_4, [0]_4)$ by 1, 1 and 2 electron-hole pair(s). 
A similar observation holds for the charge excitations: a hole excitation $N=N_M-1$ or an electron excitation $N=N_M+1$ (see App.~\ref{ed:app:excitationEDnu3} and Fig.~\ref{ed:fig:charge_excitation_4x2} for $N_M=4\times2$), which indicate the charge $\pm1$ excitations differ from the ground state wavefunctions by only an electron (hole) plus a few electron-hole pairs. 

By feeding the model parameters and system sizes used here into the scattering matrix method for exactly solvable charge neutral excitations introduced in Ref. \cite{ourpaper5}, we find that the energies of the lowest exact charge neutral excitations in Ref. \cite{ourpaper5} match those of the excited states with the irreps $([7,1]_4,[0]_4)$ and $([7]_4,[1]_4)$ here to machine precision in both Fig.~\ref{ed:fig:irrep_spectrum_4x2}  $\lambda=0, \lambda=1$. Since these lowest neutral excitations are proved  in Ref.~\cite{ourpaper5} to be the Goldstone mode branches, which connect to the gapless Goldstone modes for sufficiently small momentum (not attainable in our finte-size calculation), we identify the states $([7,1]_4,[0]_4)$ and $([7]_4,[1]_4)$ in Fig.\ref{ed:fig:irrep_spectrum_4x2} (which are one electron-hole pair from the ground state) as the single Goldstone branch excitations. They also have the correct representations for the Goldstone branches (see Ref. \cite{ourpaper5}).
However, the excited state with the irrep $([6,2]_4,[0]_4)$, which corresponds to two electron-hole pairs, cannot be obtained from the scattering matrix method in Ref.~\cite{ourpaper5}, which only applies to one electron-hole pair charge neutral excitations.

It is therefore reasonable to expect that the wavefunctions of the lowest few charge $\pm1$ or neutral excitations will only differ from the ground state wavefunctions (which occupy the maximally symmetric irreps) by a few electron-hole pairs. This hypothesis allows us to examine the low energy excitations of larger system sizes with ED. 
Therefore, we focus on these irreps and study the size effect of the low energy charge $\pm1$ excitations. 

To do this, for charge +1 (-1) excitations, we perform ED in sub-Hilbert space sectors which are at most one electron-hole pair plus one electron (hole) different from the ground states (i.e., the sub-Hilbert space of states $c_i^\dag |\Psi\rangle$ and $c_i^\dag c_{i'}^\dag c_{i''}|\Psi\rangle$ for charge $+1$ excitations of ground state $|\Psi\rangle$).  Focusing on these sectors allows us to reach larger system sizes, which is important in order to validate our full calculations at small sizes and to see the possible differences from the small sizes towards the thermodynamic limit.

\begin{figure}
    \centering
    \includegraphics[width=\linewidth]{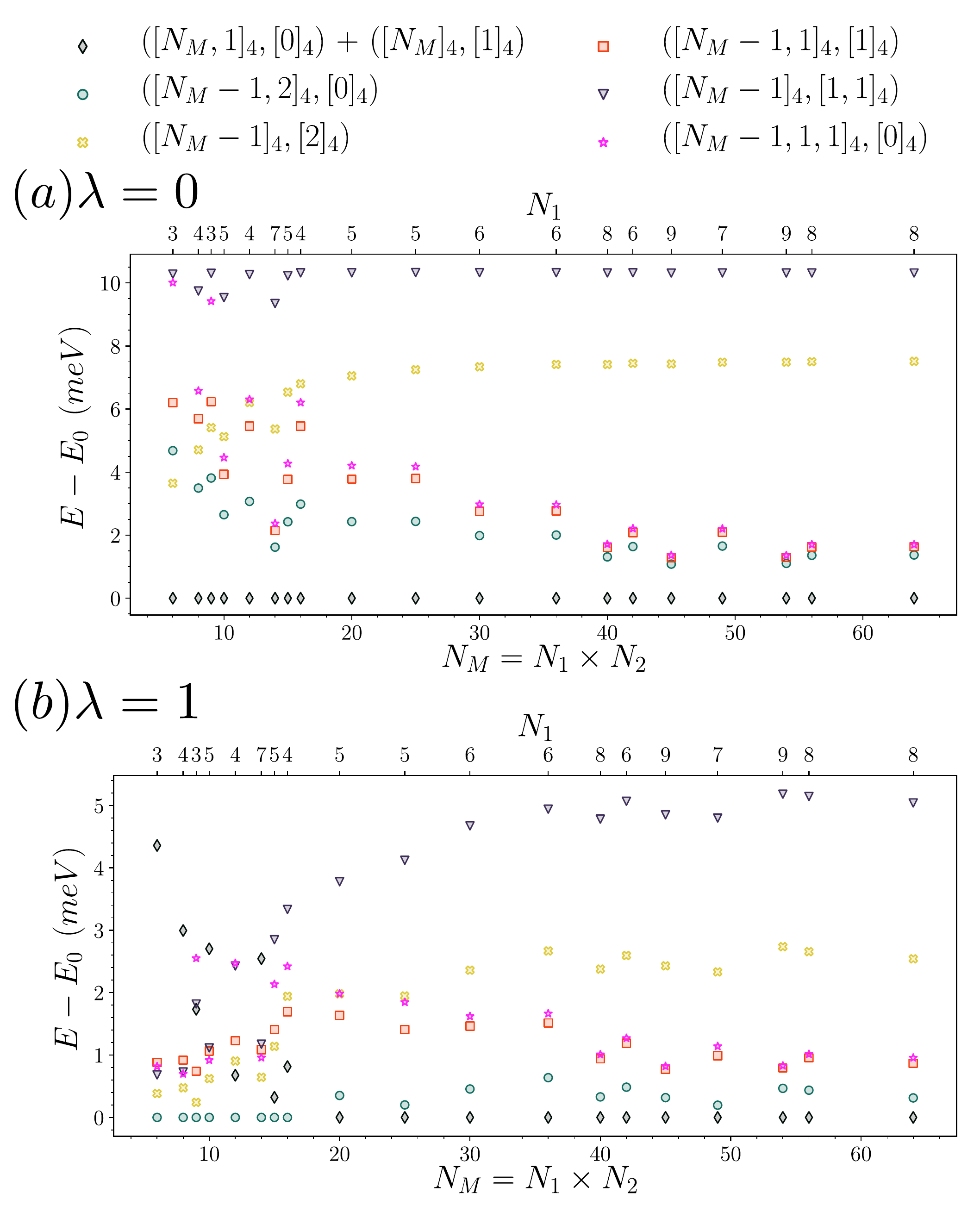}
    \caption{Charge $+1$ (electron) excitation at $\nu=-3$  (a) with the FMC ($\lambda=0$), and (b) without the FMC ($\lambda=1$). The system size is $N_M=N_1 \times N_2$. All energy levels have been shifted by the lowest energy $E_0$ in the charge +1 sub-Hilbert space sector of the corresponding system size. 
    The energies of the proposed ground state at filling factor $\nu=-3$ with one additional electron, along with the states with have an additional U(4)$\times$U(4) excitation have been calculated. We use the notation "+" between two irreps, like $([N_M,1]_4,[0]_4) + ([N_M]_4,[1]_4)$, when these irreps always appear with an exact degeneracy.}\label{ed:fig:excitations_nu-3}
\end{figure}

\begin{figure}
    \centering
    \includegraphics[width=\linewidth]{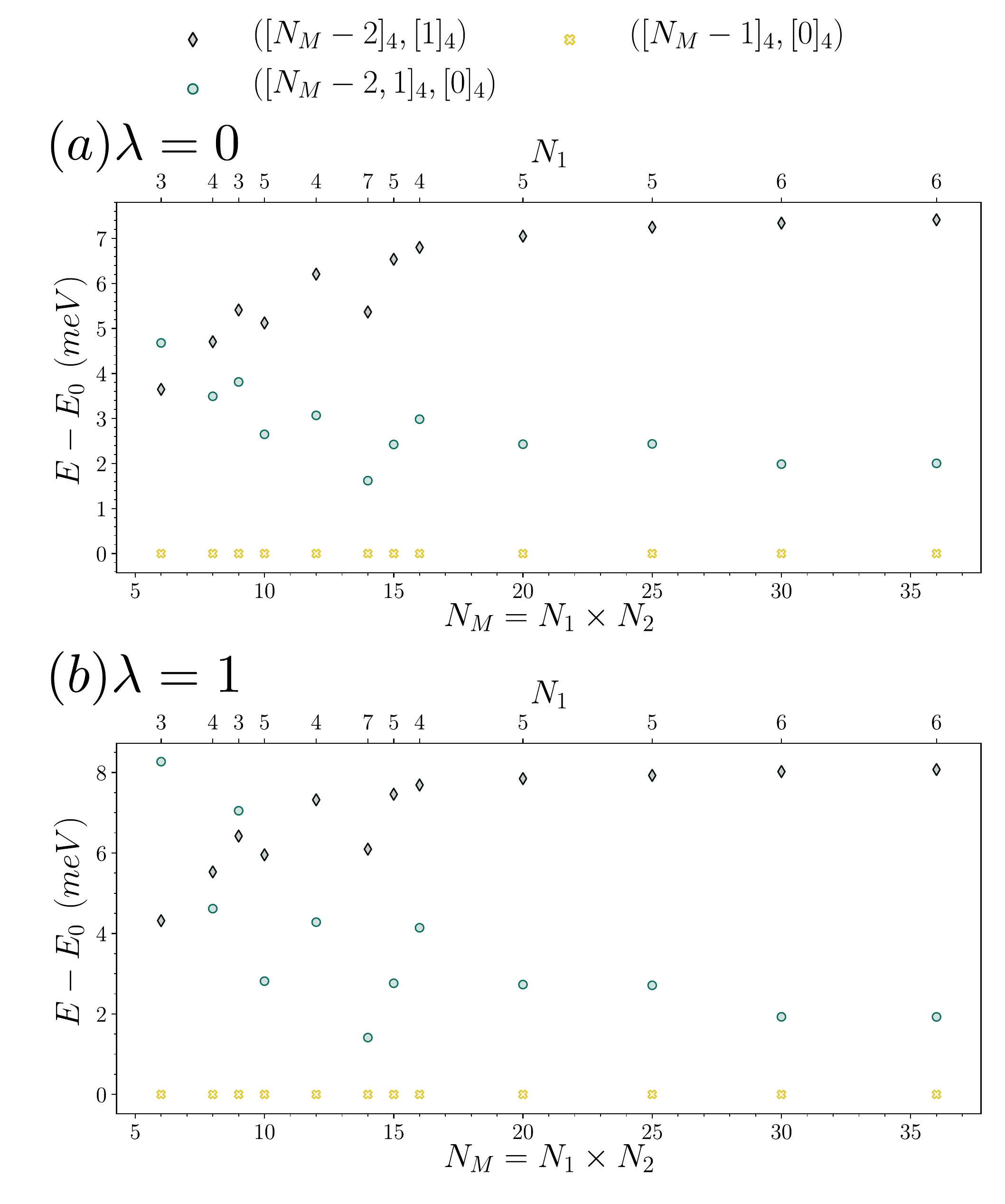}
    \caption{Charge $-1$ (hole) excitations at $\nu = -3$ with $\lambda = 0$ (a) and $\lambda = 1$ (b). $N_M = N_1 \times N_2$. All energies have been shifted by the lowest energy $E_0$ among all calculated states in the charge $-1$ sub-Hilbert space sector at the corresponding system size. The lowest hole excitation's representation is clearly $([N_M -1]_4, [0]_4)$, with the system size only affecting the gap to the second excited state. This validates our results on relatively small lattice sizes for the hole excitations. Indeed the strong similarity between the plots supports the use of the FMC model for hole excitations as a simplified approximation to the (first) chiral-limit Hamiltonian.}
    \label{ed:fig:hole_nu_-3}
\end{figure}

The energies of the charge $+1$ excitations for several slightly depolarized irreps are shown in Fig.~\ref{ed:fig:excitations_nu-3}a for $\lambda=0$ and Fig.~\ref{ed:fig:excitations_nu-3}b for $\lambda=1$. More precisely, we provide the lowest energy state in each irrep sector irrespective of its total momentum (we provide a momentum-resolved discussion in App.~\ref{ed:app:excitationEDnu3}). For $\lambda=0$, the overall lowest electron excitations correspond to the irreps $([N_M,1]_4,[0]_4)$ and $([N_M]_4,[1]_4)$, which in Ref.~\cite{ourpaper5} was proved to be an exact excitation, but not necessarily the \emph{lowest} energy excitations above the ground state (with or without the FMC). 
Physically, the excitations can be understood as adding an electron in a band with the same Chern number (for $([N_M,1]_4,[0]_4)$), or with the opposite Chern number (for $([N_M]_4,[1]_4)$) as the filled band, generating exactly one state per total momentum depending on the additional electron's momentum. Similar to the discussion about the irrep $([N_M]_4,[0]_4)$, the sector of $([N_M,1]_4,[0]_4)$ (as well as $([N_M]_4,[1]_4)$) is of dimension one (up to the irrep multiplicity) once we fix the total momentum. Thus it is always an exact eigenstate in the chiral-flat limit, irrespective of $\lambda$. Note that $([N_M]_4,[1]_4)$ and $([N_M,1]_4,[0]_4)$ are degenerate in energy in the chiral-flat limit, as shown in Ref.~\cite{ourpaper5}. Our numerical results show that for $\lambda=0$ the charge excitations with irreps $([N_M,1]_4, [0]_4)$  and $([N_M]_4,[1]_4)$ \emph{are the lowest ones}, irrespective  of the system size. For $\lambda=1$,  they only become the lowest electron excitation when $N_M \geq 20$. Note that this method focusing on irreps close to the "fully Chern band polarized" irrep, allow us to reach much larger sizes (up to 8$\times$8 moir\'e unit cells).
Despite the low energy landscape being not as clearly separated for $\lambda=1$ compared to $\lambda=0$, the two spectra are qualitatively remarkably similar. 
For example, at $N_M=64$, the order of the irreps with $\lambda=0$ (Fig.~\ref{ed:fig:excitations_nu-3}a) are the same as the order of the irreps with $\lambda=1$ (Fig.~\ref{ed:fig:excitations_nu-3}b). Remarkably, we see that in this case, at $\lambda=1$ (but not at $\lambda=0$) small sizes are misleading, as they would suggest the (first) chiral-flat limit has different charge excitations than the  simplified FMC Hamiltonian in the (first) chiral-limit. 
However, by going to the largest sizes possible, we show that they have, however, the same irreps for lowest excited states, showing that the FMC is appropriate in the (first) chiral limit.  
This similarity is even more acute when considering the one hole excitations (see Fig.~\ref{ed:fig:hole_nu_-3}). Note that, similar to  $([N_M,1,0]_4,[0]_4)$, $([N_M-1]_4,[0]_4)$ is also an exact eigenstate in the chiral-flat limit. We find (see Fig.~\ref{ed:fig:hole_nu_-3}) that it is the lowest energy hole excitation irrespective of the system size.

\subsection{Phase diagrams in the nonchiral-nonflat cases} \label{sec:phasediagramnu3}

\subsubsection{All symmetry sectors}\label{sec:allsymsectorsnu3}

We have provided evidence that the Chern insulator ground state (and its charge excitations) is robust in the (first) chiral-flat limit, which represent analytical results for the FMC $\lambda=0$ model \cite{ourpaper4}, even when we relax the flat metric condition Eq.~(\ref{ed:eq:FMCham}) towards the chiral-flat Hamiltonian $\lambda=1$ Eq.~ \ref{eq:tunableham}.
Next, we study the robustness of the insulating phase with more realistic values for $t$ and $w_0$. By adding kinetic energy ($t>0$), or by moving away from the first chiral limit ($w_0>0$), we break the U(4)$\times$U(4) symmetry according to the discussion of Sec.~\ref{sec:symmetries}. Therefore, the electron numbers in each Chern band basis $e_Y=\pm1$ are not conserved, 
and the Chern insulating wavefunctions are no longer exact eigenstates of the Hamiltonian (irrespective of $\lambda$).

The perturbation in $t$ and $w_0$ will split the chiral-flat U(4)$\times$U(4) ground state multiplet (manifold) $([N_M]_4,[0]_4)$ and $([0]_4,[N_M]_4)$ into a series of either U(4) irreps (in the nonchiral-flat limit or in the chiral-nonflat limit) or U(2)$\times$U(2) irreps (in the most generic case of nonchiral-nonflat limit). We denote the energy of the lowest (highest) states of the chiral-flat ground state manifold after splitting as $E_{0,N_M}$ ($E_{0,N_M}+\delta$), thus $\delta\ge0$ characterizes the energy spread of the U(4)$\times$U(4) multiplet ($\delta=0$ in the chiral-flat limit). For perturbations not too strong, we expect the ground states to be the lowest states with energy $E_{0,N_M}$ from the chiral-flat manifold $([N_M]_4,[0]_4)$ and $([0]_4,[N_M]_4)$ after splitting. However, as the perturbations grow, phase transitions to other phases may happen, which may be due to either the softening of neutral excitations (gapped Goldstone modes, other higher energy excitations, etc) at zero momentum (e.g., 1st order transition to another translationaly invariant insulator) or at finite momenta (e.g., into translation breaking phases), or the vanishing of Goldstone mode stiffness (e.g., into a metallic phase). To examine this possibility, we also calculate the energy difference $\Delta=E_{1,N_M}'-E_{0,N_M}$ which we call the \emph{finite size gap}, where $E_{1,N_M}'$ is the energy of the lowest $N_M$ electron state (irrespective of its total momentum) \emph{not} adiabatically connected to the chiral-flat multiplet $([N_M]_4,[0]_4)$ and $([0]_4,[N_M]_4)$. Due to the finite system size, the lowest Goldstone branch energy near zero momentum (which have quadratic dispersions \cite{ourpaper5}) are expected to have an energy $m|\mathbf{b}_{M1}|^2/2N_M$, where $m$ is the Goldstone mode stiffness, computed in Ref.~\cite{ourpaper5}. Therefore, we expect either $\Delta\sim m|\mathbf{b}_{M1}|^2/2N_M$ if the energy level $E_{1,N_M}'$ is near zero momentum (which is not attainable with our finite size calculations), or $\Delta$ to be determined by certain finite momentum softened neutral excitations $E_{1,N_M}'$ - for example finite momentum Goldstone branches gone soft. Therefore, a vanishing finite size gap $\Delta$ in our calculation would imply either vanishing Goldstone stiffness or softening of some other neutral excited states (possibly part of the finite momentum goldstone branch) and hence the possible transition to other ground states.

\begin{figure*}[!t]
    \centering
    \includegraphics[width=\linewidth]{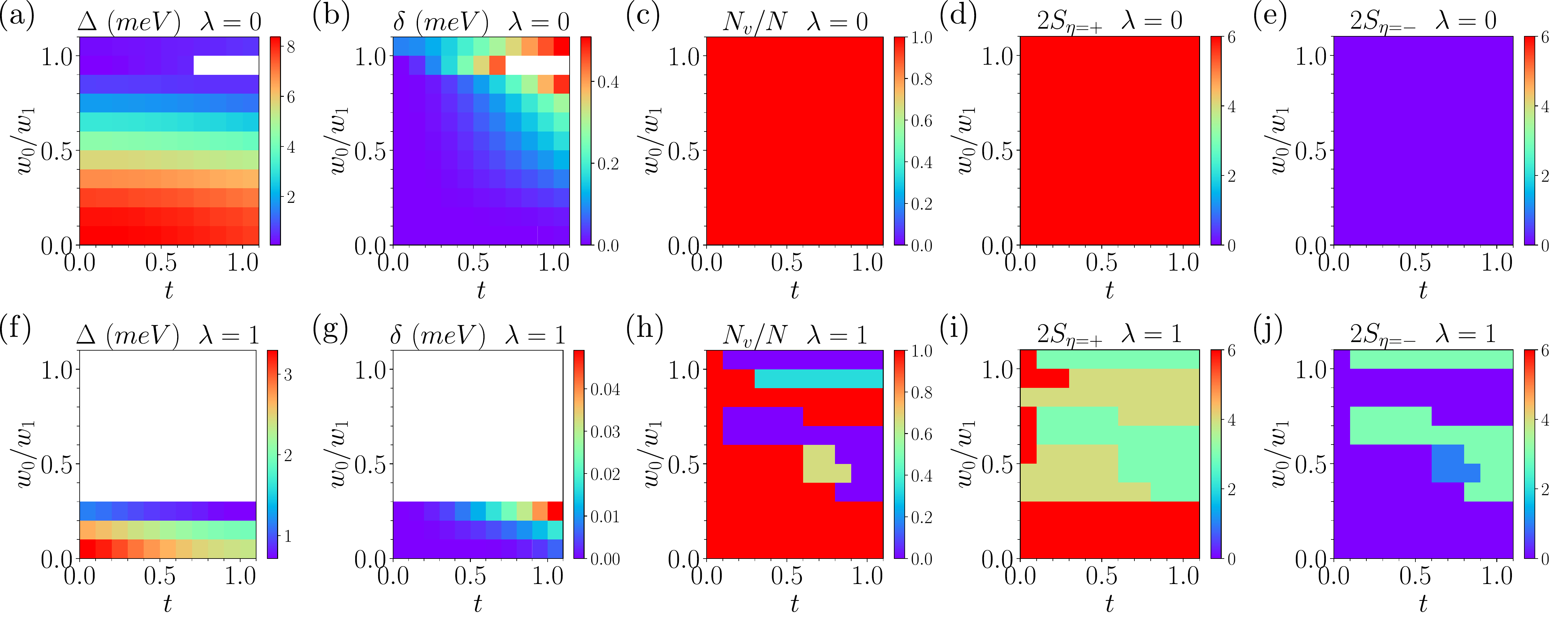}
    \caption{The phase diagrams at filling $\nu=-3$ evaluated on $3\times 2$ lattice with all symmetry sectors considered. We assume $\lambda=0$ in  subfigures (a-e) and $\lambda=1$ in subfigures (f-j). In subfigures (a, f), $\Delta$ is the finite size gap between the excited states and the lowest energy in the ground manifold. In subfigures (b, g), $\delta$ is the spread of the ground state manifold after we move away from chiral-flat limit. Subfigures (c) and (h) demonstrate the $N_v$ values in the entire phase diagram. Subfigures (d-e) and (i-j) show the biggest possible spin quantum numbers $S_1$ and $S_2$ in valleys $\eta=+$ and $\eta=-$ of the lowest energy state, respectively. Due to the $C_{2z}$ symmetry, the spectra of $N_v$ and $-N_v$ are identical, therefore we only show the positive value. Note that the system is always spin and valley polarized in the Chern insulator phase.
    }
    \label{ed:fig:phase_diagram_3x2}
\end{figure*}

First, we have computed the phase diagrams for $\lambda=0$ and $\lambda=1$ with respect to $w_0$ and $t$ covering all the symmetry sectors for a rather small system size $3\times 2$, as shown in Fig.~\ref{ed:fig:phase_diagram_3x2}. We immediately see that the FMC model has a larger finite size gap $\Delta$ over a wider parameter range than the full TBG model (Figs.~\ref{ed:fig:phase_diagram_3x2}a and~\ref{ed:fig:phase_diagram_3x2}f). As we have discussed, a vanishing $\Delta$ hints an unstable ground state and thus a possible phase transition. For the FMC model, this transition happens at around $w_0/w_1 \simeq 0.9$, where $\Delta$ becomes vanishingly small which is at zero momentum (Fig.~\ref{fig:momentum_spectrum_3x3}), implying possible first order or nematic transition  into other zero-momentum phases.
Meanwhile, for the model with $\lambda=1$, this transition happens at a much smaller $w_0/w_1\simeq 0.3$ at nonzero momentum (Fig.~\ref{fig:momentum_spectrum_3x3}), implying a softening of Goldstone stiffness or some collective modes at finite momenta, which may drive the system into metallic or translation breaking phases. This is qualitatively in support of the nematic metal or stripe phases at $\nu=-3$ with relatively large $w_0/w_1$ found in recent DMRG simulations \cite{kang_nonabelian_2020,soejima2020efficient}. In App.~\ref{ed:app:phasediagramlambdanu3}, Tables \ref{ed:tab:momentumlambda0nu-3nx3ny3}-\ref{ed:tab:momentumlambda1nu-3nx4ny3} further shows the ground state momentum (relative to the ground state in the chiral-flat limit which is Chern number $\pm1$ insulator as shown in Sec. \ref{sec:fullypolsectorsnu3} and Ref.~\cite{ourpaper4}) for various parameters at $\nu=-3$, where we observe that for $\lambda=1$, $w_0/w_1\gtrsim 0.3$ or $\lambda=0$, $w_0/w_1\gtrsim 0.9$, the ground state momenta occur near $\Gamma_M$, $M_M$ or $K_M$ at different system sizes/parameters, which suggest possible competing nematic, stripe or momentum $K_M$ CDW orders.
The ground state manifold spread $\delta$, shown in Figs.~\ref{ed:fig:phase_diagram_3x2}b and~\ref{ed:fig:phase_diagram_3x2}g, are always small in both cases when compared with the finite size gap $\Delta$. In particular, for larger lattices that we will discuss in the next subsection (see Sec.~\ref{sec:fullypolsectorsnu3} and App.~\ref{ed:app:additionalsizesnu3}), we notice: for $3\times 3$, $4\times 3$ and $5\times 3$ moir\'e lattice with the FMC, there grounds state changes from the Chern insulator to another  type of ground state at $w_0/w_1\gtrsim 0.9$ but does \emph{not} change momentum sector. For $\lambda=1$, without the FMC condition, the situation is more complicated.  On  a $3\times 3$ lattice, the Chern insulator is a ground state at $\Gamma_M$ up to $w_0/w_1 \le 0.3$.  For $0.3\le w_0/w_1 \le 0.8$ the ground state changes momentum to $K_M$, indicating a CDW.  For $0.8\le w_0/w_1 \le 1$ the ground state momentum changes again to $(1,0)$, close to the $\Gamma_M$ point, indicating a possible nematic transition.  We note that the $3\times 3$ lattice does not have a mometnum mesh that touches the $M_M$ point. For $4\times 3$ sites, the Chern ground state, stable for $0\le w_0/w_1\le 0.4$ is at momentum $(2,0)$ - or the $M_M$ point, due to the finite size of the system. Since we know that in the infinite size limit, the Chern insulating states will be at zero momentum, we measure all the momenta \emph{from} that of the Chern insulator ground state. The system then has a phase transition at $w_0/w_1 \approx 0.6$ to a CDW with momentum $M_M$, while for larger ratios, it seems to favor lower momenta ground states, probably towards zero.

We also notice that the effect of the kinetic term controlled by $t$ is relatively smaller than $w_0/w_1$, as predicted by Ref. \cite{ourpaper4}. Remarkably, we did not observe a vanishing finite size gap for any $t < 1$ in both $\lambda=0$ and $\lambda = 1$ cases.

Focusing on the properties of the absolute lowest energy state, we compute the valley polarization defined as the ratio between $N_v$, the difference of the electrons numbers in valley $+$ and $-$ (a conserved quantity), and the total number of electrons. The valley polarization is shown in Figs.~\ref{ed:fig:phase_diagram_3x2}c and~\ref{ed:fig:phase_diagram_3x2}h. 
First we see that for $w_0/w_1 \lesssim 0.3$, the splitting of the symmetry broken U(4)$\times$U(4) ground state favors the fully valley polarized states for both $\lambda=0$ and $\lambda = 1$ cases. This is in agreement with our perturbation calculations at $\nu=-3$ in \cite{ourpaper4}. For $\lambda=0$, this is actually valid over the whole phase diagram. But for $\lambda=1$, the system undergoes many level crossing involving different valley polarizations if $w_0/w_1$ goes beyond  0.3.
In Figs.~\ref{ed:fig:phase_diagram_3x2}d-e and Figs.~\ref{ed:fig:phase_diagram_3x2}i-j, we also provide the total spin in each valley $S_{\eta=+}$ and $S_{\eta=-}$ for the absolute lowest energy state with $N_v \ge 0$. (Due to the $C_{2z}$ symmetry, the spectra of $N_v$ and $-N_v$ are identical therefore only non-negative $N_v$ values are shown.). While the FMC model, the absolute lowest energy state is  spin and valley fully polarized irrespective of the values of $t\in[0, 1]$ and $w_0/w_1 \in [0, 1]$, the full TBG model display a more diverse spin polarization once the Chern insulator phase is washed out. Due to the small system size, a strong conclusion about the physics in this region would be too speculative.

\subsubsection{Fully polarized sectors}\label{sec:fullypolsectorsnu3}

\begin{figure*}[!t]
    \centering
    \includegraphics[width=0.9\linewidth]{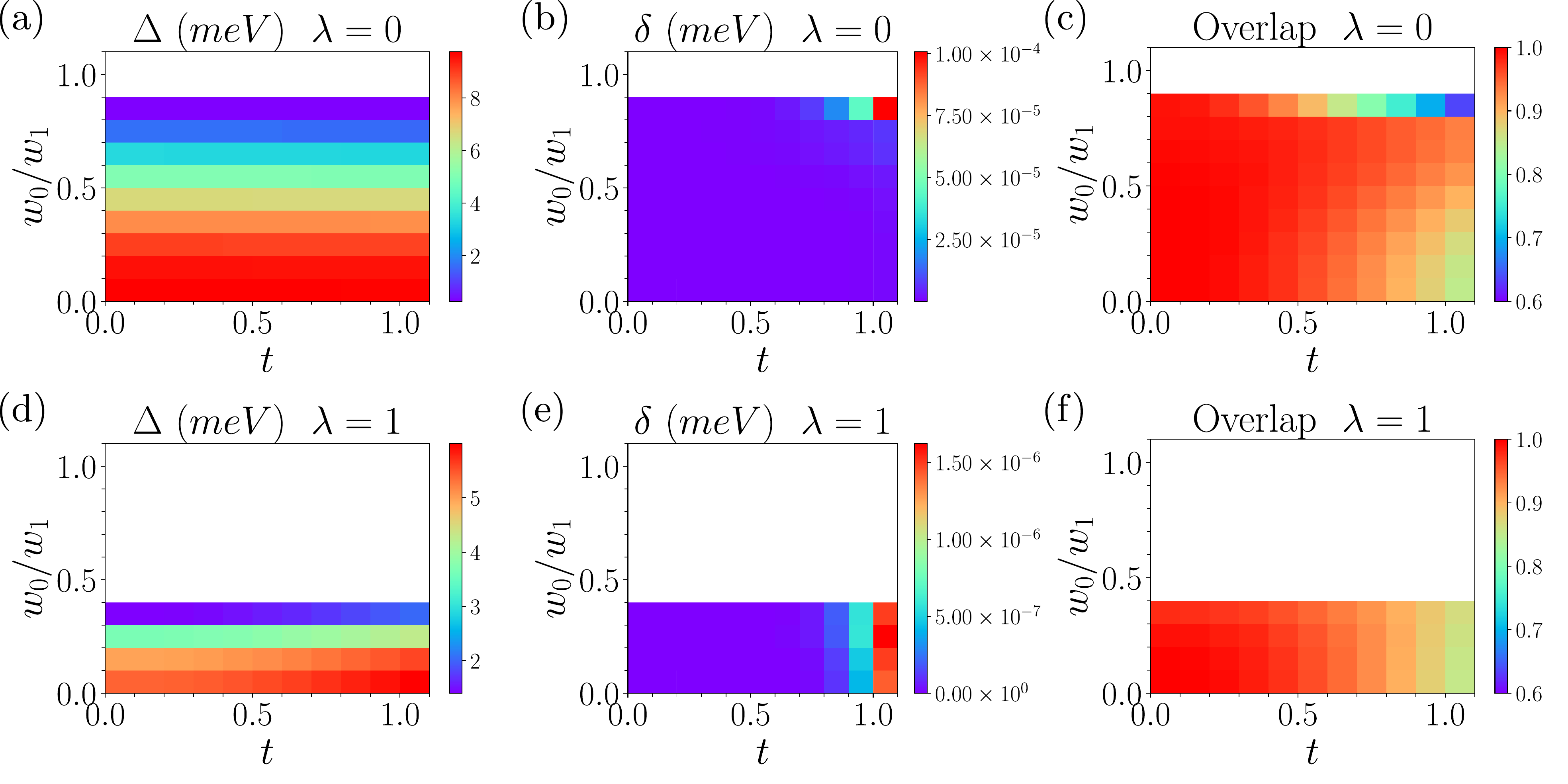}
    \caption{The phase diagram at filling $\nu=-3$ on $4\times 3$ lattice in spin and valley polarized symmetry sectors with $\lambda = 0$ and $\lambda= 1$. The finite size gap (a, d), the spread between the two lowest states (b, e) and the overlap between the two lowest states and Chern insulator states (c, f) are shown by color. We choose $\lambda = 0$ in subfigures (a-c) and $\lambda = 1$ in subfigures (d-f). The white regions are beyond the Chern insulator phase, in which the overlap between the ED ground states and Chern insulator states is zero. Note that the overlap scale starts at $0.6$. Overall, the overlap is never smaller than 0.9 in above $80\%$ of the area in the Chern insulator phase.}
    \label{ed:fig:phase_diagram_t_w0}
\end{figure*}

To reach bigger system sizes in the nonchiral ($w_0>0$) and nonflat ($t>0$) limit, we can focus on a specific symmetry sector of U(2)$\times$U(2): the fully valley and spin polarized sector. As we discussed previously,  the assumption that the ground state is the valley and spin polarized sector would break down for $\lambda=1$ when the system transitions away for the Chern insulator phase at around $w_0/w_1\simeq 0.3$. 
More precisely, by focusing on one symmetry sector, one might miss phase transitions in other sectors. Hence any phase boundary obtained by focusing on one symmetry sector only can over-estimate the stability regime of the phase - a phase transition might have already happened in a different symmetry sector. Still, it provides some valuable insight on the system size influence on the many-body spectrum. In Fig.~\ref{ed:fig:phase_diagram_t_w0}, we present phase diagrams for a $N_1=4$, $N_2=3$ system in the fully valley and spin polarized sector. Starting with the gap (Figs.~\ref{ed:fig:phase_diagram_t_w0}a and~\ref{ed:fig:phase_diagram_t_w0}d) and spread (Figs.~\ref{ed:fig:phase_diagram_t_w0}b and~\ref{ed:fig:phase_diagram_t_w0}e), we see that, for $\lambda=1$, the results of the fully polarized calculation barely changes  when compared with the phase diagrams from the full symmetry sector calculation in Fig.~\ref{ed:fig:phase_diagram_3x2} (just a slightly higher transition value around $w_0/w_1 \simeq 0.4$). In this symmetry sector, the gap  $\Delta$ between the Chern insulator ground state for $\lambda=0$ and the next energy level starts considerably diminishing only at a much larger value $w_0/w_1 \simeq 0.9$ than in the full spectrum.

We observe that the transition away from the Chern insulator phase mostly occurs by a level crossing with states at finite momentum (i.e. a total momentum not invariant under $C_{2z}$) for $\lambda=1$, as opposed to $\lambda=0$, where it never changes momentum. As long as the system is in the Chern insulator phase, the splitting between the two Chern states $\nu_C=\pm1$ is barely noticeable.  Note that on momentum lattice with $C_{3z}$ symmetry (such as the similar phase diagrams  on a $3 \times 3$ and $5 \times 3$ lattices provided in App.~\ref{ed:app:phasediagramlambdanu3}) the Chern states $\nu_C=\pm 1$ are also $C_{3z}$ eigenstates with different eigenvalues, are thus exactly degenerate.

Besides computing the many-body finite size gap and spread, we can  also rely on wavefunction overlaps to quantify how close the ground state is from a Chern insulator state. As discussed in App.~\ref{ed:appsubsec:chern}, the Chern band basis, suitable for this task, is well-defined for each given value of $w_0$ \cite{ourpaper2}. The corresponding Chern insulator wavefunctions are given by:
\begin{equation}
    |\Phi_{\nu=-3}^{\pm1}(w_0)\rangle = \prod_{\vk}d^\dagger_{\vk, e_Y=\pm1, +, \uparrow}|0\rangle\,. \label{ed:eq:chernstatew0nu3}
\end{equation}
Note that these Fock states at $w_0>0$ are different from the Fock states Eqs.~(\ref{ed:eq:nu3chern1}) and~(\ref{ed:eq:nu3chernm1}) in the chiral-flat limit $w_0= 0$, since the single-particle wavefunctions are different for different $w_0$. Although they have the same expression with Eqs.~(\ref{ed:eq:nu3chern1}) and~(\ref{ed:eq:nu3chernm1}), the operators that create the state are the Chern basis in the non-chiral limit.

By ED, we obtain the wavefunctions of the two lowest states in the spin and valley fully polarized sector as a function  of $t$, $w_0$ and $\lambda$ $|\psi_{\rm ED}^j(t, w_0, \lambda)\rangle$ with $j=1,2$ the index of the two lowest states. We define the overlap between the two lowest states in the ED spin and valley fully polarized sector and the Chern insulator states by
\begin{equation}
    \textrm{Overlap} = \frac12\sum_{j=1}^2\sum_{\nu_C=\pm1}|\langle \Phi_{\nu=-3}^{\nu_C}(w_0) |  \psi_{\rm ED}^j(t, w_0, \lambda) \rangle|^2\,.
\end{equation}
This overlap is unity when the two states $|\psi_{\rm ED}^j(t, w_0, \lambda)\rangle$ span the same subspace generated by Eq.~\ref{ed:eq:chernstatew0nu3}. In Figs.~\ref{ed:fig:phase_diagram_t_w0}c and f, we provide the overlap as a function of $w_0$ and $t$ with ($\lambda=0$) and without ($\lambda=1$) the FMC. In the regions where the Chern insulator description is expected to be good, we obtain an overlap on the order of $0.9$ or higher which drops quickly only in the vicinity of the transition. This high overlap shows that the Chern insulator states are, to a good approximation, close to non-interacting Slater determinants.

To provide a more complete picture, we have also computed the phase diagram as a function of $\lambda$ and $t$ with $w_0 = 0$ and the phase diagram as a function of $\lambda$ and $w_0$ with $t = 0$ in App.~\ref{ed:app:phasediagramlambdanu3}.  
The interpolation shows that the transition point of $w_0/w_1$ decreases smoothly when $\lambda > 0.5$.

\section{Numerical results at filling factor \texorpdfstring{$\nu = -2$}{nu=-2}}\label{sec:nu2}

\subsection{Chiral-flat limit}\label{sec:chiralflatbandnu2}

In Ref.~\cite{ourpaper4}, it is proved that in the chiral-flat limit, at $\nu=-2$, and with the FMC that the following Chern insulator states of Chern number $\nu_C$ are ground states: 
\begin{eqnarray}
    |\Psi^{1,1}_{\nu=-2}\rangle &=& \prod_{\vk}d^\dagger_{\vk,+1,+,\uparrow}d^\dagger_{\vk,-1,+,\uparrow}|0\rangle,~~\nu_C = 0\label{ed:eq:chern0nu2}\\
    |\Psi^{2, 0}_{\nu=-2}\rangle &=& \prod_\vk d^\dagger_{\vk,+1,+,\uparrow}d^\dagger_{\vk,+1,+,\downarrow}|0\rangle,~~\nu_C = 2\label{ed:eq:chern2nu2}\\
    |\Psi^{0, 2}_{\nu=-2}\rangle &=& \prod_\vk d^\dagger_{\vk, -1, +, \uparrow}d^\dagger_{\vk, -1, +, \downarrow}|0\rangle,~~\nu_C=-2,\label{ed:eq:chern-2nu2}
\end{eqnarray}
all of which are degenerate. All the U(4)$\times$U(4) rotations of these states give the ground state manifold. We note that without the FMC, these states are still eigenstates of the Hamiltonian; even with FMC, additional ground states are not excluded.
The multiplet of the $\nu_C=0$ state $|\Psi^{1,1}_{\nu=-2}\rangle$ is spin and valley polarized in each U(4) sector (i.e., Chern basis $e_Y$  sector). 
The other two states $|\Psi^{2, 0}_{\nu=-2}\rangle, |\Psi^{0, 2}_{\nu=-2}\rangle$ with Chern numbers $\nu_C=\pm2$ have all electrons occupying one Chern basis sector; within the occupied Chern basis sector, they can be either a spin polarized valley singlet, or a valley polarized spin singlet. 
The U(4)$\times$U(4) irreps of these states are $([N_M]_4,[N_M]_4)$ for $\nu_C=0$, $([N_M, N_M]_4,[0]_4)$ for $\nu_C=2$ and $([0]_4, [N_M, N_M]_4)$ for $\nu_C=-2$.  Given their irreps and conserved charges, these 3  wavefunctions are the only ones that can be built from the Hilbert space of $N=2N_M$ electrons and $N_M$ moir\'e unit cells (similar to the $\nu=-3$ situation where the irrep $([N_M]_4,[0]_4)$ was unique in the Hilbert space of $N=N_M$ electrons and $N_M$ moir\'e unit cells). Thus the states of Eqs.~(\ref{ed:eq:chern0nu2})-(\ref{ed:eq:chern-2nu2}) are always eigenstates of the Hamiltonian in the chiral-flat limit, but are not guaranteed to be ground states unless the FMC is satisfied.

\begin{figure}
    \centering
    \includegraphics[width=0.9\linewidth]{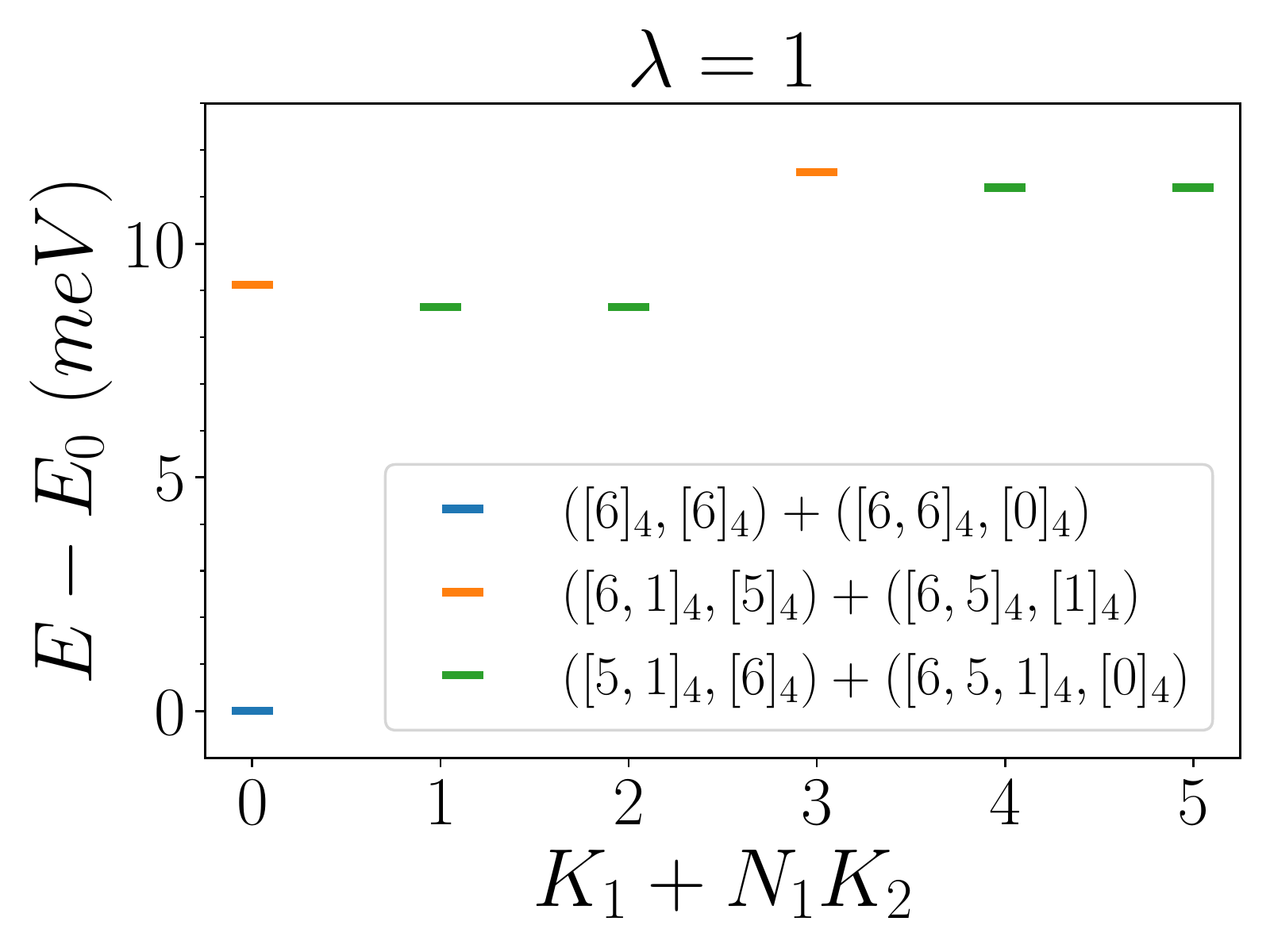}
    \caption{The low energy spectrum and the corresponding irreps at $\nu=-2$ filling on $3\times 2$ lattice at chiral-flat limit with $\lambda = 1$. The twisting angle is $\theta = 1.1014^\circ$. We use the notation ``+'' between two irreps if the states with these two irreps have the same energy.}
    \label{ed:fig:irrep_3x2_lambda=1_N=12}
\end{figure}

In Fig \ref{ed:fig:irrep_3x2_lambda=1_N=12}, we show the low energy spectrum for the full TBG model ($\lambda=1$) on a $3\times 2$ lattice in the (first) chiral-flat limit confirms, as predicted, the ground state manifold is made of the irreps $([6, 6]_4, [0]_4)$, $([0]_4, [6,6]_4)$ (not shown here) and $([6]_4, [6]_4)$. This confirms that, in the first chiral limit, the FMC - the condition under which we can prove that eigenstates Eq.~(\ref{ed:eq:chern-2nu2}) are in fact ground states, is a good approximation and no other ground states are present. Similar to the $\nu=-3$ case, the charge neutral excitation irreps, including $([6,1]_4,[5]_4)$, $([5, 1]_4, [6]_4)$, $([6,5]_4, [1]_4)$ and $([6, 5, 1]_4, [0]_4)$, can be interpreted as moving one electron from the fully-filled Chern insulator ground state to other energy bands (i.e., creating one electron-hole pair), and are thus close (as defined in Sec.~\ref{sec:chiralflatbandnu3}) to the irreps of the ground state manifold.

\begin{figure}
    \centering
    \includegraphics[width=\linewidth]{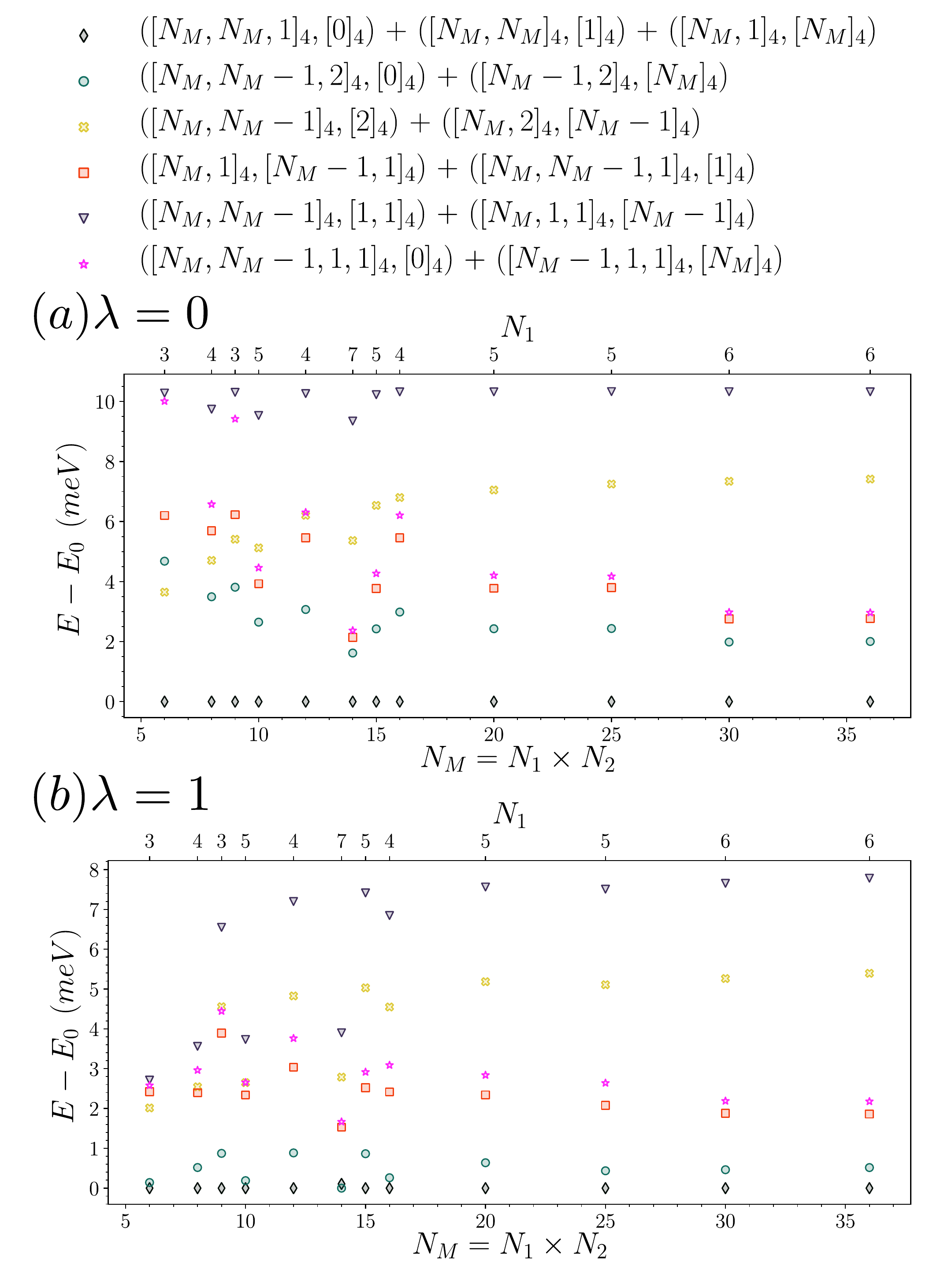}
    \caption{Charge $+1$ (electron) excitation at $\nu = -2$ with $\lambda = 0$ (a) and $\lambda = 1$ (b). All energies have been shifted by the lowest energy $E_0$ at the corresponding system size. Again after $N_M \approx 16$ the system settles towards its thermodynamic properties with the $([N_M, N_M, 1]_4, [0]_4), ([N_M, 1]_4, [N_M]_4), ([N_M, N_M]_4, [1]_4)$ as the irreps of the lowest state. In this the $\lambda =0$ and $\lambda = 1$ plots agree though the gap is much greater for the $\lambda = 0$ case. We use the notation ``+" between irreps when they always appear with an exact degeneracy.}
    \label{ed:fig:excitation_nu_-2}
\end{figure}

\begin{figure}
    \centering
    \includegraphics[width=\linewidth]{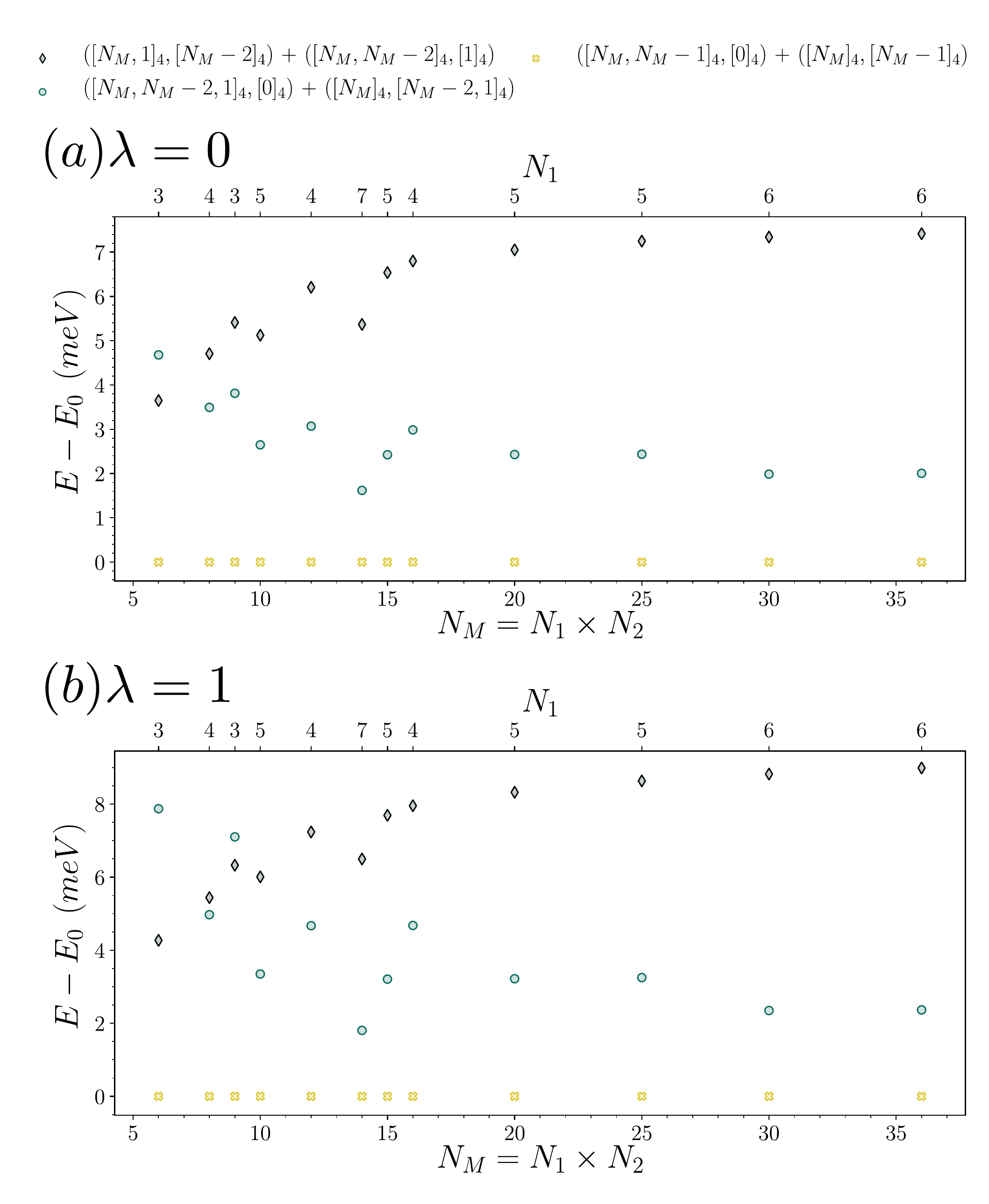}
    \caption{Charge $-1$ (hole) excitations at $\nu=-2$ with $\lambda =0$ (a) and $\lambda = 1$ (b). Again, all energies have been shifted by the lowest energy $E_0$ at the corresponding system size. As is typical for the hole excitations there is not a strong size effect and the ground state is the irreps $([N_M, N_M-1]_4, [0]_4)$ and $([N_M]_4, [N_M-1]_4)$ for with or without the FMC.  We use the notation "+" between irreps when they always appear with an exact degeneracy.}
    \label{ed:fig:hole_nu_-2}
\end{figure}

Ref.~\cite{ourpaper5} introduced neutral and charge excitations on top of the eigenstates Eq.~(\ref{ed:eq:chern-2nu2}). The excitations are eigenstates of the TBG Hamiltonian, but analytically one cannot prove that they are the lowest energy eigenstates, even with the FMC satisfied. Based on the fact that the irreps of the low-energy excitations should be (and are in the analytic model) close to the irreps of the Chern insulator ground state, we study the charge $\pm1$ excitations for both the FMC model and the full interacting TBG model in chiral-flat limit in the sub-Hilbert space of states which differ from the ground states in Eqs.~(\ref{ed:eq:chern0nu2}-\ref{ed:eq:chern-2nu2}) by at most one electron (hole) plus one electron-hole pair (similar to what we did for $\nu=-3$). The results are given in Fig.~\ref{ed:fig:excitation_nu_-2} for the charge $+1$ excitation and Fig.~\ref{ed:fig:hole_nu_-2} for the charge $-1$ excitation, respectively. The charge $+1$ excitation with irreps $([N_M, N_M, 1]_4, [0]_4)$ and $([N_M, 1]_4, [N_M]_4)$ are favored energetically for both $\lambda=0$ and $\lambda=1$ cases, even on rather small lattice sizes (as opposed to the charge excitations at $\nu=-3$, which stabilize at large sizes $N_M\ge 20$).
Similarly, for the hole excitations, the state with irrep $([N_M, N_M - 1]_4, [0]_4)$ and $([N_M]_4,[N_M - 1]_4)$ have the lowest energies for all the system sizes we studied. Notice that spectra at  $\lambda=0$ and $\lambda=1$ in  both Fig.~\ref{ed:fig:excitation_nu_-2} (charge $+1$ excitation) and Fig.~\ref{ed:fig:hole_nu_-2} (charge $-1$ excitation) contain the same energy order of the irreps in the spectra, showing that the FMC ($\lambda=0$) and the first chiral-flat limit without the FMC condition ($\lambda=1$) have the same qualitative spectra. In particular, these lowest charge excitations we found here are exactly the analytic charge excitations obtained in Ref.~\cite{ourpaper5}. For example, in Fig.~\ref{ed:fig:hole_nu_-2} (charge $-1$ excitation),  the state with irrep $([N_M, N_M - 1]_4, [0]_4)$, identical to the analytic charge excitations in Ref.~\cite{ourpaper5}, are the lowest. 

\subsection{Phase diagrams in the nonchiral-nonflat case}\label{sec:phasedigramnu-2}

Due to the large  number of electrons, the dimensions of the symmetry sectors  at $\nu=-2$ are much bigger than at $\nu=-3$ (see App.~\ref{ed:appsec:symhamiltonian} and Table.~\ref{table:nu2}). Therefore, we limit our phase diagram calculation to either valley or spin polarized sectors on a $3\times 2$ lattice.

\subsubsection{Valley polarized phase diagrams}\label{sec:phasedigramnu-2valley}

We first consider the valley polarized sectors, setting $\eta=+$.  We subduce the U(4)$\times$U(4) irreps built from the three Chern insulator states of Eqs.~(\ref{ed:eq:chern0nu2}),~(\ref{ed:eq:chern2nu2}) and~(\ref{ed:eq:chern-2nu2}) into U(2)$\times$U(2) irreps;  some of the subduced irreps  will appear in the fully valley polarized sectors $\eta=+$ (those with no particle in the second U(2)). In the valley $+$ polarized sectors, we only have one conserved total spin, the total spin $S_{\eta=+}$ of valley $+$. The total spin for the valley polarized Chern insulator states with $\nu_C=\pm 2$ can only be $S_{\eta=+}=0$, as they correspond to filling one valley, both spins,  with the same Chern number. However, the states with Chern number $\nu_C = 0$ can have different spin quantum numbers $S_{\eta=+} = 0, 1, \cdots, N/2$. 
To summarize, close to the chiral-flat limit in the valley polarized sector we expect to see 3 $S_{\eta=+}=0$ states (one $\nu_C=0$, one $\nu_C=+2$ and one $\nu_C=-2$), and a set of spin multiplets with $S_{\eta=+} = 1, \cdots, N/2$ and Chern number $\nu_C=0$.

\begin{figure*}[t]
    \centering
    \includegraphics[width=\linewidth]{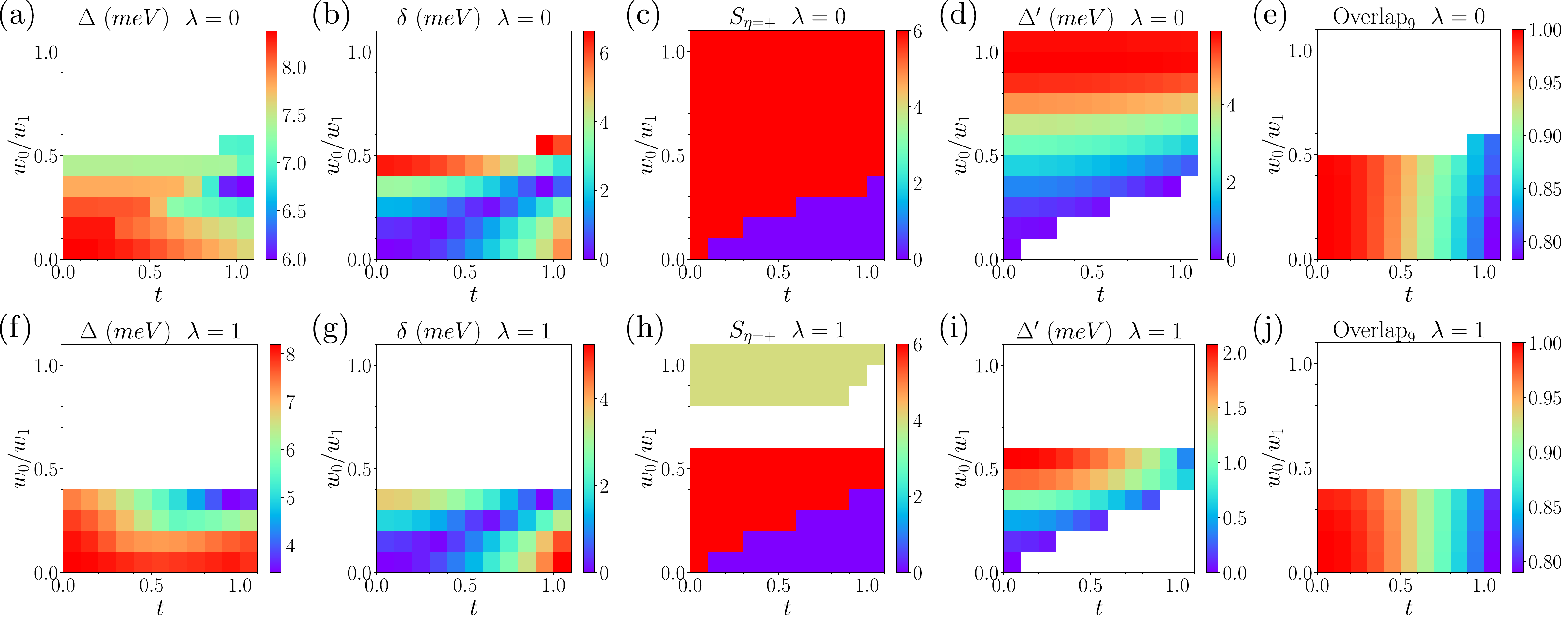}
    \caption{The phase diagrams at $\nu = -2$ on $3\times 2$ lattice with $\lambda=0$ and $1$. The calculation is done in valley polarized symmetry sectors. At the chiral-flat limit, the ground state contains $\nu_C = 0$ and $\nu_C = \pm 2$ Chern insulator states, and these states will split into multiple states with different total spins when moving away from chiral-flat limit. 
    The finite size gap $\Delta$ shown in subfigures (a, f) is defined by the distance between the lowest charge neutral excitation energy and the lowest energy in the ground state manifold, and the split $\delta$ in subfigures (b, g) is the separation between the highest and the lowest state in the ground state manifold.
    We also show the spin quantum number $S_{\eta=+}$ of the lowest state in the ground state manifold in subfigures (c, h). Note that in the ferromagnetic region we also provide the gap $\Delta^\prime$ in subfigures (d, i), which measures the energy difference between the $S_{\eta=+}=6$ ground state and the state above it. This ground state always has a total momentum $K_1=K_2=0$ except in the white region of (i).
    The overlap between the ground state manifold states and the Chern insulator states with $\nu_C = \pm 2$ in subfigures (e, j) is above $80\%$ in the entire phase, which indicates that the two $\nu_C=\pm2$ states are in these low energy states.} \label{ed:fig:phase_diagram_t_w0_3x2_-2}
\end{figure*}

Similar to Sec.~\ref{sec:phasediagramnu3} for $\nu=-3$, we now consider the phase diagrams as a function of $w_0$ and $t$ for both the FMC model $\lambda=0$ and for the full TBG model $\lambda=1$. The results are provided in Fig.~\ref{ed:fig:phase_diagram_t_w0_3x2_-2}. Figs.~\ref{ed:fig:phase_diagram_t_w0_3x2_-2}a and f show the finite size charge neutral gap $\Delta$ while Figs.~\ref{ed:fig:phase_diagram_t_w0_3x2_-2}b and g give the spread $\delta$ (both defined in Sec.~\ref{sec:phasediagramnu3}). 
Again, for stable ground state, the Goldstone branches will have a gap $\sim m|\mathbf{b}_{M1}|^2/2N_M$ for finite systems. Hence $\Delta\to0$ implies either the vanishing of Goldstone stiffness ($m\to0$) or softening of some collective modes (including possibly finite momentum Goldstone branches) at finite momenta, leading to an instability of the ground state.
The ground states manifold that we have considered here consists of all the Chern states discussed above: 3 $S_{\eta=+}=0$ states (one $\nu_C=0$, one $\nu_C=+2$ and one $\nu_C=-2$) and a $N/2$ spin multiplet with $S_{\eta=+} = 1, \cdots, N/2$ and $\nu_C=0$. Interestingly the FMC model $\lambda=0$ does not differ much from the full TBG $\lambda=1$ model. In particular, and as opposed to $\nu=-3$, the full Chern insulator phase for $\lambda=0$ and $\lambda=1$ disappears roughly at the same value of $w_0/w_1$ ($w_0/w_1\simeq 0.5$ for $\lambda=0$ and $w_0/w_1\simeq 0.4$ for $\lambda=1$). $t$ increases the spread $\delta$, and reduces the finite-size gap $\Delta$ but is never able to close the later. Since we are considering here only the valley polarized sector,  the  values of the parameters for which the finite size gap closes are thus only the upper bounds of what a fully unpolarized calculation would give.

The behaviour of the splitting of ground state manifold made of $|\Psi_{-2}^{2,0}\rangle$, $|\Psi_{-2}^{1,1}\rangle$, $|\Psi_{-2}^{0,2}\rangle$ at $\nu=-2$ is also  different from the splitting at $\nu=-3$. 
At $\nu=-2$, the spread $\delta$ (the energy difference between highest and lowest states of the Chern number $0,\pm2$ chiral-flat ground states after splitting)
is overall larger than that the $\nu=-3$ case, except along a line $w_0/w_1\approx 0.5t$ (see Fig.~\ref{ed:fig:phase_diagram_t_w0_3x2_-2}b,g) To probe this region in more detail, we have computed the spin quantum number $S_{\eta=+}$ of the absolute ground state in valley fully polarized sectors,
which can be found in Figs.~\ref{ed:fig:phase_diagram_t_w0_3x2_-2}c and h. These plots show that the insulating phase in the valley polarized sector can be separated into two phases with different magnetic orders. The region dominated by the nonchiral-flat limit prefers the largest possible spin polarization (ferromagnetic), while the region dominated by the  chiral-nonflat  limit favors the spin singlet.

The phase boundary between the ferromagnetic phase and spin singlet phase can be seen clearly in both Figs.~\ref{ed:fig:phase_diagram_t_w0_3x2_-2}c and~\ref{ed:fig:phase_diagram_t_w0_3x2_-2}h. This boundary matches well with the low spread $\delta$ line $w_0/w_1\approx 0.5t$ in Fig.~\ref{ed:fig:phase_diagram_t_w0_3x2_-2}b and g. Our numerical results validate the exact/perturbative approach in Ref.~\cite{ourpaper4}, where it is shown that $\nu=-2$ in the nonchiral-flat limit prefers to fully occupy one spin-valley flavor (thus is a spin-valley ferromagnet), while in the chiral-nonflat limit it prefers to half-occupy two different spin-valley flavors (thus spin singlet when valley is polarized). We also note that in the nonchiral-nonflat case, it is proposed by earlier HF studies \cite{liu2020nematic,zhang_HF_2020} as well as perturbation theory \cite{bultinck_ground_2020,ourpaper4} that an intervalley-coherent state may be the ground state. Such a state, however, which has valley quantum number $N_v=0$, demands a Hilbert space dimension of ED far beyond our computational power, thus will not be discussed here.

As we have mentioned earlier, the Chern insulator states Eqs.~(\ref{ed:eq:chern0nu2}-\ref{ed:eq:chern-2nu2}) with $\nu_C = \pm 2$ always have zero spin in valley polarized sectors, therefore the ground state in the ferromagnetic phase can only carry zero Chern number. However, both the states with $\nu_C=0$ and $\nu_C=\pm 2$ can have zero spin. Thus, we use the wavefunction overlap to probe the Chern number of the preferred ground state in the valley-polarized spin singlet phase near the chiral-nonflat limit
In the valley polarized sectors, the wavefunctions of Chern insulator states with $\nu_C = \pm2$ are 
\begin{equation}
    |\Phi^{\pm2}_{\nu=-2}(w_0)\rangle = \prod_\vk d^\dagger_{\vk, \pm1, +1, \uparrow}d^\dagger_{\vk, \pm1, +1, \downarrow}|0\rangle\,.\label{ed:eq:model_state-2}
\end{equation} 
We focus on the low energy states with spin $z$ component $S_{z,\eta=+} = 0$ and full valley polarization. The ground state manifold has $9$ states in this symmetry sector on $3\times 2$ lattice: two of them are the Chern insulator states with $\nu_C = \pm 2$ and the other 7 are the spin $z$-component zero states of the total spin $S_{\eta=+} = 0,1,\cdots,6$ phases. We can obtain the exact wavefunctions of the low energy states in the valley polarized sector with given values of $t, w_0$ and $\lambda$ by performing ED. We call these states $|\psi^j_{\rm ED}(t, w_0, \lambda)\rangle$, and the wavefunction overlap between the two Chern insulator states and the lowest $n$ states can be defined as shown:
\begin{equation}
    \textrm{Overlap}_n = \frac{1}{2} \sum_{\nu_C=\pm2}\sum_{j=1}^n|\langle \Phi^{\nu_C}_{\nu=-2}(w_0) | \psi^j_{\rm ED}(t, w_0, \lambda) \rangle|^2.\label{ed:eqn:overlap_-2}
\end{equation}
This overlap measures whether the Chern insulator Fock states in Eq.~(\ref{ed:eq:model_state-2}) are close to the lowest $n$ states obtained by numerical calculation. If the overlap is equal to one, the two Chern insulator states must be inside the Hilbert space spanned by these $n$ wavefunctions. When we choose $n=2$, we focus on the two lowest energy states, and the largest overlap away from chiral-flat limit in the insulating phases is around $3.7\%$. This result indicates the lowest two states in ground state manifold (all the states in) after splitting are never the nonzero Chern number states when either near the nonchiral-flat limit or near the chiral-nonflat limit. We also study the wavefunction overlap when $n=9$. The results are shown in Figs.~\ref{ed:fig:phase_diagram_t_w0_3x2_-2}e and \ref{ed:fig:phase_diagram_t_w0_3x2_-2}j. This overlap is above $80\%$ at almost everywhere in the insulating phase, which confirms that there are states carrying nonzero Chern numbers in the ground state manifold, although they are not favored energetically by a nonchiral-nonflat Hamiltonian in valley polarized sector. 

Another overlap that we can easily evaluate is the overlap between the ferromagnetic state (the spin $z$-component $S_{z,\eta=+}=6$ state with total spin $S_{\eta=+}=6$, and Chern number $0$) of the ground state manifold in numerical calculation and the ferromagnetic Fock state one can write down at $w_0$ (analogous to Eq.~(\ref{ed:eq:model_state-2})).
Since there is only one such a state with the given quantum numbers in the whole Hilbert space, if we see that the absolute ground state has this total spin, we are guarantee that the overlap is $100\%$ (i.e. the red regions in Figs.~\ref{ed:fig:phase_diagram_t_w0_3x2_-2}c and \ref{ed:fig:phase_diagram_t_w0_3x2_-2}h). For sake of completeness, we provide the finite size gap $\Delta^\prime$ above the ferromagnetic state when it becomes the system ground state ($\Delta'$ is defined as the energy difference between the ferromagnetic ground state and the next level either in or not in the ground state manifold, see  Figs.~\ref{ed:fig:phase_diagram_t_w0_3x2_-2}d and \ref{ed:fig:phase_diagram_t_w0_3x2_-2}i).

Interestingly, for the FMC model, once the ground state manifold at chiral-flat band limit (corresponding to the states in Eqs.~(\ref{ed:eq:chern0nu2})-(\ref{ed:eq:chern-2nu2}) and other states related by $U(4) \times U(4)$ symmetry operations) 
has been washed out (for $w_0/w_1 > 0.5$, where states of other U(4)$\times$U(4) irreps move down and the finite size gap $\Delta$ is smaller than $\delta$, see Fig.~\ref{ed:fig:phase_diagram_t_w0_3x2_-2}(a)), a  substantial gap of at least $3 \rm meV$ 
above the ferromagnetic state appears, indicating that the system has become a Chern $\nu_C=0$ insulator.

\begin{figure}
    \centering
    \includegraphics[width=\linewidth]{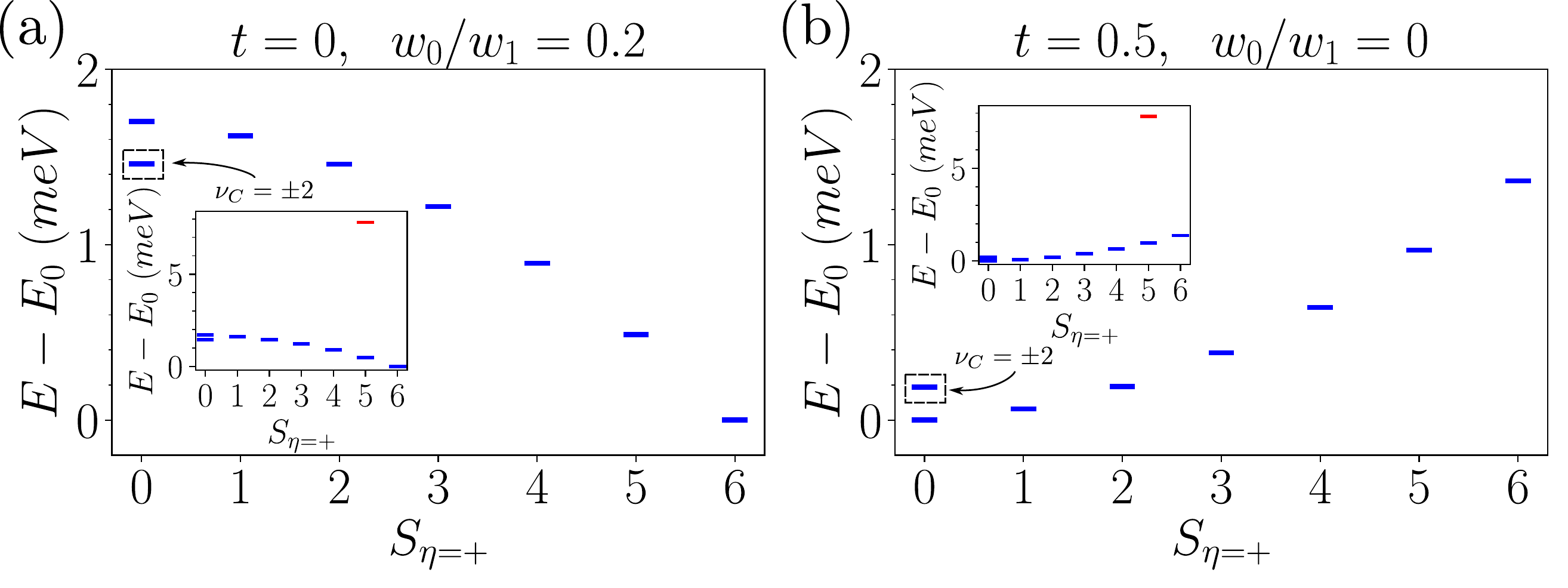}
    \caption{The valley polarized spectrum on $3\times 2$ lattice at filling factor $\nu=-2$. We choose $t = 0$, $w_0/w_1 = 0.2$ for subfigure (a) and $t = 0.5$, $w_0 = 0$ for subfigure (b), with $\lambda = 1$. We use blue dashes to label the states with total momentum $(K_1, K_2) = (0, 0)$. The insets zoom out to unveil the first state (in red) not belonging to the ground state manifold. The system prefers a spin singlet ground state if we add some band dispersion, and it prefers a ferromagnetic ground state if we move away from the chiral limit. When both $t$ and $w_0$ are turned on, the competition will lead to a phase transition within the valley polarized sectors.}
    \label{ed:fig:spin_split}
\end{figure}

To illustrate more clearly the dominance of the $\nu_C=0$ insulating phase and the ferromagnetic/spin singlet phases we show in Fig.~\ref{ed:fig:spin_split}a and~\ref{ed:fig:spin_split}b, typical cases of the ground state manifold splitting in each phase. With nonchiral-flat limit (Fig.~\ref{ed:fig:spin_split}a), the states with largest total spin are favored. For chiral-nonflat limit, the spin singlet state is favored. In both cases, the two states with $S_{\eta=+}=0$ and $\nu_C=\pm2$ are part of the ground state manifold but they are never the lowest energy states. Similar to our analysis for the $\nu=-3$ case (App.~\ref{ed:app:phasediagramlambdanu3}),
we also studied the interpolation phase diagram between $\lambda = 0$ and $\lambda = 1$ (see App.~\ref{ed:appsec:numerical-othernu} and Fig.~\ref{ed:fig:phase_diagram_lambda_t_lambda_w0_3x2_-2}). All the quantities that we probed show rather smooth dependence on $\lambda$. 

\subsubsection{Spin polarized phase diagrams}\label{sec:phasedigramnu-2spin}

\begin{figure}
    \centering
    \includegraphics[width=0.9\linewidth]{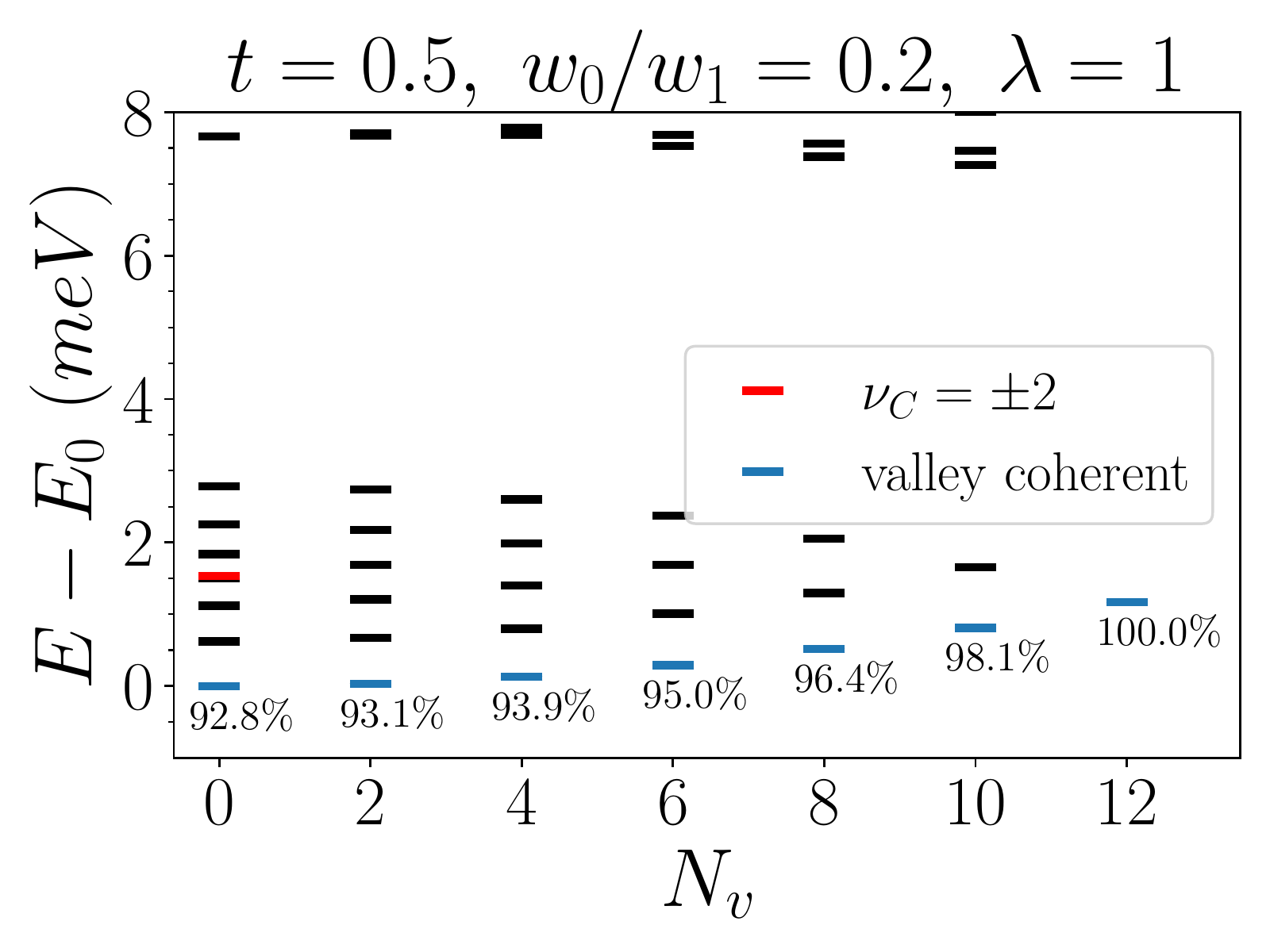}
    \caption{The spin polarized energy spectrum of $3\times 2$ lattice at $\nu=-2$ filling with $t=0.5, w_0/w_1=0.2$ and $\lambda = 1$. The states labeled by red symbol are the states carrying a Chern number $\nu_C = \pm2$. In each $N_v$ sector, the lowest energy states are indicated by blue symbols, and their overlaps with the model states $|\psi_{\rm K-IVC}(N_v)\rangle$ are written next to each level. We can see that the wavefunctions obtained from the ED have large overlaps over $90\%$.}
    \label{fig:spectrum_sp_coherent_overlaps}
\end{figure}

We now turn to the spin polarized sector, setting $s=\uparrow$.
When the system has U(4)$\times$U(4) symmetry at chiral flat band limit, the valley polarized and valley coherent states are degenerate. As predicted in Refs.~\cite{ourpaper4, kang_symmetry_2018, zhang_HF_2020, bultinck_ground_2020}, the ground state will be an inter valley coherent state if both $t$ and $w_0$ are nonzero. However in finite size exact diagonalization where no spontaneous symmetry breaking can occur, the states we obtained are always eigenstates of the valley polarization $N_v$. We start from the expression of the inter valley coherent state provided in Ref.~\cite{ourpaper4}
\begin{align}
|\Psi^{\mathrm{K-IVC}}_{\nu=-2}\rangle &= \prod_{\substack{\mathbf{k}\in{\rm MBZ}\\e_Y = \pm1}}\frac{\left(e^{\frac{-i\gamma}{2}}d^\dagger_{\mathbf{k},e_Y, +,\uparrow} + e^{\frac{i\gamma}{2}}e_Yd^\dagger_{\mathbf{k},e_Y,-,\uparrow}\right)}{\sqrt{2}}|0\rangle\label{eq:KIVCdefinition}
\end{align}
where $\gamma$ is an angle free parameter. This state can be decomposed as
\begin{align}
|\Psi^{\mathrm{K-IVC}}_{\nu=-2}\rangle&=\sum_{N_v=-N_M}^{N_M}\exp{\left(-\frac{i\gamma N_v}{2}\right)} \mathcal{N}_{N_v} |\psi_{\rm K-IVC}(N_v)\rangle\label{eq:KIVCdecomposition}
\end{align}
in which $\mathcal{N}_{N_v}$ is a normalization factor and $|\psi_{\rm K-IVC}(N_v)\rangle$ is the normalized component in the $N_v$ symmetry sector. Note that all the $\gamma$ dependence is encoded in the phase factors. In order to determine whether this state is a good approximation, we compute the overlap between the lowest energy state in each $N_v$ sector obtained by ED, and the model state wavefunction $|\phi^{\mathrm{K-IVC}}_{N_v}\rangle$, namely
\begin{equation}
	\mathrm{Overlap}(N_v) = |\langle \psi_{ED}(N_v) | \psi^{\mathrm{K-IVC}}_{\rm VC}(N_v) \rangle |^2.
\end{equation}
As an example, we consider the spin polarized Hamiltonian at $t=0.5$, $w_0/w_1=0.2$, i.e., away from the chiral flat limit, and $\lambda=1$. The low energy spectrum and overlaps are given in Fig.~\ref{fig:spectrum_sp_coherent_overlaps}. There we show that the ED low energy states in each $N_v$ sectors agree well with the model states $|\psi_{\rm K-IVC}(N_v) \rangle$, with overlaps above $90\%$.

\begin{figure}
    \centering
    \includegraphics[width=\linewidth]{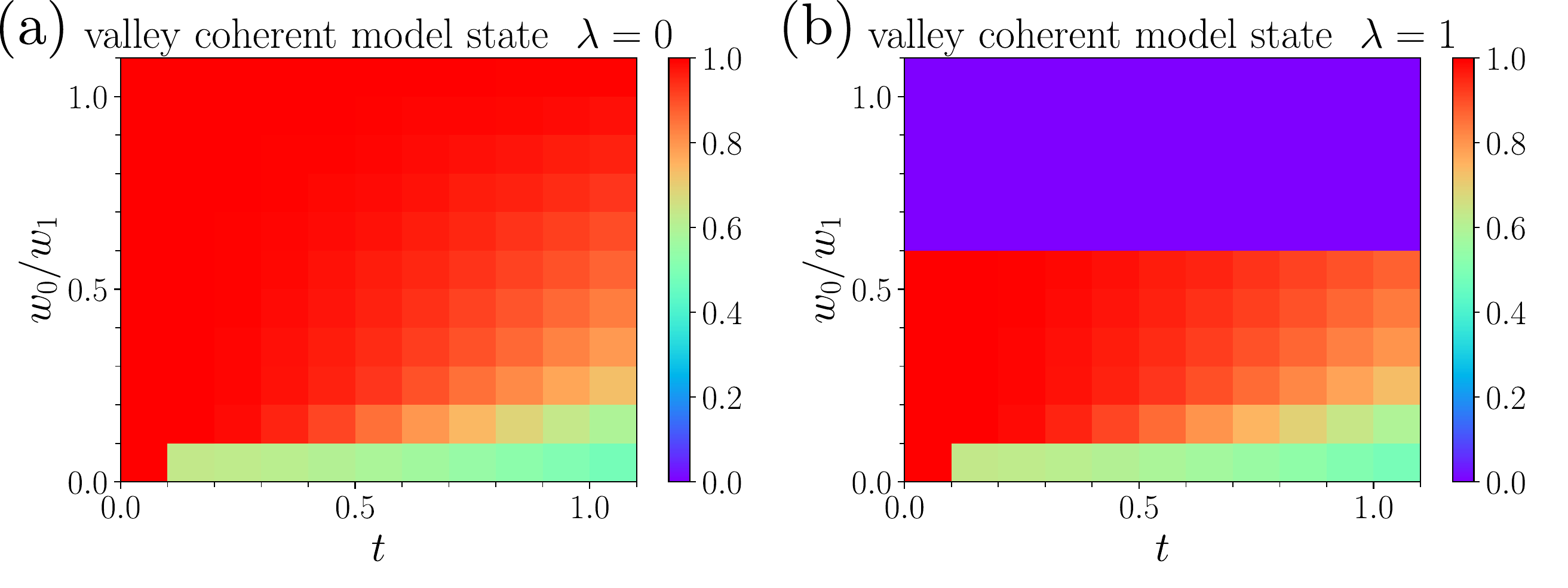}
    \caption{The phase diagrams at $\nu=-2$ filling calculated on $3\times 2$ lattice in the spin polarized sector without (a) and with (b) the flat metric condition. The color code represents the overlap between the ED ground state at $N_V=0$ and the model state $|\psi_{\rm K-IVC}(N_v=0)\rangle$.}
    \label{fig:coherent_phase_diagram}
\end{figure}

To probe how the intervalley coherent wavefunction approximation depends on kinetic energy and nonchiral contributions, we calculate the overlap in $N_v = 0$ sector as a function of $t$ and $w_0$ with and without the FMC in Fig.~\ref{fig:coherent_phase_diagram}. As can be seen in Fig.~\ref{fig:spectrum_sp_coherent_overlaps}, focusing on the $N_v = 0$ sector captures the worst case scenario for the overlap. Our numerical results show that the spin polarized ground states always have a decent overlap $>80\%$ with the model state $|\psi_{\rm K-IVC}(0)\rangle$ in most of the phase diagram if $\lambda=0$. Similarly, if $\lambda=1$, the overlap between the ED ground state and the K-IVC state is close to unity when $w_0/w_1 \lesssim 0.6$. This result implies that the ground state obtained by ED can be well approximated by the K-IVC Slater determinant model state. However, we note that the overlap drops around chiral nonflat limit, and is smaller than $70\%$ when $w_0 = 0$. This steams from the higher symmetry (the chiral-nonflat U(4) symmetry \cite{ourpaper3,bultinck_ground_2020}) in the chiral nonflat limit, which no longer pins the ground state to be intervalley coherent. We provide a detailed explanation in App.~\ref{ed:appsec:nu2spin}. From the higher symmetry in the chiral-nonflat limit, we also build a valley SU$(2)$ singlet model state, which has a large overlap $> 75 \%$ with the ED ground states in the chiral-nonflat limit (see Fig.~\ref{ed:fig:phase_diagram_lambda_t_lambda_w0_3x2_-2}).

When $t>0$ and $w_0>0$, we generically find the ground state energy in the fully spin polarized sector is lower than that in the fully valley polarized sector (with or without FMC). This agrees with the predictions in Refs.~\cite{ourpaper4, bultinck_ground_2020} that the $\nu=-2$ ground state is an intervalley coherent insulator (for small $w_0/w_1$ without FMC). As an example, at $t=1$ and $w_0/w_1=0.3$, the ground state in the fully spin polarized sector is $0.275$ meV/electron lower than that in the fully valley polarized sector, in agreement with the perturbation theory 
estimations in Refs.~\cite{ourpaper4, bultinck_ground_2020}.

\section{Numerical results at filling factor \texorpdfstring{$\nu = -1$}{nu=-1}}\label{sec:nu1}
\begin{figure}
    \centering
    \includegraphics[width=0.9\linewidth]{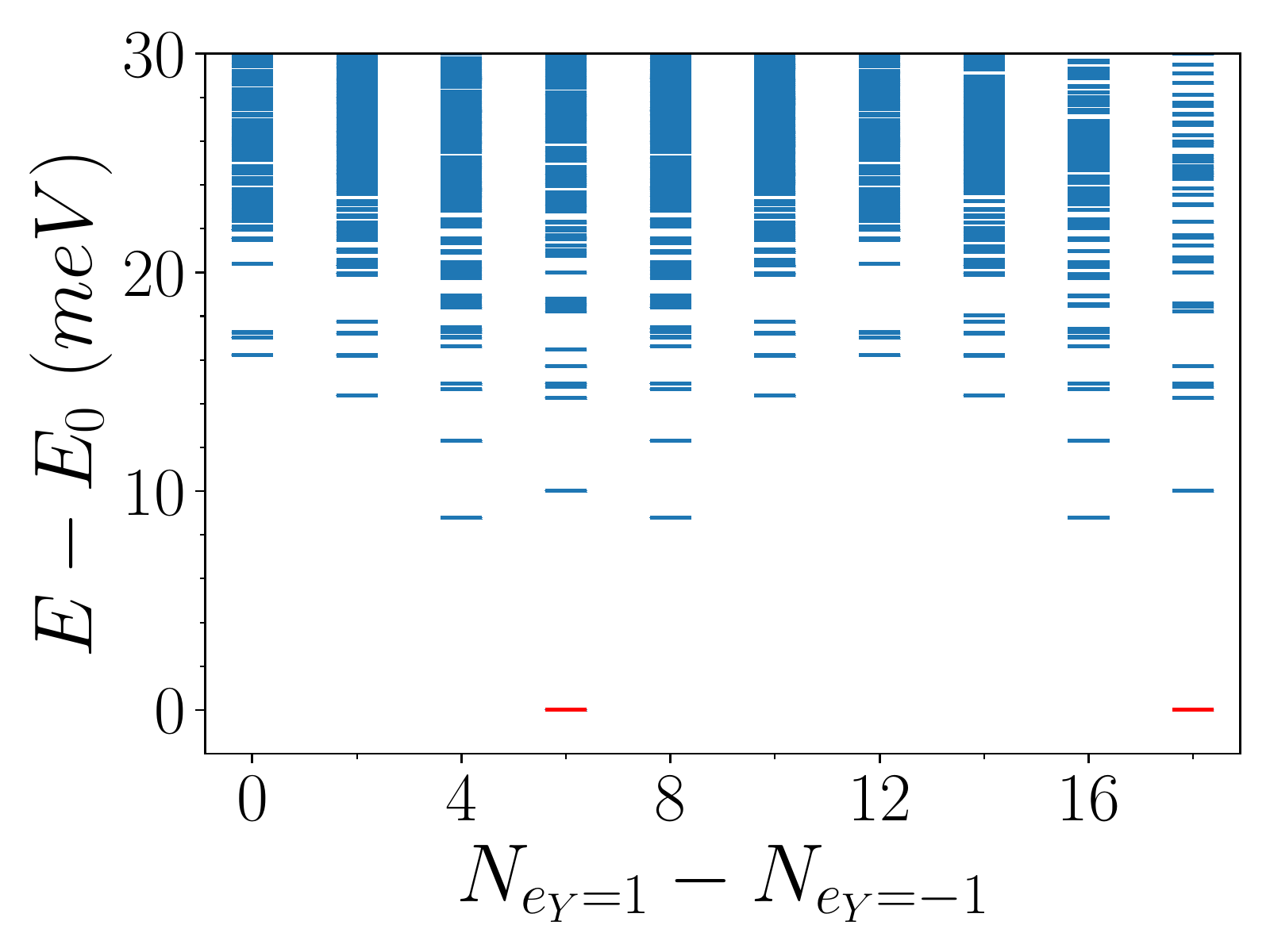}
    \caption{The low-lying states on $3\times 2$ lattice at the chiral-flat limit $\lambda=1$ with filling factors $\nu = -1$ ($N=18$). We only calculated the spectra of symmetry sectors whose dimension is below $10^6$ irrespective of their quantum numbers (at least 2 states per sector). The spectrum is plotted versus the Chern band polarization, where $N_{e_Y}$ is the electron numbers in the band with Chern number $e_Y$ (thanks to the $C_{2z}T$ symmetry, we only consider $N_{e_Y=+1} - N_{e_Y=-1} \ge 0$). The states labeled by red dashes are the Slater determinants, which corresponds to the exact Chern insulator states with $\nu_C = 1$ (at $N_{e_Y=+1} - N_{e_Y=-1}=N_M=6$) or $\nu_C = 3$ (at $N_{e_Y=+1} - N_{e_Y=-1}=3N_M=18$).}
    \label{ed:fig:spectrum3x2_-1_0}
\end{figure}

\begin{figure}
    \centering
    \includegraphics[width=\linewidth]{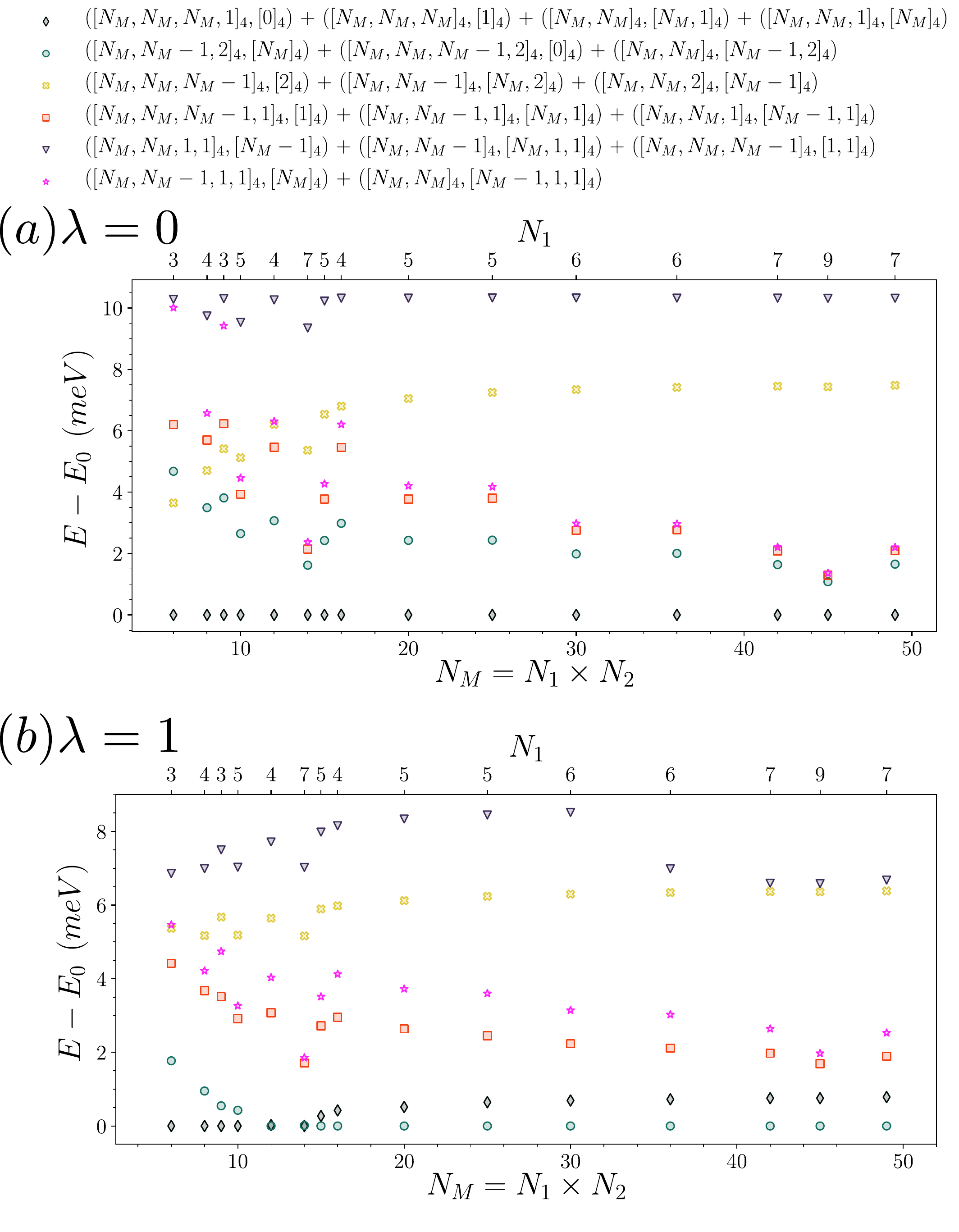}
    \caption{Charge $+1$ (electron) excitation at $\nu=-1$ for the FMC model $\lambda=0$ (a) and the full model $\lambda=1$ (b). $N_M=N_1 \times N_2$. All energies have been shifted by the lowest energy $E_0$ of the corresponding to the given system size. Here we can see a difference between the Hamiltonian with and without FMC in that the irreducible representation of the lowest calculated state differs among the two conditions. We use the notation "+" between irreps when they always appear with an exact degeneracy.}\label{ed:fig:excitations_nu-1}
\end{figure}

\begin{figure}
    \centering
    \includegraphics[width=\linewidth]{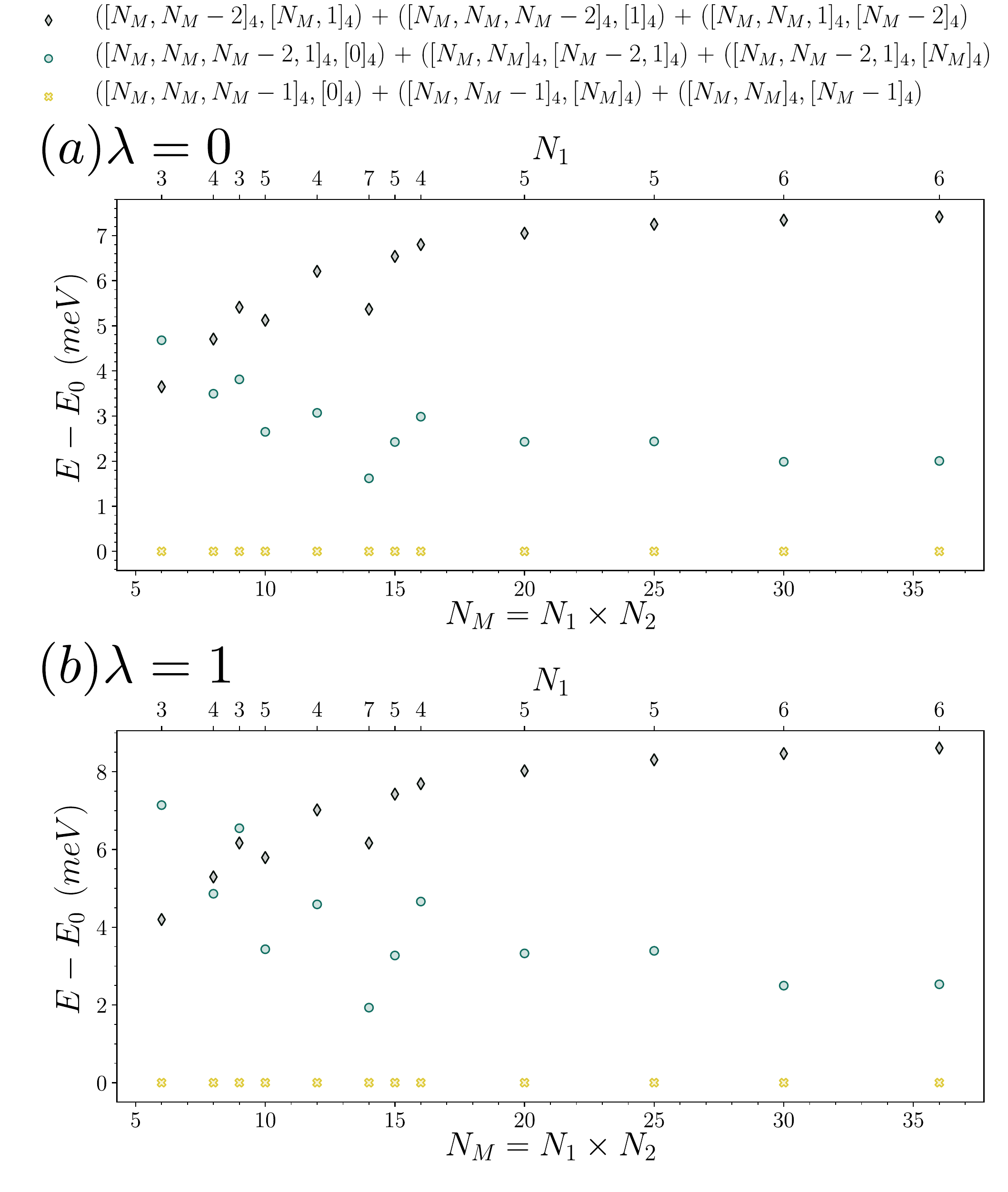}
    \caption{Charge $-1$ (hole) excitations at $\nu = -1$ with $\lambda = 0$ (a) and $\lambda = 1$ (b). All energies have been shifted by the lowest energy $E_0$ at the corresponding system size. The irreps of the lowest charge $-1$ excitations are $([N_M, N_M, N_M -1 ]_4, [0]_4)$, $([N_M, N_M],[ N_M -1 ]_4)$, $([N_M, N_M, N_M -1 ]_4, [0]_4)$ and $([N_M, N_M, - 1]_4, [N_M]_4)$ for both the $\lambda=0$ and $\lambda = 1$ models and for all system sizes we have checked. We use the notation "+" between irreps when they always appear with an exact degeneracy.}
    \label{ed:fig:hole_nu_-1}
\end{figure}

Due to the huge Hilbert space dimensions at filling factor $\nu=-1$ (see App.~\ref{ed:appsec:symhamiltonian} and Table.~\ref{table:nu3} therein), we solely focus on the (first) chiral-flat limit with U(4)$\times$U(4) symmetry. Just like the other integer filling factors, the FMC model has Chern insulator states as exact ground states \cite{ourpaper4}. At $\nu=-1$, the Chern insulating ground states of the FMC model are:
{\small
\begin{align}
    |\Psi_{\nu=-1}^{2,1}\rangle &= \prod_\vk d^\dagger_{\vk, +1, +, \uparrow}d^\dagger_{\vk, -1, +, \uparrow}d^\dagger_{\vk, +1, +, \downarrow}|0\rangle,~\nu_C=1,\label{ed:eqn:chern1nu-1}\\
    |\Psi_{\nu=-1}^{1,2}\rangle &= \prod_\vk d^\dagger_{\vk, +1, +, \uparrow}d^\dagger_{\vk, -1, +, \uparrow}d^\dagger_{\vk, -1, +, \downarrow}|0\rangle,~\nu_C=-1,\label{ed:eqn:chern-1nu-1}\\
    |\Psi_{\nu=-1}^{3,0}\rangle &= \prod_\vk d^\dagger_{\vk, +1, +, \uparrow}d^\dagger_{\vk, +1, +, \downarrow}d^\dagger_{\vk, +1, -, \uparrow}|0\rangle,~\nu_C=3,\label{ed:eqn:chern3nu-1}\\
    |\Psi_{\nu=-1}^{0,3}\rangle &= \prod_\vk d^\dagger_{\vk, -1, +, \uparrow}d^\dagger_{\vk, -1, +, \downarrow}d^\dagger_{\vk, -1, -, \uparrow}|0\rangle,~\nu_C=-3.\label{ed:eqn:chern-3nu-1}
\end{align}}
The above four states belong to the U(4)$\times$U(4) irreps $([N_M, N_M]_4, [N_M]_4)$ ($\nu_C = 1$), $([N_M]_4, [N_M, N_M]_4)$ ($\nu_C=-1$), $([N_M, N_M, N_M]_4, [0]_4)$ ($\nu_C = 3$) and $([0]_4, [N_M, N_M, N_M]_4)$ ($\nu_C = -3$). For the same reasons that we have mentioned in Secs.~\ref{sec:chiralflatbandnu3} and~\ref{sec:chiralflatbandnu2}, these states are the only states which can form these irreps and conserved charges up to U(4)$\times$U(4) transformations, and consequently they must be eigenstates in the (first) chiral-flat limit, but not necessarily the ground states away from the FMC model $\lambda=0$. In this respect, they are similar to the $\nu=-3$ states, which are also not eigenstates away from the chiral limit; they are  unlike the $\nu=-2$ states, which remain eigenstates in the non-chiral limit.

The spectrum for the valley polarized and some slightly depolarized symmetry sectors at this filling factor can be found in Fig.~\ref{ed:fig:spectrum3x2_-1_0}. Here we only consider the full TBG model $\lambda=1$. In the Chern band basis, these Chern insulator states defined in Eqs.~(\ref{ed:eqn:chern1nu-1})-(\ref{ed:eqn:chern-3nu-1}) are in symmetry sectors of dimension one. Therefore we can easily find them by the quantum numbers. The energy spectrum plot shows that these states have the same  energy value, although they carry different Chern numbers. Among the symmetry sectors we have studied in Fig.~\ref{ed:fig:spectrum3x2_-1_0}, these Chern insulator states have the lowest energy, which support the validity of FMC model with non-zero $\lambda$.

Focusing on the irreps close to those of the ground state manifold of states in Eqs.~(\ref{ed:eqn:chern1nu-1})-(\ref{ed:eqn:chern-3nu-1}), 
we can study the energy of charge excitations. The results are displayed in Fig.~\ref{ed:fig:excitations_nu-1}  (for the charge $+1$ excitation) and Fig.~ \ref{ed:fig:hole_nu_-1} (for the charge $-1$ excitation). The charge $+1$ excitation with the lowest energy has the degenerate irreps $([N_M, N_M, N_M, 1]_4, [0]_4)$, $([N_M, N_M, 1]_4, [N_M]_4)$, $([N_M, N_M]_4, [N_M, 1]_4)$, and $([N_M, N_M, N_M]_4, [1]_4)$  for the FMC model, while the model with $\lambda=1$ prefers $([N_M, N_M, N_M - 1,2]_4, [0]_4)$, $([N_M, N_M - 1,2]_4, [N_M]_4)$, and $([N_M, N_M]_4, [N_M - 1,2]_4)$ when the system size gets bigger. On the charge $-1$ excitation side, both the FMC model and $\lambda=1$ model favor the excitation with irreps $([N_M, N_M, N_M -1]_4, [0]_4)$, $([N_M, N_M -1]_4, [N_M]_4)$, and $([N_M, N_M]_4, [N_M -1]_4)$ irrespective of the system sizes we choose. These results are closer to those of $\nu=-3$ (with odd Chern numbers) rather than those of $\nu=-2$ (with even Chern numbers): the difference for the lowest charge $+1$ excitation between the two models might be a more important size effect at $\nu=-1$ than $\nu=-3$ (we can only reach up to $7\times 7$ for $\nu=-1$, while we were able to go up to $8\times 8$ for $\nu=-3$ to have the finite size effect under control). 

Finally, we address the question of the filling factor $\nu=0$. This is by far the most demanding case (see App.~\ref{ed:appsec:symhamiltonian} and Table.~\ref{table:nu4} therein). On the other hand, this is also the filling factor where properties can be derived analytically as discussed in Refs.~\cite{ourpaper4,ourpaper5} even beyond the various limits. For that reason, $\nu=0$ will not be discussed in this article (with the exception of App.~\ref{ed:appsec:phs}).

\section{Conclusion}

We performed an ED study of the phases of first  magic angle TBG with Coulomb interactions at integer fillings. We employ the momentum space interacting Hamiltonian projected into the lowest 8 flat bands (2 per spin and per valley) of the BM continuum model \cite{bistritzer_moire_2011,ourpaper1}, which is shown to have a positive semidefinite interaction Hamiltonian (analogous to that found by Kang and Vafek \cite{kang_strong_2019}) and is explicitly gauge fixed in Ref.~\cite{ourpaper3}. For integer fillings $\nu=-3,-2,-1$ (relative to the CNP), we explore the ground states and excitations in the parameter space of $w_0/w_1\in[0,1]$ (the ratio between $AA$ and $AB$ stacking hoppings), single-particle bandwidth $t\in[0,1]$ (dimensionless, $t=1$ corresponds to the bandwidth of the BM model), and a parameter $\lambda\in[0,1]$ which interpolates the Hamiltonian between having the FMC Eq.~(\ref{ed:eq:FMC}) ($\lambda=0$) and realistic parameters without the FMC ($\lambda=1$). As shown in Ref.~\cite{ourpaper4}, the FMC is a weak condition that allows us to analytically find exact ground states (but potentially not all) at integer fillings $\nu$. In particular, for any $\lambda$, the Hamiltonian enjoys a U(4)$\times$U(4) symmetry in the first chiral-flat limit ($w_0=0$, $t=0$), and have a reduced U(4) symmetry in either the nonchiral-flat limit ($w_0>0,t=0$) or the chiral-nonflat limit ($w_0=0,t>0$) (which are different U(4)'s), as revealed in Refs.~\cite{bultinck_ground_2020,ourpaper3,kang_strong_2019,seo_ferro_2019}. We therefore also study the U(4)$\times$U(4) or U(4) irreps of the ground states and excitations in these limits. The symmetry of the Hamiltonian reduces into U(2)$\times$U(2) in the physical chiral-nonflat case.

For $\nu=-3$, our calculations show the ground state is uniquely the spin and valley polarized Chern insulator with $\nu_C=\pm1$ when $w_0/w_1\lesssim0.9$ with the FMC ($\lambda=0$), and when $w_0/w_1\lesssim0.3$ without the FMC ($\lambda=1$). The phase has almost no dependence on the bandwidth $t\in[0,1]$. This conclusion is independent of the system size (up to the maximal size $5\times3$), and is in agreement with our conclusion in \cite{ourpaper4} from analytical perturbation calculations. In the chiral-flat limit, such a Chern insulator with Chern number $\nu=\pm1$ becomes an analytical exact ground state \cite{ourpaper4}. By restricting to sub-Hilbert spaces close to the ground state, we numerically verified that the exactly solvable charge $\pm1$ excitations found in Ref.~\cite{ourpaper5} are the lowest charge excitations up to a system size $8\times8$ in the chiral-flat limit, with or without the FMC. When $w_0/w_1\gtrsim0.9$ with the FMC ($\lambda=0$) or when $w_0/w_1\gtrsim 0.3$ without the FMC ($\lambda=1$), the finite-size gap $\Delta$ to the charge neutral excitations vanishes (due to either a vanishing Goldstone stiffness or a softening of other neutral excited states), which leads us to conjecture a phase transition into metallic or translation breaking phases in these parameter ranges. This qualitatively agrees with the recent DMRG studies for $\nu=-3$ \cite{bultinck_ground_2020,kang_nonabelian_2020}, which found a transition from Chern insulator to nematic semimetal or stripe phase near $w_0/w_1=0.8$. Our further analysis of the ground state momentum sectors suggests a competition between among (nematic) metal, $M_M$ ($\pi$ momentum) stripe and $K_M$-CDW orders in the large $w_0/w_1$ regime.

We also examined the phase diagram at $\nu=-2$ in the fully valley polarized sector with all electrons in one valley, or the fully spin polarized sector with all electrons in spin up, in a $3\times2$ momentum lattice. We find the following results when the FMC holds ($\lambda=0$), or when the FMC is absent ($\lambda=1$) and $w_0/w_1\lesssim 0.6$: (1) in the fully valley polarized sector, we find a spin ferromagnetic phase when $w_0/w_1\gtrsim 0.5 t$, and a spin singlet phase when $w_0/w_1\lesssim 0.5 t$, both of which have Chern number $0$. (2) In the fully spin polarized sector, we find the intervalley coherent state is always favored, which is always lower in energy than the ground state in the fully valley polarized sector. This agrees with the exact and perturbation analysis in Ref.~\cite{ourpaper4} (see a similar analysis without FMC in Ref.~\cite{bultinck_ground_2020}), where it is shown that with the FMC, the nonchiral-flat limit has a U(4) ferromagnetic exact ground state, while the chiral-nonflat limit prefers half-occupying different spin-valley flavors (up to further U(4) rotations), which together favors an intervalley coherent ground state in the nonchiral-nonflat case. Importantly, while other ground states cannot be excluded in Ref.~\cite{ourpaper4} at filling $\nu=-2$ (with the FMC), we showed here the Chern number $0$ state at $\nu=-2$ is the unique ground state in the chiral-nonflat and nonchiral-flat limits. When the FMC is absent ($\lambda=1$), we find the ground state changes for $w_0/w_1\gtrsim 0.6$, which indicates a possible phase transition (into metallic phases, etc). Moreover, in the chiral-flat limit, we show that the exact charge $\pm1$ excitations found in Ref.~\cite{ourpaper5} are the lowest charge excitations at $\nu=-2$ with or without the FMC (in restricted Hilbert spaces up to system size $6\times6$). Lastly, we note that it is shown by perturbation theory \cite{ourpaper4,bultinck_ground_2020} that $\nu=-2$ may favor an intervalley coherent ground state with valley polarization $N_v=0$. The investigation of such a state is, however, beyond our computational ability due to the enormous Hilbert space dimension needed, and we leave it to future studies.

The last filling we explored is $\nu=-1$, where we are limited to the study of the chiral-flat limit (where a U(4)$\times$U(4) symmetry emerge) in nearly valley polarized sectors due to limitation of Hilbert space dimensions. While the Chern number $\nu_C=\pm1,\pm3$ insulators are proved to be ground states at $\nu=-1$ with FMC in Ref.~\cite{ourpaper4} but not necessarily the only ground states, our numerical result does not find any other states which have lower energy than these Chern insulator states in symmetry sectors whose dimension is not larger than $10^6$, and therefore the Chern number $\nu_C = \pm1, \pm3$ states are likely to be the only ground states. Furthermore, we show that the exact charge excitations given in Ref.~\cite{ourpaper5} are the lowest charge excitations at $\nu=-1$ except for charge $+1$ excitations without FMC (in restricted Hilbert spaces of up to system size $7\times 7$).

Our work verified the validity of the exact/perturbative ground states and charge excitations at nonzero integer fillings in our earlier studies \cite{ourpaper4,ourpaper5}, and has proved the utility of enhanced U(4) and U(4)$\times$U(4) symmetries in various limits \cite{bultinck_ground_2020,ourpaper3,kang_strong_2019,seo_ferro_2019} useful for identifying the phases in magic angle TBG. Beyond the regime where our analytic states are ground states, our work further suggests the possible existence of $C_{3z}$ and/or translation breaking new phases at large $w_0/w_1$, which we will investigate in the future.

\begin{acknowledgments}
We thank Michael Zaletel, Allan MacDonald, Christophe Mora and Oskar Vafek for fruitful discussions. This work was supported by the DOE Grant No. DE-SC0016239, the Schmidt Fund for Innovative Research, Simons Investigator Grant No. 404513, and the Packard Foundation. Further support was provided by the NSF-EAGER No. DMR 1643312, NSF-MRSEC No. DMR-1420541 and DMR-2011750, ONR No. N00014-20-1-2303, Gordon and Betty Moore Foundation through Grant GBMF8685 towards the Princeton theory program, BSF Israel US foundation No. 2018226, and the Princeton Global Network Funds. B.L. acknowledge the support of Princeton Center for Theoretical Science at Princeton University in the early stage of this work. N.R. was also supported by Grant No. ANR-16-CE30-0025.
\end{acknowledgments}

\bibliography{TBLGHexalogy.bib,HexalogyInternalRefs.bib}

\begin{thebibliography}{115}%
\makeatletter
\providecommand \@ifxundefined [1]{%
 \@ifx{#1\undefined}
}%
\providecommand \@ifnum [1]{%
 \ifnum #1\expandafter \@firstoftwo
 \else \expandafter \@secondoftwo
 \fi
}%
\providecommand \@ifx [1]{%
 \ifx #1\expandafter \@firstoftwo
 \else \expandafter \@secondoftwo
 \fi
}%
\providecommand \natexlab [1]{#1}%
\providecommand \enquote  [1]{``#1''}%
\providecommand \bibnamefont  [1]{#1}%
\providecommand \bibfnamefont [1]{#1}%
\providecommand \citenamefont [1]{#1}%
\providecommand \href@noop [0]{\@secondoftwo}%
\providecommand \href [0]{\begingroup \@sanitize@url \@href}%
\providecommand \@href[1]{\@@startlink{#1}\@@href}%
\providecommand \@@href[1]{\endgroup#1\@@endlink}%
\providecommand \@sanitize@url [0]{\catcode `\\12\catcode `\$12\catcode
  `\&12\catcode `\#12\catcode `\^12\catcode `\_12\catcode `\%12\relax}%
\providecommand \@@startlink[1]{}%
\providecommand \@@endlink[0]{}%
\providecommand \url  [0]{\begingroup\@sanitize@url \@url }%
\providecommand \@url [1]{\endgroup\@href {#1}{\urlprefix }}%
\providecommand \urlprefix  [0]{URL }%
\providecommand \Eprint [0]{\href }%
\providecommand \doibase [0]{https://doi.org/}%
\providecommand \selectlanguage [0]{\@gobble}%
\providecommand \bibinfo  [0]{\@secondoftwo}%
\providecommand \bibfield  [0]{\@secondoftwo}%
\providecommand \translation [1]{[#1]}%
\providecommand \BibitemOpen [0]{}%
\providecommand \bibitemStop [0]{}%
\providecommand \bibitemNoStop [0]{.\EOS\space}%
\providecommand \EOS [0]{\spacefactor3000\relax}%
\providecommand \BibitemShut  [1]{\csname bibitem#1\endcsname}%
\let\auto@bib@innerbib\@empty
\bibitem [{\citenamefont {Bistritzer}\ and\ \citenamefont
  {MacDonald}(2011)}]{bistritzer_moire_2011}%
  \BibitemOpen
  \bibfield  {author} {\bibinfo {author} {\bibfnamefont {R.}~\bibnamefont
  {Bistritzer}}\ and\ \bibinfo {author} {\bibfnamefont {A.~H.}\ \bibnamefont
  {MacDonald}},\ }\href {https://doi.org/10.1073/pnas.1108174108} {\bibfield
  {journal} {\bibinfo  {journal} {Proceedings of the National Academy of
  Sciences}\ }\textbf {\bibinfo {volume} {108}},\ \bibinfo {pages} {12233}
  (\bibinfo {year} {2011})}\BibitemShut {NoStop}%
\bibitem [{\citenamefont {Cao}\ \emph {et~al.}(2018{\natexlab{a}})\citenamefont
  {Cao}, \citenamefont {Fatemi}, \citenamefont {Demir}, \citenamefont {Fang},
  \citenamefont {Tomarken}, \citenamefont {Luo}, \citenamefont
  {Sanchez-Yamagishi}, \citenamefont {Watanabe}, \citenamefont {Taniguchi},
  \citenamefont {Kaxiras}, \citenamefont {Ashoori},\ and\ \citenamefont
  {Jarillo-Herrero}}]{cao_correlated_2018}%
  \BibitemOpen
  \bibfield  {author} {\bibinfo {author} {\bibfnamefont {Y.}~\bibnamefont
  {Cao}}, \bibinfo {author} {\bibfnamefont {V.}~\bibnamefont {Fatemi}},
  \bibinfo {author} {\bibfnamefont {A.}~\bibnamefont {Demir}}, \bibinfo
  {author} {\bibfnamefont {S.}~\bibnamefont {Fang}}, \bibinfo {author}
  {\bibfnamefont {S.~L.}\ \bibnamefont {Tomarken}}, \bibinfo {author}
  {\bibfnamefont {J.~Y.}\ \bibnamefont {Luo}}, \bibinfo {author} {\bibfnamefont
  {J.~D.}\ \bibnamefont {Sanchez-Yamagishi}}, \bibinfo {author} {\bibfnamefont
  {K.}~\bibnamefont {Watanabe}}, \bibinfo {author} {\bibfnamefont
  {T.}~\bibnamefont {Taniguchi}}, \bibinfo {author} {\bibfnamefont
  {E.}~\bibnamefont {Kaxiras}}, \bibinfo {author} {\bibfnamefont {R.~C.}\
  \bibnamefont {Ashoori}},\ and\ \bibinfo {author} {\bibfnamefont
  {P.}~\bibnamefont {Jarillo-Herrero}},\ }\href
  {https://doi.org/10.1038/nature26154} {\bibfield  {journal} {\bibinfo
  {journal} {Nature}\ }\textbf {\bibinfo {volume} {556}},\ \bibinfo {pages}
  {80} (\bibinfo {year} {2018}{\natexlab{a}})}\BibitemShut {NoStop}%
\bibitem [{\citenamefont {Cao}\ \emph {et~al.}(2018{\natexlab{b}})\citenamefont
  {Cao}, \citenamefont {Fatemi}, \citenamefont {Fang}, \citenamefont
  {Watanabe}, \citenamefont {Taniguchi}, \citenamefont {Kaxiras},\ and\
  \citenamefont {Jarillo-Herrero}}]{cao_unconventional_2018}%
  \BibitemOpen
  \bibfield  {author} {\bibinfo {author} {\bibfnamefont {Y.}~\bibnamefont
  {Cao}}, \bibinfo {author} {\bibfnamefont {V.}~\bibnamefont {Fatemi}},
  \bibinfo {author} {\bibfnamefont {S.}~\bibnamefont {Fang}}, \bibinfo {author}
  {\bibfnamefont {K.}~\bibnamefont {Watanabe}}, \bibinfo {author}
  {\bibfnamefont {T.}~\bibnamefont {Taniguchi}}, \bibinfo {author}
  {\bibfnamefont {E.}~\bibnamefont {Kaxiras}},\ and\ \bibinfo {author}
  {\bibfnamefont {P.}~\bibnamefont {Jarillo-Herrero}},\ }\href
  {https://doi.org/10.1038/nature26160} {\bibfield  {journal} {\bibinfo
  {journal} {Nature}\ }\textbf {\bibinfo {volume} {556}},\ \bibinfo {pages}
  {43} (\bibinfo {year} {2018}{\natexlab{b}})}\BibitemShut {NoStop}%
\bibitem [{\citenamefont {Lu}\ \emph {et~al.}(2019)\citenamefont {Lu},
  \citenamefont {Stepanov}, \citenamefont {Yang}, \citenamefont {Xie},
  \citenamefont {Aamir}, \citenamefont {Das}, \citenamefont {Urgell},
  \citenamefont {Watanabe}, \citenamefont {Taniguchi}, \citenamefont {Zhang}
  \emph {et~al.}}]{lu2019superconductors}%
  \BibitemOpen
  \bibfield  {author} {\bibinfo {author} {\bibfnamefont {X.}~\bibnamefont
  {Lu}}, \bibinfo {author} {\bibfnamefont {P.}~\bibnamefont {Stepanov}},
  \bibinfo {author} {\bibfnamefont {W.}~\bibnamefont {Yang}}, \bibinfo {author}
  {\bibfnamefont {M.}~\bibnamefont {Xie}}, \bibinfo {author} {\bibfnamefont
  {M.~A.}\ \bibnamefont {Aamir}}, \bibinfo {author} {\bibfnamefont
  {I.}~\bibnamefont {Das}}, \bibinfo {author} {\bibfnamefont {C.}~\bibnamefont
  {Urgell}}, \bibinfo {author} {\bibfnamefont {K.}~\bibnamefont {Watanabe}},
  \bibinfo {author} {\bibfnamefont {T.}~\bibnamefont {Taniguchi}}, \bibinfo
  {author} {\bibfnamefont {G.}~\bibnamefont {Zhang}}, \emph {et~al.},\ }\href
  {https://www.nature.com/articles/s41586-019-1695-0} {\bibfield  {journal}
  {\bibinfo  {journal} {Nature}\ }\textbf {\bibinfo {volume} {574}},\ \bibinfo
  {pages} {653} (\bibinfo {year} {2019})}\BibitemShut {NoStop}%
\bibitem [{\citenamefont {Yankowitz}\ \emph {et~al.}(2019)\citenamefont
  {Yankowitz}, \citenamefont {Chen}, \citenamefont {Polshyn}, \citenamefont
  {Zhang}, \citenamefont {Watanabe}, \citenamefont {Taniguchi}, \citenamefont
  {Graf}, \citenamefont {Young},\ and\ \citenamefont
  {Dean}}]{yankowitz2019tuning}%
  \BibitemOpen
  \bibfield  {author} {\bibinfo {author} {\bibfnamefont {M.}~\bibnamefont
  {Yankowitz}}, \bibinfo {author} {\bibfnamefont {S.}~\bibnamefont {Chen}},
  \bibinfo {author} {\bibfnamefont {H.}~\bibnamefont {Polshyn}}, \bibinfo
  {author} {\bibfnamefont {Y.}~\bibnamefont {Zhang}}, \bibinfo {author}
  {\bibfnamefont {K.}~\bibnamefont {Watanabe}}, \bibinfo {author}
  {\bibfnamefont {T.}~\bibnamefont {Taniguchi}}, \bibinfo {author}
  {\bibfnamefont {D.}~\bibnamefont {Graf}}, \bibinfo {author} {\bibfnamefont
  {A.~F.}\ \bibnamefont {Young}},\ and\ \bibinfo {author} {\bibfnamefont
  {C.~R.}\ \bibnamefont {Dean}},\ }\href
  {https://science.sciencemag.org/content/363/6431/1059.abstract?casa_token=lunm_WUYs2YAAAAA:VYMQ9xKAP9yNieasreqWu0I0g8sN82wxfevMxLMfsegLO9RZtKOt45kmqcsGZAKERIiy2VDY21ejfWs}
  {\bibfield  {journal} {\bibinfo  {journal} {Science}\ }\textbf {\bibinfo
  {volume} {363}},\ \bibinfo {pages} {1059} (\bibinfo {year}
  {2019})}\BibitemShut {NoStop}%
\bibitem [{\citenamefont {Sharpe}\ \emph {et~al.}(2019)\citenamefont {Sharpe},
  \citenamefont {Fox}, \citenamefont {Barnard}, \citenamefont {Finney},
  \citenamefont {Watanabe}, \citenamefont {Taniguchi}, \citenamefont
  {Kastner},\ and\ \citenamefont {Goldhaber-Gordon}}]{sharpe_emergent_2019}%
  \BibitemOpen
  \bibfield  {author} {\bibinfo {author} {\bibfnamefont {A.~L.}\ \bibnamefont
  {Sharpe}}, \bibinfo {author} {\bibfnamefont {E.~J.}\ \bibnamefont {Fox}},
  \bibinfo {author} {\bibfnamefont {A.~W.}\ \bibnamefont {Barnard}}, \bibinfo
  {author} {\bibfnamefont {J.}~\bibnamefont {Finney}}, \bibinfo {author}
  {\bibfnamefont {K.}~\bibnamefont {Watanabe}}, \bibinfo {author}
  {\bibfnamefont {T.}~\bibnamefont {Taniguchi}}, \bibinfo {author}
  {\bibfnamefont {M.~A.}\ \bibnamefont {Kastner}},\ and\ \bibinfo {author}
  {\bibfnamefont {D.}~\bibnamefont {Goldhaber-Gordon}},\ }\href
  {https://doi.org/10.1126/science.aaw3780} {\bibfield  {journal} {\bibinfo
  {journal} {Science}\ }\textbf {\bibinfo {volume} {365}},\ \bibinfo {pages}
  {605–608} (\bibinfo {year} {2019})}\BibitemShut {NoStop}%
\bibitem [{\citenamefont {Saito}\ \emph {et~al.}(2020)\citenamefont {Saito},
  \citenamefont {Ge}, \citenamefont {Watanabe}, \citenamefont {Taniguchi},\
  and\ \citenamefont {Young}}]{saito_independent_2020}%
  \BibitemOpen
  \bibfield  {author} {\bibinfo {author} {\bibfnamefont {Y.}~\bibnamefont
  {Saito}}, \bibinfo {author} {\bibfnamefont {J.}~\bibnamefont {Ge}}, \bibinfo
  {author} {\bibfnamefont {K.}~\bibnamefont {Watanabe}}, \bibinfo {author}
  {\bibfnamefont {T.}~\bibnamefont {Taniguchi}},\ and\ \bibinfo {author}
  {\bibfnamefont {A.~F.}\ \bibnamefont {Young}},\ }\href
  {https://doi.org/10.1038/s41567-020-0928-3} {\bibfield  {journal} {\bibinfo
  {journal} {Nature Physics}\ }\textbf {\bibinfo {volume} {16}},\ \bibinfo
  {pages} {926–930} (\bibinfo {year} {2020})}\BibitemShut {NoStop}%
\bibitem [{\citenamefont {Stepanov}\ \emph {et~al.}(2020)\citenamefont
  {Stepanov}, \citenamefont {Das}, \citenamefont {Lu}, \citenamefont
  {Fahimniya}, \citenamefont {Watanabe}, \citenamefont {Taniguchi},
  \citenamefont {Koppens}, \citenamefont {Lischner}, \citenamefont {Levitov},\
  and\ \citenamefont {Efetov}}]{stepanov_interplay_2020}%
  \BibitemOpen
  \bibfield  {author} {\bibinfo {author} {\bibfnamefont {P.}~\bibnamefont
  {Stepanov}}, \bibinfo {author} {\bibfnamefont {I.}~\bibnamefont {Das}},
  \bibinfo {author} {\bibfnamefont {X.}~\bibnamefont {Lu}}, \bibinfo {author}
  {\bibfnamefont {A.}~\bibnamefont {Fahimniya}}, \bibinfo {author}
  {\bibfnamefont {K.}~\bibnamefont {Watanabe}}, \bibinfo {author}
  {\bibfnamefont {T.}~\bibnamefont {Taniguchi}}, \bibinfo {author}
  {\bibfnamefont {F.~H.~L.}\ \bibnamefont {Koppens}}, \bibinfo {author}
  {\bibfnamefont {J.}~\bibnamefont {Lischner}}, \bibinfo {author}
  {\bibfnamefont {L.}~\bibnamefont {Levitov}},\ and\ \bibinfo {author}
  {\bibfnamefont {D.~K.}\ \bibnamefont {Efetov}},\ }\href
  {https://doi.org/10.1038/s41586-020-2459-6} {\bibfield  {journal} {\bibinfo
  {journal} {Nature}\ }\textbf {\bibinfo {volume} {583}},\ \bibinfo {pages}
  {375–378} (\bibinfo {year} {2020})}\BibitemShut {NoStop}%
\bibitem [{\citenamefont {Liu}\ \emph {et~al.}(2021{\natexlab{a}})\citenamefont
  {Liu}, \citenamefont {Wang}, \citenamefont {Watanabe}, \citenamefont
  {Taniguchi}, \citenamefont {Vafek},\ and\ \citenamefont
  {Li}}]{liu2020tuning}%
  \BibitemOpen
  \bibfield  {author} {\bibinfo {author} {\bibfnamefont {X.}~\bibnamefont
  {Liu}}, \bibinfo {author} {\bibfnamefont {Z.}~\bibnamefont {Wang}}, \bibinfo
  {author} {\bibfnamefont {K.}~\bibnamefont {Watanabe}}, \bibinfo {author}
  {\bibfnamefont {T.}~\bibnamefont {Taniguchi}}, \bibinfo {author}
  {\bibfnamefont {O.}~\bibnamefont {Vafek}},\ and\ \bibinfo {author}
  {\bibfnamefont {J.~I.~A.}\ \bibnamefont {Li}},\ }\href
  {https://doi.org/10.1126/science.abb8754} {\bibfield  {journal} {\bibinfo
  {journal} {Science}\ }\textbf {\bibinfo {volume} {371}},\ \bibinfo {pages}
  {1261} (\bibinfo {year} {2021}{\natexlab{a}})},\ \Eprint
  {https://arxiv.org/abs/https://www.science.org/doi/pdf/10.1126/science.abb8754}
  {https://www.science.org/doi/pdf/10.1126/science.abb8754} \BibitemShut
  {NoStop}%
\bibitem [{\citenamefont {Arora}\ \emph {et~al.}(2020)\citenamefont {Arora},
  \citenamefont {Polski}, \citenamefont {Zhang}, \citenamefont {Thomson},
  \citenamefont {Choi}, \citenamefont {Kim}, \citenamefont {Lin}, \citenamefont
  {Wilson}, \citenamefont {Xu}, \citenamefont {Chu},\ and\ \citenamefont
  {et~al.}}]{arora_2020}%
  \BibitemOpen
  \bibfield  {author} {\bibinfo {author} {\bibfnamefont {H.~S.}\ \bibnamefont
  {Arora}}, \bibinfo {author} {\bibfnamefont {R.}~\bibnamefont {Polski}},
  \bibinfo {author} {\bibfnamefont {Y.}~\bibnamefont {Zhang}}, \bibinfo
  {author} {\bibfnamefont {A.}~\bibnamefont {Thomson}}, \bibinfo {author}
  {\bibfnamefont {Y.}~\bibnamefont {Choi}}, \bibinfo {author} {\bibfnamefont
  {H.}~\bibnamefont {Kim}}, \bibinfo {author} {\bibfnamefont {Z.}~\bibnamefont
  {Lin}}, \bibinfo {author} {\bibfnamefont {I.~Z.}\ \bibnamefont {Wilson}},
  \bibinfo {author} {\bibfnamefont {X.}~\bibnamefont {Xu}}, \bibinfo {author}
  {\bibfnamefont {J.-H.}\ \bibnamefont {Chu}},\ and\ \bibinfo {author}
  {\bibnamefont {et~al.}},\ }\href {https://doi.org/10.1038/s41586-020-2473-8}
  {\bibfield  {journal} {\bibinfo  {journal} {Nature}\ }\textbf {\bibinfo
  {volume} {583}},\ \bibinfo {pages} {379–384} (\bibinfo {year}
  {2020})}\BibitemShut {NoStop}%
\bibitem [{\citenamefont {Serlin}\ \emph {et~al.}(2019)\citenamefont {Serlin},
  \citenamefont {Tschirhart}, \citenamefont {Polshyn}, \citenamefont {Zhang},
  \citenamefont {Zhu}, \citenamefont {Watanabe}, \citenamefont {Taniguchi},
  \citenamefont {Balents},\ and\ \citenamefont {Young}}]{serlin_QAH_2019}%
  \BibitemOpen
  \bibfield  {author} {\bibinfo {author} {\bibfnamefont {M.}~\bibnamefont
  {Serlin}}, \bibinfo {author} {\bibfnamefont {C.~L.}\ \bibnamefont
  {Tschirhart}}, \bibinfo {author} {\bibfnamefont {H.}~\bibnamefont {Polshyn}},
  \bibinfo {author} {\bibfnamefont {Y.}~\bibnamefont {Zhang}}, \bibinfo
  {author} {\bibfnamefont {J.}~\bibnamefont {Zhu}}, \bibinfo {author}
  {\bibfnamefont {K.}~\bibnamefont {Watanabe}}, \bibinfo {author}
  {\bibfnamefont {T.}~\bibnamefont {Taniguchi}}, \bibinfo {author}
  {\bibfnamefont {L.}~\bibnamefont {Balents}},\ and\ \bibinfo {author}
  {\bibfnamefont {A.~F.}\ \bibnamefont {Young}},\ }\href
  {https://doi.org/10.1126/science.aay5533} {\bibfield  {journal} {\bibinfo
  {journal} {Science}\ }\textbf {\bibinfo {volume} {367}},\ \bibinfo {pages}
  {900–903} (\bibinfo {year} {2019})}\BibitemShut {NoStop}%
\bibitem [{\citenamefont {Cao}\ \emph {et~al.}(2020{\natexlab{a}})\citenamefont
  {Cao}, \citenamefont {Chowdhury}, \citenamefont {Rodan-Legrain},
  \citenamefont {Rubies-Bigorda}, \citenamefont {Watanabe}, \citenamefont
  {Taniguchi}, \citenamefont {Senthil},\ and\ \citenamefont
  {Jarillo-Herrero}}]{cao_strange_2020}%
  \BibitemOpen
  \bibfield  {author} {\bibinfo {author} {\bibfnamefont {Y.}~\bibnamefont
  {Cao}}, \bibinfo {author} {\bibfnamefont {D.}~\bibnamefont {Chowdhury}},
  \bibinfo {author} {\bibfnamefont {D.}~\bibnamefont {Rodan-Legrain}}, \bibinfo
  {author} {\bibfnamefont {O.}~\bibnamefont {Rubies-Bigorda}}, \bibinfo
  {author} {\bibfnamefont {K.}~\bibnamefont {Watanabe}}, \bibinfo {author}
  {\bibfnamefont {T.}~\bibnamefont {Taniguchi}}, \bibinfo {author}
  {\bibfnamefont {T.}~\bibnamefont {Senthil}},\ and\ \bibinfo {author}
  {\bibfnamefont {P.}~\bibnamefont {Jarillo-Herrero}},\ }\href
  {https://doi.org/10.1103/PhysRevLett.124.076801} {\bibfield  {journal}
  {\bibinfo  {journal} {Phys. Rev. Lett.}\ }\textbf {\bibinfo {volume} {124}},\
  \bibinfo {pages} {076801} (\bibinfo {year} {2020}{\natexlab{a}})}\BibitemShut
  {NoStop}%
\bibitem [{\citenamefont {Polshyn}\ \emph {et~al.}(2019)\citenamefont
  {Polshyn}, \citenamefont {Yankowitz}, \citenamefont {Chen}, \citenamefont
  {Zhang}, \citenamefont {Watanabe}, \citenamefont {Taniguchi}, \citenamefont
  {Dean},\ and\ \citenamefont {Young}}]{polshyn_linear_2019}%
  \BibitemOpen
  \bibfield  {author} {\bibinfo {author} {\bibfnamefont {H.}~\bibnamefont
  {Polshyn}}, \bibinfo {author} {\bibfnamefont {M.}~\bibnamefont {Yankowitz}},
  \bibinfo {author} {\bibfnamefont {S.}~\bibnamefont {Chen}}, \bibinfo {author}
  {\bibfnamefont {Y.}~\bibnamefont {Zhang}}, \bibinfo {author} {\bibfnamefont
  {K.}~\bibnamefont {Watanabe}}, \bibinfo {author} {\bibfnamefont
  {T.}~\bibnamefont {Taniguchi}}, \bibinfo {author} {\bibfnamefont {C.~R.}\
  \bibnamefont {Dean}},\ and\ \bibinfo {author} {\bibfnamefont {A.~F.}\
  \bibnamefont {Young}},\ }\href {https://doi.org/10.1038/s41567-019-0596-3}
  {\bibfield  {journal} {\bibinfo  {journal} {Nature Physics}\ }\textbf
  {\bibinfo {volume} {15}},\ \bibinfo {pages} {1011–1016} (\bibinfo {year}
  {2019})}\BibitemShut {NoStop}%
\bibitem [{\citenamefont {Xie}\ \emph {et~al.}(2019)\citenamefont {Xie},
  \citenamefont {Lian}, \citenamefont {J{\"a}ck}, \citenamefont {Liu},
  \citenamefont {Chiu}, \citenamefont {Watanabe}, \citenamefont {Taniguchi},
  \citenamefont {Bernevig},\ and\ \citenamefont
  {Yazdani}}]{xie2019spectroscopic}%
  \BibitemOpen
  \bibfield  {author} {\bibinfo {author} {\bibfnamefont {Y.}~\bibnamefont
  {Xie}}, \bibinfo {author} {\bibfnamefont {B.}~\bibnamefont {Lian}}, \bibinfo
  {author} {\bibfnamefont {B.}~\bibnamefont {J{\"a}ck}}, \bibinfo {author}
  {\bibfnamefont {X.}~\bibnamefont {Liu}}, \bibinfo {author} {\bibfnamefont
  {C.-L.}\ \bibnamefont {Chiu}}, \bibinfo {author} {\bibfnamefont
  {K.}~\bibnamefont {Watanabe}}, \bibinfo {author} {\bibfnamefont
  {T.}~\bibnamefont {Taniguchi}}, \bibinfo {author} {\bibfnamefont {B.~A.}\
  \bibnamefont {Bernevig}},\ and\ \bibinfo {author} {\bibfnamefont
  {A.}~\bibnamefont {Yazdani}},\ }\href
  {https://www.nature.com/articles/s41586-019-1422-x} {\bibfield  {journal}
  {\bibinfo  {journal} {Nature}\ }\textbf {\bibinfo {volume} {572}},\ \bibinfo
  {pages} {101} (\bibinfo {year} {2019})}\BibitemShut {NoStop}%
\bibitem [{\citenamefont {Choi}\ \emph {et~al.}(2019)\citenamefont {Choi},
  \citenamefont {Kemmer}, \citenamefont {Peng}, \citenamefont {Thomson},
  \citenamefont {Arora}, \citenamefont {Polski}, \citenamefont {Zhang},
  \citenamefont {Ren}, \citenamefont {Alicea}, \citenamefont {Refael},\ and\
  \citenamefont {et~al.}}]{choi_imaging_2019}%
  \BibitemOpen
  \bibfield  {author} {\bibinfo {author} {\bibfnamefont {Y.}~\bibnamefont
  {Choi}}, \bibinfo {author} {\bibfnamefont {J.}~\bibnamefont {Kemmer}},
  \bibinfo {author} {\bibfnamefont {Y.}~\bibnamefont {Peng}}, \bibinfo {author}
  {\bibfnamefont {A.}~\bibnamefont {Thomson}}, \bibinfo {author} {\bibfnamefont
  {H.}~\bibnamefont {Arora}}, \bibinfo {author} {\bibfnamefont
  {R.}~\bibnamefont {Polski}}, \bibinfo {author} {\bibfnamefont
  {Y.}~\bibnamefont {Zhang}}, \bibinfo {author} {\bibfnamefont
  {H.}~\bibnamefont {Ren}}, \bibinfo {author} {\bibfnamefont {J.}~\bibnamefont
  {Alicea}}, \bibinfo {author} {\bibfnamefont {G.}~\bibnamefont {Refael}},\
  and\ \bibinfo {author} {\bibnamefont {et~al.}},\ }\href
  {https://doi.org/10.1038/s41567-019-0606-5} {\bibfield  {journal} {\bibinfo
  {journal} {Nature Physics}\ }\textbf {\bibinfo {volume} {15}},\ \bibinfo
  {pages} {1174–1180} (\bibinfo {year} {2019})}\BibitemShut {NoStop}%
\bibitem [{\citenamefont {Kerelsky}\ \emph {et~al.}(2019)\citenamefont
  {Kerelsky}, \citenamefont {McGilly}, \citenamefont {Kennes}, \citenamefont
  {Xian}, \citenamefont {Yankowitz}, \citenamefont {Chen}, \citenamefont
  {Watanabe}, \citenamefont {Taniguchi}, \citenamefont {Hone}, \citenamefont
  {Dean},\ and\ \citenamefont {et~al.}}]{kerelsky_2019_stm}%
  \BibitemOpen
  \bibfield  {author} {\bibinfo {author} {\bibfnamefont {A.}~\bibnamefont
  {Kerelsky}}, \bibinfo {author} {\bibfnamefont {L.~J.}\ \bibnamefont
  {McGilly}}, \bibinfo {author} {\bibfnamefont {D.~M.}\ \bibnamefont {Kennes}},
  \bibinfo {author} {\bibfnamefont {L.}~\bibnamefont {Xian}}, \bibinfo {author}
  {\bibfnamefont {M.}~\bibnamefont {Yankowitz}}, \bibinfo {author}
  {\bibfnamefont {S.}~\bibnamefont {Chen}}, \bibinfo {author} {\bibfnamefont
  {K.}~\bibnamefont {Watanabe}}, \bibinfo {author} {\bibfnamefont
  {T.}~\bibnamefont {Taniguchi}}, \bibinfo {author} {\bibfnamefont
  {J.}~\bibnamefont {Hone}}, \bibinfo {author} {\bibfnamefont {C.}~\bibnamefont
  {Dean}},\ and\ \bibinfo {author} {\bibnamefont {et~al.}},\ }\href
  {https://doi.org/10.1038/s41586-019-1431-9} {\bibfield  {journal} {\bibinfo
  {journal} {Nature}\ }\textbf {\bibinfo {volume} {572}},\ \bibinfo {pages}
  {95–100} (\bibinfo {year} {2019})}\BibitemShut {NoStop}%
\bibitem [{\citenamefont {Jiang}\ \emph {et~al.}(2019)\citenamefont {Jiang},
  \citenamefont {Lai}, \citenamefont {Watanabe}, \citenamefont {Taniguchi},
  \citenamefont {Haule}, \citenamefont {Mao},\ and\ \citenamefont
  {Andrei}}]{jiang_charge_2019}%
  \BibitemOpen
  \bibfield  {author} {\bibinfo {author} {\bibfnamefont {Y.}~\bibnamefont
  {Jiang}}, \bibinfo {author} {\bibfnamefont {X.}~\bibnamefont {Lai}}, \bibinfo
  {author} {\bibfnamefont {K.}~\bibnamefont {Watanabe}}, \bibinfo {author}
  {\bibfnamefont {T.}~\bibnamefont {Taniguchi}}, \bibinfo {author}
  {\bibfnamefont {K.}~\bibnamefont {Haule}}, \bibinfo {author} {\bibfnamefont
  {J.}~\bibnamefont {Mao}},\ and\ \bibinfo {author} {\bibfnamefont {E.~Y.}\
  \bibnamefont {Andrei}},\ }\href {https://doi.org/10.1038/s41586-019-1460-4}
  {\bibfield  {journal} {\bibinfo  {journal} {Nature}\ }\textbf {\bibinfo
  {volume} {573}},\ \bibinfo {pages} {91–95} (\bibinfo {year}
  {2019})}\BibitemShut {NoStop}%
\bibitem [{\citenamefont {Wong}\ \emph {et~al.}(2020)\citenamefont {Wong},
  \citenamefont {Nuckolls}, \citenamefont {Oh}, \citenamefont {Lian},
  \citenamefont {Xie}, \citenamefont {Jeon}, \citenamefont {Watanabe},
  \citenamefont {Taniguchi}, \citenamefont {Bernevig},\ and\ \citenamefont
  {Yazdani}}]{wong_cascade_2020}%
  \BibitemOpen
  \bibfield  {author} {\bibinfo {author} {\bibfnamefont {D.}~\bibnamefont
  {Wong}}, \bibinfo {author} {\bibfnamefont {K.~P.}\ \bibnamefont {Nuckolls}},
  \bibinfo {author} {\bibfnamefont {M.}~\bibnamefont {Oh}}, \bibinfo {author}
  {\bibfnamefont {B.}~\bibnamefont {Lian}}, \bibinfo {author} {\bibfnamefont
  {Y.}~\bibnamefont {Xie}}, \bibinfo {author} {\bibfnamefont {S.}~\bibnamefont
  {Jeon}}, \bibinfo {author} {\bibfnamefont {K.}~\bibnamefont {Watanabe}},
  \bibinfo {author} {\bibfnamefont {T.}~\bibnamefont {Taniguchi}}, \bibinfo
  {author} {\bibfnamefont {B.~A.}\ \bibnamefont {Bernevig}},\ and\ \bibinfo
  {author} {\bibfnamefont {A.}~\bibnamefont {Yazdani}},\ }\href
  {https://doi.org/10.1038/s41586-020-2339-0} {\bibfield  {journal} {\bibinfo
  {journal} {Nature}\ }\textbf {\bibinfo {volume} {582}},\ \bibinfo {pages}
  {198–202} (\bibinfo {year} {2020})}\BibitemShut {NoStop}%
\bibitem [{\citenamefont {Zondiner}\ \emph {et~al.}(2020)\citenamefont
  {Zondiner}, \citenamefont {Rozen}, \citenamefont {Rodan-Legrain},
  \citenamefont {Cao}, \citenamefont {Queiroz}, \citenamefont {Taniguchi},
  \citenamefont {Watanabe}, \citenamefont {Oreg}, \citenamefont {von Oppen},
  \citenamefont {Stern},\ and\ \citenamefont {et~al.}}]{zondiner_cascade_2020}%
  \BibitemOpen
  \bibfield  {author} {\bibinfo {author} {\bibfnamefont {U.}~\bibnamefont
  {Zondiner}}, \bibinfo {author} {\bibfnamefont {A.}~\bibnamefont {Rozen}},
  \bibinfo {author} {\bibfnamefont {D.}~\bibnamefont {Rodan-Legrain}}, \bibinfo
  {author} {\bibfnamefont {Y.}~\bibnamefont {Cao}}, \bibinfo {author}
  {\bibfnamefont {R.}~\bibnamefont {Queiroz}}, \bibinfo {author} {\bibfnamefont
  {T.}~\bibnamefont {Taniguchi}}, \bibinfo {author} {\bibfnamefont
  {K.}~\bibnamefont {Watanabe}}, \bibinfo {author} {\bibfnamefont
  {Y.}~\bibnamefont {Oreg}}, \bibinfo {author} {\bibfnamefont {F.}~\bibnamefont
  {von Oppen}}, \bibinfo {author} {\bibfnamefont {A.}~\bibnamefont {Stern}},\
  and\ \bibinfo {author} {\bibnamefont {et~al.}},\ }\href
  {https://doi.org/10.1038/s41586-020-2373-y} {\bibfield  {journal} {\bibinfo
  {journal} {Nature}\ }\textbf {\bibinfo {volume} {582}},\ \bibinfo {pages}
  {203–208} (\bibinfo {year} {2020})}\BibitemShut {NoStop}%
\bibitem [{\citenamefont {Nuckolls}\ \emph {et~al.}(2020)\citenamefont
  {Nuckolls}, \citenamefont {Oh}, \citenamefont {Wong}, \citenamefont {Lian},
  \citenamefont {Watanabe}, \citenamefont {Taniguchi}, \citenamefont
  {Bernevig},\ and\ \citenamefont {Yazdani}}]{nuckolls_chern_2020}%
  \BibitemOpen
  \bibfield  {author} {\bibinfo {author} {\bibfnamefont {K.~P.}\ \bibnamefont
  {Nuckolls}}, \bibinfo {author} {\bibfnamefont {M.}~\bibnamefont {Oh}},
  \bibinfo {author} {\bibfnamefont {D.}~\bibnamefont {Wong}}, \bibinfo {author}
  {\bibfnamefont {B.}~\bibnamefont {Lian}}, \bibinfo {author} {\bibfnamefont
  {K.}~\bibnamefont {Watanabe}}, \bibinfo {author} {\bibfnamefont
  {T.}~\bibnamefont {Taniguchi}}, \bibinfo {author} {\bibfnamefont {B.~A.}\
  \bibnamefont {Bernevig}},\ and\ \bibinfo {author} {\bibfnamefont
  {A.}~\bibnamefont {Yazdani}},\ }\href
  {https://doi.org/10.1038/s41586-020-3028-8} {\bibfield  {journal} {\bibinfo
  {journal} {Nature}\ }\textbf {\bibinfo {volume} {588}},\ \bibinfo {pages}
  {610} (\bibinfo {year} {2020})}\BibitemShut {NoStop}%
\bibitem [{\citenamefont {Choi}\ \emph {et~al.}(2021)\citenamefont {Choi},
  \citenamefont {Kim}, \citenamefont {Peng}, \citenamefont {Thomson},
  \citenamefont {Lewandowski}, \citenamefont {Polski}, \citenamefont {Zhang},
  \citenamefont {Arora}, \citenamefont {Watanabe}, \citenamefont {Taniguchi},
  \citenamefont {Alicea},\ and\ \citenamefont {Nadj-Perge}}]{choi2020tracing}%
  \BibitemOpen
  \bibfield  {author} {\bibinfo {author} {\bibfnamefont {Y.}~\bibnamefont
  {Choi}}, \bibinfo {author} {\bibfnamefont {H.}~\bibnamefont {Kim}}, \bibinfo
  {author} {\bibfnamefont {Y.}~\bibnamefont {Peng}}, \bibinfo {author}
  {\bibfnamefont {A.}~\bibnamefont {Thomson}}, \bibinfo {author} {\bibfnamefont
  {C.}~\bibnamefont {Lewandowski}}, \bibinfo {author} {\bibfnamefont
  {R.}~\bibnamefont {Polski}}, \bibinfo {author} {\bibfnamefont
  {Y.}~\bibnamefont {Zhang}}, \bibinfo {author} {\bibfnamefont {H.~S.}\
  \bibnamefont {Arora}}, \bibinfo {author} {\bibfnamefont {K.}~\bibnamefont
  {Watanabe}}, \bibinfo {author} {\bibfnamefont {T.}~\bibnamefont {Taniguchi}},
  \bibinfo {author} {\bibfnamefont {J.}~\bibnamefont {Alicea}},\ and\ \bibinfo
  {author} {\bibfnamefont {S.}~\bibnamefont {Nadj-Perge}},\ }\href@noop {}
  {\bibfield  {journal} {\bibinfo  {journal} {Nature}\ }\textbf {\bibinfo
  {volume} {589}},\ \bibinfo {pages} {536} (\bibinfo {year} {2021})},\ \Eprint
  {https://arxiv.org/abs/2008.11746} {arXiv:2008.11746 [cond-mat.str-el]}
  \BibitemShut {NoStop}%
\bibitem [{\citenamefont {Saito}\ \emph
  {et~al.}(2021{\natexlab{a}})\citenamefont {Saito}, \citenamefont {Ge},
  \citenamefont {Rademaker}, \citenamefont {Watanabe}, \citenamefont
  {Taniguchi}, \citenamefont {Abanin},\ and\ \citenamefont
  {Young}}]{saito2020}%
  \BibitemOpen
  \bibfield  {author} {\bibinfo {author} {\bibfnamefont {Y.}~\bibnamefont
  {Saito}}, \bibinfo {author} {\bibfnamefont {J.}~\bibnamefont {Ge}}, \bibinfo
  {author} {\bibfnamefont {L.}~\bibnamefont {Rademaker}}, \bibinfo {author}
  {\bibfnamefont {K.}~\bibnamefont {Watanabe}}, \bibinfo {author}
  {\bibfnamefont {T.}~\bibnamefont {Taniguchi}}, \bibinfo {author}
  {\bibfnamefont {D.~A.}\ \bibnamefont {Abanin}},\ and\ \bibinfo {author}
  {\bibfnamefont {A.~F.}\ \bibnamefont {Young}},\ }\href
  {https://doi.org/10.1038/s41567-020-01129-4} {\bibfield  {journal} {\bibinfo
  {journal} {Nature Physics}\ }\textbf {\bibinfo {volume} {17}},\ \bibinfo
  {pages} {478} (\bibinfo {year} {2021}{\natexlab{a}})}\BibitemShut {NoStop}%
\bibitem [{\citenamefont {Das}\ \emph {et~al.}(2021)\citenamefont {Das},
  \citenamefont {Lu}, \citenamefont {Herzog-Arbeitman}, \citenamefont {Song},
  \citenamefont {Watanabe}, \citenamefont {Taniguchi}, \citenamefont
  {Bernevig},\ and\ \citenamefont {Efetov}}]{das2020symmetry}%
  \BibitemOpen
  \bibfield  {author} {\bibinfo {author} {\bibfnamefont {I.}~\bibnamefont
  {Das}}, \bibinfo {author} {\bibfnamefont {X.}~\bibnamefont {Lu}}, \bibinfo
  {author} {\bibfnamefont {J.}~\bibnamefont {Herzog-Arbeitman}}, \bibinfo
  {author} {\bibfnamefont {Z.-D.}\ \bibnamefont {Song}}, \bibinfo {author}
  {\bibfnamefont {K.}~\bibnamefont {Watanabe}}, \bibinfo {author}
  {\bibfnamefont {T.}~\bibnamefont {Taniguchi}}, \bibinfo {author}
  {\bibfnamefont {B.~A.}\ \bibnamefont {Bernevig}},\ and\ \bibinfo {author}
  {\bibfnamefont {D.~K.}\ \bibnamefont {Efetov}},\ }\href
  {https://doi.org/10.1038/s41567-021-01186-3} {\bibfield  {journal} {\bibinfo
  {journal} {Nat. Phys.}\ } (\bibinfo {year} {2021})}\BibitemShut {NoStop}%
\bibitem [{\citenamefont {Wu}\ \emph {et~al.}(2021)\citenamefont {Wu},
  \citenamefont {Zhang}, \citenamefont {Watanabe}, \citenamefont {Taniguchi},\
  and\ \citenamefont {Andrei}}]{wu_chern_2020}%
  \BibitemOpen
  \bibfield  {author} {\bibinfo {author} {\bibfnamefont {S.}~\bibnamefont
  {Wu}}, \bibinfo {author} {\bibfnamefont {Z.}~\bibnamefont {Zhang}}, \bibinfo
  {author} {\bibfnamefont {K.}~\bibnamefont {Watanabe}}, \bibinfo {author}
  {\bibfnamefont {T.}~\bibnamefont {Taniguchi}},\ and\ \bibinfo {author}
  {\bibfnamefont {E.~Y.}\ \bibnamefont {Andrei}},\ }\href
  {https://doi.org/10.1038/s41563-020-00911-2} {\bibfield  {journal} {\bibinfo
  {journal} {Nature Materials}\ }\textbf {\bibinfo {volume} {20}},\ \bibinfo
  {pages} {488} (\bibinfo {year} {2021})}\BibitemShut {NoStop}%
\bibitem [{\citenamefont {Park}\ \emph {et~al.}(2021)\citenamefont {Park},
  \citenamefont {Cao}, \citenamefont {Watanabe}, \citenamefont {Taniguchi},\
  and\ \citenamefont {Jarillo-Herrero}}]{park2020flavour}%
  \BibitemOpen
  \bibfield  {author} {\bibinfo {author} {\bibfnamefont {J.~M.}\ \bibnamefont
  {Park}}, \bibinfo {author} {\bibfnamefont {Y.}~\bibnamefont {Cao}}, \bibinfo
  {author} {\bibfnamefont {K.}~\bibnamefont {Watanabe}}, \bibinfo {author}
  {\bibfnamefont {T.}~\bibnamefont {Taniguchi}},\ and\ \bibinfo {author}
  {\bibfnamefont {P.}~\bibnamefont {Jarillo-Herrero}},\ }\href
  {https://doi.org/10.1038/s41586-021-03366-w} {\bibfield  {journal} {\bibinfo
  {journal} {Nature}\ }\textbf {\bibinfo {volume} {592}},\ \bibinfo {pages}
  {43} (\bibinfo {year} {2021})}\BibitemShut {NoStop}%
\bibitem [{\citenamefont {Saito}\ \emph
  {et~al.}(2021{\natexlab{b}})\citenamefont {Saito}, \citenamefont {Yang},
  \citenamefont {Ge}, \citenamefont {Liu}, \citenamefont {Taniguchi},
  \citenamefont {Watanabe}, \citenamefont {Li}, \citenamefont {Berg},\ and\
  \citenamefont {Young}}]{saito2020isospin}%
  \BibitemOpen
  \bibfield  {author} {\bibinfo {author} {\bibfnamefont {Y.}~\bibnamefont
  {Saito}}, \bibinfo {author} {\bibfnamefont {F.}~\bibnamefont {Yang}},
  \bibinfo {author} {\bibfnamefont {J.}~\bibnamefont {Ge}}, \bibinfo {author}
  {\bibfnamefont {X.}~\bibnamefont {Liu}}, \bibinfo {author} {\bibfnamefont
  {T.}~\bibnamefont {Taniguchi}}, \bibinfo {author} {\bibfnamefont
  {K.}~\bibnamefont {Watanabe}}, \bibinfo {author} {\bibfnamefont {J.~I.~A.}\
  \bibnamefont {Li}}, \bibinfo {author} {\bibfnamefont {E.}~\bibnamefont
  {Berg}},\ and\ \bibinfo {author} {\bibfnamefont {A.~F.}\ \bibnamefont
  {Young}},\ }\href {https://doi.org/10.1038/s41586-021-03409-2} {\bibfield
  {journal} {\bibinfo  {journal} {Nature}\ }\textbf {\bibinfo {volume} {592}},\
  \bibinfo {pages} {220} (\bibinfo {year} {2021}{\natexlab{b}})}\BibitemShut
  {NoStop}%
\bibitem [{\citenamefont {Rozen}\ \emph {et~al.}(2021)\citenamefont {Rozen},
  \citenamefont {Park}, \citenamefont {Zondiner}, \citenamefont {Cao},
  \citenamefont {Rodan-Legrain}, \citenamefont {Taniguchi}, \citenamefont
  {Watanabe}, \citenamefont {Oreg}, \citenamefont {Stern}, \citenamefont
  {Berg}, \citenamefont {Jarillo-Herrero},\ and\ \citenamefont
  {Ilani}}]{rozen2020entropic}%
  \BibitemOpen
  \bibfield  {author} {\bibinfo {author} {\bibfnamefont {A.}~\bibnamefont
  {Rozen}}, \bibinfo {author} {\bibfnamefont {J.~M.}\ \bibnamefont {Park}},
  \bibinfo {author} {\bibfnamefont {U.}~\bibnamefont {Zondiner}}, \bibinfo
  {author} {\bibfnamefont {Y.}~\bibnamefont {Cao}}, \bibinfo {author}
  {\bibfnamefont {D.}~\bibnamefont {Rodan-Legrain}}, \bibinfo {author}
  {\bibfnamefont {T.}~\bibnamefont {Taniguchi}}, \bibinfo {author}
  {\bibfnamefont {K.}~\bibnamefont {Watanabe}}, \bibinfo {author}
  {\bibfnamefont {Y.}~\bibnamefont {Oreg}}, \bibinfo {author} {\bibfnamefont
  {A.}~\bibnamefont {Stern}}, \bibinfo {author} {\bibfnamefont
  {E.}~\bibnamefont {Berg}}, \bibinfo {author} {\bibfnamefont {P.}~\bibnamefont
  {Jarillo-Herrero}},\ and\ \bibinfo {author} {\bibfnamefont {S.}~\bibnamefont
  {Ilani}},\ }\href {https://doi.org/10.1038/s41586-021-03319-3} {\bibfield
  {journal} {\bibinfo  {journal} {Nature}\ }\textbf {\bibinfo {volume} {592}},\
  \bibinfo {pages} {214} (\bibinfo {year} {2021})}\BibitemShut {NoStop}%
\bibitem [{\citenamefont {Lu}\ \emph {et~al.}(2021)\citenamefont {Lu},
  \citenamefont {Lian}, \citenamefont {Chaudhary}, \citenamefont {Piot},
  \citenamefont {Romagnoli}, \citenamefont {Watanabe}, \citenamefont
  {Taniguchi}, \citenamefont {Poggio}, \citenamefont {MacDonald}, \citenamefont
  {Bernevig},\ and\ \citenamefont {Efetov}}]{lu2020fingerprints}%
  \BibitemOpen
  \bibfield  {author} {\bibinfo {author} {\bibfnamefont {X.}~\bibnamefont
  {Lu}}, \bibinfo {author} {\bibfnamefont {B.}~\bibnamefont {Lian}}, \bibinfo
  {author} {\bibfnamefont {G.}~\bibnamefont {Chaudhary}}, \bibinfo {author}
  {\bibfnamefont {B.~A.}\ \bibnamefont {Piot}}, \bibinfo {author}
  {\bibfnamefont {G.}~\bibnamefont {Romagnoli}}, \bibinfo {author}
  {\bibfnamefont {K.}~\bibnamefont {Watanabe}}, \bibinfo {author}
  {\bibfnamefont {T.}~\bibnamefont {Taniguchi}}, \bibinfo {author}
  {\bibfnamefont {M.}~\bibnamefont {Poggio}}, \bibinfo {author} {\bibfnamefont
  {A.~H.}\ \bibnamefont {MacDonald}}, \bibinfo {author} {\bibfnamefont {B.~A.}\
  \bibnamefont {Bernevig}},\ and\ \bibinfo {author} {\bibfnamefont {D.~K.}\
  \bibnamefont {Efetov}},\ }\href {https://doi.org/10.1073/pnas.2100006118}
  {\bibinfo {title} {Multiple flat bands and topological hofstadter butterfly
  in twisted bilayer graphene close to the second magic angle}} (\bibinfo
  {year} {2021}),\ \Eprint
  {https://arxiv.org/abs/https://www.pnas.org/doi/pdf/10.1073/pnas.2100006118}
  {https://www.pnas.org/doi/pdf/10.1073/pnas.2100006118} \BibitemShut {NoStop}%
\bibitem [{\citenamefont {Burg}\ \emph {et~al.}(2019)\citenamefont {Burg},
  \citenamefont {Zhu}, \citenamefont {Taniguchi}, \citenamefont {Watanabe},
  \citenamefont {MacDonald},\ and\ \citenamefont
  {Tutuc}}]{burg_correlated_2019}%
  \BibitemOpen
  \bibfield  {author} {\bibinfo {author} {\bibfnamefont {G.~W.}\ \bibnamefont
  {Burg}}, \bibinfo {author} {\bibfnamefont {J.}~\bibnamefont {Zhu}}, \bibinfo
  {author} {\bibfnamefont {T.}~\bibnamefont {Taniguchi}}, \bibinfo {author}
  {\bibfnamefont {K.}~\bibnamefont {Watanabe}}, \bibinfo {author}
  {\bibfnamefont {A.~H.}\ \bibnamefont {MacDonald}},\ and\ \bibinfo {author}
  {\bibfnamefont {E.}~\bibnamefont {Tutuc}},\ }\href
  {https://doi.org/10.1103/PhysRevLett.123.197702} {\bibfield  {journal}
  {\bibinfo  {journal} {Phys. Rev. Lett.}\ }\textbf {\bibinfo {volume} {123}},\
  \bibinfo {pages} {197702} (\bibinfo {year} {2019})}\BibitemShut {NoStop}%
\bibitem [{\citenamefont {Shen}\ \emph {et~al.}(2020)\citenamefont {Shen},
  \citenamefont {Chu}, \citenamefont {Wu}, \citenamefont {Li}, \citenamefont
  {Wang}, \citenamefont {Zhao}, \citenamefont {Tang}, \citenamefont {Liu},
  \citenamefont {Tian}, \citenamefont {Watanabe}, \citenamefont {Taniguchi},
  \citenamefont {Yang}, \citenamefont {Meng}, \citenamefont {Shi},
  \citenamefont {Yazyev},\ and\ \citenamefont {Zhang}}]{shen_correlated_2020}%
  \BibitemOpen
  \bibfield  {author} {\bibinfo {author} {\bibfnamefont {C.}~\bibnamefont
  {Shen}}, \bibinfo {author} {\bibfnamefont {Y.}~\bibnamefont {Chu}}, \bibinfo
  {author} {\bibfnamefont {Q.}~\bibnamefont {Wu}}, \bibinfo {author}
  {\bibfnamefont {N.}~\bibnamefont {Li}}, \bibinfo {author} {\bibfnamefont
  {S.}~\bibnamefont {Wang}}, \bibinfo {author} {\bibfnamefont {Y.}~\bibnamefont
  {Zhao}}, \bibinfo {author} {\bibfnamefont {J.}~\bibnamefont {Tang}}, \bibinfo
  {author} {\bibfnamefont {J.}~\bibnamefont {Liu}}, \bibinfo {author}
  {\bibfnamefont {J.}~\bibnamefont {Tian}}, \bibinfo {author} {\bibfnamefont
  {K.}~\bibnamefont {Watanabe}}, \bibinfo {author} {\bibfnamefont
  {T.}~\bibnamefont {Taniguchi}}, \bibinfo {author} {\bibfnamefont
  {R.}~\bibnamefont {Yang}}, \bibinfo {author} {\bibfnamefont {Z.~Y.}\
  \bibnamefont {Meng}}, \bibinfo {author} {\bibfnamefont {D.}~\bibnamefont
  {Shi}}, \bibinfo {author} {\bibfnamefont {O.~V.}\ \bibnamefont {Yazyev}},\
  and\ \bibinfo {author} {\bibfnamefont {G.}~\bibnamefont {Zhang}},\ }\href
  {https://doi.org/10.1038/s41567-020-0825-9} {\bibfield  {journal} {\bibinfo
  {journal} {Nature Physics}\ }\textbf {\bibinfo {volume} {16}},\ \bibinfo
  {pages} {520} (\bibinfo {year} {2020})}\BibitemShut {NoStop}%
\bibitem [{\citenamefont {Cao}\ \emph {et~al.}(2020{\natexlab{b}})\citenamefont
  {Cao}, \citenamefont {Rodan-Legrain}, \citenamefont {Rubies-Bigorda},
  \citenamefont {Park}, \citenamefont {Watanabe}, \citenamefont {Taniguchi},\
  and\ \citenamefont {Jarillo-Herrero}}]{cao_tunable_2020}%
  \BibitemOpen
  \bibfield  {author} {\bibinfo {author} {\bibfnamefont {Y.}~\bibnamefont
  {Cao}}, \bibinfo {author} {\bibfnamefont {D.}~\bibnamefont {Rodan-Legrain}},
  \bibinfo {author} {\bibfnamefont {O.}~\bibnamefont {Rubies-Bigorda}},
  \bibinfo {author} {\bibfnamefont {J.~M.}\ \bibnamefont {Park}}, \bibinfo
  {author} {\bibfnamefont {K.}~\bibnamefont {Watanabe}}, \bibinfo {author}
  {\bibfnamefont {T.}~\bibnamefont {Taniguchi}},\ and\ \bibinfo {author}
  {\bibfnamefont {P.}~\bibnamefont {Jarillo-Herrero}},\ }\href
  {https://doi.org/10.1038/s41586-020-2260-6} {\bibfield  {journal} {\bibinfo
  {journal} {Nature}\ }\textbf {\bibinfo {volume} {583}},\ \bibinfo {pages}
  {215} (\bibinfo {year} {2020}{\natexlab{b}})}\BibitemShut {NoStop}%
\bibitem [{\citenamefont {Liu}\ \emph {et~al.}(2020)\citenamefont {Liu},
  \citenamefont {Hao}, \citenamefont {Khalaf}, \citenamefont {Lee},
  \citenamefont {Ronen}, \citenamefont {Yoo}, \citenamefont {Haei~Najafabadi},
  \citenamefont {Watanabe}, \citenamefont {Taniguchi}, \citenamefont
  {Vishwanath},\ and\ \citenamefont {Kim}}]{liu_spin-polarized_2019}%
  \BibitemOpen
  \bibfield  {author} {\bibinfo {author} {\bibfnamefont {X.}~\bibnamefont
  {Liu}}, \bibinfo {author} {\bibfnamefont {Z.}~\bibnamefont {Hao}}, \bibinfo
  {author} {\bibfnamefont {E.}~\bibnamefont {Khalaf}}, \bibinfo {author}
  {\bibfnamefont {J.~Y.}\ \bibnamefont {Lee}}, \bibinfo {author} {\bibfnamefont
  {Y.}~\bibnamefont {Ronen}}, \bibinfo {author} {\bibfnamefont
  {H.}~\bibnamefont {Yoo}}, \bibinfo {author} {\bibfnamefont {D.}~\bibnamefont
  {Haei~Najafabadi}}, \bibinfo {author} {\bibfnamefont {K.}~\bibnamefont
  {Watanabe}}, \bibinfo {author} {\bibfnamefont {T.}~\bibnamefont {Taniguchi}},
  \bibinfo {author} {\bibfnamefont {A.}~\bibnamefont {Vishwanath}},\ and\
  \bibinfo {author} {\bibfnamefont {P.}~\bibnamefont {Kim}},\ }\href
  {https://doi.org/10.1038/s41586-020-2458-7} {\bibfield  {journal} {\bibinfo
  {journal} {Nature}\ }\textbf {\bibinfo {volume} {583}},\ \bibinfo {pages}
  {221} (\bibinfo {year} {2020})}\BibitemShut {NoStop}%
\bibitem [{\citenamefont {Chen}\ \emph
  {et~al.}(2019{\natexlab{a}})\citenamefont {Chen}, \citenamefont {Jiang},
  \citenamefont {Wu}, \citenamefont {Lyu}, \citenamefont {Li}, \citenamefont
  {Chittari}, \citenamefont {Watanabe}, \citenamefont {Taniguchi},
  \citenamefont {Shi}, \citenamefont {Jung}, \citenamefont {Zhang},\ and\
  \citenamefont {Wang}}]{chen_evidence_2019}%
  \BibitemOpen
  \bibfield  {author} {\bibinfo {author} {\bibfnamefont {G.}~\bibnamefont
  {Chen}}, \bibinfo {author} {\bibfnamefont {L.}~\bibnamefont {Jiang}},
  \bibinfo {author} {\bibfnamefont {S.}~\bibnamefont {Wu}}, \bibinfo {author}
  {\bibfnamefont {B.}~\bibnamefont {Lyu}}, \bibinfo {author} {\bibfnamefont
  {H.}~\bibnamefont {Li}}, \bibinfo {author} {\bibfnamefont {B.~L.}\
  \bibnamefont {Chittari}}, \bibinfo {author} {\bibfnamefont {K.}~\bibnamefont
  {Watanabe}}, \bibinfo {author} {\bibfnamefont {T.}~\bibnamefont {Taniguchi}},
  \bibinfo {author} {\bibfnamefont {Z.}~\bibnamefont {Shi}}, \bibinfo {author}
  {\bibfnamefont {J.}~\bibnamefont {Jung}}, \bibinfo {author} {\bibfnamefont
  {Y.}~\bibnamefont {Zhang}},\ and\ \bibinfo {author} {\bibfnamefont
  {F.}~\bibnamefont {Wang}},\ }\href
  {https://doi.org/10.1038/s41567-018-0387-2} {\bibfield  {journal} {\bibinfo
  {journal} {Nature Physics}\ }\textbf {\bibinfo {volume} {15}},\ \bibinfo
  {pages} {237} (\bibinfo {year} {2019}{\natexlab{a}})}\BibitemShut {NoStop}%
\bibitem [{\citenamefont {Chen}\ \emph
  {et~al.}(2019{\natexlab{b}})\citenamefont {Chen}, \citenamefont {Sharpe},
  \citenamefont {Gallagher}, \citenamefont {Rosen}, \citenamefont {Fox},
  \citenamefont {Jiang}, \citenamefont {Lyu}, \citenamefont {Li}, \citenamefont
  {Watanabe}, \citenamefont {Taniguchi}, \citenamefont {Jung}, \citenamefont
  {Shi}, \citenamefont {Goldhaber-Gordon}, \citenamefont {Zhang},\ and\
  \citenamefont {Wang}}]{chen_signatures_2019}%
  \BibitemOpen
  \bibfield  {author} {\bibinfo {author} {\bibfnamefont {G.}~\bibnamefont
  {Chen}}, \bibinfo {author} {\bibfnamefont {A.~L.}\ \bibnamefont {Sharpe}},
  \bibinfo {author} {\bibfnamefont {P.}~\bibnamefont {Gallagher}}, \bibinfo
  {author} {\bibfnamefont {I.~T.}\ \bibnamefont {Rosen}}, \bibinfo {author}
  {\bibfnamefont {E.~J.}\ \bibnamefont {Fox}}, \bibinfo {author} {\bibfnamefont
  {L.}~\bibnamefont {Jiang}}, \bibinfo {author} {\bibfnamefont
  {B.}~\bibnamefont {Lyu}}, \bibinfo {author} {\bibfnamefont {H.}~\bibnamefont
  {Li}}, \bibinfo {author} {\bibfnamefont {K.}~\bibnamefont {Watanabe}},
  \bibinfo {author} {\bibfnamefont {T.}~\bibnamefont {Taniguchi}}, \bibinfo
  {author} {\bibfnamefont {J.}~\bibnamefont {Jung}}, \bibinfo {author}
  {\bibfnamefont {Z.}~\bibnamefont {Shi}}, \bibinfo {author} {\bibfnamefont
  {D.}~\bibnamefont {Goldhaber-Gordon}}, \bibinfo {author} {\bibfnamefont
  {Y.}~\bibnamefont {Zhang}},\ and\ \bibinfo {author} {\bibfnamefont
  {F.}~\bibnamefont {Wang}},\ }\href
  {https://doi.org/10.1038/s41586-019-1393-y} {\bibfield  {journal} {\bibinfo
  {journal} {Nature}\ }\textbf {\bibinfo {volume} {572}},\ \bibinfo {pages}
  {215} (\bibinfo {year} {2019}{\natexlab{b}})}\BibitemShut {NoStop}%
\bibitem [{\citenamefont {Chen}\ \emph {et~al.}(2020)\citenamefont {Chen},
  \citenamefont {Sharpe}, \citenamefont {Fox}, \citenamefont {Zhang},
  \citenamefont {Wang}, \citenamefont {Jiang}, \citenamefont {Lyu},
  \citenamefont {Li}, \citenamefont {Watanabe}, \citenamefont {Taniguchi},
  \citenamefont {Shi}, \citenamefont {Senthil}, \citenamefont
  {Goldhaber-Gordon}, \citenamefont {Zhang},\ and\ \citenamefont
  {Wang}}]{chen_tunable_2020}%
  \BibitemOpen
  \bibfield  {author} {\bibinfo {author} {\bibfnamefont {G.}~\bibnamefont
  {Chen}}, \bibinfo {author} {\bibfnamefont {A.~L.}\ \bibnamefont {Sharpe}},
  \bibinfo {author} {\bibfnamefont {E.~J.}\ \bibnamefont {Fox}}, \bibinfo
  {author} {\bibfnamefont {Y.-H.}\ \bibnamefont {Zhang}}, \bibinfo {author}
  {\bibfnamefont {S.}~\bibnamefont {Wang}}, \bibinfo {author} {\bibfnamefont
  {L.}~\bibnamefont {Jiang}}, \bibinfo {author} {\bibfnamefont
  {B.}~\bibnamefont {Lyu}}, \bibinfo {author} {\bibfnamefont {H.}~\bibnamefont
  {Li}}, \bibinfo {author} {\bibfnamefont {K.}~\bibnamefont {Watanabe}},
  \bibinfo {author} {\bibfnamefont {T.}~\bibnamefont {Taniguchi}}, \bibinfo
  {author} {\bibfnamefont {Z.}~\bibnamefont {Shi}}, \bibinfo {author}
  {\bibfnamefont {T.}~\bibnamefont {Senthil}}, \bibinfo {author} {\bibfnamefont
  {D.}~\bibnamefont {Goldhaber-Gordon}}, \bibinfo {author} {\bibfnamefont
  {Y.}~\bibnamefont {Zhang}},\ and\ \bibinfo {author} {\bibfnamefont
  {F.}~\bibnamefont {Wang}},\ }\href
  {https://doi.org/10.1038/s41586-020-2049-7} {\bibfield  {journal} {\bibinfo
  {journal} {Nature}\ }\textbf {\bibinfo {volume} {579}},\ \bibinfo {pages}
  {56} (\bibinfo {year} {2020})}\BibitemShut {NoStop}%
\bibitem [{\citenamefont {Burg}\ \emph {et~al.}(2020)\citenamefont {Burg},
  \citenamefont {Lian}, \citenamefont {Taniguchi}, \citenamefont {Watanabe},
  \citenamefont {Bernevig},\ and\ \citenamefont {Tutuc}}]{burg2020evidence}%
  \BibitemOpen
  \bibfield  {author} {\bibinfo {author} {\bibfnamefont {G.~W.}\ \bibnamefont
  {Burg}}, \bibinfo {author} {\bibfnamefont {B.}~\bibnamefont {Lian}}, \bibinfo
  {author} {\bibfnamefont {T.}~\bibnamefont {Taniguchi}}, \bibinfo {author}
  {\bibfnamefont {K.}~\bibnamefont {Watanabe}}, \bibinfo {author}
  {\bibfnamefont {B.~A.}\ \bibnamefont {Bernevig}},\ and\ \bibinfo {author}
  {\bibfnamefont {E.}~\bibnamefont {Tutuc}},\ }\href@noop {} {\bibinfo {title}
  {Evidence of emergent symmetry and valley chern number in twisted
  double-bilayer graphene}} (\bibinfo {year} {2020}),\ \Eprint
  {https://arxiv.org/abs/2006.14000} {arXiv:2006.14000 [cond-mat.mes-hall]}
  \BibitemShut {NoStop}%
\bibitem [{\citenamefont {Tarnopolsky}\ \emph {et~al.}(2019)\citenamefont
  {Tarnopolsky}, \citenamefont {Kruchkov},\ and\ \citenamefont
  {Vishwanath}}]{tarnopolsky_origin_2019}%
  \BibitemOpen
  \bibfield  {author} {\bibinfo {author} {\bibfnamefont {G.}~\bibnamefont
  {Tarnopolsky}}, \bibinfo {author} {\bibfnamefont {A.~J.}\ \bibnamefont
  {Kruchkov}},\ and\ \bibinfo {author} {\bibfnamefont {A.}~\bibnamefont
  {Vishwanath}},\ }\href {https://doi.org/10.1103/PhysRevLett.122.106405}
  {\bibfield  {journal} {\bibinfo  {journal} {Physical Review Letters}\
  }\textbf {\bibinfo {volume} {122}},\ \bibinfo {pages} {106405} (\bibinfo
  {year} {2019})}\BibitemShut {NoStop}%
\bibitem [{\citenamefont {Zou}\ \emph {et~al.}(2018)\citenamefont {Zou},
  \citenamefont {Po}, \citenamefont {Vishwanath},\ and\ \citenamefont
  {Senthil}}]{zou2018}%
  \BibitemOpen
  \bibfield  {author} {\bibinfo {author} {\bibfnamefont {L.}~\bibnamefont
  {Zou}}, \bibinfo {author} {\bibfnamefont {H.~C.}\ \bibnamefont {Po}},
  \bibinfo {author} {\bibfnamefont {A.}~\bibnamefont {Vishwanath}},\ and\
  \bibinfo {author} {\bibfnamefont {T.}~\bibnamefont {Senthil}},\ }\href
  {https://doi.org/10.1103/PhysRevB.98.085435} {\bibfield  {journal} {\bibinfo
  {journal} {Phys. Rev. B}\ }\textbf {\bibinfo {volume} {98}},\ \bibinfo
  {pages} {085435} (\bibinfo {year} {2018})}\BibitemShut {NoStop}%
\bibitem [{\citenamefont {Fu}\ \emph {et~al.}(2020)\citenamefont {Fu},
  \citenamefont {K{\"o}nig}, \citenamefont {Wilson}, \citenamefont {Chou},\
  and\ \citenamefont {Pixley}}]{fu2018magicangle}%
  \BibitemOpen
  \bibfield  {author} {\bibinfo {author} {\bibfnamefont {Y.}~\bibnamefont
  {Fu}}, \bibinfo {author} {\bibfnamefont {E.~J.}\ \bibnamefont {K{\"o}nig}},
  \bibinfo {author} {\bibfnamefont {J.~H.}\ \bibnamefont {Wilson}}, \bibinfo
  {author} {\bibfnamefont {Y.-Z.}\ \bibnamefont {Chou}},\ and\ \bibinfo
  {author} {\bibfnamefont {J.~H.}\ \bibnamefont {Pixley}},\ }\href
  {https://doi.org/10.1038/s41535-020-00271-9} {\bibinfo {title} {Magic-angle
  semimetals}} (\bibinfo {year} {2020})\BibitemShut {NoStop}%
\bibitem [{\citenamefont {Liu}\ \emph {et~al.}(2019{\natexlab{a}})\citenamefont
  {Liu}, \citenamefont {Liu},\ and\ \citenamefont {Dai}}]{liu2019pseudo}%
  \BibitemOpen
  \bibfield  {author} {\bibinfo {author} {\bibfnamefont {J.}~\bibnamefont
  {Liu}}, \bibinfo {author} {\bibfnamefont {J.}~\bibnamefont {Liu}},\ and\
  \bibinfo {author} {\bibfnamefont {X.}~\bibnamefont {Dai}},\ }\href
  {https://journals.aps.org/prb/abstract/10.1103/PhysRevB.99.155415} {\bibfield
   {journal} {\bibinfo  {journal} {Physical Review B}\ }\textbf {\bibinfo
  {volume} {99}},\ \bibinfo {pages} {155415} (\bibinfo {year}
  {2019}{\natexlab{a}})}\BibitemShut {NoStop}%
\bibitem [{\citenamefont {Efimkin}\ and\ \citenamefont
  {MacDonald}(2018)}]{Efimkin2018TBG}%
  \BibitemOpen
  \bibfield  {author} {\bibinfo {author} {\bibfnamefont {D.~K.}\ \bibnamefont
  {Efimkin}}\ and\ \bibinfo {author} {\bibfnamefont {A.~H.}\ \bibnamefont
  {MacDonald}},\ }\href {https://doi.org/10.1103/PhysRevB.98.035404} {\bibfield
   {journal} {\bibinfo  {journal} {Phys. Rev. B}\ }\textbf {\bibinfo {volume}
  {98}},\ \bibinfo {pages} {035404} (\bibinfo {year} {2018})}\BibitemShut
  {NoStop}%
\bibitem [{\citenamefont {Kang}\ and\ \citenamefont
  {Vafek}(2018)}]{kang_symmetry_2018}%
  \BibitemOpen
  \bibfield  {author} {\bibinfo {author} {\bibfnamefont {J.}~\bibnamefont
  {Kang}}\ and\ \bibinfo {author} {\bibfnamefont {O.}~\bibnamefont {Vafek}},\
  }\href {https://doi.org/10.1103/PhysRevX.8.031088} {\bibfield  {journal}
  {\bibinfo  {journal} {Phys. Rev. X}\ }\textbf {\bibinfo {volume} {8}},\
  \bibinfo {pages} {031088} (\bibinfo {year} {2018})}\BibitemShut {NoStop}%
\bibitem [{\citenamefont {Song}\ \emph {et~al.}(2019)\citenamefont {Song},
  \citenamefont {Wang}, \citenamefont {Shi}, \citenamefont {Li}, \citenamefont
  {Fang},\ and\ \citenamefont {Bernevig}}]{song_all_2019}%
  \BibitemOpen
  \bibfield  {author} {\bibinfo {author} {\bibfnamefont {Z.}~\bibnamefont
  {Song}}, \bibinfo {author} {\bibfnamefont {Z.}~\bibnamefont {Wang}}, \bibinfo
  {author} {\bibfnamefont {W.}~\bibnamefont {Shi}}, \bibinfo {author}
  {\bibfnamefont {G.}~\bibnamefont {Li}}, \bibinfo {author} {\bibfnamefont
  {C.}~\bibnamefont {Fang}},\ and\ \bibinfo {author} {\bibfnamefont {B.~A.}\
  \bibnamefont {Bernevig}},\ }\href
  {https://doi.org/10.1103/PhysRevLett.123.036401} {\bibfield  {journal}
  {\bibinfo  {journal} {Physical Review Letters}\ }\textbf {\bibinfo {volume}
  {123}},\ \bibinfo {pages} {036401} (\bibinfo {year} {2019})}\BibitemShut
  {NoStop}%
\bibitem [{\citenamefont {Po}\ \emph {et~al.}(2019)\citenamefont {Po},
  \citenamefont {Zou}, \citenamefont {Senthil},\ and\ \citenamefont
  {Vishwanath}}]{po_faithful_2019}%
  \BibitemOpen
  \bibfield  {author} {\bibinfo {author} {\bibfnamefont {H.~C.}\ \bibnamefont
  {Po}}, \bibinfo {author} {\bibfnamefont {L.}~\bibnamefont {Zou}}, \bibinfo
  {author} {\bibfnamefont {T.}~\bibnamefont {Senthil}},\ and\ \bibinfo {author}
  {\bibfnamefont {A.}~\bibnamefont {Vishwanath}},\ }\href
  {https://doi.org/10.1103/PhysRevB.99.195455} {\bibfield  {journal} {\bibinfo
  {journal} {Physical Review B}\ }\textbf {\bibinfo {volume} {99}},\ \bibinfo
  {pages} {195455} (\bibinfo {year} {2019})}\BibitemShut {NoStop}%
\bibitem [{\citenamefont {Ahn}\ \emph {et~al.}(2019)\citenamefont {Ahn},
  \citenamefont {Park},\ and\ \citenamefont {Yang}}]{ahn_failure_2019}%
  \BibitemOpen
  \bibfield  {author} {\bibinfo {author} {\bibfnamefont {J.}~\bibnamefont
  {Ahn}}, \bibinfo {author} {\bibfnamefont {S.}~\bibnamefont {Park}},\ and\
  \bibinfo {author} {\bibfnamefont {B.-J.}\ \bibnamefont {Yang}},\ }\href
  {https://doi.org/10.1103/PhysRevX.9.021013} {\bibfield  {journal} {\bibinfo
  {journal} {Physical Review X}\ }\textbf {\bibinfo {volume} {9}},\ \bibinfo
  {pages} {021013} (\bibinfo {year} {2019})}\BibitemShut {NoStop}%
\bibitem [{\citenamefont {Bouhon}\ \emph {et~al.}(2019)\citenamefont {Bouhon},
  \citenamefont {Black-Schaffer},\ and\ \citenamefont {Slager}}]{Slager2019WL}%
  \BibitemOpen
  \bibfield  {author} {\bibinfo {author} {\bibfnamefont {A.}~\bibnamefont
  {Bouhon}}, \bibinfo {author} {\bibfnamefont {A.~M.}\ \bibnamefont
  {Black-Schaffer}},\ and\ \bibinfo {author} {\bibfnamefont {R.-J.}\
  \bibnamefont {Slager}},\ }\href {https://doi.org/10.1103/PhysRevB.100.195135}
  {\bibfield  {journal} {\bibinfo  {journal} {Phys. Rev. B}\ }\textbf {\bibinfo
  {volume} {100}},\ \bibinfo {pages} {195135} (\bibinfo {year}
  {2019})}\BibitemShut {NoStop}%
\bibitem [{\citenamefont {Hejazi}\ \emph
  {et~al.}(2019{\natexlab{a}})\citenamefont {Hejazi}, \citenamefont {Liu},
  \citenamefont {Shapourian}, \citenamefont {Chen},\ and\ \citenamefont
  {Balents}}]{hejazi_multiple_2019}%
  \BibitemOpen
  \bibfield  {author} {\bibinfo {author} {\bibfnamefont {K.}~\bibnamefont
  {Hejazi}}, \bibinfo {author} {\bibfnamefont {C.}~\bibnamefont {Liu}},
  \bibinfo {author} {\bibfnamefont {H.}~\bibnamefont {Shapourian}}, \bibinfo
  {author} {\bibfnamefont {X.}~\bibnamefont {Chen}},\ and\ \bibinfo {author}
  {\bibfnamefont {L.}~\bibnamefont {Balents}},\ }\href
  {https://doi.org/10.1103/PhysRevB.99.035111} {\bibfield  {journal} {\bibinfo
  {journal} {Phys. Rev. B}\ }\textbf {\bibinfo {volume} {99}},\ \bibinfo
  {pages} {035111} (\bibinfo {year} {2019}{\natexlab{a}})}\BibitemShut
  {NoStop}%
\bibitem [{\citenamefont {Lian}\ \emph {et~al.}(2020)\citenamefont {Lian},
  \citenamefont {Xie},\ and\ \citenamefont {Bernevig}}]{lian2020}%
  \BibitemOpen
  \bibfield  {author} {\bibinfo {author} {\bibfnamefont {B.}~\bibnamefont
  {Lian}}, \bibinfo {author} {\bibfnamefont {F.}~\bibnamefont {Xie}},\ and\
  \bibinfo {author} {\bibfnamefont {B.~A.}\ \bibnamefont {Bernevig}},\ }\href
  {https://doi.org/10.1103/PhysRevB.102.041402} {\bibfield  {journal} {\bibinfo
   {journal} {Phys. Rev. B}\ }\textbf {\bibinfo {volume} {102}},\ \bibinfo
  {pages} {041402} (\bibinfo {year} {2020})}\BibitemShut {NoStop}%
\bibitem [{\citenamefont {Hejazi}\ \emph
  {et~al.}(2019{\natexlab{b}})\citenamefont {Hejazi}, \citenamefont {Liu},\
  and\ \citenamefont {Balents}}]{hejazi_landau_2019}%
  \BibitemOpen
  \bibfield  {author} {\bibinfo {author} {\bibfnamefont {K.}~\bibnamefont
  {Hejazi}}, \bibinfo {author} {\bibfnamefont {C.}~\bibnamefont {Liu}},\ and\
  \bibinfo {author} {\bibfnamefont {L.}~\bibnamefont {Balents}},\ }\bibfield
  {journal} {\bibinfo  {journal} {Physical Review B}\ }\textbf {\bibinfo
  {volume} {100}},\ \href {https://doi.org/10.1103/physrevb.100.035115}
  {10.1103/physrevb.100.035115} (\bibinfo {year}
  {2019}{\natexlab{b}})\BibitemShut {NoStop}%
\bibitem [{\citenamefont {Padhi}\ \emph {et~al.}(2020)\citenamefont {Padhi},
  \citenamefont {Tiwari}, \citenamefont {Neupert},\ and\ \citenamefont
  {Ryu}}]{padhi2020transport}%
  \BibitemOpen
  \bibfield  {author} {\bibinfo {author} {\bibfnamefont {B.}~\bibnamefont
  {Padhi}}, \bibinfo {author} {\bibfnamefont {A.}~\bibnamefont {Tiwari}},
  \bibinfo {author} {\bibfnamefont {T.}~\bibnamefont {Neupert}},\ and\ \bibinfo
  {author} {\bibfnamefont {S.}~\bibnamefont {Ryu}},\ }\href
  {https://doi.org/10.1103/PhysRevResearch.2.033458} {\bibinfo {title}
  {Transport across twist angle domains in moir\'e graphene}} (\bibinfo {year}
  {2020})\BibitemShut {NoStop}%
\bibitem [{\citenamefont {Xu}\ and\ \citenamefont
  {Balents}(2018)}]{xu2018topological}%
  \BibitemOpen
  \bibfield  {author} {\bibinfo {author} {\bibfnamefont {C.}~\bibnamefont
  {Xu}}\ and\ \bibinfo {author} {\bibfnamefont {L.}~\bibnamefont {Balents}},\
  }\href {https://journals.aps.org/prl/abstract/10.1103/PhysRevLett.121.087001}
  {\bibfield  {journal} {\bibinfo  {journal} {Physical review letters}\
  }\textbf {\bibinfo {volume} {121}},\ \bibinfo {pages} {087001} (\bibinfo
  {year} {2018})}\BibitemShut {NoStop}%
\bibitem [{\citenamefont {Koshino}\ \emph {et~al.}(2018)\citenamefont
  {Koshino}, \citenamefont {Yuan}, \citenamefont {Koretsune}, \citenamefont
  {Ochi}, \citenamefont {Kuroki},\ and\ \citenamefont
  {Fu}}]{koshino_maximally_2018}%
  \BibitemOpen
  \bibfield  {author} {\bibinfo {author} {\bibfnamefont {M.}~\bibnamefont
  {Koshino}}, \bibinfo {author} {\bibfnamefont {N.~F.~Q.}\ \bibnamefont
  {Yuan}}, \bibinfo {author} {\bibfnamefont {T.}~\bibnamefont {Koretsune}},
  \bibinfo {author} {\bibfnamefont {M.}~\bibnamefont {Ochi}}, \bibinfo {author}
  {\bibfnamefont {K.}~\bibnamefont {Kuroki}},\ and\ \bibinfo {author}
  {\bibfnamefont {L.}~\bibnamefont {Fu}},\ }\href
  {https://doi.org/10.1103/PhysRevX.8.031087} {\bibfield  {journal} {\bibinfo
  {journal} {Phys. Rev. X}\ }\textbf {\bibinfo {volume} {8}},\ \bibinfo {pages}
  {031087} (\bibinfo {year} {2018})}\BibitemShut {NoStop}%
\bibitem [{\citenamefont {Ochi}\ \emph {et~al.}(2018)\citenamefont {Ochi},
  \citenamefont {Koshino},\ and\ \citenamefont {Kuroki}}]{ochi_possible_2018}%
  \BibitemOpen
  \bibfield  {author} {\bibinfo {author} {\bibfnamefont {M.}~\bibnamefont
  {Ochi}}, \bibinfo {author} {\bibfnamefont {M.}~\bibnamefont {Koshino}},\ and\
  \bibinfo {author} {\bibfnamefont {K.}~\bibnamefont {Kuroki}},\ }\href
  {https://doi.org/10.1103/PhysRevB.98.081102} {\bibfield  {journal} {\bibinfo
  {journal} {Phys. Rev. B}\ }\textbf {\bibinfo {volume} {98}},\ \bibinfo
  {pages} {081102} (\bibinfo {year} {2018})}\BibitemShut {NoStop}%
\bibitem [{\citenamefont {Xu}\ \emph {et~al.}(2018)\citenamefont {Xu},
  \citenamefont {Law},\ and\ \citenamefont {Lee}}]{xux2018}%
  \BibitemOpen
  \bibfield  {author} {\bibinfo {author} {\bibfnamefont {X.~Y.}\ \bibnamefont
  {Xu}}, \bibinfo {author} {\bibfnamefont {K.~T.}\ \bibnamefont {Law}},\ and\
  \bibinfo {author} {\bibfnamefont {P.~A.}\ \bibnamefont {Lee}},\ }\href
  {https://doi.org/10.1103/PhysRevB.98.121406} {\bibfield  {journal} {\bibinfo
  {journal} {Phys. Rev. B}\ }\textbf {\bibinfo {volume} {98}},\ \bibinfo
  {pages} {121406} (\bibinfo {year} {2018})}\BibitemShut {NoStop}%
\bibitem [{\citenamefont {Guinea}\ and\ \citenamefont
  {Walet}(2018)}]{guinea2018}%
  \BibitemOpen
  \bibfield  {author} {\bibinfo {author} {\bibfnamefont {F.}~\bibnamefont
  {Guinea}}\ and\ \bibinfo {author} {\bibfnamefont {N.~R.}\ \bibnamefont
  {Walet}},\ }\href {https://doi.org/10.1073/pnas.1810947115} {\bibfield
  {journal} {\bibinfo  {journal} {Proceedings of the National Academy of
  Sciences}\ }\textbf {\bibinfo {volume} {115}},\ \bibinfo {pages} {13174}
  (\bibinfo {year} {2018})}\BibitemShut {NoStop}%
\bibitem [{\citenamefont {Venderbos}\ and\ \citenamefont
  {Fernandes}(2018)}]{venderbos2018}%
  \BibitemOpen
  \bibfield  {author} {\bibinfo {author} {\bibfnamefont {J.~W.~F.}\
  \bibnamefont {Venderbos}}\ and\ \bibinfo {author} {\bibfnamefont {R.~M.}\
  \bibnamefont {Fernandes}},\ }\href
  {https://doi.org/10.1103/PhysRevB.98.245103} {\bibfield  {journal} {\bibinfo
  {journal} {Phys. Rev. B}\ }\textbf {\bibinfo {volume} {98}},\ \bibinfo
  {pages} {245103} (\bibinfo {year} {2018})}\BibitemShut {NoStop}%
\bibitem [{\citenamefont {{You}}\ and\ \citenamefont
  {{Vishwanath}}(2019)}]{you2019}%
  \BibitemOpen
  \bibfield  {author} {\bibinfo {author} {\bibfnamefont {Y.-Z.}\ \bibnamefont
  {{You}}}\ and\ \bibinfo {author} {\bibfnamefont {A.}~\bibnamefont
  {{Vishwanath}}},\ }\href {https://doi.org/10.1038/s41535-019-0153-4}
  {\bibfield  {journal} {\bibinfo  {journal} {npj Quantum Materials}\ }\textbf
  {\bibinfo {volume} {4}},\ \bibinfo {pages} {16} (\bibinfo {year}
  {2019})}\BibitemShut {NoStop}%
\bibitem [{\citenamefont {Wu}\ and\ \citenamefont
  {Das~Sarma}(2020)}]{wu_collective_2020}%
  \BibitemOpen
  \bibfield  {author} {\bibinfo {author} {\bibfnamefont {F.}~\bibnamefont
  {Wu}}\ and\ \bibinfo {author} {\bibfnamefont {S.}~\bibnamefont {Das~Sarma}},\
  }\bibfield  {journal} {\bibinfo  {journal} {Physical Review Letters}\
  }\textbf {\bibinfo {volume} {124}},\ \href
  {https://doi.org/10.1103/physrevlett.124.046403}
  {10.1103/physrevlett.124.046403} (\bibinfo {year} {2020})\BibitemShut
  {NoStop}%
\bibitem [{\citenamefont {Lian}\ \emph {et~al.}(2019)\citenamefont {Lian},
  \citenamefont {Wang},\ and\ \citenamefont {Bernevig}}]{Lian2019TBG}%
  \BibitemOpen
  \bibfield  {author} {\bibinfo {author} {\bibfnamefont {B.}~\bibnamefont
  {Lian}}, \bibinfo {author} {\bibfnamefont {Z.}~\bibnamefont {Wang}},\ and\
  \bibinfo {author} {\bibfnamefont {B.~A.}\ \bibnamefont {Bernevig}},\ }\href
  {https://doi.org/10.1103/PhysRevLett.122.257002} {\bibfield  {journal}
  {\bibinfo  {journal} {Phys. Rev. Lett.}\ }\textbf {\bibinfo {volume} {122}},\
  \bibinfo {pages} {257002} (\bibinfo {year} {2019})}\BibitemShut {NoStop}%
\bibitem [{\citenamefont {Wu}\ \emph {et~al.}(2018)\citenamefont {Wu},
  \citenamefont {MacDonald},\ and\ \citenamefont {Martin}}]{Wu2018TBG-BCS}%
  \BibitemOpen
  \bibfield  {author} {\bibinfo {author} {\bibfnamefont {F.}~\bibnamefont
  {Wu}}, \bibinfo {author} {\bibfnamefont {A.~H.}\ \bibnamefont {MacDonald}},\
  and\ \bibinfo {author} {\bibfnamefont {I.}~\bibnamefont {Martin}},\ }\href
  {https://doi.org/10.1103/PhysRevLett.121.257001} {\bibfield  {journal}
  {\bibinfo  {journal} {Phys. Rev. Lett.}\ }\textbf {\bibinfo {volume} {121}},\
  \bibinfo {pages} {257001} (\bibinfo {year} {2018})}\BibitemShut {NoStop}%
\bibitem [{\citenamefont {Isobe}\ \emph {et~al.}(2018)\citenamefont {Isobe},
  \citenamefont {Yuan},\ and\ \citenamefont {Fu}}]{isobe2018unconventional}%
  \BibitemOpen
  \bibfield  {author} {\bibinfo {author} {\bibfnamefont {H.}~\bibnamefont
  {Isobe}}, \bibinfo {author} {\bibfnamefont {N.~F.}\ \bibnamefont {Yuan}},\
  and\ \bibinfo {author} {\bibfnamefont {L.}~\bibnamefont {Fu}},\ }\href@noop
  {} {\bibfield  {journal} {\bibinfo  {journal} {Physical Review X}\ }\textbf
  {\bibinfo {volume} {8}},\ \bibinfo {pages} {041041} (\bibinfo {year}
  {2018})}\BibitemShut {NoStop}%
\bibitem [{\citenamefont {Liu}\ \emph {et~al.}(2018)\citenamefont {Liu},
  \citenamefont {Zhang}, \citenamefont {Chen},\ and\ \citenamefont
  {Yang}}]{liu2018chiral}%
  \BibitemOpen
  \bibfield  {author} {\bibinfo {author} {\bibfnamefont {C.-C.}\ \bibnamefont
  {Liu}}, \bibinfo {author} {\bibfnamefont {L.-D.}\ \bibnamefont {Zhang}},
  \bibinfo {author} {\bibfnamefont {W.-Q.}\ \bibnamefont {Chen}},\ and\
  \bibinfo {author} {\bibfnamefont {F.}~\bibnamefont {Yang}},\ }\href
  {https://journals.aps.org/prl/abstract/10.1103/PhysRevLett.121.217001}
  {\bibfield  {journal} {\bibinfo  {journal} {Physical review letters}\
  }\textbf {\bibinfo {volume} {121}},\ \bibinfo {pages} {217001} (\bibinfo
  {year} {2018})}\BibitemShut {NoStop}%
\bibitem [{\citenamefont {Bultinck}\ \emph
  {et~al.}(2020{\natexlab{a}})\citenamefont {Bultinck}, \citenamefont
  {Chatterjee},\ and\ \citenamefont {Zaletel}}]{bultinck2020}%
  \BibitemOpen
  \bibfield  {author} {\bibinfo {author} {\bibfnamefont {N.}~\bibnamefont
  {Bultinck}}, \bibinfo {author} {\bibfnamefont {S.}~\bibnamefont
  {Chatterjee}},\ and\ \bibinfo {author} {\bibfnamefont {M.~P.}\ \bibnamefont
  {Zaletel}},\ }\href {https://doi.org/10.1103/PhysRevLett.124.166601}
  {\bibfield  {journal} {\bibinfo  {journal} {Phys. Rev. Lett.}\ }\textbf
  {\bibinfo {volume} {124}},\ \bibinfo {pages} {166601} (\bibinfo {year}
  {2020}{\natexlab{a}})}\BibitemShut {NoStop}%
\bibitem [{\citenamefont {Zhang}\ \emph {et~al.}(2019)\citenamefont {Zhang},
  \citenamefont {Mao}, \citenamefont {Cao}, \citenamefont {Jarillo-Herrero},\
  and\ \citenamefont {Senthil}}]{zhang2019nearly}%
  \BibitemOpen
  \bibfield  {author} {\bibinfo {author} {\bibfnamefont {Y.-H.}\ \bibnamefont
  {Zhang}}, \bibinfo {author} {\bibfnamefont {D.}~\bibnamefont {Mao}}, \bibinfo
  {author} {\bibfnamefont {Y.}~\bibnamefont {Cao}}, \bibinfo {author}
  {\bibfnamefont {P.}~\bibnamefont {Jarillo-Herrero}},\ and\ \bibinfo {author}
  {\bibfnamefont {T.}~\bibnamefont {Senthil}},\ }\href
  {https://journals.aps.org/prb/abstract/10.1103/PhysRevB.99.075127} {\bibfield
   {journal} {\bibinfo  {journal} {Physical Review B}\ }\textbf {\bibinfo
  {volume} {99}},\ \bibinfo {pages} {075127} (\bibinfo {year}
  {2019})}\BibitemShut {NoStop}%
\bibitem [{\citenamefont {Liu}\ \emph {et~al.}(2019{\natexlab{b}})\citenamefont
  {Liu}, \citenamefont {Ma}, \citenamefont {Gao},\ and\ \citenamefont
  {Dai}}]{liu2019quantum}%
  \BibitemOpen
  \bibfield  {author} {\bibinfo {author} {\bibfnamefont {J.}~\bibnamefont
  {Liu}}, \bibinfo {author} {\bibfnamefont {Z.}~\bibnamefont {Ma}}, \bibinfo
  {author} {\bibfnamefont {J.}~\bibnamefont {Gao}},\ and\ \bibinfo {author}
  {\bibfnamefont {X.}~\bibnamefont {Dai}},\ }\href
  {https://journals.aps.org/prx/abstract/10.1103/PhysRevX.9.031021} {\bibfield
  {journal} {\bibinfo  {journal} {Physical Review X}\ }\textbf {\bibinfo
  {volume} {9}},\ \bibinfo {pages} {031021} (\bibinfo {year}
  {2019}{\natexlab{b}})}\BibitemShut {NoStop}%
\bibitem [{\citenamefont {Wu}\ \emph {et~al.}(2019{\natexlab{a}})\citenamefont
  {Wu}, \citenamefont {Jian},\ and\ \citenamefont {Xu}}]{wux2018b}%
  \BibitemOpen
  \bibfield  {author} {\bibinfo {author} {\bibfnamefont {X.-C.}\ \bibnamefont
  {Wu}}, \bibinfo {author} {\bibfnamefont {C.-M.}\ \bibnamefont {Jian}},\ and\
  \bibinfo {author} {\bibfnamefont {C.}~\bibnamefont {Xu}},\ }\bibfield
  {journal} {\bibinfo  {journal} {Physical Review B}\ }\textbf {\bibinfo
  {volume} {99}},\ \href {https://doi.org/10.1103/physrevb.99.161405}
  {10.1103/physrevb.99.161405} (\bibinfo {year}
  {2019}{\natexlab{a}})\BibitemShut {NoStop}%
\bibitem [{\citenamefont {Thomson}\ \emph {et~al.}(2018)\citenamefont
  {Thomson}, \citenamefont {Chatterjee}, \citenamefont {Sachdev},\ and\
  \citenamefont {Scheurer}}]{thomson2018triangular}%
  \BibitemOpen
  \bibfield  {author} {\bibinfo {author} {\bibfnamefont {A.}~\bibnamefont
  {Thomson}}, \bibinfo {author} {\bibfnamefont {S.}~\bibnamefont {Chatterjee}},
  \bibinfo {author} {\bibfnamefont {S.}~\bibnamefont {Sachdev}},\ and\ \bibinfo
  {author} {\bibfnamefont {M.~S.}\ \bibnamefont {Scheurer}},\ }\bibfield
  {journal} {\bibinfo  {journal} {Physical Review B}\ }\textbf {\bibinfo
  {volume} {98}},\ \href {https://doi.org/10.1103/physrevb.98.075109}
  {10.1103/physrevb.98.075109} (\bibinfo {year} {2018})\BibitemShut {NoStop}%
\bibitem [{\citenamefont {Dodaro}\ \emph {et~al.}(2018)\citenamefont {Dodaro},
  \citenamefont {Kivelson}, \citenamefont {Schattner}, \citenamefont {Sun},\
  and\ \citenamefont {Wang}}]{dodaro2018phases}%
  \BibitemOpen
  \bibfield  {author} {\bibinfo {author} {\bibfnamefont {J.~F.}\ \bibnamefont
  {Dodaro}}, \bibinfo {author} {\bibfnamefont {S.~A.}\ \bibnamefont
  {Kivelson}}, \bibinfo {author} {\bibfnamefont {Y.}~\bibnamefont {Schattner}},
  \bibinfo {author} {\bibfnamefont {X.-Q.}\ \bibnamefont {Sun}},\ and\ \bibinfo
  {author} {\bibfnamefont {C.}~\bibnamefont {Wang}},\ }\href
  {https://journals.aps.org/prb/abstract/10.1103/PhysRevB.98.075154} {\bibfield
   {journal} {\bibinfo  {journal} {Physical Review B}\ }\textbf {\bibinfo
  {volume} {98}},\ \bibinfo {pages} {075154} (\bibinfo {year}
  {2018})}\BibitemShut {NoStop}%
\bibitem [{\citenamefont {Gonzalez}\ and\ \citenamefont
  {Stauber}(2019)}]{gonzalez2019kohn}%
  \BibitemOpen
  \bibfield  {author} {\bibinfo {author} {\bibfnamefont {J.}~\bibnamefont
  {Gonzalez}}\ and\ \bibinfo {author} {\bibfnamefont {T.}~\bibnamefont
  {Stauber}},\ }\href
  {https://journals.aps.org/prl/abstract/10.1103/PhysRevLett.122.026801}
  {\bibfield  {journal} {\bibinfo  {journal} {Physical review letters}\
  }\textbf {\bibinfo {volume} {122}},\ \bibinfo {pages} {026801} (\bibinfo
  {year} {2019})}\BibitemShut {NoStop}%
\bibitem [{\citenamefont {Yuan}\ and\ \citenamefont
  {Fu}(2018)}]{yuan2018model}%
  \BibitemOpen
  \bibfield  {author} {\bibinfo {author} {\bibfnamefont {N.~F.}\ \bibnamefont
  {Yuan}}\ and\ \bibinfo {author} {\bibfnamefont {L.}~\bibnamefont {Fu}},\
  }\href {https://journals.aps.org/prb/abstract/10.1103/PhysRevB.98.045103}
  {\bibfield  {journal} {\bibinfo  {journal} {Physical Review B}\ }\textbf
  {\bibinfo {volume} {98}},\ \bibinfo {pages} {045103} (\bibinfo {year}
  {2018})}\BibitemShut {NoStop}%
\bibitem [{\citenamefont {Kang}\ and\ \citenamefont
  {Vafek}(2019)}]{kang_strong_2019}%
  \BibitemOpen
  \bibfield  {author} {\bibinfo {author} {\bibfnamefont {J.}~\bibnamefont
  {Kang}}\ and\ \bibinfo {author} {\bibfnamefont {O.}~\bibnamefont {Vafek}},\
  }\href {https://doi.org/10.1103/PhysRevLett.122.246401} {\bibfield  {journal}
  {\bibinfo  {journal} {Physical Review Letters}\ }\textbf {\bibinfo {volume}
  {122}},\ \bibinfo {pages} {246401} (\bibinfo {year} {2019})}\BibitemShut
  {NoStop}%
\bibitem [{\citenamefont {Bultinck}\ \emph
  {et~al.}(2020{\natexlab{b}})\citenamefont {Bultinck}, \citenamefont {Khalaf},
  \citenamefont {Liu}, \citenamefont {Chatterjee}, \citenamefont {Vishwanath},\
  and\ \citenamefont {Zaletel}}]{bultinck_ground_2020}%
  \BibitemOpen
  \bibfield  {author} {\bibinfo {author} {\bibfnamefont {N.}~\bibnamefont
  {Bultinck}}, \bibinfo {author} {\bibfnamefont {E.}~\bibnamefont {Khalaf}},
  \bibinfo {author} {\bibfnamefont {S.}~\bibnamefont {Liu}}, \bibinfo {author}
  {\bibfnamefont {S.}~\bibnamefont {Chatterjee}}, \bibinfo {author}
  {\bibfnamefont {A.}~\bibnamefont {Vishwanath}},\ and\ \bibinfo {author}
  {\bibfnamefont {M.~P.}\ \bibnamefont {Zaletel}},\ }\href
  {https://doi.org/10.1103/PhysRevX.10.031034} {\bibfield  {journal} {\bibinfo
  {journal} {Phys. Rev. X}\ }\textbf {\bibinfo {volume} {10}},\ \bibinfo
  {pages} {031034} (\bibinfo {year} {2020}{\natexlab{b}})}\BibitemShut
  {NoStop}%
\bibitem [{\citenamefont {Seo}\ \emph {et~al.}(2019)\citenamefont {Seo},
  \citenamefont {Kotov},\ and\ \citenamefont {Uchoa}}]{seo_ferro_2019}%
  \BibitemOpen
  \bibfield  {author} {\bibinfo {author} {\bibfnamefont {K.}~\bibnamefont
  {Seo}}, \bibinfo {author} {\bibfnamefont {V.~N.}\ \bibnamefont {Kotov}},\
  and\ \bibinfo {author} {\bibfnamefont {B.}~\bibnamefont {Uchoa}},\ }\href
  {https://doi.org/10.1103/PhysRevLett.122.246402} {\bibfield  {journal}
  {\bibinfo  {journal} {Phys. Rev. Lett.}\ }\textbf {\bibinfo {volume} {122}},\
  \bibinfo {pages} {246402} (\bibinfo {year} {2019})}\BibitemShut {NoStop}%
\bibitem [{\citenamefont {Hejazi}\ \emph {et~al.}(2021)\citenamefont {Hejazi},
  \citenamefont {Chen},\ and\ \citenamefont {Balents}}]{hejazi2020hybrid}%
  \BibitemOpen
  \bibfield  {author} {\bibinfo {author} {\bibfnamefont {K.}~\bibnamefont
  {Hejazi}}, \bibinfo {author} {\bibfnamefont {X.}~\bibnamefont {Chen}},\ and\
  \bibinfo {author} {\bibfnamefont {L.}~\bibnamefont {Balents}},\ }\href
  {https://doi.org/10.1103/PhysRevResearch.3.013242} {\bibinfo {title} {Hybrid
  wannier chern bands in magic angle twisted bilayer graphene and the quantized
  anomalous hall effect}} (\bibinfo {year} {2021})\BibitemShut {NoStop}%
\bibitem [{\citenamefont {Khalaf}\ \emph {et~al.}(2020)\citenamefont {Khalaf},
  \citenamefont {Chatterjee}, \citenamefont {Bultinck}, \citenamefont
  {Zaletel},\ and\ \citenamefont {Vishwanath}}]{khalaf_charged_2020}%
  \BibitemOpen
  \bibfield  {author} {\bibinfo {author} {\bibfnamefont {E.}~\bibnamefont
  {Khalaf}}, \bibinfo {author} {\bibfnamefont {S.}~\bibnamefont {Chatterjee}},
  \bibinfo {author} {\bibfnamefont {N.}~\bibnamefont {Bultinck}}, \bibinfo
  {author} {\bibfnamefont {M.~P.}\ \bibnamefont {Zaletel}},\ and\ \bibinfo
  {author} {\bibfnamefont {A.}~\bibnamefont {Vishwanath}},\ }\href@noop {}
  {\bibinfo {title} {Charged skyrmions and topological origin of
  superconductivity in magic angle graphene}} (\bibinfo {year} {2020}),\
  \Eprint {https://arxiv.org/abs/2004.00638} {arXiv:2004.00638
  [cond-mat.str-el]} \BibitemShut {NoStop}%
\bibitem [{\citenamefont {Po}\ \emph {et~al.}(2018)\citenamefont {Po},
  \citenamefont {Zou}, \citenamefont {Vishwanath},\ and\ \citenamefont
  {Senthil}}]{po_origin_2018}%
  \BibitemOpen
  \bibfield  {author} {\bibinfo {author} {\bibfnamefont {H.~C.}\ \bibnamefont
  {Po}}, \bibinfo {author} {\bibfnamefont {L.}~\bibnamefont {Zou}}, \bibinfo
  {author} {\bibfnamefont {A.}~\bibnamefont {Vishwanath}},\ and\ \bibinfo
  {author} {\bibfnamefont {T.}~\bibnamefont {Senthil}},\ }\href
  {https://doi.org/10.1103/PhysRevX.8.031089} {\bibfield  {journal} {\bibinfo
  {journal} {Physical Review X}\ }\textbf {\bibinfo {volume} {8}},\ \bibinfo
  {pages} {031089} (\bibinfo {year} {2018})}\BibitemShut {NoStop}%
\bibitem [{\citenamefont {Xie}\ \emph {et~al.}(2020)\citenamefont {Xie},
  \citenamefont {Song}, \citenamefont {Lian},\ and\ \citenamefont
  {Bernevig}}]{xie_superfluid_2020}%
  \BibitemOpen
  \bibfield  {author} {\bibinfo {author} {\bibfnamefont {F.}~\bibnamefont
  {Xie}}, \bibinfo {author} {\bibfnamefont {Z.}~\bibnamefont {Song}}, \bibinfo
  {author} {\bibfnamefont {B.}~\bibnamefont {Lian}},\ and\ \bibinfo {author}
  {\bibfnamefont {B.~A.}\ \bibnamefont {Bernevig}},\ }\href
  {https://doi.org/10.1103/PhysRevLett.124.167002} {\bibfield  {journal}
  {\bibinfo  {journal} {Phys. Rev. Lett.}\ }\textbf {\bibinfo {volume} {124}},\
  \bibinfo {pages} {167002} (\bibinfo {year} {2020})}\BibitemShut {NoStop}%
\bibitem [{\citenamefont {Julku}\ \emph {et~al.}(2020)\citenamefont {Julku},
  \citenamefont {Peltonen}, \citenamefont {Liang}, \citenamefont {Heikkilä},\
  and\ \citenamefont {Törmä}}]{julku_superfluid_2020}%
  \BibitemOpen
  \bibfield  {author} {\bibinfo {author} {\bibfnamefont {A.}~\bibnamefont
  {Julku}}, \bibinfo {author} {\bibfnamefont {T.~J.}\ \bibnamefont {Peltonen}},
  \bibinfo {author} {\bibfnamefont {L.}~\bibnamefont {Liang}}, \bibinfo
  {author} {\bibfnamefont {T.~T.}\ \bibnamefont {Heikkilä}},\ and\ \bibinfo
  {author} {\bibfnamefont {P.}~\bibnamefont {Törmä}},\ }\bibfield  {journal}
  {\bibinfo  {journal} {Physical Review B}\ }\textbf {\bibinfo {volume}
  {101}},\ \href {https://doi.org/10.1103/physrevb.101.060505}
  {10.1103/physrevb.101.060505} (\bibinfo {year} {2020})\BibitemShut {NoStop}%
\bibitem [{\citenamefont {Hu}\ \emph {et~al.}(2019)\citenamefont {Hu},
  \citenamefont {Hyart}, \citenamefont {Pikulin},\ and\ \citenamefont
  {Rossi}}]{hu2019_superfluid}%
  \BibitemOpen
  \bibfield  {author} {\bibinfo {author} {\bibfnamefont {X.}~\bibnamefont
  {Hu}}, \bibinfo {author} {\bibfnamefont {T.}~\bibnamefont {Hyart}}, \bibinfo
  {author} {\bibfnamefont {D.~I.}\ \bibnamefont {Pikulin}},\ and\ \bibinfo
  {author} {\bibfnamefont {E.}~\bibnamefont {Rossi}},\ }\href
  {https://doi.org/10.1103/PhysRevLett.123.237002} {\bibfield  {journal}
  {\bibinfo  {journal} {Phys. Rev. Lett.}\ }\textbf {\bibinfo {volume} {123}},\
  \bibinfo {pages} {237002} (\bibinfo {year} {2019})}\BibitemShut {NoStop}%
\bibitem [{\citenamefont {Kang}\ and\ \citenamefont
  {Vafek}(2020)}]{kang_nonabelian_2020}%
  \BibitemOpen
  \bibfield  {author} {\bibinfo {author} {\bibfnamefont {J.}~\bibnamefont
  {Kang}}\ and\ \bibinfo {author} {\bibfnamefont {O.}~\bibnamefont {Vafek}},\
  }\href {https://doi.org/10.1103/PhysRevB.102.035161} {\bibfield  {journal}
  {\bibinfo  {journal} {Phys. Rev. B}\ }\textbf {\bibinfo {volume} {102}},\
  \bibinfo {pages} {035161} (\bibinfo {year} {2020})}\BibitemShut {NoStop}%
\bibitem [{\citenamefont {Soejima}\ \emph {et~al.}(2020)\citenamefont
  {Soejima}, \citenamefont {Parker}, \citenamefont {Bultinck}, \citenamefont
  {Hauschild},\ and\ \citenamefont {Zaletel}}]{soejima2020efficient}%
  \BibitemOpen
  \bibfield  {author} {\bibinfo {author} {\bibfnamefont {T.}~\bibnamefont
  {Soejima}}, \bibinfo {author} {\bibfnamefont {D.~E.}\ \bibnamefont {Parker}},
  \bibinfo {author} {\bibfnamefont {N.}~\bibnamefont {Bultinck}}, \bibinfo
  {author} {\bibfnamefont {J.}~\bibnamefont {Hauschild}},\ and\ \bibinfo
  {author} {\bibfnamefont {M.~P.}\ \bibnamefont {Zaletel}},\ }\href
  {https://doi.org/10.1103/PhysRevB.102.205111} {\bibinfo {title} {Efficient
  simulation of moir\'e materials using the density matrix renormalization
  group}} (\bibinfo {year} {2020})\BibitemShut {NoStop}%
\bibitem [{\citenamefont {Pixley}\ and\ \citenamefont
  {Andrei}(2019)}]{pixley2019}%
  \BibitemOpen
  \bibfield  {author} {\bibinfo {author} {\bibfnamefont {J.~H.}\ \bibnamefont
  {Pixley}}\ and\ \bibinfo {author} {\bibfnamefont {E.~Y.}\ \bibnamefont
  {Andrei}},\ }\href {https://doi.org/10.1126/science.aay3409} {\bibfield
  {journal} {\bibinfo  {journal} {Science}\ }\textbf {\bibinfo {volume}
  {365}},\ \bibinfo {pages} {543} (\bibinfo {year} {2019})},\ \Eprint
  {https://arxiv.org/abs/https://science.sciencemag.org/content/365/6453/543.full.pdf}
  {https://science.sciencemag.org/content/365/6453/543.full.pdf} \BibitemShut
  {NoStop}%
\bibitem [{\citenamefont {K\"onig}\ \emph {et~al.}(2020)\citenamefont
  {K\"onig}, \citenamefont {Coleman},\ and\ \citenamefont
  {Tsvelik}}]{knig2020spin}%
  \BibitemOpen
  \bibfield  {author} {\bibinfo {author} {\bibfnamefont {E.~J.}\ \bibnamefont
  {K\"onig}}, \bibinfo {author} {\bibfnamefont {P.}~\bibnamefont {Coleman}},\
  and\ \bibinfo {author} {\bibfnamefont {A.~M.}\ \bibnamefont {Tsvelik}},\
  }\href {https://doi.org/10.1103/PhysRevB.102.104514} {\bibinfo {title} {Spin
  magnetometry as a probe of stripe superconductivity in twisted bilayer
  graphene}} (\bibinfo {year} {2020})\BibitemShut {NoStop}%
\bibitem [{\citenamefont {Christos}\ \emph {et~al.}(2020)\citenamefont
  {Christos}, \citenamefont {Sachdev},\ and\ \citenamefont
  {Scheurer}}]{christos2020superconductivity}%
  \BibitemOpen
  \bibfield  {author} {\bibinfo {author} {\bibfnamefont {M.}~\bibnamefont
  {Christos}}, \bibinfo {author} {\bibfnamefont {S.}~\bibnamefont {Sachdev}},\
  and\ \bibinfo {author} {\bibfnamefont {M.~S.}\ \bibnamefont {Scheurer}},\
  }\href {https://doi.org/10.1073/pnas.2014691117} {\bibinfo {title}
  {Superconductivity, correlated insulators, and wess-zumino-witten terms in
  twisted bilayer graphene}} (\bibinfo {year} {2020}),\ \Eprint
  {https://arxiv.org/abs/https://www.pnas.org/doi/pdf/10.1073/pnas.2014691117}
  {https://www.pnas.org/doi/pdf/10.1073/pnas.2014691117} \BibitemShut {NoStop}%
\bibitem [{\citenamefont {Lewandowski}\ \emph {et~al.}(2021)\citenamefont
  {Lewandowski}, \citenamefont {Chowdhury},\ and\ \citenamefont
  {Ruhman}}]{lewandowski2020pairing}%
  \BibitemOpen
  \bibfield  {author} {\bibinfo {author} {\bibfnamefont {C.}~\bibnamefont
  {Lewandowski}}, \bibinfo {author} {\bibfnamefont {D.}~\bibnamefont
  {Chowdhury}},\ and\ \bibinfo {author} {\bibfnamefont {J.}~\bibnamefont
  {Ruhman}},\ }\href {https://doi.org/10.1103/PhysRevB.103.235401} {\bibinfo
  {title} {Pairing in magic-angle twisted bilayer graphene: Role of phonon and
  plasmon umklapp}} (\bibinfo {year} {2021})\BibitemShut {NoStop}%
\bibitem [{\citenamefont {Xie}\ and\ \citenamefont
  {MacDonald}(2020)}]{xie_HF_2020}%
  \BibitemOpen
  \bibfield  {author} {\bibinfo {author} {\bibfnamefont {M.}~\bibnamefont
  {Xie}}\ and\ \bibinfo {author} {\bibfnamefont {A.~H.}\ \bibnamefont
  {MacDonald}},\ }\href {https://doi.org/10.1103/PhysRevLett.124.097601}
  {\bibfield  {journal} {\bibinfo  {journal} {Phys. Rev. Lett.}\ }\textbf
  {\bibinfo {volume} {124}},\ \bibinfo {pages} {097601} (\bibinfo {year}
  {2020})}\BibitemShut {NoStop}%
\bibitem [{\citenamefont {Liu}\ and\ \citenamefont
  {Dai}(2021)}]{liu2020theories}%
  \BibitemOpen
  \bibfield  {author} {\bibinfo {author} {\bibfnamefont {J.}~\bibnamefont
  {Liu}}\ and\ \bibinfo {author} {\bibfnamefont {X.}~\bibnamefont {Dai}},\
  }\href {https://doi.org/10.1103/PhysRevB.103.035427} {\bibinfo {title}
  {Theories for the correlated insulating states and quantum anomalous hall
  effect phenomena in twisted bilayer graphene}} (\bibinfo {year}
  {2021})\BibitemShut {NoStop}%
\bibitem [{\citenamefont {Cea}\ and\ \citenamefont
  {Guinea}(2020)}]{cea_band_2020}%
  \BibitemOpen
  \bibfield  {author} {\bibinfo {author} {\bibfnamefont {T.}~\bibnamefont
  {Cea}}\ and\ \bibinfo {author} {\bibfnamefont {F.}~\bibnamefont {Guinea}},\
  }\href {https://doi.org/10.1103/PhysRevB.102.045107} {\bibfield  {journal}
  {\bibinfo  {journal} {Phys. Rev. B}\ }\textbf {\bibinfo {volume} {102}},\
  \bibinfo {pages} {045107} (\bibinfo {year} {2020})}\BibitemShut {NoStop}%
\bibitem [{\citenamefont {Zhang}\ \emph {et~al.}(2020)\citenamefont {Zhang},
  \citenamefont {Jiang}, \citenamefont {Wang},\ and\ \citenamefont
  {Zhang}}]{zhang_HF_2020}%
  \BibitemOpen
  \bibfield  {author} {\bibinfo {author} {\bibfnamefont {Y.}~\bibnamefont
  {Zhang}}, \bibinfo {author} {\bibfnamefont {K.}~\bibnamefont {Jiang}},
  \bibinfo {author} {\bibfnamefont {Z.}~\bibnamefont {Wang}},\ and\ \bibinfo
  {author} {\bibfnamefont {F.}~\bibnamefont {Zhang}},\ }\href
  {https://doi.org/10.1103/PhysRevB.102.035136} {\bibfield  {journal} {\bibinfo
   {journal} {Phys. Rev. B}\ }\textbf {\bibinfo {volume} {102}},\ \bibinfo
  {pages} {035136} (\bibinfo {year} {2020})}\BibitemShut {NoStop}%
\bibitem [{\citenamefont {Liu}\ \emph {et~al.}(2021{\natexlab{b}})\citenamefont
  {Liu}, \citenamefont {Khalaf}, \citenamefont {Lee},\ and\ \citenamefont
  {Vishwanath}}]{liu2020nematic}%
  \BibitemOpen
  \bibfield  {author} {\bibinfo {author} {\bibfnamefont {S.}~\bibnamefont
  {Liu}}, \bibinfo {author} {\bibfnamefont {E.}~\bibnamefont {Khalaf}},
  \bibinfo {author} {\bibfnamefont {J.~Y.}\ \bibnamefont {Lee}},\ and\ \bibinfo
  {author} {\bibfnamefont {A.}~\bibnamefont {Vishwanath}},\ }\href
  {https://doi.org/10.1103/PhysRevResearch.3.013033} {\bibinfo {title} {Nematic
  topological semimetal and insulator in magic-angle bilayer graphene at charge
  neutrality}} (\bibinfo {year} {2021}{\natexlab{b}})\BibitemShut {NoStop}%
\bibitem [{\citenamefont {Da~Liao}\ \emph {et~al.}(2019)\citenamefont
  {Da~Liao}, \citenamefont {Meng},\ and\ \citenamefont {Xu}}]{daliao_VBO_2019}%
  \BibitemOpen
  \bibfield  {author} {\bibinfo {author} {\bibfnamefont {Y.}~\bibnamefont
  {Da~Liao}}, \bibinfo {author} {\bibfnamefont {Z.~Y.}\ \bibnamefont {Meng}},\
  and\ \bibinfo {author} {\bibfnamefont {X.~Y.}\ \bibnamefont {Xu}},\ }\href
  {https://doi.org/10.1103/PhysRevLett.123.157601} {\bibfield  {journal}
  {\bibinfo  {journal} {Phys. Rev. Lett.}\ }\textbf {\bibinfo {volume} {123}},\
  \bibinfo {pages} {157601} (\bibinfo {year} {2019})}\BibitemShut {NoStop}%
\bibitem [{\citenamefont {Da~Liao}\ \emph {et~al.}(2021)\citenamefont
  {Da~Liao}, \citenamefont {Kang}, \citenamefont {Brei\o{}}, \citenamefont
  {Xu}, \citenamefont {Wu}, \citenamefont {Andersen}, \citenamefont
  {Fernandes},\ and\ \citenamefont {Meng}}]{daliao2020correlation}%
  \BibitemOpen
  \bibfield  {author} {\bibinfo {author} {\bibfnamefont {Y.}~\bibnamefont
  {Da~Liao}}, \bibinfo {author} {\bibfnamefont {J.}~\bibnamefont {Kang}},
  \bibinfo {author} {\bibfnamefont {C.~N.}\ \bibnamefont {Brei\o{}}}, \bibinfo
  {author} {\bibfnamefont {X.~Y.}\ \bibnamefont {Xu}}, \bibinfo {author}
  {\bibfnamefont {H.-Q.}\ \bibnamefont {Wu}}, \bibinfo {author} {\bibfnamefont
  {B.~M.}\ \bibnamefont {Andersen}}, \bibinfo {author} {\bibfnamefont {R.~M.}\
  \bibnamefont {Fernandes}},\ and\ \bibinfo {author} {\bibfnamefont {Z.~Y.}\
  \bibnamefont {Meng}},\ }\href {https://doi.org/10.1103/PhysRevX.11.011014}
  {\bibinfo {title} {Correlation-induced insulating topological phases at
  charge neutrality in twisted bilayer graphene}} (\bibinfo {year}
  {2021})\BibitemShut {NoStop}%
\bibitem [{\citenamefont {Classen}\ \emph {et~al.}(2019)\citenamefont
  {Classen}, \citenamefont {Honerkamp},\ and\ \citenamefont
  {Scherer}}]{classen2019competing}%
  \BibitemOpen
  \bibfield  {author} {\bibinfo {author} {\bibfnamefont {L.}~\bibnamefont
  {Classen}}, \bibinfo {author} {\bibfnamefont {C.}~\bibnamefont {Honerkamp}},\
  and\ \bibinfo {author} {\bibfnamefont {M.~M.}\ \bibnamefont {Scherer}},\
  }\href@noop {} {\bibfield  {journal} {\bibinfo  {journal} {Physical Review
  B}\ }\textbf {\bibinfo {volume} {99}},\ \bibinfo {pages} {195120} (\bibinfo
  {year} {2019})}\BibitemShut {NoStop}%
\bibitem [{\citenamefont {Kennes}\ \emph {et~al.}(2018)\citenamefont {Kennes},
  \citenamefont {Lischner},\ and\ \citenamefont {Karrasch}}]{kennes2018strong}%
  \BibitemOpen
  \bibfield  {author} {\bibinfo {author} {\bibfnamefont {D.~M.}\ \bibnamefont
  {Kennes}}, \bibinfo {author} {\bibfnamefont {J.}~\bibnamefont {Lischner}},\
  and\ \bibinfo {author} {\bibfnamefont {C.}~\bibnamefont {Karrasch}},\
  }\href@noop {} {\bibfield  {journal} {\bibinfo  {journal} {Physical Review
  B}\ }\textbf {\bibinfo {volume} {98}},\ \bibinfo {pages} {241407} (\bibinfo
  {year} {2018})}\BibitemShut {NoStop}%
\bibitem [{\citenamefont {Eugenio}\ and\ \citenamefont
  {Dağ}(2020)}]{eugenio2020dmrg}%
  \BibitemOpen
  \bibfield  {author} {\bibinfo {author} {\bibfnamefont {P.~M.}\ \bibnamefont
  {Eugenio}}\ and\ \bibinfo {author} {\bibfnamefont {C.~B.}\ \bibnamefont
  {Dağ}},\ }\href {https://doi.org/10.21468/SciPostPhysCore.3.2.015}
  {\bibfield  {journal} {\bibinfo  {journal} {SciPost Phys. Core}\ }\textbf
  {\bibinfo {volume} {3}},\ \bibinfo {pages} {15} (\bibinfo {year}
  {2020})}\BibitemShut {NoStop}%
\bibitem [{\citenamefont {Huang}\ \emph {et~al.}(2020)\citenamefont {Huang},
  \citenamefont {Hosur},\ and\ \citenamefont {Pal}}]{huang2020deconstructing}%
  \BibitemOpen
  \bibfield  {author} {\bibinfo {author} {\bibfnamefont {Y.}~\bibnamefont
  {Huang}}, \bibinfo {author} {\bibfnamefont {P.}~\bibnamefont {Hosur}},\ and\
  \bibinfo {author} {\bibfnamefont {H.~K.}\ \bibnamefont {Pal}},\ }\href
  {https://doi.org/10.1103/PhysRevB.102.155429} {\bibfield  {journal} {\bibinfo
   {journal} {Phys. Rev. B}\ }\textbf {\bibinfo {volume} {102}},\ \bibinfo
  {pages} {155429} (\bibinfo {year} {2020})}\BibitemShut {NoStop}%
\bibitem [{\citenamefont {Huang}\ \emph {et~al.}(2019)\citenamefont {Huang},
  \citenamefont {Zhang},\ and\ \citenamefont
  {Ma}}]{huang2019antiferromagnetically}%
  \BibitemOpen
  \bibfield  {author} {\bibinfo {author} {\bibfnamefont {T.}~\bibnamefont
  {Huang}}, \bibinfo {author} {\bibfnamefont {L.}~\bibnamefont {Zhang}},\ and\
  \bibinfo {author} {\bibfnamefont {T.}~\bibnamefont {Ma}},\ }\href@noop {}
  {\bibfield  {journal} {\bibinfo  {journal} {Science Bulletin}\ }\textbf
  {\bibinfo {volume} {64}},\ \bibinfo {pages} {310} (\bibinfo {year}
  {2019})}\BibitemShut {NoStop}%
\bibitem [{\citenamefont {Guo}\ \emph {et~al.}(2018)\citenamefont {Guo},
  \citenamefont {Zhu}, \citenamefont {Feng},\ and\ \citenamefont
  {Scalettar}}]{guo2018pairing}%
  \BibitemOpen
  \bibfield  {author} {\bibinfo {author} {\bibfnamefont {H.}~\bibnamefont
  {Guo}}, \bibinfo {author} {\bibfnamefont {X.}~\bibnamefont {Zhu}}, \bibinfo
  {author} {\bibfnamefont {S.}~\bibnamefont {Feng}},\ and\ \bibinfo {author}
  {\bibfnamefont {R.~T.}\ \bibnamefont {Scalettar}},\ }\href@noop {} {\bibfield
   {journal} {\bibinfo  {journal} {Physical Review B}\ }\textbf {\bibinfo
  {volume} {97}},\ \bibinfo {pages} {235453} (\bibinfo {year}
  {2018})}\BibitemShut {NoStop}%
\bibitem [{\citenamefont {Ledwith}\ \emph {et~al.}(2020)\citenamefont
  {Ledwith}, \citenamefont {Tarnopolsky}, \citenamefont {Khalaf},\ and\
  \citenamefont {Vishwanath}}]{ledwith2020}%
  \BibitemOpen
  \bibfield  {author} {\bibinfo {author} {\bibfnamefont {P.~J.}\ \bibnamefont
  {Ledwith}}, \bibinfo {author} {\bibfnamefont {G.}~\bibnamefont
  {Tarnopolsky}}, \bibinfo {author} {\bibfnamefont {E.}~\bibnamefont
  {Khalaf}},\ and\ \bibinfo {author} {\bibfnamefont {A.}~\bibnamefont
  {Vishwanath}},\ }\href {https://doi.org/10.1103/PhysRevResearch.2.023237}
  {\bibfield  {journal} {\bibinfo  {journal} {Phys. Rev. Research}\ }\textbf
  {\bibinfo {volume} {2}},\ \bibinfo {pages} {023237} (\bibinfo {year}
  {2020})}\BibitemShut {NoStop}%
\bibitem [{\citenamefont {Repellin}\ \emph {et~al.}(2020)\citenamefont
  {Repellin}, \citenamefont {Dong}, \citenamefont {Zhang},\ and\ \citenamefont
  {Senthil}}]{repellin_EDDMRG_2020}%
  \BibitemOpen
  \bibfield  {author} {\bibinfo {author} {\bibfnamefont {C.}~\bibnamefont
  {Repellin}}, \bibinfo {author} {\bibfnamefont {Z.}~\bibnamefont {Dong}},
  \bibinfo {author} {\bibfnamefont {Y.-H.}\ \bibnamefont {Zhang}},\ and\
  \bibinfo {author} {\bibfnamefont {T.}~\bibnamefont {Senthil}},\ }\href
  {https://doi.org/10.1103/PhysRevLett.124.187601} {\bibfield  {journal}
  {\bibinfo  {journal} {Phys. Rev. Lett.}\ }\textbf {\bibinfo {volume} {124}},\
  \bibinfo {pages} {187601} (\bibinfo {year} {2020})}\BibitemShut {NoStop}%
\bibitem [{\citenamefont {Abouelkomsan}\ \emph {et~al.}(2020)\citenamefont
  {Abouelkomsan}, \citenamefont {Liu},\ and\ \citenamefont
  {Bergholtz}}]{abouelkomsan2020}%
  \BibitemOpen
  \bibfield  {author} {\bibinfo {author} {\bibfnamefont {A.}~\bibnamefont
  {Abouelkomsan}}, \bibinfo {author} {\bibfnamefont {Z.}~\bibnamefont {Liu}},\
  and\ \bibinfo {author} {\bibfnamefont {E.~J.}\ \bibnamefont {Bergholtz}},\
  }\href {https://doi.org/10.1103/PhysRevLett.124.106803} {\bibfield  {journal}
  {\bibinfo  {journal} {Phys. Rev. Lett.}\ }\textbf {\bibinfo {volume} {124}},\
  \bibinfo {pages} {106803} (\bibinfo {year} {2020})}\BibitemShut {NoStop}%
\bibitem [{\citenamefont {Repellin}\ and\ \citenamefont
  {Senthil}(2020)}]{repellin_FCI_2020}%
  \BibitemOpen
  \bibfield  {author} {\bibinfo {author} {\bibfnamefont {C.}~\bibnamefont
  {Repellin}}\ and\ \bibinfo {author} {\bibfnamefont {T.}~\bibnamefont
  {Senthil}},\ }\href {https://doi.org/10.1103/PhysRevResearch.2.023238}
  {\bibfield  {journal} {\bibinfo  {journal} {Phys. Rev. Research}\ }\textbf
  {\bibinfo {volume} {2}},\ \bibinfo {pages} {023238} (\bibinfo {year}
  {2020})}\BibitemShut {NoStop}%
\bibitem [{\citenamefont {Vafek}\ and\ \citenamefont
  {Kang}(2020)}]{vafek2020hidden}%
  \BibitemOpen
  \bibfield  {author} {\bibinfo {author} {\bibfnamefont {O.}~\bibnamefont
  {Vafek}}\ and\ \bibinfo {author} {\bibfnamefont {J.}~\bibnamefont {Kang}},\
  }\href@noop {} {\bibinfo {title} {Towards the hidden symmetry in coulomb
  interacting twisted bilayer graphene: renormalization group approach}}
  (\bibinfo {year} {2020}),\ \Eprint {https://arxiv.org/abs/2009.09413}
  {arXiv:2009.09413 [cond-mat.str-el]} \BibitemShut {NoStop}%
\bibitem [{\citenamefont {Fernandes}\ and\ \citenamefont
  {Venderbos}(2020)}]{fernandes_nematic_2020}%
  \BibitemOpen
  \bibfield  {author} {\bibinfo {author} {\bibfnamefont {R.~M.}\ \bibnamefont
  {Fernandes}}\ and\ \bibinfo {author} {\bibfnamefont {J.~W.~F.}\ \bibnamefont
  {Venderbos}},\ }\bibfield  {journal} {\bibinfo  {journal} {Science Advances}\
  }\textbf {\bibinfo {volume} {6}},\ \href
  {https://doi.org/10.1126/sciadv.aba8834} {10.1126/sciadv.aba8834} (\bibinfo
  {year} {2020}),\ \Eprint
  {https://arxiv.org/abs/https://advances.sciencemag.org/content/6/32/eaba8834.full.pdf}
  {https://advances.sciencemag.org/content/6/32/eaba8834.full.pdf} \BibitemShut
  {NoStop}%
\bibitem [{\citenamefont {Wilson}\ \emph {et~al.}(2020)\citenamefont {Wilson},
  \citenamefont {Fu}, \citenamefont {Das~Sarma},\ and\ \citenamefont
  {Pixley}}]{Wilson2020TBG}%
  \BibitemOpen
  \bibfield  {author} {\bibinfo {author} {\bibfnamefont {J.~H.}\ \bibnamefont
  {Wilson}}, \bibinfo {author} {\bibfnamefont {Y.}~\bibnamefont {Fu}}, \bibinfo
  {author} {\bibfnamefont {S.}~\bibnamefont {Das~Sarma}},\ and\ \bibinfo
  {author} {\bibfnamefont {J.~H.}\ \bibnamefont {Pixley}},\ }\href
  {https://doi.org/10.1103/PhysRevResearch.2.023325} {\bibfield  {journal}
  {\bibinfo  {journal} {Phys. Rev. Research}\ }\textbf {\bibinfo {volume}
  {2}},\ \bibinfo {pages} {023325} (\bibinfo {year} {2020})}\BibitemShut
  {NoStop}%
\bibitem [{\citenamefont {Wang}\ \emph {et~al.}(2021)\citenamefont {Wang},
  \citenamefont {Zheng}, \citenamefont {Millis},\ and\ \citenamefont
  {Cano}}]{wang2020chiral}%
  \BibitemOpen
  \bibfield  {author} {\bibinfo {author} {\bibfnamefont {J.}~\bibnamefont
  {Wang}}, \bibinfo {author} {\bibfnamefont {Y.}~\bibnamefont {Zheng}},
  \bibinfo {author} {\bibfnamefont {A.~J.}\ \bibnamefont {Millis}},\ and\
  \bibinfo {author} {\bibfnamefont {J.}~\bibnamefont {Cano}},\ }\href
  {https://doi.org/10.1103/PhysRevResearch.3.023155} {\bibinfo {title} {Chiral
  approximation to twisted bilayer graphene: Exact intravalley inversion
  symmetry, nodal structure, and implications for higher magic angles}}
  (\bibinfo {year} {2021})\BibitemShut {NoStop}%
\bibitem [{\citenamefont {Bernevig}\ \emph
  {et~al.}(2021{\natexlab{a}})\citenamefont {Bernevig}, \citenamefont {Song},
  \citenamefont {Regnault},\ and\ \citenamefont {Lian}}]{ourpaper1}%
  \BibitemOpen
  \bibfield  {author} {\bibinfo {author} {\bibfnamefont {B.~A.}\ \bibnamefont
  {Bernevig}}, \bibinfo {author} {\bibfnamefont {Z.-D.}\ \bibnamefont {Song}},
  \bibinfo {author} {\bibfnamefont {N.}~\bibnamefont {Regnault}},\ and\
  \bibinfo {author} {\bibfnamefont {B.}~\bibnamefont {Lian}},\ }\bibfield
  {journal} {\bibinfo  {journal} {Physical Review B}\ }\textbf {\bibinfo
  {volume} {103}},\ \href {https://doi.org/10.1103/physrevb.103.205411}
  {10.1103/physrevb.103.205411} (\bibinfo {year}
  {2021}{\natexlab{a}})\BibitemShut {NoStop}%
\bibitem [{\citenamefont {Song}\ \emph {et~al.}(2021)\citenamefont {Song},
  \citenamefont {Lian}, \citenamefont {Regnault},\ and\ \citenamefont
  {Bernevig}}]{ourpaper2}%
  \BibitemOpen
  \bibfield  {author} {\bibinfo {author} {\bibfnamefont {Z.-D.}\ \bibnamefont
  {Song}}, \bibinfo {author} {\bibfnamefont {B.}~\bibnamefont {Lian}}, \bibinfo
  {author} {\bibfnamefont {N.}~\bibnamefont {Regnault}},\ and\ \bibinfo
  {author} {\bibfnamefont {B.~A.}\ \bibnamefont {Bernevig}},\ }\bibfield
  {journal} {\bibinfo  {journal} {Physical Review B}\ }\textbf {\bibinfo
  {volume} {103}},\ \href {https://doi.org/10.1103/physrevb.103.205412}
  {10.1103/physrevb.103.205412} (\bibinfo {year} {2021})\BibitemShut {NoStop}%
\bibitem [{\citenamefont {Bernevig}\ \emph
  {et~al.}(2021{\natexlab{b}})\citenamefont {Bernevig}, \citenamefont {Song},
  \citenamefont {Regnault},\ and\ \citenamefont {Lian}}]{ourpaper3}%
  \BibitemOpen
  \bibfield  {author} {\bibinfo {author} {\bibfnamefont {B.~A.}\ \bibnamefont
  {Bernevig}}, \bibinfo {author} {\bibfnamefont {Z.-D.}\ \bibnamefont {Song}},
  \bibinfo {author} {\bibfnamefont {N.}~\bibnamefont {Regnault}},\ and\
  \bibinfo {author} {\bibfnamefont {B.}~\bibnamefont {Lian}},\ }\bibfield
  {journal} {\bibinfo  {journal} {Physical Review B}\ }\textbf {\bibinfo
  {volume} {103}},\ \href {https://doi.org/10.1103/physrevb.103.205413}
  {10.1103/physrevb.103.205413} (\bibinfo {year}
  {2021}{\natexlab{b}})\BibitemShut {NoStop}%
\bibitem [{\citenamefont {Lian}\ \emph {et~al.}(2021)\citenamefont {Lian},
  \citenamefont {Song}, \citenamefont {Regnault}, \citenamefont {Efetov},
  \citenamefont {Yazdani},\ and\ \citenamefont {Bernevig}}]{ourpaper4}%
  \BibitemOpen
  \bibfield  {author} {\bibinfo {author} {\bibfnamefont {B.}~\bibnamefont
  {Lian}}, \bibinfo {author} {\bibfnamefont {Z.-D.}\ \bibnamefont {Song}},
  \bibinfo {author} {\bibfnamefont {N.}~\bibnamefont {Regnault}}, \bibinfo
  {author} {\bibfnamefont {D.~K.}\ \bibnamefont {Efetov}}, \bibinfo {author}
  {\bibfnamefont {A.}~\bibnamefont {Yazdani}},\ and\ \bibinfo {author}
  {\bibfnamefont {B.~A.}\ \bibnamefont {Bernevig}},\ }\bibfield  {journal}
  {\bibinfo  {journal} {Physical Review B}\ }\textbf {\bibinfo {volume}
  {103}},\ \href {https://doi.org/10.1103/physrevb.103.205414}
  {10.1103/physrevb.103.205414} (\bibinfo {year} {2021})\BibitemShut {NoStop}%
\bibitem [{\citenamefont {Bernevig}\ \emph
  {et~al.}(2021{\natexlab{c}})\citenamefont {Bernevig}, \citenamefont {Lian},
  \citenamefont {Cowsik}, \citenamefont {Xie}, \citenamefont {Regnault},\ and\
  \citenamefont {Song}}]{ourpaper5}%
  \BibitemOpen
  \bibfield  {author} {\bibinfo {author} {\bibfnamefont {B.~A.}\ \bibnamefont
  {Bernevig}}, \bibinfo {author} {\bibfnamefont {B.}~\bibnamefont {Lian}},
  \bibinfo {author} {\bibfnamefont {A.}~\bibnamefont {Cowsik}}, \bibinfo
  {author} {\bibfnamefont {F.}~\bibnamefont {Xie}}, \bibinfo {author}
  {\bibfnamefont {N.}~\bibnamefont {Regnault}},\ and\ \bibinfo {author}
  {\bibfnamefont {Z.-D.}\ \bibnamefont {Song}},\ }\bibfield  {journal}
  {\bibinfo  {journal} {Physical Review B}\ }\textbf {\bibinfo {volume}
  {103}},\ \href {https://doi.org/10.1103/physrevb.103.205415}
  {10.1103/physrevb.103.205415} (\bibinfo {year}
  {2021}{\natexlab{c}})\BibitemShut {NoStop}%
\bibitem [{\citenamefont {Wu}\ \emph {et~al.}(2019{\natexlab{b}})\citenamefont
  {Wu}, \citenamefont {Hwang},\ and\ \citenamefont
  {Das~Sarma}}]{wu_phonon_linearT2019}%
  \BibitemOpen
  \bibfield  {author} {\bibinfo {author} {\bibfnamefont {F.}~\bibnamefont
  {Wu}}, \bibinfo {author} {\bibfnamefont {E.}~\bibnamefont {Hwang}},\ and\
  \bibinfo {author} {\bibfnamefont {S.}~\bibnamefont {Das~Sarma}},\ }\href
  {https://doi.org/10.1103/PhysRevB.99.165112} {\bibfield  {journal} {\bibinfo
  {journal} {Phys. Rev. B}\ }\textbf {\bibinfo {volume} {99}},\ \bibinfo
  {pages} {165112} (\bibinfo {year} {2019}{\natexlab{b}})}\BibitemShut
  {NoStop}%
\bibitem [{\citenamefont {Fang}\ \emph {et~al.}(2012)\citenamefont {Fang},
  \citenamefont {Gilbert},\ and\ \citenamefont {Bernevig}}]{fang_bulk_2012}%
  \BibitemOpen
  \bibfield  {author} {\bibinfo {author} {\bibfnamefont {C.}~\bibnamefont
  {Fang}}, \bibinfo {author} {\bibfnamefont {M.~J.}\ \bibnamefont {Gilbert}},\
  and\ \bibinfo {author} {\bibfnamefont {B.~A.}\ \bibnamefont {Bernevig}},\
  }\href {https://doi.org/10.1103/PhysRevB.86.115112} {\bibfield  {journal}
  {\bibinfo  {journal} {Phys. Rev. B}\ }\textbf {\bibinfo {volume} {86}},\
  \bibinfo {pages} {115112} (\bibinfo {year} {2012})}\BibitemShut {NoStop}%
\bibitem [{\citenamefont {Bernevig}\ and\ \citenamefont
  {Regnault}(2012)}]{Bernevig-PhysRevB.85.075128}%
  \BibitemOpen
  \bibfield  {author} {\bibinfo {author} {\bibfnamefont {B.~A.}\ \bibnamefont
  {Bernevig}}\ and\ \bibinfo {author} {\bibfnamefont {N.}~\bibnamefont
  {Regnault}},\ }\href {https://doi.org/10.1103/PhysRevB.85.075128} {\bibfield
  {journal} {\bibinfo  {journal} {Phys. Rev. B}\ }\textbf {\bibinfo {volume}
  {85}},\ \bibinfo {pages} {075128} (\bibinfo {year} {2012})}\BibitemShut
  {NoStop}%
\bibitem [{\citenamefont {Liu}\ \emph {et~al.}(2012)\citenamefont {Liu},
  \citenamefont {Bergholtz}, \citenamefont {Fan},\ and\ \citenamefont
  {L\"auchli}}]{Zhao-PhysRevLett.109.186805}%
  \BibitemOpen
  \bibfield  {author} {\bibinfo {author} {\bibfnamefont {Z.}~\bibnamefont
  {Liu}}, \bibinfo {author} {\bibfnamefont {E.~J.}\ \bibnamefont {Bergholtz}},
  \bibinfo {author} {\bibfnamefont {H.}~\bibnamefont {Fan}},\ and\ \bibinfo
  {author} {\bibfnamefont {A.~M.}\ \bibnamefont {L\"auchli}},\ }\href
  {https://doi.org/10.1103/PhysRevLett.109.186805} {\bibfield  {journal}
  {\bibinfo  {journal} {Phys. Rev. Lett.}\ }\textbf {\bibinfo {volume} {109}},\
  \bibinfo {pages} {186805} (\bibinfo {year} {2012})}\BibitemShut {NoStop}%
\end{thebibliography}%
\bibliographystyle{apsrev4-2}

\appendix

\onecolumngrid
\tableofcontents

\section{Projected many-body Hamiltonian of TBG}\label{ed:appsec:mbhamiltonian}
In this Appendix, we briefly review the definition and symmetries of the non-interacting Hamiltonian of TBG, which was first introduced in Ref. \cite{bistritzer_moire_2011}.  We then derive the projected interacting Hamiltonian matrix elements in terms of  single particle wavefunctions. We also discuss two gauge choices which are beneficial for numerical study. Our notations are identical to the paper Ref. \cite{ourpaper3}, which also provides more detailed derivation and discussion.

\subsection{Single Particle Hamiltonian}\label{app:singleparticle}
We first define the creation operator $c^\dagger_{\vp, \alpha, s, l}$, where $\vp$ is the electron momentum measured from single layer graphene $\Gamma_M$ point, $\alpha = A, B$ is the graphene sublattice, $s = \uparrow, \downarrow$ is the electron spin and $l=\pm 1$ refers to the layer index. 
The low energy physics in TBG is mostly dominated by states around the two Dirac points $K$ and $K'$. By focusing on one valley $K$, we define vectors $\vq_{j} = C_{3z}^{j-1}(\mathbf{K}_- - \mathbf{K}_+)$, where $\mathbf{K}_l$ is the momentum of the Dirac point $K$ in layer $l$, and $|\mathbf{K}_l| = 1.703\,\textrm{\AA}^{-1}$. The reciprocal vectors of the moir\'e lattice, denoted by $\mathcal{Q}_0$, are spanned by basis vectors $\mathbf{b}_{M1} = \vq_{3} - \vq_{1}$ and $\mathbf{b}_{M2} = \vq_{3} - \vq_{2}$. 
The momenta lattices $\mathcal{Q}_\pm = \mathcal{Q}_0 \pm \vq_1$ form a hexagonal lattice in momentum space, and they stand for Dirac points of the top and bottom layers, respectively.
For convenience, we introduce the electron operators:
\begin{equation}
    c^\dagger_{\vk, \mathbf{Q}, \eta, \alpha, s} = 
    c^\dagger_{\eta\mathbf{K}_{\eta\cdot \ell} + \vk - \mathbf{Q}, \alpha, s, \eta\cdot \ell}~~~\text{if}~\mathbf{Q}\in\mathcal{Q}_{\ell}\,.
\end{equation}
Therefore the second quantized non-interacting Hamiltonian of TBG can be written as
\begin{equation}
    \hat{H}_0 = \sum_{\vk\in{\rm MBZ}}\sum_{\mathbf{Q},\mathbf{Q}'\in{\mathcal{Q}_{\pm}}}\sum_{\eta,s, \alpha,\beta}\Big[h^{(\eta)}_{\mathbf{Q} \mathbf{Q}'}(\vk)\Big]_{\alpha\beta}c^\dagger_{\vk,\mathbf{Q},\eta, \alpha, s}c_{\vk,\mathbf{Q}', \eta, \beta, s}\,,
\end{equation}
where MBZ stands for moir\'e Brillouin zone, and the ``first quantized'' single-body Hamiltonian of TBG with valley $\eta = +1$ is given by the following equation Ref.~\cite{bistritzer_moire_2011}:
\begin{equation}\label{ed:eqn:sbh}
\begin{aligned}
h^{(+1)}_{\mathbf{Q} \mathbf{Q}'}(\vk) =& v_F \bm{\sigma}\cdot(\vk - \mathbf{Q})\delta_{\mathbf{Q}, \mathbf{Q}'} + \sum_{j=1,2,3}\left(T_j \delta_{\mathbf{Q} - \mathbf{Q}',\vq_j} + T^\dagger_j \delta_{\mathbf{Q} - \mathbf{Q}', -\vq_j}\right) \\
h^{(-1)}_{\mathbf{Q} \mathbf{Q}'}(\vk) =& v_F \bm{\sigma}^*\cdot(\vk - \mathbf{Q})\delta_{\mathbf{Q}, \mathbf{Q}'} + \sum_{j=1,2,3}\left(\sigma_x T_j\sigma_x \delta_{\mathbf{Q} - \mathbf{Q}',-\vq_j} + \sigma_x T^\dagger_j\sigma_x \delta_{\mathbf{Q} - \mathbf{Q}', \vq_j}\right)
\end{aligned}
\end{equation}
where $\bm{\sigma}=(\sigma_x,\sigma_y)$, $\bm{\sigma}^*=(-\sigma_x,\sigma_y)$, with $\sigma_{0,x,y,z}$ being the 2$\times$2 identity and Pauli matrices, $v_F = 6104.5\,\mathrm{meV}\cdot \textrm{\AA}$ is the Fermi velocity of single layer graphene, and $w = 110\,\rm{meV}$ is the strength of interlayer hopping. Interlayer hopping matrices $T_j$ are given by:
\begin{equation}
    T_j = w_0 \sigma_0 + w_1 \left[\cos\left(\frac{2\pi (j - 1)}{3}\right)\sigma_x + \sin\left(\frac{2\pi (j - 1)}{3}\right)\sigma_y\right]\,.
\end{equation}
The parameters $w_0$ and $w_1$ represents the relative strength of interlayer hopping at the AA and AB stacking centers in TBG. In the original BM model \cite{bistritzer_moire_2011}, both of the two parameters $w_0 = w_1 = 110~\rm meV$. It has been shown that in reality the value of $w_0/w_1$ is smaller than 1 \cite{koshino_maximally_2018}. In this article we set $w_1 = 110~\rm meV$ and use $w_0$ as a tunable parameter.

The single-body Hamitonian at valley $\eta=-1$ is given by $h_{\mathbf{Q},\mathbf{Q}^{\prime}}^{\left(-\right)}\left(\mathbf{k}\right)=\sigma_x h_{-\mathbf{Q},-\mathbf{Q}^{\prime}}^{\left(+\right)}\left(-\mathbf{k}\right) \sigma_x$.
It can be shown that the single valley Hamiltonian is invariant under the crystalline transformations $C_{2z}T$, $C_{3z}$, $C_{2x}$  and a unitary particle hole transformation $P$. These symmetries are represented by the following matrices:
\begin{align}
    D_{\mathbf{Q}\alpha, \mathbf{Q}'\beta}(C_{2z}T) &= \left(\sigma_x\right)_{\alpha, \beta} \delta_{\mathbf{Q},\mathbf{Q}'}\\
    D_{\mathbf{Q},\mathbf{Q}'}\left(C_{3z}\right) &= e^{i\frac{2\pi}{3}\sigma_{z}}\delta_{\mathbf{Q},C_{3z}\mathbf{Q}'}\\
    D_{\mathbf{Q},\mathbf{Q}'}\left(C_{2x}\right) &= \sigma_{x}\delta_{\mathbf{Q},C_{2x}\mathbf{Q}'}\\
    D_{\mathbf{Q}\alpha, \mathbf{Q}'\beta}(P) &= \zeta_\mathbf{Q}\delta_{\mathbf{Q}, -\mathbf{Q}'}\delta_{\alpha,\beta}
\end{align}
where $\zeta_{\mathbf{Q}} = \pm 1$ for $\mathbf{Q}\in \mathcal{Q}_\pm$. It can be shown that these representation matrices satisfy the following relations:
\begin{align}
    h^{(\eta)}(\vk) &= D^\dagger(C_{2z}T)h^{(\eta) *}(\vk)D(C_{2z}T)\\
    h^{(\eta)}(\vk) &= D^\dagger(C_{3z})h^{(\eta)}(C_{3z}\vk)D(C_{3z})\\
    h^{(\eta)}(\vk) &= D^\dagger(C_{2x})h^{(\eta)}(C_{2x}\vk)D(C_{2x})\\
    h^{(\eta)}(\vk) &= -D^\dagger(P)h^{(\eta)}(-\vk)D(P)
\end{align}
Furthermore, at the first chiral limit $w_0 = 0$, the single-body Hamiltonian Eq.~(\ref{ed:eqn:sbh}) has only $\sigma_x$ and $\sigma_y$. Therefore, the chiral symmetry $C$ can be defined, and its representation $D(C)$ satisfies the following equations:
\begin{align}
    D_{\mathbf{Q}\alpha, \mathbf{Q}'\beta}(C) &=(\sigma_z)_{\alpha\beta}\delta_{\mathbf{Q},\mathbf{Q}'}\,,\\
    \left\{D(C), h^{(\eta)}(\vk)\right\} & = 0\,. 
\end{align}

The symmetries discussed in the previous paragraph do not change the valley quantum number. Another symmetry $C_{2z}$, which is represented by $D_{\mathbf{Q}\alpha, \mathbf{Q}'\beta}(C_{2z}) =_x)_{\alpha\beta}\delta_{\mathbf{Q},-\mathbf{Q}'}$, transforms the single-body Hamiltonian in Eq.~(\ref{ed:eqn:sbh}) to the other valley:
\begin{align}
D_{\mathbf{Q}\alpha,\mathbf{Q}'\beta}(C_{2z}) = & (\sigma_x)_{\alpha,\beta} \delta_{\mathbf{Q},-\mathbf{Q}'} \\
h^{(\eta)}(\vk) =& D^\dagger(C_{2z})h^{(-\eta)}(-\vk)D(C_{2z})\,.
\end{align}
The symmetries $C_{2z}$, $C_{2z}T$ and $P$ (and $C$ at the chiral limit) will be used to fix the gauge choices when deriving the matrix elements of projected interacting Hamiltonian.

By diagonalizing the single-body Hamiltonian, we can obtain the band structure $\varepsilon_{\vk, m, \eta}$ and single-body wavefunctions $u^{(\eta)}_{\mathbf{Q}\alpha, m}(\vk)$ of TBG:
\begin{equation}
    \sum_{\mathbf{Q}'\beta} h^{(\eta)}_{\mathbf{Q}\alpha,\mathbf{Q}'\beta}(\vk) u_{\mathbf{Q}' \beta, m\eta}(\vk) = \epsilon_{\vk, m, \eta}u_{\mathbf{Q} \alpha, m \eta}(\vk)\,.
\end{equation}
Here $m$ is the band index. Thus the non-interacting Hamiltonian can be brought to the following form:
\begin{equation}
    \hat{H}_0 = \sum_{\vk\in{\rm MBZ}}\sum_{\eta, s} \sum_{m\neq 0} \epsilon_{\vk,m,\eta}  c^\dagger_{\vk, m, \eta, s}c_{\vk, m, \eta, s}\,,
\end{equation}
where the electron operators in the energy band basis $c^\dagger_{\vk, m, \eta, s}$ are defined as follows:
\begin{align}
    c^\dagger_{\vk, m, \eta, s} &= \sum_{\mathbf{Q}\alpha}u_{\mathbf{Q} \alpha, m \eta}(\vk) c^\dagger_{\vk, \mathbf{Q}, \eta, \alpha, s}\,,\label{ed:eqn:band_to_planewave}\\
    c^\dagger_{\vk, \mathbf{Q}, \eta, \alpha, s} &= \sum_{m} u^*_{\mathbf{Q}, \alpha, m\eta}(\vk) c^\dagger_{\vk, m, \eta, s} \,.\label{ed:eqn:planewave_to_band}
\end{align} 

As shown in the earlier studies of Bistritzer and MacDonald  \cite{bistritzer_moire_2011}, there are two flat bands whose band width can be smaller than $10\,\rm meV$ for each valley and spin. Projecting into the two flat bands around charge neutral point labeled by $m = \pm 1$, we obtain the kinetic term of the Hamiltonian in the main text:
\begin{equation}\label{ed:eqn:H_kin}
    H_0 = \sum_{\vk\in{\rm MBZ}}\sum_{m=\pm 1}\sum_{\eta, s}\epsilon_{\vk, m, \eta} c^\dagger_{\vk, m, \eta, s}c_{\vk, m, \eta, s}\,.
\end{equation}
In the following subsection, we will derive the interacting Hamiltonian projected onto these flat bands.

\subsection{Projected Coulomb Interaction and Real Gauge Fixing}\label{ed:appsubsec:real}
We assume that the interaction between the electrons is the screened Coulomb potential, whose Fourier transformation is given by:
\begin{equation}
    V(\vq) = \pi \xi^2 U_\xi\frac{\tanh(\xi q/2)}{\xi q/2}\,.
\end{equation}
Here $\xi = 10\,\rm nm$ is the distance between the metal gates, and $U_\xi = e^2/4\pi\epsilon\xi \approx 24\rm\,meV$ is the strength of the interaction. Before projected into the flat bands, the two body interacting Hamiltonian has the following form:
\begin{equation}
    \hat{H}_{I} = \frac{1}{2\Omega_{\rm tot}}\sum_{\vq,\mathbf{G}}V(\vq + \mathbf{G})\delta\rho_{\vq + \mathbf{G}} \delta\rho_{-\vq - \mathbf{G}}
\end{equation}
where $\Omega_{\rm tot}$ is the total area of the moir\'e lattice, and the relative electron density $\delta \rho_{\vq + \mathbf{G}}$ is given by
\begin{equation}
    \delta\rho_{\vq + \mathbf{G}} = \sum_{\vk\in{\rm MBZ}}\sum_{\eta, s}\sum_{\mathbf{Q}\in\mathcal{Q}_\pm}\sum_{\alpha}\left(c^\dagger_{\vk + \vq, \mathbf{Q} - \mathbf{G}, \alpha, \eta, s}c_{\vk, \mathbf{Q}, \alpha, \eta, s} - \frac{1}{2}\delta_{\vq, 0}\delta_{\mathbf{G}, 0}\right)\,.
\end{equation}
We use Eq.~(\ref{ed:eqn:planewave_to_band}) to rewrite the density operator in the energy band basis:
\begin{align}
    \delta\rho_{\vq + \mathbf{G}} &= \sum_{\vk\in{\rm MBZ}}\sum_{\eta, s}\sum_{\mathbf{Q}\in\mathcal{Q}_\pm}\sum_{\alpha}\left(\sum_{m,n}u^*_{\mathbf{Q}-\mathbf{G},\alpha m\eta}(\vk + \vq)u_{\mathbf{Q},\alpha n\eta}(\vk)c^\dagger_{\vk + \vq, m, \eta, s}c_{\vk, n, \eta, s} - \frac{1}{2}\delta_{\vq, 0}\delta_{\mathbf{G}, 0}\right)\nonumber\\
    & =\sum_{\vk\in{\rm MBZ}}\sum_{\eta, s}\sum_{m,n}\left(\sum_{\mathbf{Q}\in\mathcal{Q}_\pm}\sum_{\alpha}u^*_{\mathbf{Q}-\mathbf{G},\alpha m\eta}(\vk + \vq)u_{\mathbf{Q}, \alpha n\eta}(\vk)\right)\left(c^\dagger_{\vk + \vq, m, \eta, s}c_{\vk, n, \eta, s} - \frac{1}{2}\delta_{\vq, 0}\delta_{m, n}\right)\,,
\end{align}
where we use the following unitarity condition of single-body wavefunctions to get the second line from the first line:
\begin{equation}
     \sum_{m}u^*_{\mathbf{Q}-\mathbf{G}, \alpha m\eta}(\vk)u_{\mathbf{Q}, \alpha m\eta}(\vk) = \delta_{\mathbf{G}, 0}
\end{equation}
Then we can obtain the projected density operator around the first magic angle by only keeping $m, n = \pm 1$ terms. For convenience, we define the form factor (overlap) matrix 
\begin{equation}
    M^{(\eta)}_{mn}(\vk, \vq + \mathbf{G}) = \sum_{\mathbf{Q}\in\mathcal{Q}_\pm}\sum_{\alpha}u^*_{\mathbf{Q}-\mathbf{G}, \alpha m\eta}(\vk + \vq)u_{\mathbf{Q}, \alpha n\eta}(\vk)\,,\label{ed:app:eqn:formfactormatrix}
\end{equation}
and the projected density operator will have the following form:
\begin{equation}
    \overline{\delta\rho}_{\vq + \mathbf{G}} = \sum_{\vk \in {\rm MBZ}}\sum_{\eta, s}\sum_{m, n = \pm 1}M^{(\eta)}_{mn}(\vk, \vq + \mathbf{G})\left(c^\dagger_{\vk + \vq, m, \eta, s}c_{\vk, n, \eta, s} - \frac{1}{2}\delta_{\vq, 0}\delta_{m, n}\right)\,.
\end{equation}
Consequently, the projected two body Hamiltonian can be written as:
\begin{align}
    H_{I} =& \frac{1}{2\Omega_{\rm tot}}\sum_{\vk,\vk',\vq\in{\rm MBZ}}\sum_{\eta\eta'ss'}\sum_{mn;m'n'}U^{(\eta\eta')}_{mn;m'n'}(\vq;\vk,\vk')\nonumber\\
    & \times\left(c^\dagger_{\vk + \vq, m, \eta, s}c_{\vk, n, \eta, s} - \frac{1}{2}\delta_{\vq, 0}\delta_{m, n}\right)\left(c^\dagger_{\vk' - \vq, m', \eta', s'}c_{\vk, n', \eta', s'} - \frac{1}{2}\delta_{\vq, 0}\delta_{m', n'}\right)\,,\label{ed:eqn:int_ham}\\
    U^{(\eta\eta')}_{mn;m'n'}(\vq;\vk, \vk') =& \sum_{\mathbf{G}\in\mathcal{Q}_0}V(\vq + \mathbf{G})M^{(\eta)}_{mn}(\vk, \vq + \mathbf{G})M^{(\eta')}_{m'n'}(\vk', -\vq - \mathbf{G})\,.
\end{align}
This is the projected interacting Hamiltonian appeared in Eq.~(\ref{ed:eqn:four-fermion-ham}) in the main text. 

All the matrix elements in the interacting Hamiltonian $U^{(\eta\eta')}_{mn;m'n'}(\vq;\vk,\vk')$ can be obtained from single particle wavefunctions, no matter which gauge is chosen. However, carefully choosing a specific gauge is highly beneficial for analyzing the symmetry and for the numerical calculation. We choose the phase of the wavefunction at a given $\vk$ by fixing the sewing matrices of $C_{2z}T$ and $C_{2z}P$. Once we have obtained  the wavefunctions $u^{(\eta)}_{\mathbf{Q}\alpha, m}(\vk)$ of valley $\eta$ and band $m$, we first fix the sewing matrix of $C_{2z} T$ symmetry. If the two flat bands are non-degenerate at momentum $\vk$, then $C_{2z}T$ symmetry will give us a phase:
\begin{equation}\label{ed:eqn:C2Tsewing_nondegenerate}
    \sum_{\mathbf{Q}'\beta}D_{\mathbf{Q}\alpha, \mathbf{Q}'\beta}(C_{2z}T)u^*_{\mathbf{Q}',\beta m\eta}(\vk) = e^{i\varphi_\vk}u_{\mathbf{Q}, \alpha\eta m}(\vk)\,.
\end{equation}
By doing the following gauge transformation, the phase factor in Eq.~(\ref{ed:eqn:C2Tsewing_nondegenerate}) will disappear:
\begin{equation}\label{ed:eqn:non_degenerate_gauge_transformation}
    u_{\mathbf{Q}, \alpha m\eta}(\vk)\rightarrow e^{i\frac{\varphi_\vk}{2}} u_{\mathbf{Q},\alpha m\eta}(\vk)\,.
\end{equation}
If the flat bands are degenerate at this momentum, then in general the wavefunctions will transform under $C_{2z}T$ as shown:
\begin{equation}\label{ed:eqn:C2Tsewing_degenerate}
    \sum_{\mathbf{Q}'\beta}D_{\mathbf{Q}\alpha,\mathbf{Q}'\beta}(C_{2z}T)u^*_{\mathbf{Q}', \beta m\eta}(\vk) = \sum_{n}u_{\mathbf{Q},\alpha\eta n}(\vk)B^{C_{2z}T}_{nm}(\vk)\,,
\end{equation}
where the sewing matrix of $C_{2z}T$ symmetry $B^{C_{2z}T}(\vk)$ is defined by
\begin{equation}
    B^{C_{2z}T}_{nm}(\vk) = \sum_{\mathbf{Q}\alpha}u^*_{\mathbf{Q},\alpha\eta m}D_{\mathbf{Q}\alpha,\mathbf{Q}'\beta}(C_{2z}T)u^*_{\mathbf{Q}',\beta\eta m}(\vk)\,.
\end{equation}
For this case, we can apply a unitary gauge transformation $O_{nm}(\vk)\in U(2)$ to the wavefunction:
\begin{equation}\label{ed:eqn:degenerate_gauge_fixing}
    u_{\mathbf{Q}, \alpha m\eta} \rightarrow \sum_{n}u_{\mathbf{Q},\alpha n\eta}O_{nm}(\vk)
\end{equation}
which satisfies:
\begin{equation}\label{ed:eqn:C2T_fixing_degenerate}
        B_{nm}^{C_{2z}T}(\vk) = \sum_{m'}O_{nm'}(\vk)O_{mm'}(\vk)\,.
\end{equation}
The wavefunction after we apply this gauge transformation will also have $B^{C_{2z}T}_{nm}(\vk) = \delta_{nm}$. After fixing the sewing matrix of $C_{2z}T$ at every point in moir\'e Brillouin zone (MBZ), we can prove that all form factors satisfy 
\begin{equation}
M^{(\eta)}_{mn}(\vk, \vq + \mathbf{G}) = M^{(\eta)*}_{mn}(\vk, \vq + \mathbf{G})\,, 
\end{equation}
which means that all the matrix elements in the many-body Hamiltonian are real. 

However, when fixing the $C_{2z}T$ gauge, we still have some arbitrariness. At non-degenerate points, the gauge transformation with an extra minus sign $e^{i\frac{\varphi_{\vk}}{2}}\rightarrow -e^{i\frac{\varphi_{\vk}}{2}}$ also satisfies the gauge fixing condition. Similarly, at degenerate points, Eq.~(\ref{ed:eqn:C2T_fixing_degenerate}) can also be satisfied by the gauge transformation with an extra $O'(\vk)\in$ O(2) transformation $O(\vk)\rightarrow O(\vk)O'(\vk)$. The additional freedom is not an important issue if we keep using the real Hamiltonians. But it requires more careful attention when we use the Chern band basis, as we will discuss in App. \ref{ed:appsubsec:chern}.

We can also use $C_{2z}P$ to fix the relative phases between the two valleys. The associated transformation $C_{2z}P$ is represented by $D_{\mathbf{Q}\alpha, \mathbf{Q}'\beta}(C_{2z}P)$, and it satisfies:
\begin{align}
    D_{\mathbf{Q}\alpha, \mathbf{Q}'\beta}(C_{2z}P) &= (\sigma_{x})_{\alpha\beta}\zeta_{\mathbf{Q}}\delta_{\mathbf{Q},\mathbf{Q}'}\\
    D^\dagger(C_{2z}P)h^{(\eta)}(\vk)D(C_{2z}P) &= -h^{(-\eta)}(\vk)
\end{align}
That means the transformation $C_{2z}P$ will flip the valley and band index, but keep the momentum of the state unchanged. Therefore, this symmetry can be used for generating the single-body wavefunctions at valley $\eta = -1$ from wavefunctions at $\eta = + 1$.

In summary, the gauge choice of single body wavefunctions can be determined by following these steps:
\begin{itemize}
    \item Use the Hamiltonian for valley $\eta = +1$ to obtain the wavefunctions $u_{\mathbf{Q}, \alpha n+}(\vk)$ on a given momentum lattice in the first moir\'e Brillouin zone. 
    \item Perform gauge transformations discussed in Eq.~(\ref{ed:eqn:non_degenerate_gauge_transformation}) and Eq.~(\ref{ed:eqn:degenerate_gauge_fixing}) to fix the $C_{2z}T$ sewing matrix to be the identity.
    \item Use $C_{2z}P$ transformation to get the wavefunctions in valley $\eta = -1$: 
    \begin{equation}\label{ed:eqn:C2Pgaugefixing}
        u_{\mathbf{Q},\alpha n-}(\vk) = n\cdot\sum_{\mathbf{Q}'\beta}D_{\mathbf{Q}\alpha, \mathbf{Q}'\beta}(C_{2z}P) u_{\mathbf{Q}',\beta, -n,+}(\vk)\,.
    \end{equation}
    \item For momentum $\vk$ beyond the first MBZ, we can use the embedding matrix to shift $\vk$ back into the 1st MBZ:
    \begin{equation}
        u_{\mathbf{Q},\alpha m\eta}(\vk + \mathbf{G}) = u_{\mathbf{Q}-\mathbf{G},\alpha m\eta}(\vk)\,.
    \end{equation}
\end{itemize}

The interacting Hamiltonian Eq.~(\ref{ed:eqn:int_ham}) can also be reorganized into the summation of a normal-ordered two body term and a quadratic term:
\begin{align}
    H_{I} =& \frac{1}{2\Omega_{\rm tot}}\sum_{\vk,\vk',\vq\in{\rm MBZ}}\sum_{\eta\eta'ss'}\sum_{mn;m'n'}U^{(\eta\eta')}_{mn;m'n'}(\vq;\vk,\vk')c^\dagger_{\vk + \vq, m, \eta, s}c^\dagger_{\vk' - \vq, m', \eta', s'}c_{\vk, n', \eta', s'}c_{\vk, n, \eta, s}\nonumber\\
    &+ \sum_{\vk,m,n,\eta,s}\mathcal{E}^{\rm HF}_{\vk,m,n,\eta,s}c^\dagger_{\vk,m,\eta,s}c_{\vk,n,\eta,s}\,,\\
    \mathcal{E}^{\rm HF}_{\vk,m,n,\eta,s} &= \frac{1}{2\Omega_{\rm tot}}\left(\sum_{\vq, n'}U^{(\eta\eta)}_{mn';n'n}(\vq;\vk-\vq, \vk) - 2\sum_{\vk',\eta',m'}U^{(\eta\eta')}_{mn;m'm'}(0;\vk,\vk')\right)\,.\label{ed:eqn:def_hf}
\end{align}
In fact, it is shown in Ref.~\cite{ourpaper3} that Eq.~(\ref{ed:eqn:def_hf}) matches the ``Hartree-Fock'' effects from the filled bands below the flat bands. The interacting Hamiltonian before the projection into flat bands commutes with $\mathcal{P}_c$, which is a many-body charge-conjugation transformation. This transformation can transform a state at filling factor $\nu$ to $-\nu$. Therefore it is reasonable to have a projected Hamiltonian which satisfies this symmetry. Neither the normal-ordered four-fermion Hamiltonian nor the quadratic terms are invariant under this transformation, but their summation satisfies $\mathcal{P}_c$ symmetry. Thus the quadratic term can help the interacting Hamiltonian preserve the charge-conjugation symmetry, which is supported by experiments. The effect of the quadratic term will be studied numerically.

\begin{figure}
\centering
\includegraphics[width=0.9\linewidth]{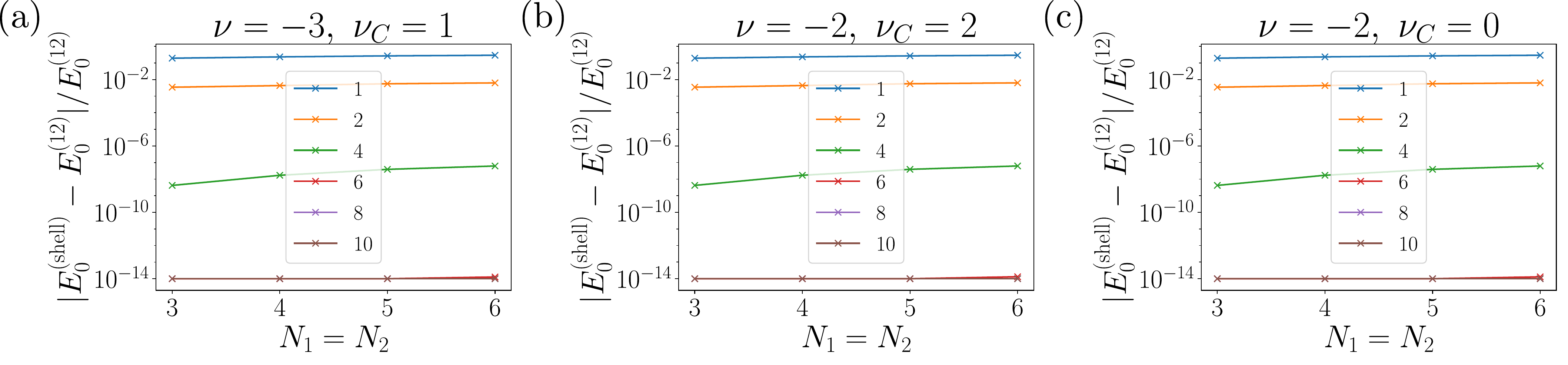}
\caption{The convergence of the energy of Chern insulator states with different number of momentum shells $N_{\rm shell}$ on different lattice mesh sizes $N_1=N_2$. We use the chiral-flat band limit without FMC (i.e., $\lambda=1$). The filling factors and Chern numbers are $\nu=-3$, $\nu_C = 1$ in subfigure (a), $\nu=-2$, $\nu_C=2$ in subfigure (b) and $\nu=-2$, $\nu_C=0$ in subfigure (c). We choose $N_{\rm shell}$ equal to 12 as a reference point. The vertical axis shows the relative difference of ground state energy using log scale. A relative difference below $10^{-14}$ is shown as $10^{-14}$ in order to eliminate the fluctuation due to numerical precision. Data with different numbers of shells is labeled by different colors. The $\mathcal{Q}_{\pm}$ lattice with $N_{\rm shell}=1$ is identical to $(A1 + B1) $in Fig.~4a of Ref.~\cite{ourpaper1}, and $N_{\rm shell}=2$ is identical to $(A1 + B1 + A2 + B2)$. From the results provided here, we can see that with only shell number equals 2, the numerical error is already smaller than $1\%$ for any of the three Chern insulator states. We also find that $N_{\rm shell}=6$ is already enough for convergence to machine precision (double accuracy).}
\label{ed:fig:shell}
\end{figure}

Finally, we comment on the summation over $\mathcal{Q}_{\pm}$ in Eq.~(\ref{ed:app:eqn:formfactormatrix}). Numerically, we have to choose a finite $\mathcal{Q}_{\pm}$ lattice. The effect of this finite truncation has been discussed in details in Ref.~\cite{ourpaper1}, especially on the form factor matrix. Here, we will provide another quantitative evidence of the exponential convergence with the truncation, focusing on the energy of the Chern insulator states. The lattice $\mathcal{Q}_{\pm}$ is given by $m_1 \vq_1 + m_2 \vq_2 + m_3 \vq_3$ where $m_1 + m_2 + m_3 = \pm 1$. We introduce the cutoff by the following constraint ${\rm max}(m_i) \leq N_{\rm shell}$. The $\mathcal{Q}_{\pm}$ lattice with $N_{\rm shell}=1$ is identical to $(A1 + B1)$ in Fig.~4a of Ref.~\cite{ourpaper1}, and $N_{\rm shell}=2$ is identical to $(A1 + B1 + A2 + B2)$. In order to show how the many-body energy converges with this cutoff, we evaluated the energy of some Chern insulator states with different cutoff and different lattice sizes at chiral-flat band limit without FMC. The results for the relative error on the energy can be found in Fig.~\ref{ed:fig:shell}. From these, we can find that the lattice shown in Fig.~4a of Ref.~\cite{ourpaper1} is already enough to get the ground state energy with a relative error lower than $1\%$, and $N_{\rm shell} = 6$ is enough for convergence at machine accuracy. Note that all the numerical calculations presented in this article have been performed with a large number of shells ($N_{\rm shell} = 12$), guaranteeing the convergence of our results with $N_{\rm shell}$.

\subsection{Chern Band Basis}\label{ed:appsubsec:chern}
In last subsection we derived the projected interacting Hamiltonian, and by $C_{2z}T$ gauge fixing, we find a basis in which the Hamiltonian matrix elements can be all real. 
Here we present another single particle basis $d^\dagger_{\vk, e_Y, \eta, s}$ where each band carries a non-zero Chern number $e_Y$ (see Ref. \cite{ourpaper2} for proof):
\begin{equation}\label{ed:eqn:def_of_chern}
    d^\dagger_{\vk, e_Y, \eta, s} = \frac{c^\dagger_{\vk, 1, \eta, s} + ie_Y c^\dagger_{\vk, -1, \eta, s}}{\sqrt{2}}\,,~~~e_Y = \pm 1\,,
\end{equation}
Here the gauge of electron operators under energy band basis $c^\dagger_{\vk, n, \eta, s}$ is fixed following the prescription in last subsection. However, the arbitrary sign at non-degenerate points, and the arbitrary O(2) transformations at degenerate points can lead to an ambiguity in the definition of $d^\dagger_{\vk, e_Y, \eta, s}$. There are several possibilities:
\begin{itemize}
    \item At non-degenerate points, if we only flip one of the sign of $c^\dagger_{\vk, \pm1, \eta, s}$, then the two $d^\dagger$ operators are swapped.
    \item At non-degenerate points, if the signs of both operators $c^\dagger_{\vk, \pm1, \eta, s}$ are flipped, the $d^\dagger_{\vk, e_Y, \eta, s}$ operators will acquire an extra minus sign, but they are not swapped.
    \item At degenerate points $\vk$, if we apply a transformation $O'(\vk) \in$ SO(2) to the energy band basis, then both the $d^\dagger_{\vk, e_Y, \eta, s}$ Chern operators will acquire a phase factor without swapping.
    \item At degenerate points $\vk$, we can also apply a gauge transformation $O'(\vk)\in$ O(2) with ${\rm det}\,O'(\vk) = -1$ to the energy band basis. 
    This transformation can be decomposed into the product of an SO(2) transformation followed by $\zeta_z$, which is the Pauli $z$ matrix applied to the energy band indices. The SO(2) transformation will not swap the two $d^\dagger_{\vk, e_Y, \eta, s}$ operators, but $\zeta_z$ transformation will add a minus sign to $c^\dagger_{\vk, -1, \eta, s}$. Thus the two $d^\dagger_{\vk, e_Y, \eta, s}$ will be swapped after this transformation.
\end{itemize}
In conclusion, no matter whether $\vk$ is at a degenerate point or not, the arbitrary sign or O(2) transformation can only  either swap the two $d^\dagger_{\vk, e_Y, \eta, s}$, or simply multiply the operators by a phase factor.

In order to find a well-defined Chern band basis, which carries a non-zero Chern number, we can use the continuous condition (\ref{ed:eqn:continuouscondition}). Similar to Eq.~(\ref{ed:eqn:def_of_chern}), the wavefunctions of the Chern basis have the following form:
\begin{equation}\label{ed:eqn:def_of_chern_1st}
    u'_{\mathbf{Q},\alpha e_Y\eta}(\vk) = \frac{u_{\mathbf{Q},\alpha 1\eta}(\vk) + ie_Y u_{\mathbf{Q},\alpha, -1 \eta}(\vk)}{\sqrt2}\,.
\end{equation}
The ambiguity can be fixed by a continuous condition:
\begin{equation}\label{ed:eqn:continuouscondition}
    \lim_{\vq \rightarrow 0}\Big{|}\sum_{\mathbf{Q}\alpha} u'^\star_{\mathbf{Q}\alpha e_Y\eta}(\vk + \vq) u'_{\mathbf{Q}\alpha e_Y'\eta}(\vk)\Big{|} = \delta_{e_Y, e_Y'}\,,
\end{equation}
and the single-body wavefunctions can be obtained by following these steps:
\begin{itemize}
    \item Similar to the first step when using real basis, we start with the Chern band basis in valley $\eta = +1$. We solve the single-body wavefunctions $u_{\mathbf{Q},\alpha m+}(\vk_0)$ at some point in the MBZ $\vk = \vk_0$, with the $C_{2z}T$ sewing matrix fixed.
    \item We move to another point $\vk_1$ in momentum space, which is close to $\vk_0$. We solve the single-body wavefunctions $u_{\mathbf{Q},\alpha m+}(\vk_1)$ at this momentum and we fix its $C_{2z}T$ sewing matrix. Next we calculate the Chern band basis wavefunctions at $\vk_0$ and $\vk_1$ with $e_Y = 1$, using Eq.~(\ref{ed:eqn:def_of_chern_1st}), and we check the inner product of these two wavefunctions. If the absolute value of the inner product is close to $1$, this means the Chern band wavefunctions with $e_Y = 1$ at $\vk_0$ and $\vk_1$ are continuous; if the absolute value of the inner product is close to zero, this means the gauge choice swapped the two Chern basis at $\vk_1$. By flipping the sign of $u_{\mathbf{Q},\alpha, -1\eta}(\vk_1)$, we can swap them back, and get the Chern basis wavefunction which is still continuous.
    \item Next we move to another point $\vk_2$, which is close to $\vk_1$. By similar methods, we can make sure that the Chern basis we obtained for $\vk_2$ is continuously connected with that at $\vk_1$.
    \item Step by step, we finally obtain the well-defined continuous Chern basis wavefunctions for the momentum lattice we need in the first MBZ with $\eta = 1$.
    \item Using the $C_{2z}P$ transformation, we can obtain the Chern basis wavefunctions in valley $\eta = -1$, as discussed in Eq.~(\ref{ed:eqn:C2Pgaugefixing}).
    \item For momentum $\vk$ beyond the first MBZ, we can use the embedding matrix to shift $\vk$ back into the 1st MBZ.
\end{itemize}

\begin{figure}[!htbp]
    \centering
    \includegraphics[width=15cm]{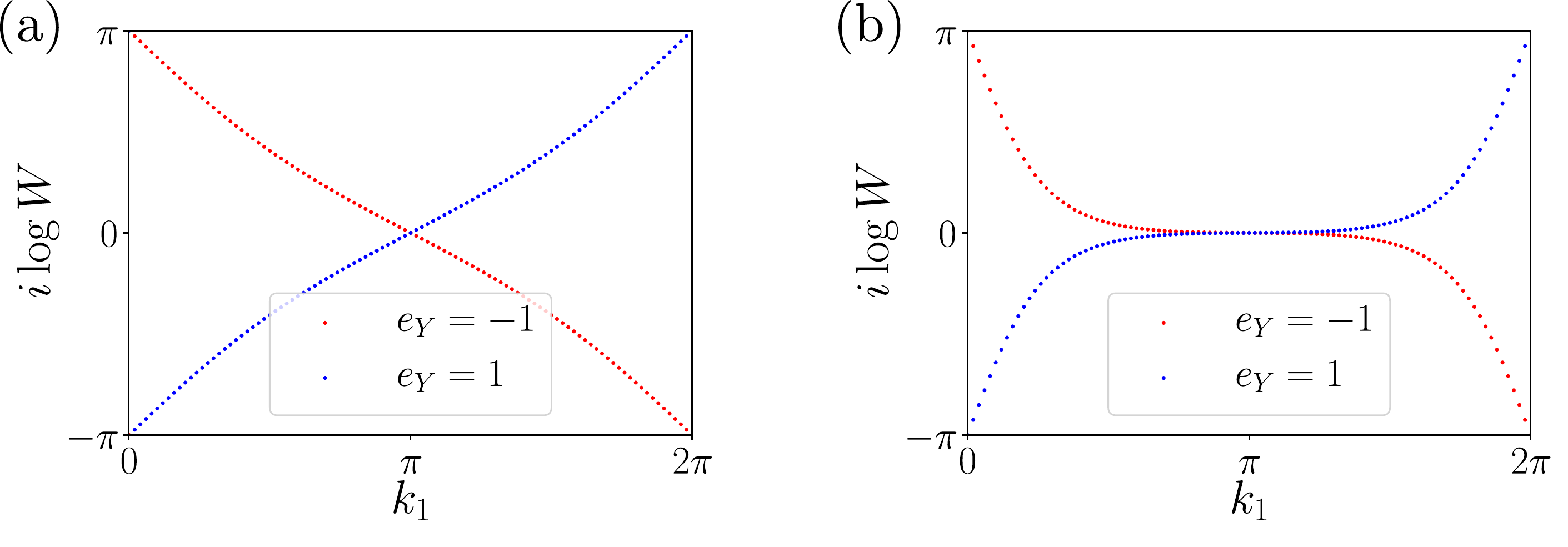}
    \caption{The Wilson loops of Chern basis in TBG at twisting angle $\theta = 1.07^\circ$ for two different values of $w_0$. a) $w_0 = 0$; b) $w_0/w_1 = 0.8$.}
    \label{ed:fig:wilsonloop}
\end{figure}

To check that the Chern basis states carry the proper Chern number, we also calculate the corresponding Wilson loops as depicted in Fig.~\ref{ed:fig:wilsonloop}. The Wilson loops with opposite $e_Y$ wind in opposite directions, with winding number equal to $\pm 1$.

The major benefit of using Chern band basis is the simple expression for wavefunction of Chern insulator states: they can be written as a single Fock state. The disadvantage is that the kinetic energy Hamiltonian will no longer be diagonal. Moreover, the Hamiltonian in the Chern basis is complex, which will be more computing intensive and memory consuming.

\subsection{Momentum Space Lattice}
For numerical calculations, the momentum space has to be discretized. By imposing periodic boundary condition, the momenta in the first moir\'e Brillouin zone of a $N_1 \times N_2$ lattice will be
\begin{equation}\label{ed:eqn:def_lattice}
    {\rm MBZ} = \left\{\vk = \frac{k_1}{N_1}\mathbf{b}_{M1} + \frac{k_2}{N_2}\mathbf{b}_{M2}\Big{|}0 \leq k_1 < N_1;~0 \leq k_2 < N_2\right\}\,.
\end{equation}
Here $k_1$ and $k_2$ are integers, and the momentum lattice MBZ  used for ED calculations is more coarse-grained than the lattice we use to do the continuous gauge fixing in \ref{ed:appsubsec:chern}. The total area of the moir\'e lattice, which appears in the prefactor of the interacting Hamiltonian, can be written as
\begin{equation}
    \Omega_{\rm tot} = N_1 N_2 \Omega_c = \frac{2\pi^2 N_1 N_2}{3\sqrt{3}|\mathbf{K}_l|^2\sin^2\frac{\theta}{2}}\,,
\end{equation}
in which $\Omega_c$ is the area of each moir\'e unit cell, $\mathbf{K}_l$ is the momentum of the Dirac point in single layer graphene, and $\theta$ is the twist angle of TBG. Therefore, once the twist angle $\theta$ and the momentum lattice in MBZ is fixed, we can obtain all the matrix elements of the interacting Hamiltonian.

\section{Implementation of the tunable Hamiltonian and its symmetry sectors}\label{ed:appsec:symhamiltonian}
\subsection{Fock Basis and the Tunable Hamiltonian}\label{ed:appsubsec:fock_ham}
To perform the ED, we define a many-body basis. We use the Fock states formed by the real basis defined in \ref{ed:appsubsec:real}. Each of these states can be labeled by a group of integers $\{n_{\vk, m, \eta, s}\}$. These integers $n_{\vk, m, \eta, s} = 0, 1$ represent the occupation number for each single-body state $c^\dagger_{\vk, m, \eta, s}$:
\begin{equation}
    |\{n_{\vk, m, \eta, s}\}\rangle = \prod_{\vk, m, \eta, s}\left(c^\dagger_{\vk, m, \eta, s}\right)^{n_{\vk, m, \eta, s}} |0\rangle\,.
\end{equation}
And as we mentioned in \ref{ed:appsubsec:real}, the many-body Hamiltonian will be real in this basis. Similarly, we can also use the Fock states formed by the Chern band basis defined in Eq.~(\ref{ed:eqn:def_of_chern}):
\begin{equation}
    |\{n_{\vk, e_Y, \eta, s}\}\rangle = \prod_{\vk, e_Y, \eta, s}\left(d^\dagger_{\vk, e_Y, \eta, s}\right)^{n_{\vk, e_Y, \eta, s}} |0\rangle
\end{equation}
We can use the Chern basis not only in the chiral-flat limit where we have a larger $U(4) \times U(4)$ symmetry but also for nonzero $w_0, t$. Even though the matrix elements are not real in this basis, it allows us to easily determine whether a state is a Chern-polarized state or not.

 We now introduce the tunable Hamiltonian. The many-body Hamiltonian has two terms, the kinetic term Eq.~(\ref{ed:eqn:H_kin}) and the two-body interacting term Eq.~(\ref{ed:eqn:int_ham}). Both the kinetic term and the two body interacting term depend on the single-body parameter $w_0$. We also add a parameter $t$ to control the amplitude of the kinetic term. Therefore, the Hamiltonian with these two parameters is given by:
\begin{equation}
    H(t, w_0) = t H_0(w_0) + H_I(w_0)\,.
\end{equation}

The flat metric condition introduced in Ref. \cite{ourpaper3} allows to derive several exact results for the interacting Hamiltonian, such as the analytical expression of the ground state and excitations at certain integer filling factors in the chiral-flat limit. It is defined by the following equation \cite{ourpaper1, ourpaper3, ourpaper4}
\begin{equation}
    M^{(\eta)}_{mn}(\vk, \mathbf{G}) = \xi({\mathbf{G}})\delta_{m,n}.\label{ed:eqn:def_fmc}
\end{equation}
By implementing this condition, the interacting Hamiltonian can be written as
\begin{align}
    H_{I,{\rm FMC}} =& \frac{1}{2\Omega_{\rm tot}}\sum_{\vq\neq 0}\sum_{\mathbf{G}}V(\vq + \mathbf{G})\overline{\delta\rho}_{\vq + \mathbf{G}}\overline{\delta\rho}_{-\vq - \mathbf{G}} \nonumber \\ 
    & +\frac{1}{2\Omega_{\rm tot}}\sum_{\vk,\vk'}\sum_{\mathbf{G}}\sum_{\eta\eta';ss'} V(\mathbf{G})\xi(\mathbf{G})\xi(-\mathbf{G})\left(c^\dagger_{\vk, m, \eta,s}c_{\vk, m,\eta, s} - \frac{1}{2}\right)\left(c^\dagger_{\vk',m',\eta',s'}c_{\vk', m', \eta', s'}-\frac12\right)
\end{align}
Now we focus on the second term. It is equal to
\begin{align}
   &\frac{1}{2\Omega_{\rm tot}}\sum_{\mathbf{G}} V(\mathbf{G})\xi(\mathbf{G})\xi(-\mathbf{G}) \sum_{\vk,\vk'} \sum_{\eta\eta';ss'} (N_{\vk, m, \eta,s} - \frac{1}{2})(N_{\vk',m',\eta',s'}-\frac12) \\
   = &\frac{1}{2\Omega_{\rm tot}}\sum_{\mathbf{G}} V(\mathbf{G})\xi(\mathbf{G})\xi(-\mathbf{G})\left(\sum_{\vk}\sum_{\eta ;s} (N_{\vk, m, \eta,s} - \frac{1}{2})\right)^2 \\
   = &\frac{1}{2\Omega_{\rm tot}}\sum_{\mathbf{G}} V(\mathbf{G})\xi(\mathbf{G})\xi(-\mathbf{G})\left(N - 2N_M\right)^2.
\end{align}
This term only depends on the total particle number $N$, and different choices of $\xi(\mathbf{G})$ can only shift the whole spectrum by a $N$ dependent constant. Since we are mostly focusing on the spectrum with a fixed total electron number, we will neglect this term. This is equivalent to removing all the terms with $\vq = 0$ from Eq.~(\ref{ed:eqn:int_ham}). We denote the interacting Hamiltonian satisfying FMC as $H_{I,\textrm{FMC}}(w_0)$, and therefore we have
\begin{equation}
    H_{\rm FMC}(t, w_0) = t H_0(w_0) + H_{I, {\rm FMC}}(w_0).
\end{equation}
It is worth studying how this FMC model $H_{\rm FMC}(t, w_0)$ is related with the exact Hamiltonian. Thus we can define a linear interpolation between $H$ and $H_{\rm FMC}$:
\begin{equation}
    H(t, w_0, \lambda) = \lambda \cdot H(t, w_0) + (1 - \lambda) \cdot H_{\rm FMC}(t, w_0)\,.
\end{equation}

\subsection{Symmetry Sectors}

When both $t$ and $w_0$ are non-zero -called the nonchiral-nonflat limit, the Hamiltonian has U(2)$\times$U(2) symmetry, because of the spin rotation symmetry and charge conservation in both valleys. If $t = 0$ and $w_0 \neq 0$, called the nonchiral-flat limit, the system has U(4) symmetry. Similarly, if $t \neq 0$ and $w_0 = 0$, called the chiral-nonflat limit, the Hamiltonian also has U(4) symmetry with different generators from the $t = 0$ and $w_0 \neq 0$ nonchiral-flat case. In all these cases, there are 4 Cartan subalgebra operators, which are the electron numbers per spin and valley. For our Fock basis, these quantum numbers have the following form:
\begin{equation}
    N_{\eta, s} = \sum_{\vk\in{\rm MBZ}}\sum_{m=\pm1}n_{\vk, m, \eta, s}\,,~~\eta = \pm 1\,, s = \uparrow,\downarrow\,.
\end{equation}
The total momentum is also conserved due to the translation symmetry. On a discrete momentum lattice defined in Eq.~(\ref{ed:eqn:def_lattice}), the total momentum  $K_1$ and $K_2$ read:
\begin{align}
    K_1 &= \left(\sum_{\vk, m, \eta, s} k_1 n_{\vk, m, \eta, s} \right) {\rm mod}\,N_1\,,~~k_1 = 0, 1, \cdots, N_1 - 1\,,\\
    K_2 &= \left(\sum_{\vk, m, \eta, s} k_2 n_{\vk, m, \eta, s} \right) {\rm mod}\,N_2\,,~~k_2 = 0, 1, \cdots, N_2 - 1\,.
\end{align}
The four quantum numbers $N_{\eta, s}$, together with the total momentum components $K_1$ and $K_2$, are the good quantum numbers we use for ED. The Hamiltonian is block diagonal in this basis, and each block can be labeled by these six quantum numbers, which we call symmetry sectors. Similarly, every many-body eigenstate can also be labeled by these quantum numbers. 

The sizes of symmetry sectors at different system sizes and fillings are given in tables \ref{table:nu1} - \ref{table:nu4}. Since the Hilbert space sizes grow super-exponentially in the system size for almost all the sectors we first focus on cases with higher symmetry like the chiral-flat limit, and then try and extend these results to the nonchiral-nonflat case, or both situations. At the chiral-flat limit, the symmetry is promoted to U(4)$\times$U(4) \cite{bultinck_ground_2020, ourpaper3}. There are hence 8 Cartan subalgebra operators. It can be shown that the form factors $M^{(\eta)}_{e_Y,e_Y'}(\vk, \vq + \mathbf{G})$ are diagonal in the Chern band basis. Thus the electron numbers in each spin, valley and band are conserved separately:
\begin{equation}
    N_{e_Y, \eta, s} = \sum_{\vk\in{\rm MBZ}} n_{\vk, e_Y, \eta, s}\,,~~e_Y = \pm 1\,, \eta = \pm 1\,, s = \uparrow, \downarrow\,.
\end{equation}
Together with the total momentum components $K_1$ and $K_2$, we have 10 good quantum numbers for the Hamiltonian $H(0, 0, \lambda)$.

Besides the Cartan subalgebra operators, we also have other symmetries which can be used to reduce the intensiveness of calculation. Because of these symmetries, symmetry sectors with different quantum numbers will have identical spectra:
\begin{itemize}
    \item When both $t$ and $w_0$ are non-zero, the spectra will not be changed if $N_{\eta, \uparrow}$ and $N_{\eta, \downarrow}$ are swapped, because of the U(2)$\times$U(2) symmetry. 
    \item When $t = 0$, $w_0 \neq 0$ or $t \neq 0$, $w_0 = 0$, any permutation of the four quantum numbers $N_{\eta, s}$ will not change the spectra, due to the U(4) symmetry.
    \item When both $t$ and $w_0$ are zero, any permutation of the four quantum numbers $N_{\eta, e_Y = 1, s}$, or any permutation of the four quantum numbers $N_{\eta, e_Y = -1, s}$ will not change the spectra. Moreover, because of the $C_{2z}T$ symmetry, the spectra will also be unchanged if we swap $N_{\eta, e_Y=1, s}$ and $N_{\eta, e_Y = -1, s}$.
\end{itemize}
Using these properties, we can compute the spectra of only a fraction of all the symmetry sectors, called the Weyl chamber. We choose a minimal subset of sectors which will allow us to generate all the sectors using these symmetry operations. Furthermore, if some sectors are invariant under a commuting subgroup of these operations, we can even split the symmetry sector into smaller blocks, which are labeled by the eigenvalues of these operators.

We only implemented the Cartan subalgebra operators instead of the full U(2)$\times$U(2), U(4) or U(4)$\times$U(4), due to the difficulty of  implementing these symmetries along with momentum conservation. Although the eigenstates we obtained in our Fock basis are not labeled by their irreps (as we have not implemented the full group, but rather only its Cartan subalgebra), it is still possible to investigate the irrep of a degenerate state by studying the degeneracy and the quantum numbers associated with the Cartan subalgebra operators. In particular each irrep corresponds to a unique set of symmetry sectors and degenearacies in these sectors. Therefore we may determine the irrep(s) of a degenerate set of states by looking at their degeneracies in the various symmetry sectors (a simple example would be, with SU(2) symmetry, having $4$ states in $S_z=3$ and $6$ states in $S_z=2$ sectors; then we would know that $6-4=2$ total spin $S=2$ states would exist in the spectrum, despite not having diagonalized the full $S^2$ operator). If we knew the irreps of our states, we would be able to determine the degeneracies of symmetry-related states -- this process can be uniquely inverted to recover the irreps from the degenerate states. Alternatively if we want to focus on a particular irrep, and there is a Cartan symmetry sector in which it is the only irrep, we may just calculate the spectrum in that sector.

For sake of completeness, we provide the largest Hilbert space dimensions that are involved for the ED calculations depending on the momentum mesh size, the symmetries and the filling factor: $\nu=-3$ (Table~\ref{table:nu1}), $\nu=-2$ (Table~\ref{table:nu2}), $\nu=-1$ (Table~\ref{table:nu3}) and $\nu=0$ (Table~\ref{table:nu4}).

\begin{table}
\begin{tabular}{lllll}
\hline
$N_1 \times N_2$   & Valley and spin polarized &  Valley or spin polarized  & Fully unpolarized  & $U(4) \times U(4)$\\
\hline
 $2\times 2$          & 22         & 208                & 1,024                   &64\\
 $3\times 2$          & 160        & 8,072              & 104,544                 &7,776\\
 $4\times 2$          & 1,638      & 414,352            & 25,921,536              &2,097,152\\
 $3 \times 3$         & 5,420      & 2,913,120          & 324,729,648             &19,131,876\\
 $5 \times 2$          & 18,504     & 24,037,408         & 4,691,556,000           &202,500,000\\
 $4\times 3$         & 225,440    & 1,509,677,768      & 1,398,494,577,664       &32,788,343,808     \\
 $5\times 3$          & 10,341,208 & 794,358,981,000    & 5,570,885,004,520,500   &140,710,042,265,625\\
 $4\times 4$          & 37,569,990 & 6,914,665,302,288  & 104,510,217,063,043,072 &2,687,385,603,145,728 \\
\hline
\end{tabular}
\caption{The largest Hilbert space dimensions for the given momentum mesh $N_1 \times N_2$ at filling factor $\nu = -3$. The Hilbert space dimensions are given for the U(2)$\times$U(2) (or equivalently the U(4) symmetry since only the Cartan subalgebra is implemented) and for the U(4)$\times$U(4) symmetry. For each system size, we focus on the quantum number sector (momentum sector and Cartan subalgebra eigenvalues, without any Weyl chamber symmetry) that gives the largest dimension. The first column is the size of the momentum lattice. The second column is the dimension of the Hilbert space with the U(2)$\times$U(2) symmetry when both the valley or spin are polarized. The third column is the Hilbert space dimension with the U(2)$\times$U(2) symmetry when either the valley or spin is polarized   with the other degree of freedom as close to unpolarized as possible. The fourth column is the Hilbert space dimension when neither are polarized. Specifically this sector happens when we divide the electrons as evenly as possible across the valleys (one valley having an extra electron for odd total numbers of electrons) and then have the smallest positive or zero $S_z$ possible in each valley. The fifth column is the Hilbert space dimension for the largest sector assuming U(4)$\times$U(4) symmetry.}
\label{table:nu1}
\end{table}

\begin{table}
\begin{tabular}{lllll}
\hline
 $N_1 \times N_2$  & Valley and Spin Polarized & Valley or Spin Polarized   & Fully Unpolarized & $U(4) \times U(4)$  \\
\hline
 $2\times 2$           &1 & 1,252           & 153,856             &16,384\\
 $3\times 2$           &1 & 142,376         & 390,426,752          &10,935,000\\
 $4\times 2$           &1 & 20,706,468      & 1,371,499,450,624     &47,225,274,368      \\
 $3\times 3$          &1 & 262,656,400      & 76,376,413,209,600    &1,706,597,351,424      \\
 $5\times 2$           &1 & 3,413,484,320    & 5,777,966,756,796,928  &85,030,560,000,000     \\
 $4\times 3$           &1 & 609,371,711,400        & 27,349,372,590,948,391,040  &457,298,946,133,344,256\\
\hline
\end{tabular}
\caption{The largest Hilbert space dimensions for the given momentum mesh $N_1 \times N_2$ at filling factor $\nu = -2$. The Hilbert space dimensions are given for the U(2)$\times$U(2) (or equivalently the U(4) symmetry since only the Cartan subalgebra is implemented) and for the U(4)$\times$U(4) symmetry. For each system size, we focus on the quantum number sector (momentum sector and Cartan subalgebra eigenvalues, without any Weyl chamber symmetry) that gives the largest dimension. The first column is the size of the momentum lattice. The second column is the dimension of the Hilbert space with the U(2)$\times$U(2) symmetry when both the valley or spin are polarized. This is always one because there are only two sites per spin, valley, and momentum and they are both filled. The third column is the Hilbert space dimension with the U(2)$\times$U(2) symmetry when either the valley or spin is polarized   with the other degree of freedom as close to unpolarized as possible. The fourth column is the Hilbert space dimension when neither are polarized. Specifically this sector happens when we divide the electrons as evenly as possible across the valleys (one valley having an extra electron for odd total numbers of electrons) and then have the smallest positive or zero $S_z$ possible in each valley. The fifth column is the Hilbert space dimension for the largest sector assuming U(4)$\times$U(4) symmetry.}
\label{table:nu2}
\end{table}

\begin{table}
\begin{tabular}{llll}
\hline
 $N_1 \times N_2$  & Spin or Valley Polarized   & Fully Unpolarized   & $U(4) \times U(4)$\\
\hline
 $2\times 2$           & 208             & 2,458,624           &82,944 \\
 $3\times 2$           & 8,072           & 25,615,893,600        &759,375,000      \\
 $4\times 2$           & 414,352         & 514,051,077,736,448     &12,089,663,946,752    \\
 $3\times 3$           & 2,913,120       & 66,480,357,719,752,704    &929,534,591,655,936  \\
 $5\times 2$           & 24,037,408      & 9,535,902,166,979,136,000   &123,503,214,240,000,000\\
 $4\times 3$           & 1,509,677,768   & N/A                         & N/A\\
\hline
\end{tabular}
\caption{The largest Hilbert space dimensions for the given momentum mesh $N_1 \times N_2$ at filling factor $\nu = -1$. The Hilbert space dimensions are given for the U(2)$\times$U(2) (or equivalently the U(4) symmetry since only the Cartan subalgebra is implemented) and for the U(4)$\times$U(4) symmetry. For each system size, we focus on the quantum number sector (momentum sector and Cartan subalgebra eigenvalues, without any Weyl chamber symmetry) that gives the largest dimension. The first column is the size of the momentum lattice. For this filling factor the system cannot be both spin and valley polarized. The second column is the Hilbert space dimension with the U(2)$\times$U(2) symmetry when either the valley or spin is polarized with the other degree of freedom as close to unpolarized as possible. The third column is the Hilbert space dimension when neither are polarized. Specifically this sector happens when we divide the electrons as evenly as possible across the valleys (one valley having an extra electron for odd total numbers of electrons) and then have the smallest positive or zero $S_z$ possible in each valley. The fourth column is the Hilbert space dimension for the largest sector assuming U(4)$\times$U(4) symmetry.}
\label{table:nu3}
\end{table}

\begin{table}
\begin{tabular}{llll}
\hline
 $N_1 \times N_2$  &   One Sector Polarized & Neither Sectors Polarized  & Largest with $U(4) \times U(4)$       \\
\hline
 $2\times 2$           &            1 & 6,003,472                  & 420,096 \\
 $3\times 2$           &            1 & 121,488,936,800            & 4,266,666,752       \\
 $4\times 2$           &            1 & 3,429,447,839,205,648          &72,060,013,129,984   \\
 $3\times 3$           &            1 & 620,893,779,148,960,000      & 7,058,653,305,387,264     \\
 $5\times 2$           &            1 & 116,518,317,397,535,713,856      & 1,626,313,721,561,225,728 \\
 $4\times 3$           &            1 & 4,456,005,538,295,087,455,458,144 & 44,278,665,537,370,662,050,560\\
\hline
\end{tabular}
\caption{The largest Hilbert space dimensions for the given momentum mesh $N_1 \times N_2$ at filling factor $\nu = 0$. The Hilbert space dimensions are given for the U(2)$\times$U(2) (or equivalently the U(4) symmetry since only the Cartan subalgebra is implemented) and for the U(4)$\times$U(4) symmetry. For each system size, we focus on the quantum number sector (momentum sector and Cartan subalgebra eigenvalues, without any Weyl chamber symmetry) that gives the largest dimension. The first column is the size of the momentum lattice. For this filling factor the system cannot be both spin and valley polarized. The second column is the Hilbert space dimension with the U(2)$\times$U(2) symmetry when either the valley or spin is polarized. This is always one since there are at most 4 sites per spin or per valley and they are all filled. The fourth column is the Hilbert space dimension for the largest sector assuming U(4)$\times$U(4) symmetry.}
\label{table:nu4}
\end{table}

\section{Numerical Results for \texorpdfstring{$\nu=-3$}{nu=-3}}\label{ed:appsec:numerical-3}

In this appendix, we provide additional numerical results for the filling factor $\nu=-3$. In particular, we address the low energy excitation dispersion and the phase diagrams with respect to interpolating parameter $\lambda$ in the three limits discussed in Sec.~\ref{sec:symmetries}.

\subsection{Charge and neutral excitations from ED}\label{ed:app:excitationEDnu3}
In addition to the data discussed in Sec.~\ref{sec:chiralflatbandnu3}, we provide a momentum-resolved discussion of the neutral and charge excitations at $\nu = -3$ and compare the results with and without the FMC.

\begin{figure}
    \centering
    \includegraphics[width=\linewidth]{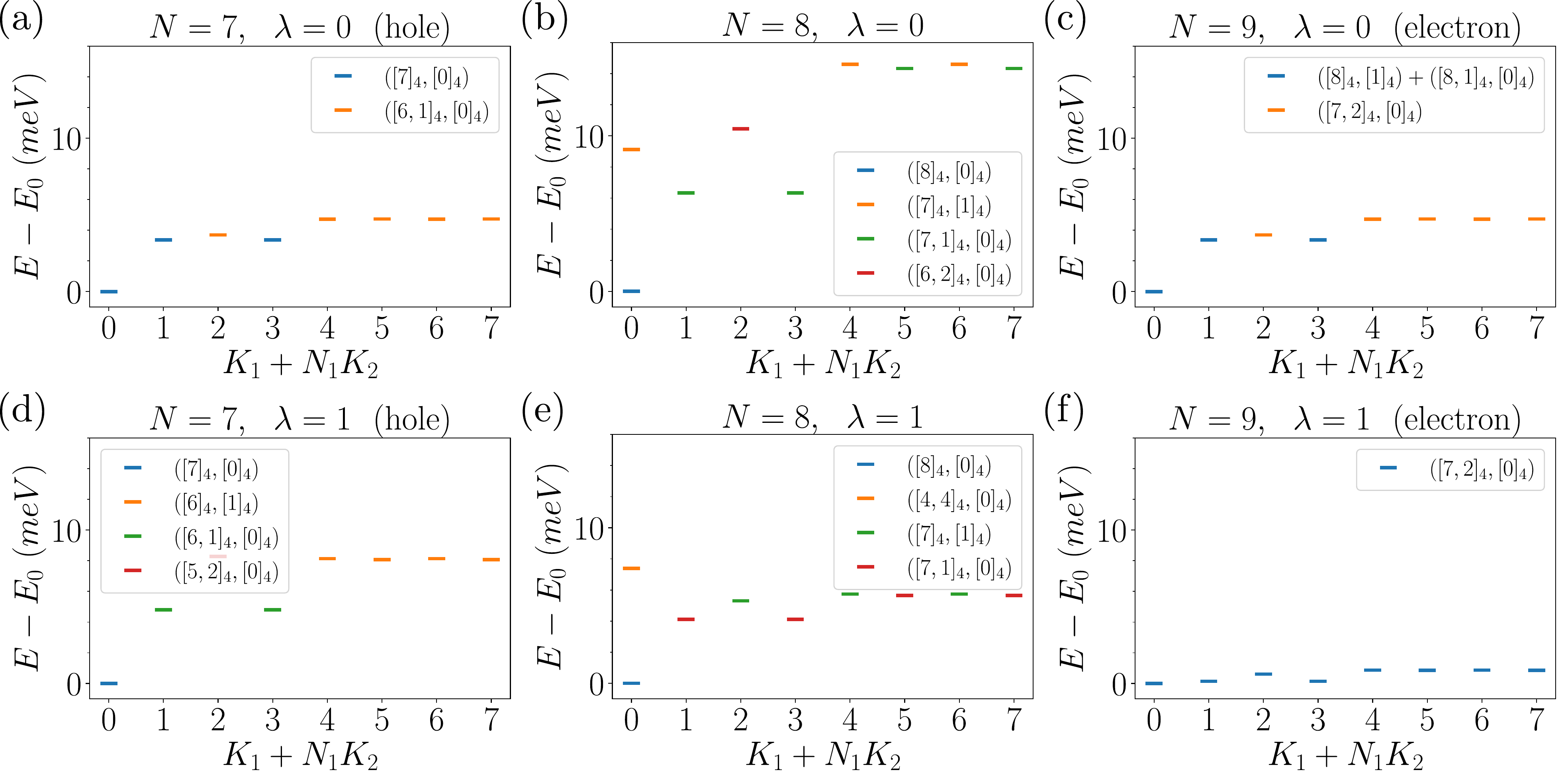}
    \caption{The low lying spectrum of $N = 7, 8$ and $9$ on $4\times 2$ lattice at twisting angle $\theta=1.1014^\circ$, for the FMC model (a-c) and the full model (d-f) at chiral-flat limit. The corresponding $U(4)\times U(4)$ irreps for each momentum sector are also shown in the plot. Note that in (c), we use the notation $([8,1]_4,[0]_4) + ([8]_4,[1]_4)$ for energy levels where the two irreps are always exactly degenerate.}
    \label{ed:fig:charge_excitation_4x2}
\end{figure}

First we consider the full (all sectors) diagonalization on a $N_1 \times N_2 = 4 \times 2$ lattice presented in Fig.~\ref{ed:fig:charge_excitation_4x2} and then move on to larger system size, but restricted to some irreps, calculations. Fig.~\ref{ed:fig:charge_excitation_4x2}a and~d show the charge excitations without and with the FMC for a charge $-1$ (hole) excitation. 
The lowest excitation is the $([7]_4, [0]_4)$ irrep at the $\Gamma_M$ point for $\lambda = 0$ and also for $\lambda = 1$, the Hamiltonian in the chiral-flat limit, $H(0, 0, \lambda)$.
This unbiased calculation without the FMC Eq.~(\ref{ed:eqn:def_fmc}) confirms that the charge $-1$ excitations analytically derived in Ref.~\cite{ourpaper5}, which includes a single hole but not $n$ holes plus $n-1$ particles, e.g., two holes plus one particle, are indeed the lowest charge $-1$ excitations.

We see that the charge $-1$ excitations at non-zero momentum are equivalent to the analytic ones $([7]_4, [0]_4)$ for  $H(0,0, 0)$ (see Fig.~\ref{ed:fig:charge_excitation_4x2}a) at momentum $(k_1,k_2)=(1,0),(3,0)$. 
At different momenta, other charged hole excitations, of different irreps from the analytic eigenstates (which still have to be exact eigenstates - the plot only shows the \emph{lowest} charge excitation per momentum sector), exhibit lower energy. 
The $\Gamma_M$-point charge excitation  $([7]_4, [0]_4)$ is the lowest (it is plotted at zero energy in Fig.~\ref{ed:fig:charge_excitation_4x2}a due to an energy substraction), which confirms the analytic result \cite{ourpaper5} that the smallest gap of the charge $-1$ excitations in the chiral-flat limit with the FMC Eq.~(\ref{ed:eqn:def_fmc}) is at the $\Gamma_M$ point. For $\nu=-3$, without the FMC, the analytic spectrum of the  hole excitation also shows minimal gap at the $\Gamma_M$ point.
(See Figs. 5d, 6d in Ref. \cite{ourpaper5}.)
Fig.~\ref{ed:fig:charge_excitation_4x2}d confirms that the analytic excitation remains the lowest excitation at the $\Gamma_M$ point in the chiral-flat limit without FMC.

Moving on to the neutral excitations we see again that the unbiased calculation performed in Figs.~\ref{ed:fig:charge_excitation_4x2}b and~e supports that the electron-hole pair excitations from analytic calculation \cite{ourpaper5} are the lowest charge neutral excitations. 
The ground state is the $([8]_4,[0]_4)$ irrep with or without the FMC. 
We see that the first neutral excitation is the $([7, 1]_4, [0]_4)$ irrep which is solely an excitation in one Chern band for both with and without the FMC. Remarkably, these excited states are actually part of the Goldstone branch analytically computed in Ref.~\cite{ourpaper5}. The gap between the ground state and the finite momentum excitations is a finite-size gap, due to the fact that the momentum $(1,0)$ and above on a $4 \times 2$ lattice is actually a large momentum relative to the thermodynamic limit - which explains the finite gap at this momentum between the Goldstone branch and the ground state. It is remarkable that we can identify the analytic Goldstone branch in the ED results on small lattices and away from the FMC.

Finally the single electron charge excitation shown in Figs.~\ref{ed:fig:charge_excitation_4x2}c and f is quite different between the $\lambda=0$ and $\lambda=1$ in the chiral-flat limit. 
The FMC Hamiltonian, $H(0,0,0)$, exhibits the analytic eigenstates as the excited states, while the chiral-flat band Hamiltonian, $H(0,0,1)$, exhibits a lowest charge excitation which is a different irrep - $([7,2]_4,[0]_4)$ - than the analytic calculation. 
However, this irrep represents a wavefunction that is obtained by dressing the analytic charge excitation with a single particle-hole pair, and we hence call it ``close in irrep space" to the analytic excitation.  
Overall the results of this type of full diagonalization - which show that the lowest excitations in the system are close in irrep space to the ground state validates our decision to restrict to the excitation sectors of irreps near the ground states of fully filled Chern band. This  provides access to larger system as  we can focus on a single symmetry sector.

The band of charge +1 excitations of $H(0,0,1)$ turns out to be rather flat in this problem, which seems also consistent with the analytic calculation of the low energy charged eigenstates of $H(0,0,0)$ without FMC.
(See Fig. 5, 6 in Ref. \cite{ourpaper5}.)
However, based on Fig.~\ref{ed:fig:excitations_nu-3}, the size of the $4 \times 2$ lattice is too small to see that the excited states are the same for the both models - for $\lambda=1$, the charge excitations with irrep $([N_M,1]_4, [0]_4)$ and $([N_M]_4,[1]_4)$ only become the lowest electron excitation when $N_M \geq 20$.

\begin{figure}
    \centering
    \includegraphics[width=\linewidth]{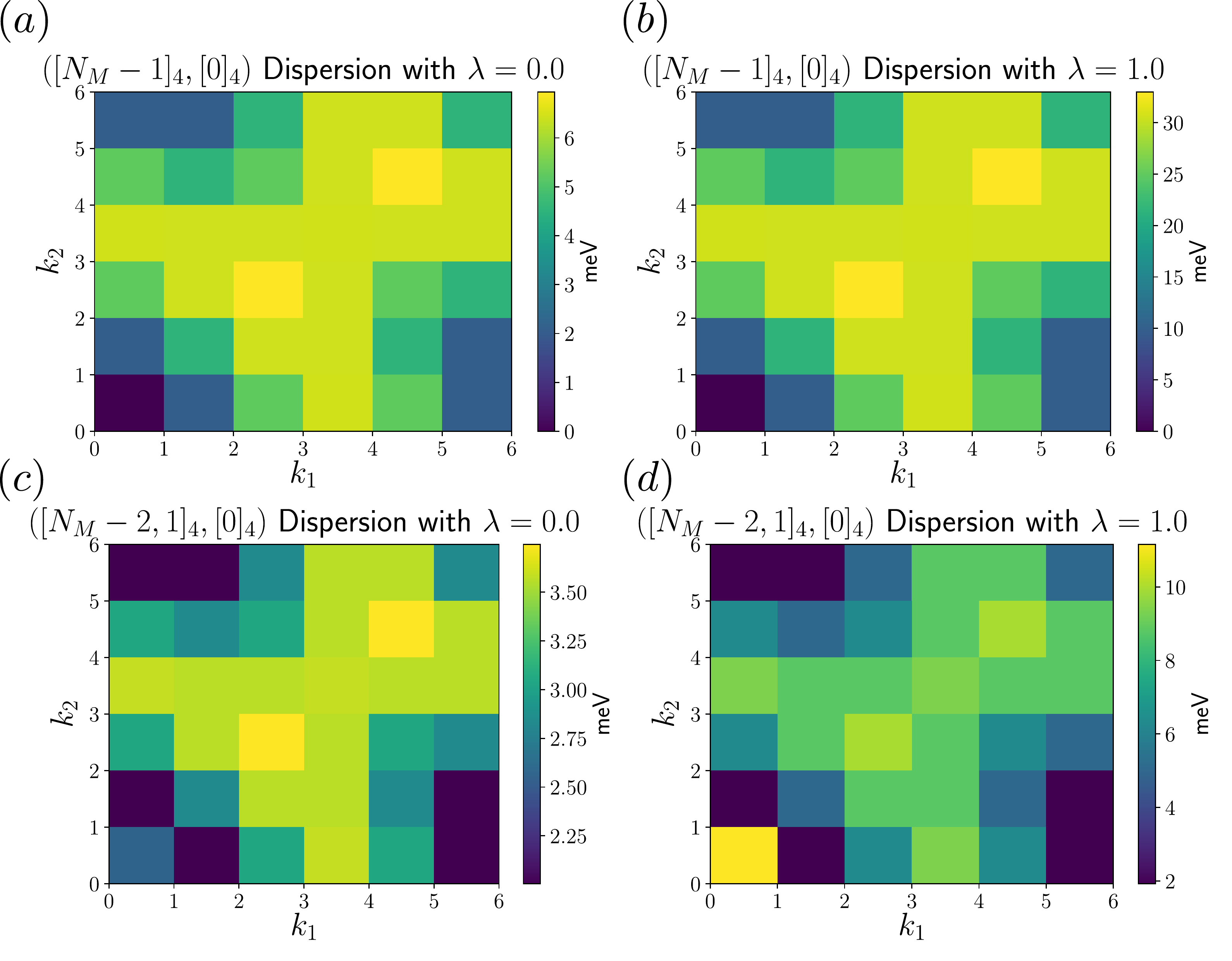}
    \caption{Energy as a function of momentum ($\vk = \frac{k_1}{N_1}\mathbf{b}_{M1} + \frac{k_2}{N_2}\mathbf{b}_{M2}$) for charge $-1$ (hole) excitation at a filling of $\nu = -3$ in the chiral-flat limit. The energies are relative to $E_0$, the minimum energy over all calculated irreps in Fig.~\ref{ed:fig:hole_nu_-3} and momentum sectors for the respective $\lambda = 0, 1$. Plots (a) and (b) are the dispersions of a hole in an otherwise filled Chern band at $\lambda = 0$ and $\lambda = 1$ respectively. Notice the remarkable similarity (up to scaling) with and without the FMC. The $U(4)$ excitations on top of the hole excitations shown in figures (c) and (d) have qualitatively similar spectra, especially for the low-lying momentum states. The largest difference occuring at the $\Gamma_M$ point.}
    \label{ed:fig:hole_dispersions_nu_-3}
\end{figure}

We further analyze the properties of the electron and hole excitations. By focusing on several representation sectors - for example not only the analytic hole excitation \cite{ourpaper5} $([N_M-1]_4, [0]_4)$ but also another excitation $([N_M-2,1]_4, [0]_4)$, close in irrep space to the analytic hole excitation - we are now able to compute their spectra in much larger system sizes, including $6\times 6$ sites. Fig.~\ref{ed:fig:hole_dispersions_nu_-3} contains the hole excitations of the aforementioned representations at $\nu=-3$. We see that, up to a rescaling, their energy dispersions are similar for both  $([N_M-1]_4, [0]_4)$ $([N_M-2,1]_4,[0]_4)$ irreps, with one deviation: the largest -and almost unique difference occurs at the $\Gamma_M$ point for the $([N_M - 2, 1]_4, [0]_4)$ irrep.

\begin{figure}
    \centering
    \includegraphics[width=\linewidth]{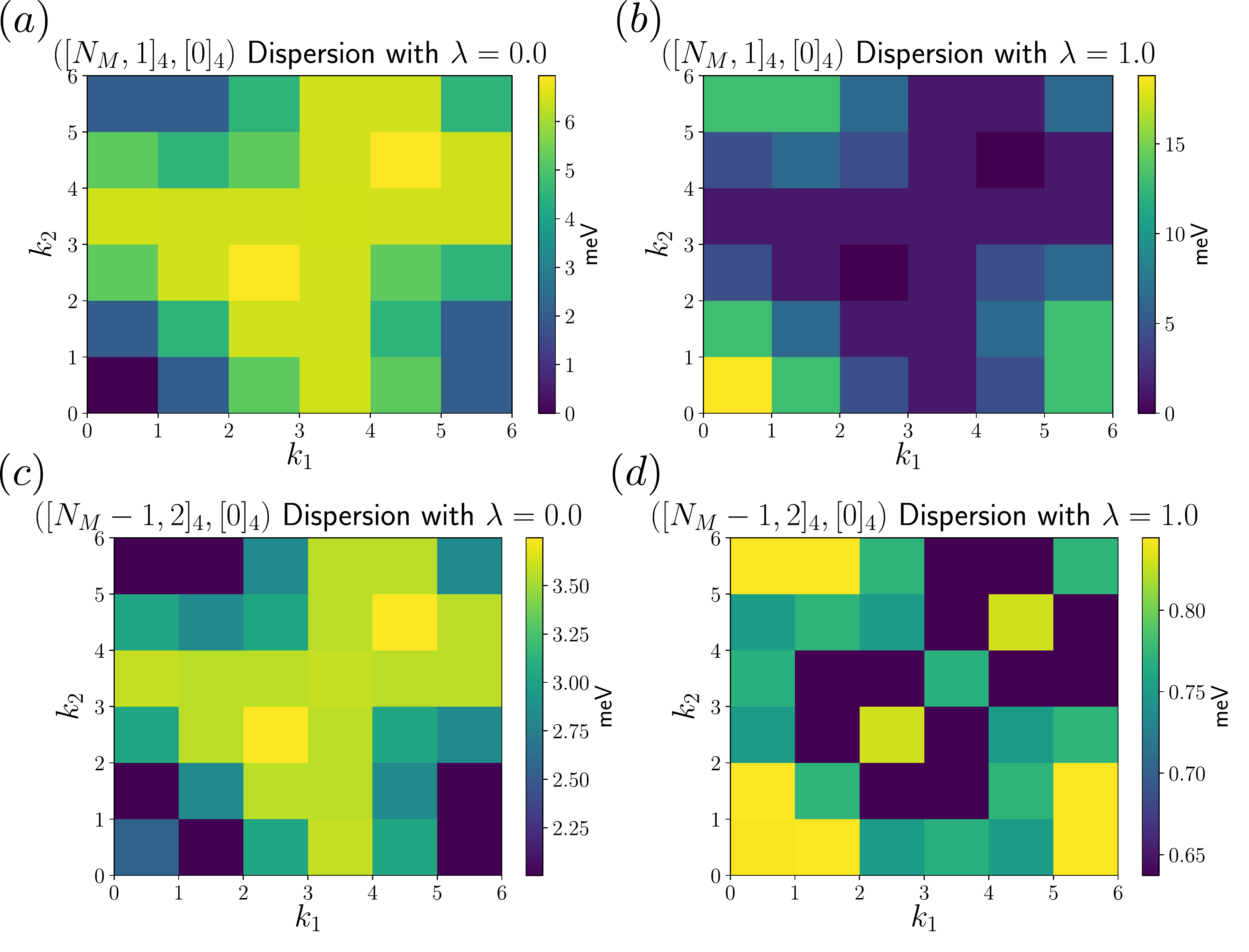}
    \caption{Energy as a function of momentum $\vk = \frac{k_1}{N_1}\mathbf{b}_{M1} + \frac{k_2}{N_2}\mathbf{b}_{M2}$ on the $6 \times 6$ lattice for a single electron excitation at $\nu =-3$ in the chiral-flat limit. The energies are relative to $E_0$, the minimum energy over all calculated irreps in Fig.~\ref{ed:fig:excitations_nu-3} and momentum sectors for the respective $\lambda = 0$ and $1$. Plots (a) and (b) are the dispersions of an additional electron (charge $+1$) on top of a filled Chern band without (a) and with (b) the FMC respectively. These plots show a distinct difference with the minimum energy at $\lambda = 0$ located at $\vk = \Gamma-M$ whereas the minimum for $\lambda = 1$ is located at $\vk = K_M$, the Dirac point of the moir\'e Billouin zone. Figures (c) and (d) are the dispersions of an additional $U(4)$ excitation on top of adding an additional electron.}
    \label{ed:fig:electron_disperions_nu_-3}
\end{figure}

On the other hand the charge excitation plots shown in Fig.~\ref{ed:fig:electron_disperions_nu_-3} show a distinct difference with and without the FMC. We also analyse two irreps of charge +1 excitations: the irrep of the analytic charge excitation \cite{ourpaper5} $([N_M,1]_4, [0]_4)$ but also another excitation $([N_M-1,2]_4, [0]_4)$, close in irrep space to the analytic hole excitation. We see large differences between $\lambda=0$ and $\lambda=1$. In particular the lowest charge excitation is at finite momentum (near the moir\'e Dirac point $K_M$) without the FMC $\lambda=1$ while it is at zero momentum with the FMC $\lambda=0$. Remarkably, this is exactly what the analytic excitation in Ref.~\cite{ourpaper5} exhibits with both FMC (see Fig.~1 of Ref.~\cite{ourpaper5}) and without the FMC  (see Figs.~5d and~6d of Ref.~\cite{ourpaper5}).

\subsection{Additional system sizes}\label{ed:app:additionalsizesnu3}

In the main text Sec. \ref{sec:fullypolsectorsnu3}, we have presented the phase diagrams of the spin and valley polarized sectors on $4\times 3$ lattice in the nonchiral-nonflat case with $\lambda = 0$ and $1$.

\begin{figure}
    \centering
    \includegraphics[width=0.9\linewidth]{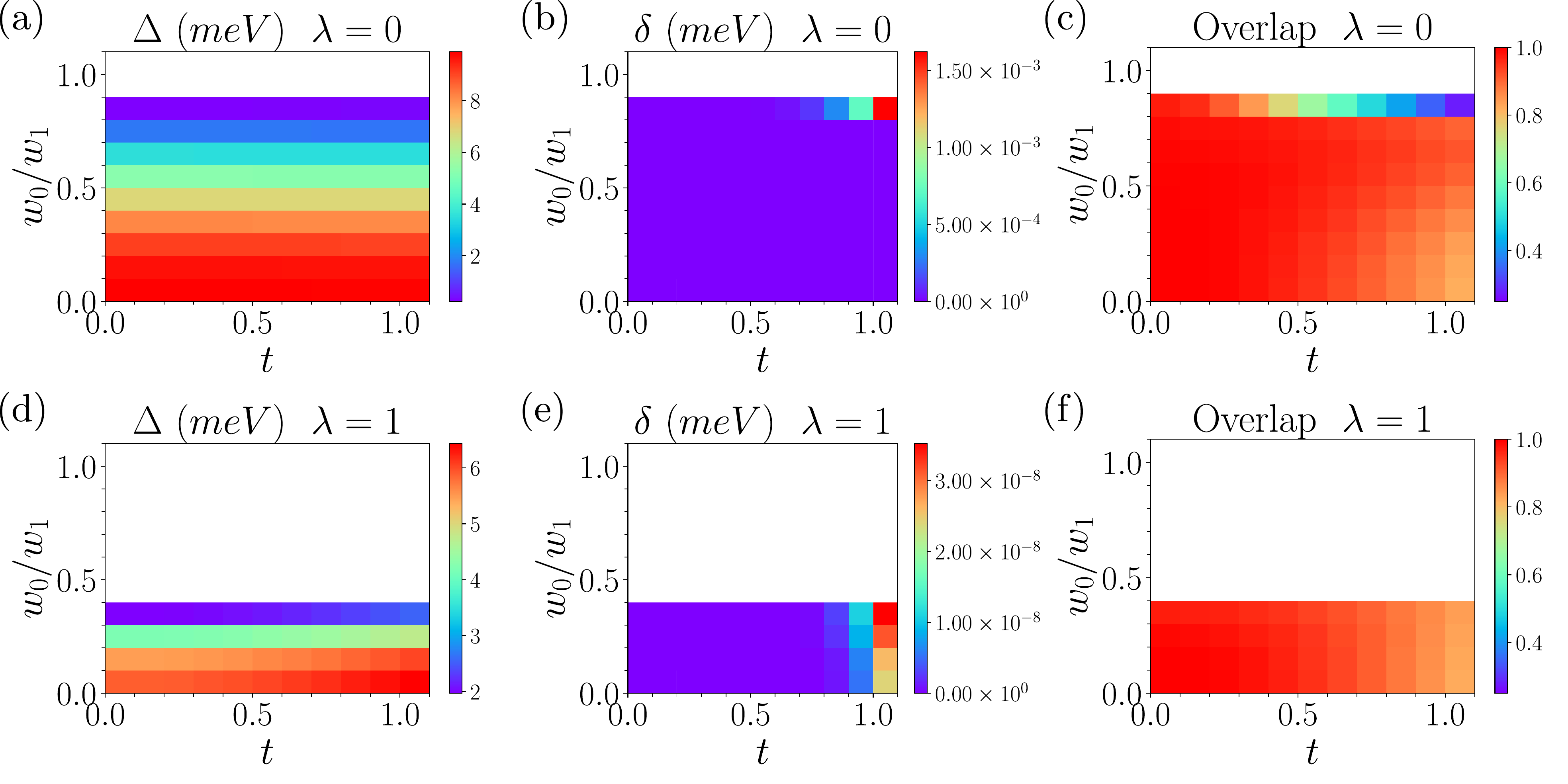}
    \caption{The phase diagram at filling $\nu=-3$ on $5 \times 3$ lattice in spin and valley polarized symmetry sectors with $\lambda = 0$ and $\lambda= 1$. The finite size gap (a, d), the spread between the two lowest states (b, e) and the overlap between the two lowest states and Chern insulator states (c, f) are shown by color. We choose $\lambda = 0$ in subfigures (a-c) and $\lambda = 1$ in subfigures (d-f). The white regions are beyond the Chern insulator phase, in which the overlap between the ED ground states and Chern insulator states is zero. Note that the overlap scale starts at $0.25$. Overall, the overlap is never smaller than 0.85 in above $85\%$ of the area in the Chern insulator phase.}
    \label{ed:fig:phase_diagram_t_w0_5x3}
\end{figure}

In this section we present the phase diagrams on a smaller lattice $3\times 3$ and a larger lattice $5\times 3$. The results for the $5\times 3$ lattice shown in Fig.~\ref{ed:fig:phase_diagram_t_w0_5x3} are almost identical to the $4\times 3$ results of Fig.~\ref{ed:fig:phase_diagram_t_w0}. We only provide them as an illustration of the small finite size effects. Note that the momentum mesh for $5\times 3$ includes neither $M_M$ nor the two Dirac points.

\begin{figure}
    \centering
    \includegraphics[width=\linewidth]{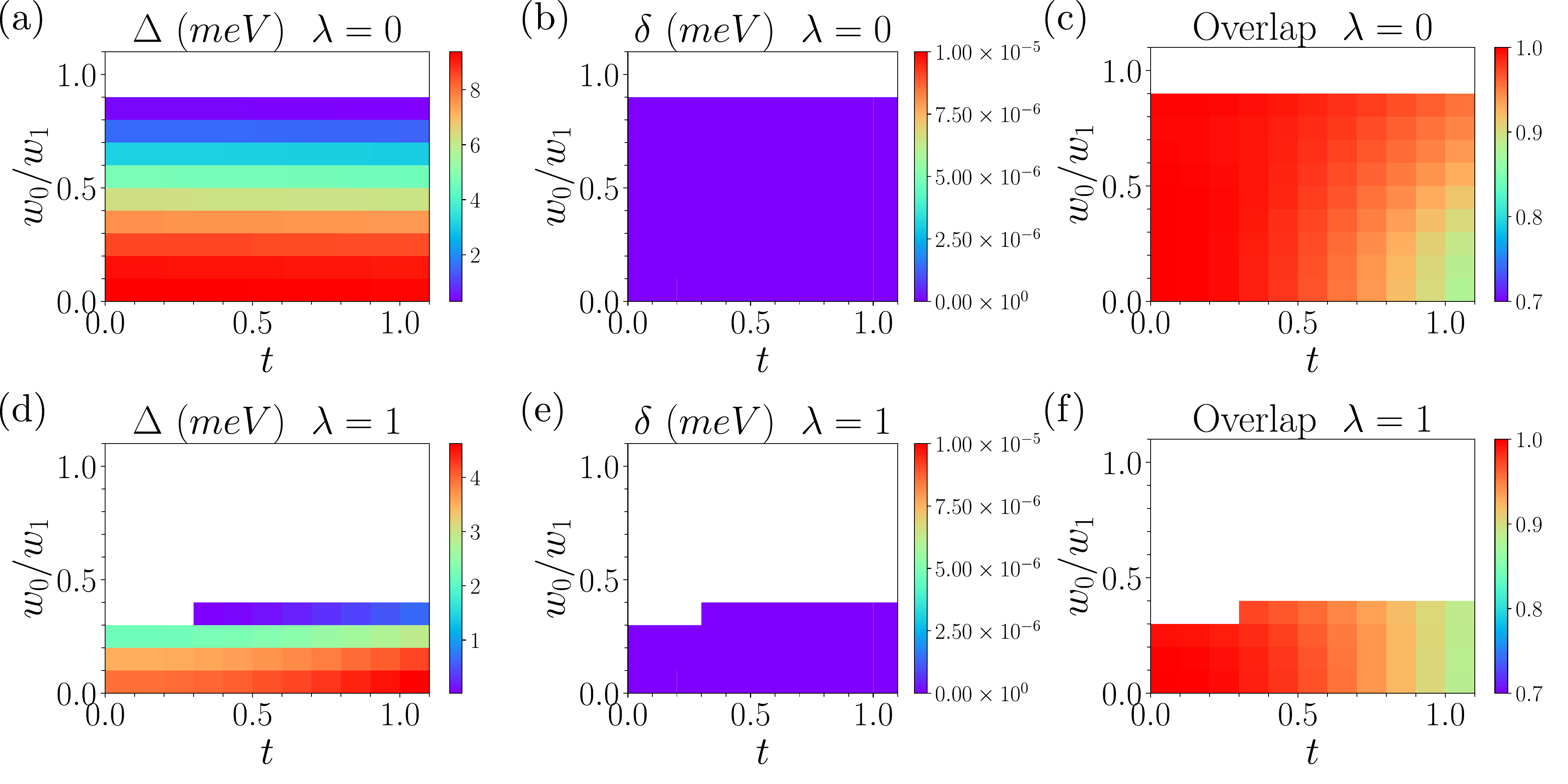}
    \caption{The phase diagrams ($\nu=-3$) in $t\sim w_0$ planes with $\lambda = 0$ and $\lambda = 1$ on $3\times 3$ lattice. The quantities shown in these plots are defined in Sec. \ref{sec:fullypolsectorsnu3}.}
    \label{ed:fig:phase_diagram_t_w0_app_3x3}
\end{figure}

Despite being smaller, the $3\times 3$ lattice MBZ contains the two Dirac points. Moreover, this lattice also satisfies the $C_{3z}$ rotation symmetry, which is absent in the many of the system sizes we have discussed in the main text. We provide the same quantities as those defined in Sec.~\ref{sec:phasediagramnu3}: the Goldstone branch finite momentum energy within the spin and valley polarized sectors, the spread between the two lowest lying states and the overlap between the ground state manifold wavefunctions and the Chern insulator wavefunctions, which are defined in Eq.~(\ref{ed:eq:chernstatew0nu3}). The results are shown in Fig.~\ref{ed:fig:phase_diagram_t_w0_app_3x3}. These phase diagrams are, overall, similar to Fig.~\ref{ed:fig:phase_diagram_t_w0}, which is calculated on $4\times 3$ lattice. The phase transition point of the FMC model ($\lambda = 0$)is at around $w_0/w_1 \simeq 0.9$, while for ($\lambda = 1$) it is around $w_0/w_1 \simeq 0.3$ for small $t$ and $w_0/w_1\simeq 0.4$ for large $t$. 

The spread between the two lowest states which carry Chern number $\nu_C = \pm 1$ in the spin and valley polarized sectors, however, is completely different from other system sizes. The $\delta$ plots in Fig.~\ref{ed:fig:phase_diagram_t_w0_app_3x3}b and e are quite small when compared with Figs.~\ref{ed:fig:phase_diagram_t_w0}b and e. In fact, these two lowest states with $\nu_C = \pm 1$ 
are degenerate within  machine precision (separated by roughly $10^{-11}~\rm meV$). This is because when the momentum lattice preserves $C_{3z}$ symmetry ($3\times 3$ here): a Chern insulator state with Chern number $\nu_C=\pm1$ have $C_{3z}$ eigenvalues $e^{\pm i 2\pi/3}$ \cite{fang_bulk_2012}, respectively, and are related by time-reversal $T$. These two states then form an irreducible representation of $T$ and $C_{3z}$, thus are exactly degenerate (even for finite sizes). For generic momentum lattices (such as $5\times3$ or $4\times3$ discussed in this article), $C_{3z}$ symmetry is broken, and the Chern numbers $\nu_C=\pm1$ are only related by time-reversal $T$. For finite system sizes, they will split into bonding and anti-bonding states which are eigenstates of $T$ (which is the reason for the finite spread $\delta$ when lattice is not $C_{3z}$ symmetric). However, in the thermodynamic limit, $\nu_C=\pm1$ spontaneously break time-reversal symmetry $T$, thus become degenerate when system size tends to infinity.

The wavefunction overlap shown in Fig.~\ref{ed:fig:phase_diagram_t_w0_app_3x3}c and f, is above $0.85$ in the Chern insulator phase for both $\lambda = 0$ and $\lambda = 1$ models. The similarity between the phase diagrams on $5\times 3$ and $3\times 3$ lattices in fully polarized sectors, and on $3\times 2$ lattice in all symmetry sectors, hints that the system size does not affect the insulating phase ground state significantly.

\begin{figure}
    \centering
    \includegraphics[width=0.9\linewidth]{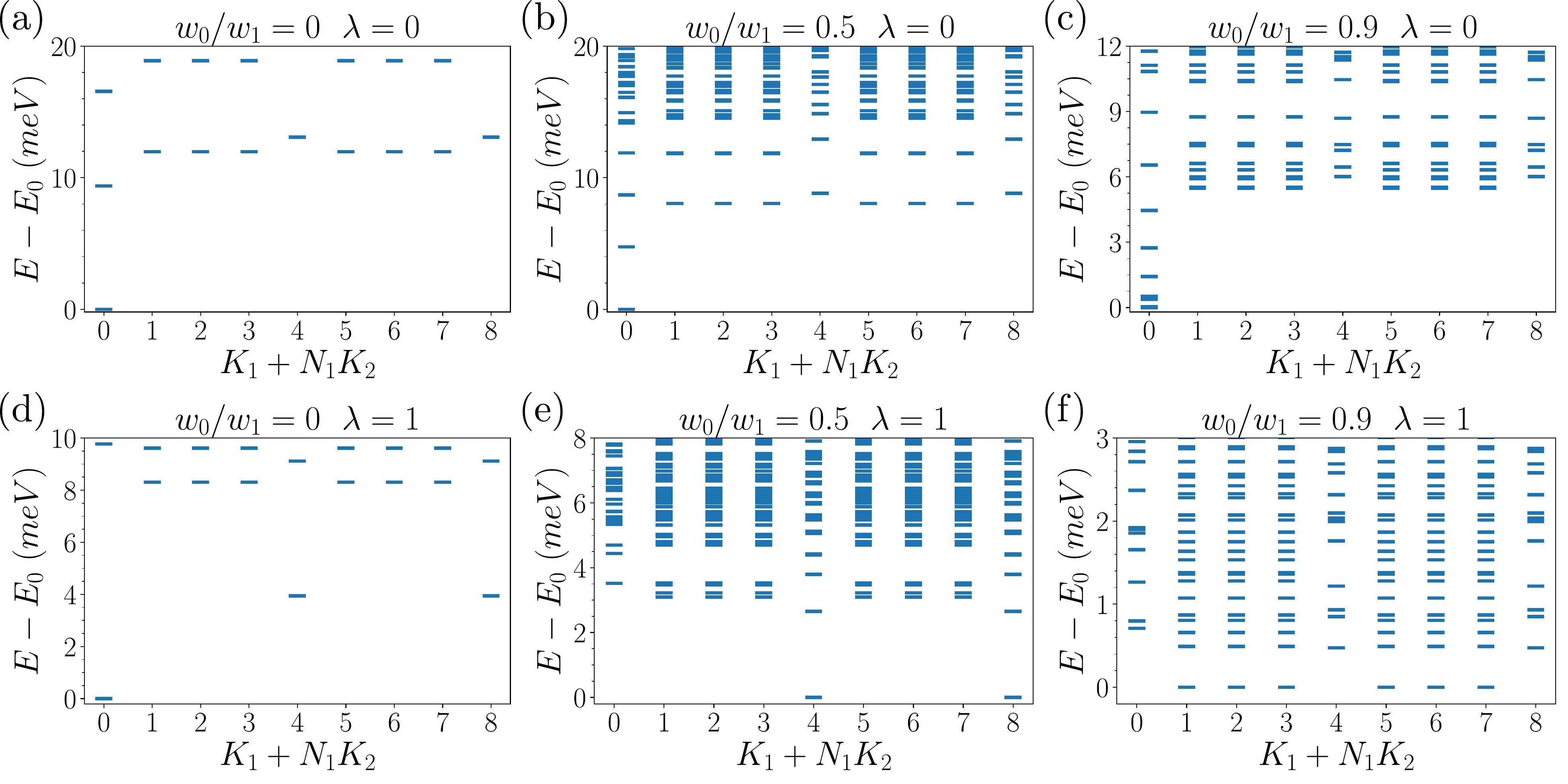}
    \caption{The energy spectra at filling factor $\nu=-3$ on a $3\times 3$ lattice in the nonchiral-flat limit and in the spin and valley polarized sectors. The upper panel (a-c) shows the spectra for $\lambda=0$, i.e., with FMC, and the lower panel (d-f) shows the spectra for $\lambda=1$ . The three different values of $w_0/w_1$, namely $w_0/w_1=0.0$, $w_0/w_1=0.5$ and $w_0/w_1=0.9$, have been selected to be representative of the different total momenta of the ground states. Note that for readability, the energy scales differs from one plot to another.}
    \label{fig:momentum_spectrum_3x3}
\end{figure}

For sake of completeness, we also provides tables giving the momenta of the lowest energy states for each point of the phase diagrams. For 3$\times$3, Table~\ref{ed:tab:momentumlambda0nu-3nx3ny3} is for $\lambda=0$ and Table~\ref{ed:tab:momentumlambda1nu-3nx3ny3} is for $\lambda=1$. 
Similarly, we give such momentum tables for the $4\times 3$ system (Tables~\ref{ed:tab:momentumlambda0nu-3nx4ny3} and \ref{ed:tab:momentumlambda1nu-3nx4ny3} for $\lambda=0$ and $\lambda=1$, respectively), and the momentum tables for the $5\times 3$ system discussed in Sec.~\ref{sec:fullypolsectorsnu3} (Tables~\ref{ed:tab:momentumlambda0nu-3nx5ny3} and~\ref{ed:tab:momentumlambda1nu-3nx5ny3} for $\lambda = 0$ and $\lambda=1$, respectively).
To exemplify these tables, we also have plotted representative momentum resolved energy spectra in Fig.~\ref{fig:momentum_spectrum_3x3} for the 3$\times$3 system in the nonchiral-flat limit for $\lambda=0$ (Figs.~\ref{fig:momentum_spectrum_3x3}a-c) and for $\lambda=1$ (Figs.~\ref{fig:momentum_spectrum_3x3}d-f).

{\tiny
\begin{table}
\begin{tabular}{|c|c|c|c|c|c|c|c|c|c|c|c|}
\hline
 & $t=0.0$ & $t=0.1$ & $t=0.2$ & $t=0.3$ & $t=0.4$ & $t=0.5$ & $t=0.6$ & $t=0.7$ & $t=0.8$ & $t=0.9$ & $t=1.0$ \\
\hline
$w_0=0.0$ & $$\rm{Ch}$$  & $$\rm{Ch}$$  & $$\rm{Ch}$$  & $$\rm{Ch}$$  & $$\rm{Ch}$$  & $$\rm{Ch}$$  & $$\rm{Ch}$$  & $$\rm{Ch}$$  & $$\rm{Ch}$$  & $$\rm{Ch}$$  & $$\rm{Ch}$$  \\ 
 \hline 
$w_0=0.1$ & $$\rm{Ch}$$  & $$\rm{Ch}$$  & $$\rm{Ch}$$  & $$\rm{Ch}$$  & $$\rm{Ch}$$  & $$\rm{Ch}$$  & $$\rm{Ch}$$  & $$\rm{Ch}$$  & $$\rm{Ch}$$  & $$\rm{Ch}$$  & $$\rm{Ch}$$  \\ 
 \hline 
$w_0=0.2$ & $$\rm{Ch}$$  & $$\rm{Ch}$$  & $$\rm{Ch}$$  & $$\rm{Ch}$$  & $$\rm{Ch}$$  & $$\rm{Ch}$$  & $$\rm{Ch}$$  & $$\rm{Ch}$$  & $$\rm{Ch}$$  & $$\rm{Ch}$$  & $$\rm{Ch}$$  \\ 
 \hline 
$w_0=0.3$ & $$\rm{Ch}$$  & $$\rm{Ch}$$  & $$\rm{Ch}$$  & $$\rm{Ch}$$  & $$\rm{Ch}$$  & $$\rm{Ch}$$  & $$\rm{Ch}$$  & $$\rm{Ch}$$  & $$\rm{Ch}$$  & $$\rm{Ch}$$  & $$\rm{Ch}$$  \\ 
 \hline 
$w_0=0.4$ & $$\rm{Ch}$$  & $$\rm{Ch}$$  & $$\rm{Ch}$$  & $$\rm{Ch}$$  & $$\rm{Ch}$$  & $$\rm{Ch}$$  & $$\rm{Ch}$$  & $$\rm{Ch}$$  & $$\rm{Ch}$$  & $$\rm{Ch}$$  & $$\rm{Ch}$$  \\ 
 \hline 
$w_0=0.5$ & $$\rm{Ch}$$  & $$\rm{Ch}$$  & $$\rm{Ch}$$  & $$\rm{Ch}$$  & $$\rm{Ch}$$  & $$\rm{Ch}$$  & $$\rm{Ch}$$  & $$\rm{Ch}$$  & $$\rm{Ch}$$  & $$\rm{Ch}$$  & $$\rm{Ch}$$  \\ 
 \hline 
$w_0=0.6$ & $$\rm{Ch}$$  & $$\rm{Ch}$$  & $$\rm{Ch}$$  & $$\rm{Ch}$$  & $$\rm{Ch}$$  & $$\rm{Ch}$$  & $$\rm{Ch}$$  & $$\rm{Ch}$$  & $$\rm{Ch}$$  & $$\rm{Ch}$$  & $$\rm{Ch}$$  \\ 
 \hline 
$w_0=0.7$ & $$\rm{Ch}$$  & $$\rm{Ch}$$  & $$\rm{Ch}$$  & $$\rm{Ch}$$  & $$\rm{Ch}$$  & $$\rm{Ch}$$  & $$\rm{Ch}$$  & $$\rm{Ch}$$  & $$\rm{Ch}$$  & $$\rm{Ch}$$  & $$\rm{Ch}$$  \\ 
 \hline 
$w_0=0.8$ & $$\rm{Ch}$$  & $$\rm{Ch}$$  & $$\rm{Ch}$$  & $$\rm{Ch}$$  & $$\rm{Ch}$$  & $$\rm{Ch}$$  & $$\rm{Ch}$$  & $$\rm{Ch}$$  & $$\rm{Ch}$$  & $$\rm{Ch}$$  & $$\rm{Ch}$$  \\ 
 \hline 
$w_0=0.9$ & ${\begin{tabular}{c}(0,0)\\(0,0)\end{tabular}}$  & ${\begin{tabular}{c}(0,0)\\(0,0)\end{tabular}}$  & ${\begin{tabular}{c}(0,0)\\(0,0)\end{tabular}}$  & ${\begin{tabular}{c}(0,0)\\(0,0)\end{tabular}}$  & ${\begin{tabular}{c}(0,0)\\(0,0)\end{tabular}}$  & ${\begin{tabular}{c}(0,0)\\(0,0)\end{tabular}}$  & ${\begin{tabular}{c}(0,0)\\(0,0)\end{tabular}}$  & ${\begin{tabular}{c}(0,0)\\(0,0)\end{tabular}}$  & ${\begin{tabular}{c}(0,0)\\(0,0)\end{tabular}}$  & ${\begin{tabular}{c}(0,0)\\(0,0)\end{tabular}}$  & ${\begin{tabular}{c}(0,0)\\(0,0)\end{tabular}}$  \\ 
 \hline 
$w_0=1.0$ & ${\begin{tabular}{c}(0,0)\\(0,0)\end{tabular}}$  & ${\begin{tabular}{c}(0,0)\\(0,0)\end{tabular}}$  & ${\begin{tabular}{c}(0,0)\\(0,0)\end{tabular}}$  & ${\begin{tabular}{c}(0,0)\\(0,0)\end{tabular}}$  & ${\begin{tabular}{c}(0,0)\\(0,0)\end{tabular}}$  & ${\begin{tabular}{c}(0,0)\\(0,0)\end{tabular}}$  & ${\begin{tabular}{c}(0,0)\\(0,0)\end{tabular}}$  & ${\begin{tabular}{c}(0,0)\\(0,0)\end{tabular}}$  & ${\begin{tabular}{c}(0,0)\\(0,0)\end{tabular}}$  & ${\begin{tabular}{c}(0,0)\\(0,0)\end{tabular}}$  & ${\begin{tabular}{c}(0,0)\\(0,0)\end{tabular}}$  \\ 
 \hline 
\hline
\end{tabular}
\caption{Momentum sectors relative to the Chern insulator state momentum, for the two lowest energy states (not related by $C_{2z}$) on a 3$\times$3 at $\nu=-3$ and $\lambda=0$ (top first lowest energy state, bottom second energy states). The system is fully spin and valley polarized. $\rm{Ch}$ indicates the Chern insulators states.}\label{ed:tab:momentumlambda0nu-3nx3ny3}
\end{table}
}

{\tiny
\begin{table}
\begin{tabular}{|c|c|c|c|c|c|c|c|c|c|c|c|}
\hline
 & $t=0.0$ & $t=0.1$ & $t=0.2$ & $t=0.3$ & $t=0.4$ & $t=0.5$ & $t=0.6$ & $t=0.7$ & $t=0.8$ & $t=0.9$ & $t=1.0$ \\
\hline
$w_0=0.0$ & $$\rm{Ch}$$  & $$\rm{Ch}$$  & $$\rm{Ch}$$  & $$\rm{Ch}$$  & $$\rm{Ch}$$  & $$\rm{Ch}$$  & $$\rm{Ch}$$  & $$\rm{Ch}$$  & $$\rm{Ch}$$  & $$\rm{Ch}$$  & $$\rm{Ch}$$  \\ 
 \hline 
$w_0=0.1$ & $$\rm{Ch}$$  & $$\rm{Ch}$$  & $$\rm{Ch}$$  & $$\rm{Ch}$$  & $$\rm{Ch}$$  & $$\rm{Ch}$$  & $$\rm{Ch}$$  & $$\rm{Ch}$$  & $$\rm{Ch}$$  & $$\rm{Ch}$$  & $$\rm{Ch}$$  \\ 
 \hline 
$w_0=0.2$ & $$\rm{Ch}$$  & $$\rm{Ch}$$  & $$\rm{Ch}$$  & $$\rm{Ch}$$  & $$\rm{Ch}$$  & $$\rm{Ch}$$  & $$\rm{Ch}$$  & $$\rm{Ch}$$  & $$\rm{Ch}$$  & $$\rm{Ch}$$  & $$\rm{Ch}$$  \\ 
 \hline 
$w_0=0.3$ & ${\begin{tabular}{c}(1,1)\\(0,0)\end{tabular}}$  & ${\begin{tabular}{c}(1,1)\\(0,0)\end{tabular}}$  & ${\begin{tabular}{c}(1,1)\\(0,0)\end{tabular}}$  & $$\rm{Ch}$$  & $$\rm{Ch}$$  & $$\rm{Ch}$$  & $$\rm{Ch}$$  & $$\rm{Ch}$$  & $$\rm{Ch}$$  & $$\rm{Ch}$$  & $$\rm{Ch}$$  \\ 
 \hline 
$w_0=0.4$ & ${\begin{tabular}{c}(1,1)\\(0,0)\end{tabular}}$  & ${\begin{tabular}{c}(1,1)\\(0,0)\end{tabular}}$  & ${\begin{tabular}{c}(1,1)\\(0,0)\end{tabular}}$  & ${\begin{tabular}{c}(1,1)\\(0,0)\end{tabular}}$  & ${\begin{tabular}{c}(1,1)\\(0,0)\end{tabular}}$  & ${\begin{tabular}{c}(1,1)\\(0,0)\end{tabular}}$  & ${\begin{tabular}{c}(1,1)\\(0,0)\end{tabular}}$  & ${\begin{tabular}{c}(1,1)\\(0,0)\end{tabular}}$  & ${\begin{tabular}{c}(1,1)\\(0,0)\end{tabular}}$  & ${\begin{tabular}{c}(1,1)\\(0,0)\end{tabular}}$  & ${\begin{tabular}{c}(1,1)\\(0,0)\end{tabular}}$  \\ 
 \hline 
$w_0=0.5$ & ${\begin{tabular}{c}(1,1)\\(1,2)\end{tabular}}$  & ${\begin{tabular}{c}(1,1)\\(1,0)\end{tabular}}$  & ${\begin{tabular}{c}(1,1)\\(1,0)\end{tabular}}$  & ${\begin{tabular}{c}(1,1)\\(0,1)\end{tabular}}$  & ${\begin{tabular}{c}(1,1)\\(1,2)\end{tabular}}$  & ${\begin{tabular}{c}(1,1)\\(0,1)\end{tabular}}$  & ${\begin{tabular}{c}(1,1)\\(0,1)\end{tabular}}$  & ${\begin{tabular}{c}(1,1)\\(0,1)\end{tabular}}$  & ${\begin{tabular}{c}(1,1)\\(1,2)\end{tabular}}$  & ${\begin{tabular}{c}(1,1)\\(0,1)\end{tabular}}$  & ${\begin{tabular}{c}(1,1)\\(1,2)\end{tabular}}$  \\ 
 \hline 
$w_0=0.6$ & ${\begin{tabular}{c}(1,1)\\(1,2)\end{tabular}}$  & ${\begin{tabular}{c}(1,1)\\(1,0)\end{tabular}}$  & ${\begin{tabular}{c}(1,1)\\(1,0)\end{tabular}}$  & ${\begin{tabular}{c}(1,1)\\(1,0)\end{tabular}}$  & ${\begin{tabular}{c}(1,1)\\(0,1)\end{tabular}}$  & ${\begin{tabular}{c}(1,1)\\(1,2)\end{tabular}}$  & ${\begin{tabular}{c}(1,1)\\(1,2)\end{tabular}}$  & ${\begin{tabular}{c}(1,1)\\(0,1)\end{tabular}}$  & ${\begin{tabular}{c}(1,1)\\(1,2)\end{tabular}}$  & ${\begin{tabular}{c}(1,1)\\(0,1)\end{tabular}}$  & ${\begin{tabular}{c}(1,1)\\(1,2)\end{tabular}}$  \\ 
 \hline 
$w_0=0.7$ & ${\begin{tabular}{c}(1,1)\\(1,2)\end{tabular}}$  & ${\begin{tabular}{c}(1,1)\\(1,2)\end{tabular}}$  & ${\begin{tabular}{c}(1,1)\\(1,2)\end{tabular}}$  & ${\begin{tabular}{c}(1,1)\\(1,2)\end{tabular}}$  & ${\begin{tabular}{c}(1,1)\\(0,1)\end{tabular}}$  & ${\begin{tabular}{c}(1,1)\\(0,1)\end{tabular}}$  & ${\begin{tabular}{c}(1,1)\\(0,1)\end{tabular}}$  & ${\begin{tabular}{c}(1,1)\\(0,1)\end{tabular}}$  & ${\begin{tabular}{c}(1,1)\\(0,1)\end{tabular}}$  & ${\begin{tabular}{c}(1,1)\\(1,2)\end{tabular}}$  & ${\begin{tabular}{c}(1,1)\\(1,2)\end{tabular}}$  \\ 
 \hline 
$w_0=0.8$ & ${\begin{tabular}{c}(1,1)\\(1,0)\end{tabular}}$  & ${\begin{tabular}{c}(1,1)\\(1,2)\end{tabular}}$  & ${\begin{tabular}{c}(1,1)\\(1,2)\end{tabular}}$  & ${\begin{tabular}{c}(1,1)\\(0,1)\end{tabular}}$  & ${\begin{tabular}{c}(1,1)\\(0,1)\end{tabular}}$  & ${\begin{tabular}{c}(1,1)\\(0,1)\end{tabular}}$  & ${\begin{tabular}{c}(1,1)\\(0,1)\end{tabular}}$  & ${\begin{tabular}{c}(1,1)\\(0,1)\end{tabular}}$  & ${\begin{tabular}{c}(1,1)\\(0,1)\end{tabular}}$  & ${\begin{tabular}{c}(1,1)\\(0,1)\end{tabular}}$  & ${\begin{tabular}{c}(1,1)\\(0,1)\end{tabular}}$  \\ 
 \hline 
$w_0=0.9$ & ${\begin{tabular}{c}(1,0)\\(1,2)\end{tabular}}$  & ${\begin{tabular}{c}(1,0)\\(0,1)\end{tabular}}$  & ${\begin{tabular}{c}(1,0)\\(0,1)\end{tabular}}$  & ${\begin{tabular}{c}(1,0)\\(1,2)\end{tabular}}$  & ${\begin{tabular}{c}(1,0)\\(0,1)\end{tabular}}$  & ${\begin{tabular}{c}(1,0)\\(0,1)\end{tabular}}$  & ${\begin{tabular}{c}(1,0)\\(1,2)\end{tabular}}$  & ${\begin{tabular}{c}(1,0)\\(1,2)\end{tabular}}$  & ${\begin{tabular}{c}(1,0)\\(1,2)\end{tabular}}$  & ${\begin{tabular}{c}(1,0)\\(0,1)\end{tabular}}$  & ${\begin{tabular}{c}(1,0)\\(1,2)\end{tabular}}$  \\ 
 \hline 
$w_0=1.0$ & ${\begin{tabular}{c}(1,2)\\(1,0)\end{tabular}}$  & ${\begin{tabular}{c}(1,0)\\(0,1)\end{tabular}}$  & ${\begin{tabular}{c}(1,0)\\(1,2)\end{tabular}}$  & ${\begin{tabular}{c}(1,0)\\(0,1)\end{tabular}}$  & ${\begin{tabular}{c}(1,0)\\(0,1)\end{tabular}}$  & ${\begin{tabular}{c}(1,0)\\(1,2)\end{tabular}}$  & ${\begin{tabular}{c}(1,0)\\(0,1)\end{tabular}}$  & ${\begin{tabular}{c}(1,0)\\(0,1)\end{tabular}}$  & ${\begin{tabular}{c}(1,0)\\(1,2)\end{tabular}}$  & ${\begin{tabular}{c}(1,0)\\(0,1)\end{tabular}}$  & ${\begin{tabular}{c}(1,0)\\(1,2)\end{tabular}}$  \\ 
 \hline
 \end{tabular}
\caption{Momentum sectors relative to the Chern insulator state momentum, for the two lowest energy states (not related by $C_{2z}$) on a 3$\times$3 at $\nu=-3$ and $\lambda=1$ (top first lowest energy state, bottom second energy states). The system is fully spin and valley polarized. $\rm{Ch}$ indicates the Chern insulators states.}\label{ed:tab:momentumlambda1nu-3nx3ny3}
\end{table}
}

{\tiny
\begin{table}
\begin{tabular}{|c|c|c|c|c|c|c|c|c|c|c|c|}
\hline
 & $t=0.0$ & $t=0.1$ & $t=0.2$ & $t=0.3$ & $t=0.4$ & $t=0.5$ & $t=0.6$ & $t=0.7$ & $t=0.8$ & $t=0.9$ & $t=1.0$ \\
\hline
$w_0=0.0$ & $$\rm{Ch}$$  & $$\rm{Ch}$$  & $$\rm{Ch}$$  & $$\rm{Ch}$$  & $$\rm{Ch}$$  & $$\rm{Ch}$$  & $$\rm{Ch}$$  & $$\rm{Ch}$$  & $$\rm{Ch}$$  & $$\rm{Ch}$$  & $$\rm{Ch}$$  \\ 
 \hline 
$w_0=0.1$ & $$\rm{Ch}$$  & $$\rm{Ch}$$  & $$\rm{Ch}$$  & $$\rm{Ch}$$  & $$\rm{Ch}$$  & $$\rm{Ch}$$  & $$\rm{Ch}$$  & $$\rm{Ch}$$  & $$\rm{Ch}$$  & $$\rm{Ch}$$  & $$\rm{Ch}$$  \\ 
 \hline 
$w_0=0.2$ & $$\rm{Ch}$$  & $$\rm{Ch}$$  & $$\rm{Ch}$$  & $$\rm{Ch}$$  & $$\rm{Ch}$$  & $$\rm{Ch}$$  & $$\rm{Ch}$$  & $$\rm{Ch}$$  & $$\rm{Ch}$$  & $$\rm{Ch}$$  & $$\rm{Ch}$$  \\ 
 \hline 
$w_0=0.3$ & $$\rm{Ch}$$  & $$\rm{Ch}$$  & $$\rm{Ch}$$  & $$\rm{Ch}$$  & $$\rm{Ch}$$  & $$\rm{Ch}$$  & $$\rm{Ch}$$  & $$\rm{Ch}$$  & $$\rm{Ch}$$  & $$\rm{Ch}$$  & $$\rm{Ch}$$  \\ 
 \hline 
$w_0=0.4$ & $$\rm{Ch}$$  & $$\rm{Ch}$$  & $$\rm{Ch}$$  & $$\rm{Ch}$$  & $$\rm{Ch}$$  & $$\rm{Ch}$$  & $$\rm{Ch}$$  & $$\rm{Ch}$$  & $$\rm{Ch}$$  & $$\rm{Ch}$$  & $$\rm{Ch}$$  \\ 
 \hline 
$w_0=0.5$ & $$\rm{Ch}$$  & $$\rm{Ch}$$  & $$\rm{Ch}$$  & $$\rm{Ch}$$  & $$\rm{Ch}$$  & $$\rm{Ch}$$  & $$\rm{Ch}$$  & $$\rm{Ch}$$  & $$\rm{Ch}$$  & $$\rm{Ch}$$  & $$\rm{Ch}$$  \\ 
 \hline 
$w_0=0.6$ & $$\rm{Ch}$$  & $$\rm{Ch}$$  & $$\rm{Ch}$$  & $$\rm{Ch}$$  & $$\rm{Ch}$$  & $$\rm{Ch}$$  & $$\rm{Ch}$$  & $$\rm{Ch}$$  & $$\rm{Ch}$$  & $$\rm{Ch}$$  & $$\rm{Ch}$$  \\ 
 \hline 
$w_0=0.7$ & $$\rm{Ch}$$  & $$\rm{Ch}$$  & $$\rm{Ch}$$  & $$\rm{Ch}$$  & $$\rm{Ch}$$  & $$\rm{Ch}$$  & $$\rm{Ch}$$  & $$\rm{Ch}$$  & $$\rm{Ch}$$  & $$\rm{Ch}$$  & $$\rm{Ch}$$  \\ 
 \hline 
$w_0=0.8$ & $$\rm{Ch}$$  & $$\rm{Ch}$$  & $$\rm{Ch}$$  & $$\rm{Ch}$$  & $$\rm{Ch}$$  & $$\rm{Ch}$$  & $$\rm{Ch}$$  & $$\rm{Ch}$$  & $$\rm{Ch}$$  & $$\rm{Ch}$$  & $$\rm{Ch}$$  \\ 
 \hline 
$w_0=0.9$ & ${\begin{tabular}{c}(0,0)\\(0,0)\end{tabular}}$  & ${\begin{tabular}{c}(0,0)\\(0,0)\end{tabular}}$  & ${\begin{tabular}{c}(0,0)\\(0,0)\end{tabular}}$  & ${\begin{tabular}{c}(0,0)\\(0,0)\end{tabular}}$  & ${\begin{tabular}{c}(0,0)\\(0,0)\end{tabular}}$  & ${\begin{tabular}{c}(0,0)\\(0,0)\end{tabular}}$  & ${\begin{tabular}{c}(0,0)\\(0,0)\end{tabular}}$  & ${\begin{tabular}{c}(0,0)\\(0,0)\end{tabular}}$  & ${\begin{tabular}{c}(0,0)\\(0,0)\end{tabular}}$  & ${\begin{tabular}{c}(0,0)\\(0,0)\end{tabular}}$  & ${\begin{tabular}{c}(0,0)\\(0,0)\end{tabular}}$  \\ 
 \hline 
$w_0=1.0$ & ${\begin{tabular}{c}(0,0)\\(0,0)\end{tabular}}$  & ${\begin{tabular}{c}(0,0)\\(0,0)\end{tabular}}$  & ${\begin{tabular}{c}(0,0)\\(0,0)\end{tabular}}$  & ${\begin{tabular}{c}(0,0)\\(0,0)\end{tabular}}$  & ${\begin{tabular}{c}(0,0)\\(0,0)\end{tabular}}$  & ${\begin{tabular}{c}(0,0)\\(0,0)\end{tabular}}$  & ${\begin{tabular}{c}(0,0)\\(0,0)\end{tabular}}$  & ${\begin{tabular}{c}(0,0)\\(0,0)\end{tabular}}$  & ${\begin{tabular}{c}(0,0)\\(0,0)\end{tabular}}$  & ${\begin{tabular}{c}(0,0)\\(0,0)\end{tabular}}$  & ${\begin{tabular}{c}(0,0)\\(0,0)\end{tabular}}$  \\ 
 \hline 
\end{tabular}
\caption{Momentum sectors relative to the Chern insulator state momentum, for the two lowest energy states (not related by $C_{2z}$) on a 4$\times$3 at $\nu=-3$ and $\lambda=0$ (top first lowest energy state, bottom second energy states). The system is fully spin and valley polarized. $\rm{Ch}$ indicates the Chern insulators states.}\label{ed:tab:momentumlambda0nu-3nx4ny3}
\end{table}
}

{\tiny
\begin{table}
\begin{tabular}{|c|c|c|c|c|c|c|c|c|c|c|c|}
\hline
 & $t=0.0$ & $t=0.1$ & $t=0.2$ & $t=0.3$ & $t=0.4$ & $t=0.5$ & $t=0.6$ & $t=0.7$ & $t=0.8$ & $t=0.9$ & $t=1.0$ \\
\hline
$w_0=0.0$ & $$\rm{Ch}$$  & $$\rm{Ch}$$  & $$\rm{Ch}$$  & $$\rm{Ch}$$  & $$\rm{Ch}$$  & $$\rm{Ch}$$  & $$\rm{Ch}$$  & $$\rm{Ch}$$  & $$\rm{Ch}$$  & $$\rm{Ch}$$  & $$\rm{Ch}$$  \\ 
 \hline 
$w_0=0.1$ & $$\rm{Ch}$$  & $$\rm{Ch}$$  & $$\rm{Ch}$$  & $$\rm{Ch}$$  & $$\rm{Ch}$$  & $$\rm{Ch}$$  & $$\rm{Ch}$$  & $$\rm{Ch}$$  & $$\rm{Ch}$$  & $$\rm{Ch}$$  & $$\rm{Ch}$$  \\ 
 \hline 
$w_0=0.2$ & $$\rm{Ch}$$  & $$\rm{Ch}$$  & $$\rm{Ch}$$  & $$\rm{Ch}$$  & $$\rm{Ch}$$  & $$\rm{Ch}$$  & $$\rm{Ch}$$  & $$\rm{Ch}$$  & $$\rm{Ch}$$  & $$\rm{Ch}$$  & $$\rm{Ch}$$  \\ 
 \hline 
$w_0=0.3$ & $$\rm{Ch}$$  & $$\rm{Ch}$$  & $$\rm{Ch}$$  & $$\rm{Ch}$$  & $$\rm{Ch}$$  & $$\rm{Ch}$$  & $$\rm{Ch}$$  & $$\rm{Ch}$$  & $$\rm{Ch}$$  & $$\rm{Ch}$$  & $$\rm{Ch}$$  \\ 
 \hline 
$w_0=0.4$ & ${\begin{tabular}{c}(1,1)\\(2,0)\end{tabular}}$  & ${\begin{tabular}{c}(1,1)\\(2,0)\end{tabular}}$  & ${\begin{tabular}{c}(1,1)\\(2,0)\end{tabular}}$  & ${\begin{tabular}{c}(1,1)\\(2,0)\end{tabular}}$  & ${\begin{tabular}{c}(1,1)\\(2,0)\end{tabular}}$  & ${\begin{tabular}{c}(1,1)\\(2,0)\end{tabular}}$  & ${\begin{tabular}{c}(1,1)\\(2,0)\end{tabular}}$  & ${\begin{tabular}{c}(1,1)\\(2,0)\end{tabular}}$  & ${\begin{tabular}{c}(1,1)\\(2,0)\end{tabular}}$  & ${\begin{tabular}{c}(1,1)\\(2,0)\end{tabular}}$  & ${\begin{tabular}{c}(1,1)\\(2,0)\end{tabular}}$  \\ 
 \hline 
$w_0=0.5$ & ${\begin{tabular}{c}(1,1)\\(2,0)\end{tabular}}$  & ${\begin{tabular}{c}(1,1)\\(2,0)\end{tabular}}$  & ${\begin{tabular}{c}(1,1)\\(2,0)\end{tabular}}$  & ${\begin{tabular}{c}(1,1)\\(2,0)\end{tabular}}$  & ${\begin{tabular}{c}(1,1)\\(2,0)\end{tabular}}$  & ${\begin{tabular}{c}(1,1)\\(2,0)\end{tabular}}$  & ${\begin{tabular}{c}(1,1)\\(2,0)\end{tabular}}$  & ${\begin{tabular}{c}(1,1)\\(2,0)\end{tabular}}$  & ${\begin{tabular}{c}(1,1)\\(2,0)\end{tabular}}$  & ${\begin{tabular}{c}(1,1)\\(2,0)\end{tabular}}$  & ${\begin{tabular}{c}(1,1)\\(2,0)\end{tabular}}$  \\ 
 \hline 
$w_0=0.6$ & ${\begin{tabular}{c}(2,0)\\(1,1)\end{tabular}}$  & ${\begin{tabular}{c}(2,0)\\(1,1)\end{tabular}}$  & ${\begin{tabular}{c}(2,0)\\(1,1)\end{tabular}}$  & ${\begin{tabular}{c}(2,0)\\(1,1)\end{tabular}}$  & ${\begin{tabular}{c}(2,0)\\(1,1)\end{tabular}}$  & ${\begin{tabular}{c}(2,0)\\(1,1)\end{tabular}}$  & ${\begin{tabular}{c}(2,0)\\(1,1)\end{tabular}}$  & ${\begin{tabular}{c}(2,0)\\(1,1)\end{tabular}}$  & ${\begin{tabular}{c}(2,0)\\(1,1)\end{tabular}}$  & ${\begin{tabular}{c}(2,0)\\(1,1)\end{tabular}}$  & ${\begin{tabular}{c}(2,0)\\(1,1)\end{tabular}}$  \\ 
 \hline 
$w_0=0.7$ & ${\begin{tabular}{c}(2,0)\\(1,1)\end{tabular}}$  & ${\begin{tabular}{c}(2,0)\\(1,1)\end{tabular}}$  & ${\begin{tabular}{c}(2,0)\\(1,1)\end{tabular}}$  & ${\begin{tabular}{c}(2,0)\\(1,1)\end{tabular}}$  & ${\begin{tabular}{c}(2,0)\\(1,1)\end{tabular}}$  & ${\begin{tabular}{c}(2,0)\\(1,1)\end{tabular}}$  & ${\begin{tabular}{c}(2,0)\\(1,1)\end{tabular}}$  & ${\begin{tabular}{c}(2,0)\\(1,1)\end{tabular}}$  & ${\begin{tabular}{c}(2,0)\\(1,1)\end{tabular}}$  & ${\begin{tabular}{c}(2,0)\\(1,1)\end{tabular}}$  & ${\begin{tabular}{c}(2,0)\\(1,1)\end{tabular}}$  \\ 
 \hline 
$w_0=0.8$ & ${\begin{tabular}{c}(2,0)\\(1,1)\end{tabular}}$  & ${\begin{tabular}{c}(2,0)\\(1,1)\end{tabular}}$  & ${\begin{tabular}{c}(2,0)\\(1,1)\end{tabular}}$  & ${\begin{tabular}{c}(2,0)\\(1,1)\end{tabular}}$  & ${\begin{tabular}{c}(2,0)\\(1,1)\end{tabular}}$  & ${\begin{tabular}{c}(2,0)\\(1,1)\end{tabular}}$  & ${\begin{tabular}{c}(2,0)\\(1,1)\end{tabular}}$  & ${\begin{tabular}{c}(2,0)\\(1,1)\end{tabular}}$  & ${\begin{tabular}{c}(2,0)\\(1,1)\end{tabular}}$  & ${\begin{tabular}{c}(2,0)\\(1,1)\end{tabular}}$  & ${\begin{tabular}{c}(2,0)\\(1,1)\end{tabular}}$  \\ 
 \hline 
$w_0=0.9$ & ${\begin{tabular}{c}(1,2)\\(1,0)\end{tabular}}$  & ${\begin{tabular}{c}(1,2)\\(1,0)\end{tabular}}$  & ${\begin{tabular}{c}(1,2)\\(1,0)\end{tabular}}$  & ${\begin{tabular}{c}(1,2)\\(1,0)\end{tabular}}$  & ${\begin{tabular}{c}(1,2)\\(1,0)\end{tabular}}$  & ${\begin{tabular}{c}(1,2)\\(1,0)\end{tabular}}$  & ${\begin{tabular}{c}(1,2)\\(1,0)\end{tabular}}$  & ${\begin{tabular}{c}(1,2)\\(1,0)\end{tabular}}$  & ${\begin{tabular}{c}(1,2)\\(1,0)\end{tabular}}$  & ${\begin{tabular}{c}(1,2)\\(2,0)\end{tabular}}$  & ${\begin{tabular}{c}(1,2)\\(2,0)\end{tabular}}$  \\ 
 \hline 
$w_0=1.0$ & ${\begin{tabular}{c}(1,0)\\(1,2)\end{tabular}}$  & ${\begin{tabular}{c}(1,0)\\(1,2)\end{tabular}}$  & ${\begin{tabular}{c}(1,0)\\(1,2)\end{tabular}}$  & ${\begin{tabular}{c}(1,2)\\(1,0)\end{tabular}}$  & ${\begin{tabular}{c}(1,2)\\(1,0)\end{tabular}}$  & ${\begin{tabular}{c}(1,2)\\(0,1)\end{tabular}}$  & ${\begin{tabular}{c}(1,2)\\(0,1)\end{tabular}}$  & ${\begin{tabular}{c}(1,2)\\(0,1)\end{tabular}}$  & ${\begin{tabular}{c}(1,2)\\(0,1)\end{tabular}}$  & ${\begin{tabular}{c}(1,2)\\(0,1)\end{tabular}}$  & ${\begin{tabular}{c}(1,2)\\(0,1)\end{tabular}}$  \\ 
 \hline 
\end{tabular}
\caption{Momentum sectors relative to the Chern insulator state momentum, for the two lowest energy states (not related by $C_{2z}$) on a 4$\times$3 at $\nu=-3$ and $\lambda=1$ (top first lowest energy state, bottom second energy states). The system is fully spin and valley polarized. $\rm{Ch}$ indicates the Chern insulators states.}\label{ed:tab:momentumlambda1nu-3nx4ny3}
\end{table}
}

{\tiny
\begin{table}
\begin{tabular}{|c|c|c|c|c|c|c|c|c|c|c|c|}
\hline
 & $t=0.0$ & $t=0.1$ & $t=0.2$ & $t=0.3$ & $t=0.4$ & $t=0.5$ & $t=0.6$ & $t=0.7$ & $t=0.8$ & $t=0.9$ & $t=1.0$ \\
\hline
$w_0=0.0$ & $$\rm{Ch}$$  & $$\rm{Ch}$$  & $$\rm{Ch}$$  & $$\rm{Ch}$$  & $$\rm{Ch}$$  & $$\rm{Ch}$$  & $$\rm{Ch}$$  & $$\rm{Ch}$$  & $$\rm{Ch}$$  & $$\rm{Ch}$$  & $$\rm{Ch}$$  \\ 
 \hline 
$w_0=0.1$ & $$\rm{Ch}$$  & $$\rm{Ch}$$  & $$\rm{Ch}$$  & $$\rm{Ch}$$  & $$\rm{Ch}$$  & $$\rm{Ch}$$  & $$\rm{Ch}$$  & $$\rm{Ch}$$  & $$\rm{Ch}$$  & $$\rm{Ch}$$  & $$\rm{Ch}$$  \\ 
 \hline 
$w_0=0.2$ & $$\rm{Ch}$$  & $$\rm{Ch}$$  & $$\rm{Ch}$$  & $$\rm{Ch}$$  & $$\rm{Ch}$$  & $$\rm{Ch}$$  & $$\rm{Ch}$$  & $$\rm{Ch}$$  & $$\rm{Ch}$$  & $$\rm{Ch}$$  & $$\rm{Ch}$$  \\ 
 \hline 
$w_0=0.3$ & $$\rm{Ch}$$  & $$\rm{Ch}$$  & $$\rm{Ch}$$  & $$\rm{Ch}$$  & $$\rm{Ch}$$  & $$\rm{Ch}$$  & $$\rm{Ch}$$  & $$\rm{Ch}$$  & $$\rm{Ch}$$  & $$\rm{Ch}$$  & $$\rm{Ch}$$  \\ 
 \hline 
$w_0=0.4$ & $$\rm{Ch}$$  & $$\rm{Ch}$$  & $$\rm{Ch}$$  & $$\rm{Ch}$$  & $$\rm{Ch}$$  & $$\rm{Ch}$$  & $$\rm{Ch}$$  & $$\rm{Ch}$$  & $$\rm{Ch}$$  & $$\rm{Ch}$$  & $$\rm{Ch}$$  \\ 
 \hline 
$w_0=0.5$ & $$\rm{Ch}$$  & $$\rm{Ch}$$  & $$\rm{Ch}$$  & $$\rm{Ch}$$  & $$\rm{Ch}$$  & $$\rm{Ch}$$  & $$\rm{Ch}$$  & $$\rm{Ch}$$  & $$\rm{Ch}$$  & $$\rm{Ch}$$  & $$\rm{Ch}$$  \\ 
 \hline 
$w_0=0.6$ & $$\rm{Ch}$$  & $$\rm{Ch}$$  & $$\rm{Ch}$$  & $$\rm{Ch}$$  & $$\rm{Ch}$$  & $$\rm{Ch}$$  & $$\rm{Ch}$$  & $$\rm{Ch}$$  & $$\rm{Ch}$$  & $$\rm{Ch}$$  & $$\rm{Ch}$$  \\ 
 \hline 
$w_0=0.7$ & $$\rm{Ch}$$  & $$\rm{Ch}$$  & $$\rm{Ch}$$  & $$\rm{Ch}$$  & $$\rm{Ch}$$  & $$\rm{Ch}$$  & $$\rm{Ch}$$  & $$\rm{Ch}$$  & $$\rm{Ch}$$  & $$\rm{Ch}$$  & $$\rm{Ch}$$  \\ 
 \hline 
$w_0=0.8$ & $$\rm{Ch}$$  & $$\rm{Ch}$$  & $$\rm{Ch}$$  & $$\rm{Ch}$$  & $$\rm{Ch}$$  & $$\rm{Ch}$$  & $$\rm{Ch}$$  & $$\rm{Ch}$$  & $$\rm{Ch}$$  & $$\rm{Ch}$$  & $$\rm{Ch}$$  \\ 
 \hline 
$w_0=0.9$ & ${\begin{tabular}{c}(0,0)\\(0,0)\end{tabular}}$  & ${\begin{tabular}{c}(0,0)\\(0,0)\end{tabular}}$  & ${\begin{tabular}{c}(0,0)\\(0,0)\end{tabular}}$  & ${\begin{tabular}{c}(0,0)\\(0,0)\end{tabular}}$  & ${\begin{tabular}{c}(0,0)\\(0,0)\end{tabular}}$  & ${\begin{tabular}{c}(0,0)\\(0,0)\end{tabular}}$  & ${\begin{tabular}{c}(0,0)\\(0,0)\end{tabular}}$  & ${\begin{tabular}{c}(0,0)\\(0,0)\end{tabular}}$  & ${\begin{tabular}{c}(0,0)\\(0,0)\end{tabular}}$  & ${\begin{tabular}{c}(0,0)\\(0,0)\end{tabular}}$  & ${\begin{tabular}{c}(0,0)\\(0,0)\end{tabular}}$  \\ 
 \hline 
$w_0=1.0$ & ${\begin{tabular}{c}(0,0)\\(0,0)\end{tabular}}$  & ${\begin{tabular}{c}(0,0)\\(0,0)\end{tabular}}$  & ${\begin{tabular}{c}(0,0)\\(0,0)\end{tabular}}$  & ${\begin{tabular}{c}(0,0)\\(0,0)\end{tabular}}$  & ${\begin{tabular}{c}(0,0)\\(0,0)\end{tabular}}$  & ${\begin{tabular}{c}(0,0)\\(0,0)\end{tabular}}$  & ${\begin{tabular}{c}(0,0)\\(0,0)\end{tabular}}$  & ${\begin{tabular}{c}(0,0)\\(0,0)\end{tabular}}$  & ${\begin{tabular}{c}(0,0)\\(0,0)\end{tabular}}$  & ${\begin{tabular}{c}(0,0)\\(0,0)\end{tabular}}$  & ${\begin{tabular}{c}(0,0)\\(0,0)\end{tabular}}$  \\ 
 \hline 
\end{tabular}
\caption{Momentum sectors relative to the Chern insulator state momentum, for the two lowest energy states (not related by $C_{2z}$) on a 5$\times$3 at $\nu=-3$ and $\lambda=0$ (top first lowest energy state, bottom second energy states). The system is fully spin and valley polarized. $\rm{Ch}$ indicates the Chern insulators states.}\label{ed:tab:momentumlambda0nu-3nx5ny3}
\end{table}
}

{\tiny
\begin{table}
\begin{tabular}{|c|c|c|c|c|c|c|c|c|c|c|c|}
\hline
 & $t=0.0$ & $t=0.1$ & $t=0.2$ & $t=0.3$ & $t=0.4$ & $t=0.5$ & $t=0.6$ & $t=0.7$ & $t=0.8$ & $t=0.9$ & $t=1.0$ \\
\hline
$w_0=0.0$ & $$\rm{Ch}$$  & $$\rm{Ch}$$  & $$\rm{Ch}$$  & $$\rm{Ch}$$  & $$\rm{Ch}$$  & $$\rm{Ch}$$  & $$\rm{Ch}$$  & $$\rm{Ch}$$  & $$\rm{Ch}$$  & $$\rm{Ch}$$  & $$\rm{Ch}$$  \\ 
 \hline 
$w_0=0.1$ & $$\rm{Ch}$$  & $$\rm{Ch}$$  & $$\rm{Ch}$$  & $$\rm{Ch}$$  & $$\rm{Ch}$$  & $$\rm{Ch}$$  & $$\rm{Ch}$$  & $$\rm{Ch}$$  & $$\rm{Ch}$$  & $$\rm{Ch}$$  & $$\rm{Ch}$$  \\ 
 \hline 
$w_0=0.2$ & $$\rm{Ch}$$  & $$\rm{Ch}$$  & $$\rm{Ch}$$  & $$\rm{Ch}$$  & $$\rm{Ch}$$  & $$\rm{Ch}$$  & $$\rm{Ch}$$  & $$\rm{Ch}$$  & $$\rm{Ch}$$  & $$\rm{Ch}$$  & $$\rm{Ch}$$  \\ 
 \hline 
$w_0=0.3$ & $$\rm{Ch}$$  & $$\rm{Ch}$$  & $$\rm{Ch}$$  & $$\rm{Ch}$$  & $$\rm{Ch}$$  & $$\rm{Ch}$$  & $$\rm{Ch}$$  & $$\rm{Ch}$$  & $$\rm{Ch}$$  & $$\rm{Ch}$$  & $$\rm{Ch}$$  \\ 
 \hline 
$w_0=0.4$ & ${\begin{tabular}{c}(2,1)\\(1,1)\end{tabular}}$  & ${\begin{tabular}{c}(2,1)\\(1,1)\end{tabular}}$  & ${\begin{tabular}{c}(2,1)\\(1,1)\end{tabular}}$  & ${\begin{tabular}{c}(2,1)\\(1,1)\end{tabular}}$  & ${\begin{tabular}{c}(2,1)\\(1,1)\end{tabular}}$  & ${\begin{tabular}{c}(2,1)\\(1,1)\end{tabular}}$  & ${\begin{tabular}{c}(2,1)\\(1,1)\end{tabular}}$  & ${\begin{tabular}{c}(2,1)\\(1,1)\end{tabular}}$  & ${\begin{tabular}{c}(2,1)\\(1,1)\end{tabular}}$  & ${\begin{tabular}{c}(2,1)\\(1,1)\end{tabular}}$  & ${\begin{tabular}{c}(2,1)\\(1,1)\end{tabular}}$  \\ 
 \hline 
$w_0=0.5$ & ${\begin{tabular}{c}(2,1)\\(1,1)\end{tabular}}$  & ${\begin{tabular}{c}(2,1)\\(1,1)\end{tabular}}$  & ${\begin{tabular}{c}(2,1)\\(1,1)\end{tabular}}$  & ${\begin{tabular}{c}(2,1)\\(1,1)\end{tabular}}$  & ${\begin{tabular}{c}(2,1)\\(1,1)\end{tabular}}$  & ${\begin{tabular}{c}(2,1)\\(1,1)\end{tabular}}$  & ${\begin{tabular}{c}(2,1)\\(1,1)\end{tabular}}$  & ${\begin{tabular}{c}(2,1)\\(1,1)\end{tabular}}$  & ${\begin{tabular}{c}(2,1)\\(1,1)\end{tabular}}$  & ${\begin{tabular}{c}(2,1)\\(1,1)\end{tabular}}$  & ${\begin{tabular}{c}(2,1)\\(1,1)\end{tabular}}$  \\ 
 \hline 
$w_0=0.6$ & ${\begin{tabular}{c}(2,1)\\(1,1)\end{tabular}}$  & ${\begin{tabular}{c}(2,1)\\(1,1)\end{tabular}}$  & ${\begin{tabular}{c}(2,1)\\(1,1)\end{tabular}}$  & ${\begin{tabular}{c}(2,1)\\(1,1)\end{tabular}}$  & ${\begin{tabular}{c}(2,1)\\(1,1)\end{tabular}}$  & ${\begin{tabular}{c}(2,1)\\(1,1)\end{tabular}}$  & ${\begin{tabular}{c}(2,1)\\(1,1)\end{tabular}}$  & ${\begin{tabular}{c}(2,1)\\(1,1)\end{tabular}}$  & ${\begin{tabular}{c}(2,1)\\(1,1)\end{tabular}}$  & ${\begin{tabular}{c}(2,1)\\(1,1)\end{tabular}}$  & ${\begin{tabular}{c}(2,1)\\(1,1)\end{tabular}}$  \\ 
 \hline 
$w_0=0.7$ & ${\begin{tabular}{c}(2,1)\\(1,1)\end{tabular}}$  & ${\begin{tabular}{c}(2,1)\\(1,1)\end{tabular}}$  & ${\begin{tabular}{c}(2,1)\\(1,1)\end{tabular}}$  & ${\begin{tabular}{c}(2,1)\\(1,1)\end{tabular}}$  & ${\begin{tabular}{c}(2,1)\\(1,1)\end{tabular}}$  & ${\begin{tabular}{c}(2,1)\\(1,1)\end{tabular}}$  & ${\begin{tabular}{c}(2,1)\\(1,1)\end{tabular}}$  & ${\begin{tabular}{c}(2,1)\\(1,1)\end{tabular}}$  & ${\begin{tabular}{c}(2,1)\\(1,1)\end{tabular}}$  & ${\begin{tabular}{c}(2,1)\\(1,1)\end{tabular}}$  & ${\begin{tabular}{c}(2,1)\\(1,1)\end{tabular}}$  \\ 
 \hline 
$w_0=0.8$ & ${\begin{tabular}{c}(2,1)\\(1,1)\end{tabular}}$  & ${\begin{tabular}{c}(2,1)\\(1,1)\end{tabular}}$  & ${\begin{tabular}{c}(2,1)\\(1,1)\end{tabular}}$  & ${\begin{tabular}{c}(2,1)\\(1,1)\end{tabular}}$  & ${\begin{tabular}{c}(2,1)\\(1,1)\end{tabular}}$  & ${\begin{tabular}{c}(2,1)\\(1,1)\end{tabular}}$  & ${\begin{tabular}{c}(2,1)\\(1,1)\end{tabular}}$  & ${\begin{tabular}{c}(2,1)\\(1,1)\end{tabular}}$  & ${\begin{tabular}{c}(2,1)\\(1,1)\end{tabular}}$  & ${\begin{tabular}{c}(2,1)\\(1,1)\end{tabular}}$  & ${\begin{tabular}{c}(2,1)\\(1,1)\end{tabular}}$  \\ 
 \hline 
$w_0=0.9$ & ${\begin{tabular}{c}(1,2)\\(2,1)\end{tabular}}$  & ${\begin{tabular}{c}(1,2)\\(2,1)\end{tabular}}$  & ${\begin{tabular}{c}(1,2)\\(2,1)\end{tabular}}$  & ${\begin{tabular}{c}(1,2)\\(2,1)\end{tabular}}$  & ${\begin{tabular}{c}(1,2)\\(2,1)\end{tabular}}$  & ${\begin{tabular}{c}(1,2)\\(2,1)\end{tabular}}$  & ${\begin{tabular}{c}(1,2)\\(2,1)\end{tabular}}$  & ${\begin{tabular}{c}(1,2)\\(2,1)\end{tabular}}$  & ${\begin{tabular}{c}(1,2)\\(2,1)\end{tabular}}$  & ${\begin{tabular}{c}(1,2)\\(2,1)\end{tabular}}$  & ${\begin{tabular}{c}(1,2)\\(2,1)\end{tabular}}$  \\ 
 \hline 
$w_0=1.0$ & ${\begin{tabular}{c}(1,0)\\(2,0)\end{tabular}}$  & ${\begin{tabular}{c}(1,0)\\(2,0)\end{tabular}}$  & ${\begin{tabular}{c}(1,0)\\(2,0)\end{tabular}}$  & ${\begin{tabular}{c}(1,0)\\(2,0)\end{tabular}}$  & ${\begin{tabular}{c}(1,0)\\(2,0)\end{tabular}}$  & ${\begin{tabular}{c}(1,0)\\(2,0)\end{tabular}}$  & ${\begin{tabular}{c}(1,0)\\(0,1)\end{tabular}}$  & ${\begin{tabular}{c}(0,0)\\(0,0)\end{tabular}}$  & ${\begin{tabular}{c}(0,0)\\(0,0)\end{tabular}}$  & ${\begin{tabular}{c}(1,0)\\(1,2)\end{tabular}}$  & ${\begin{tabular}{c}(0,0)\\(0,0)\end{tabular}}$  \\ 
 \hline 
\end{tabular}
\caption{Momentum sectors relative to the Chern insulator state momentum, for the two lowest energy states (not related by $C_{2z}$) on a 5$\times$3 at $\nu=-3$ and $\lambda=1$ (top first lowest energy state, bottom second energy states). The system is fully spin and valley polarized. $\rm{Ch}$ indicates the Chern insulators states.}\label{ed:tab:momentumlambda1nu-3nx5ny3}
\end{table}
}

\subsection{Phase diagrams and spectra for the \texorpdfstring{$\lambda$}{lambda} interpolation}\label{ed:app:phasediagramlambdanu3}

The phase diagrams and energy spectra of the four limits, i.e., chiral-flat, chiral-nonflat, nonchiral-flat, nonchiral-nonflat, shown in the main text Sec. \ref{sec:nu3} are obtained for either $\lambda=0$ or $\lambda=1$. While the $\lambda=0$ and $\lambda=1$ calculations give the same irreps of the ground state and the lowest excitations (over all momenta), they do have some differences in excitations at some momentum, and differences in the critical values of phase transitions.
For example, at chiral-flat limit, the irreps of second lowest charge neutral excitations are different (Fig.~\ref{ed:fig:irrep_spectrum_4x2}), and the transition point of $w_0/w_1$ are changed (Fig.~\ref{ed:fig:phase_diagram_3x2}). 
In this subsection we provide some spectrum and phase diagrams for the $\lambda$ interpolation to illustrate how the FMC model is connected with the full TBG model.

\begin{figure}
    \centering
    \includegraphics[width=\linewidth]{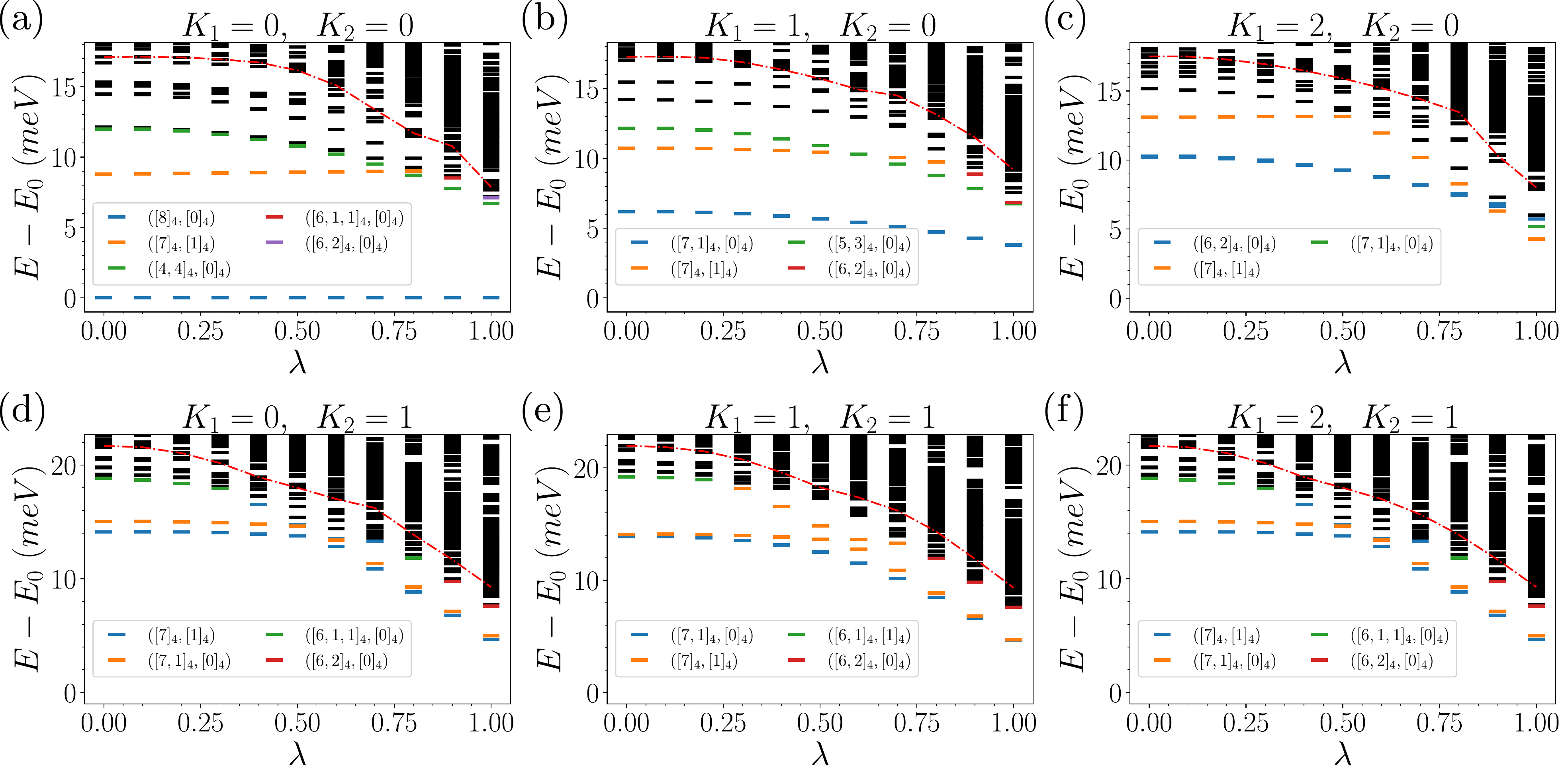}
    \caption{The energy spectrum in the chiral-flat limit on $4\times 2$ lattice with different $\lambda$ values at filling factor $\nu=-3$. Each total momentum is shown in different subfigures. The U(4)$\times$U(4) irreps of some low lying states are labeled by their color. The plots with momentum $K=(3, 0)$ and $K=(3, 1)$ are identical to the spectrum of $K=(1, 0)$ and $K=(1, 1)$ due to $C_{2z}$ symmetry, and therefore we are ignored. Because only several lowest eigenvalues are solved for each symmetry sector, the spectra above the red dashed line are incomplete. The spectra with momentum $K=(0, 1)$ and $K=(2, 1)$ are identical because of the $C_{2x}$ symmetry.}
    \label{ed:fig:lambda_sweep_spectrum}
\end{figure}

First we study how the low energy spectra at $\nu = -3$ on a $4\times 2$ lattice
change with $\lambda$ at chiral-flat limit. The irreps of low-lying states 
shown in Fig.~\ref{ed:fig:irrep_spectrum_4x2}a and b are clearly not identical. For example, the irrep of the lowest state at momentum $K_1 = 2$, $K_2 = 0$ changes from $([6,2]_4, [0]_4)$ at $\lambda=0$ to $([7]_4, [1]_4)$ at $\lambda= 1$. In Fig.~\ref{ed:fig:lambda_sweep_spectrum}, we present some low energy states with their irreps for each momentum sector with various values of $\lambda$. Although the ground state irrep is not changing, we can clearly see the level crossings between excited states. In Fig.~\ref{ed:fig:lambda_sweep_spectrum}c, we can see the level crossing between $([7]_4, [1]_4)$ state and $([6,2]_4, [0]_4)$ state. Meanwhile, an unchanged irrep cannot exclude the possibility of a level crossing, because some irreps can appear multiple times in the whole spectrum at a given filling factor, and same representations can have avoided crossings. As shown in Fig.~\ref{ed:fig:lambda_sweep_spectrum}d, the irrep of the lowest energy state with total momentum $K_1 = 0$, $K_2 = 1$ is $([7]_4, [1]_4)$ for any $0\leq\lambda\leq 1$. However we can also notice an avoided level crossing at around $\lambda \simeq 0.5$.

\begin{figure}
    \centering
    \includegraphics[width=\linewidth]{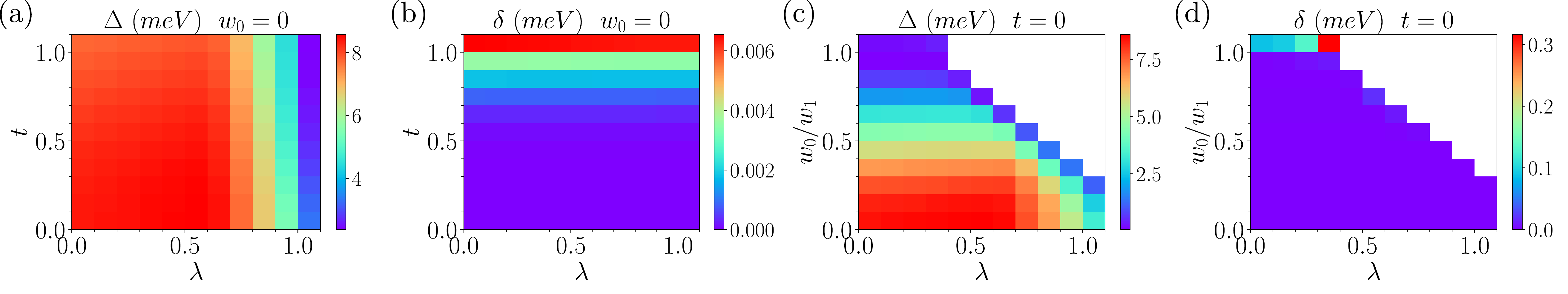}
    \caption{Phase diagrams at $\nu=-3$ on $3\times 2$ lattice in the chiral-nonflat limit (a and b) and nonchiral-flat limit (c-d), considering all the symmetry sectors. In subfigures (a and c) we show the finite size gap $\Delta$, and in subfigures (b and d) we show the spread $\delta$ as defined in Sec.~\ref{sec:phasediagramnu3}.}
    \label{fig:phase_diagram_3x2_full_sector_lambda_sweep}
\end{figure}

We also studied the phase diagram with all symmetry sectors on $3\times 2$ lattice at the (first) chiral limit with various values of $\lambda$ and $t$, and at flat band limit with various values of $\lambda$ and $w_0$. The finite size gap and the ground state manifold spread can be found in Fig.~\ref{fig:phase_diagram_3x2_full_sector_lambda_sweep}. Not surprisingly, a larger $\lambda$ leads to a significant reduction of the finite size gap $\Delta$, which implies instabilities of the ground states, as discussed in Sec. \ref{sec:nu3}, and therefore we have a decreasing $w_0/w_1$ phase transition value for increasing $\lambda$.

\begin{figure}
    \centering
    \includegraphics[width=0.9\linewidth]{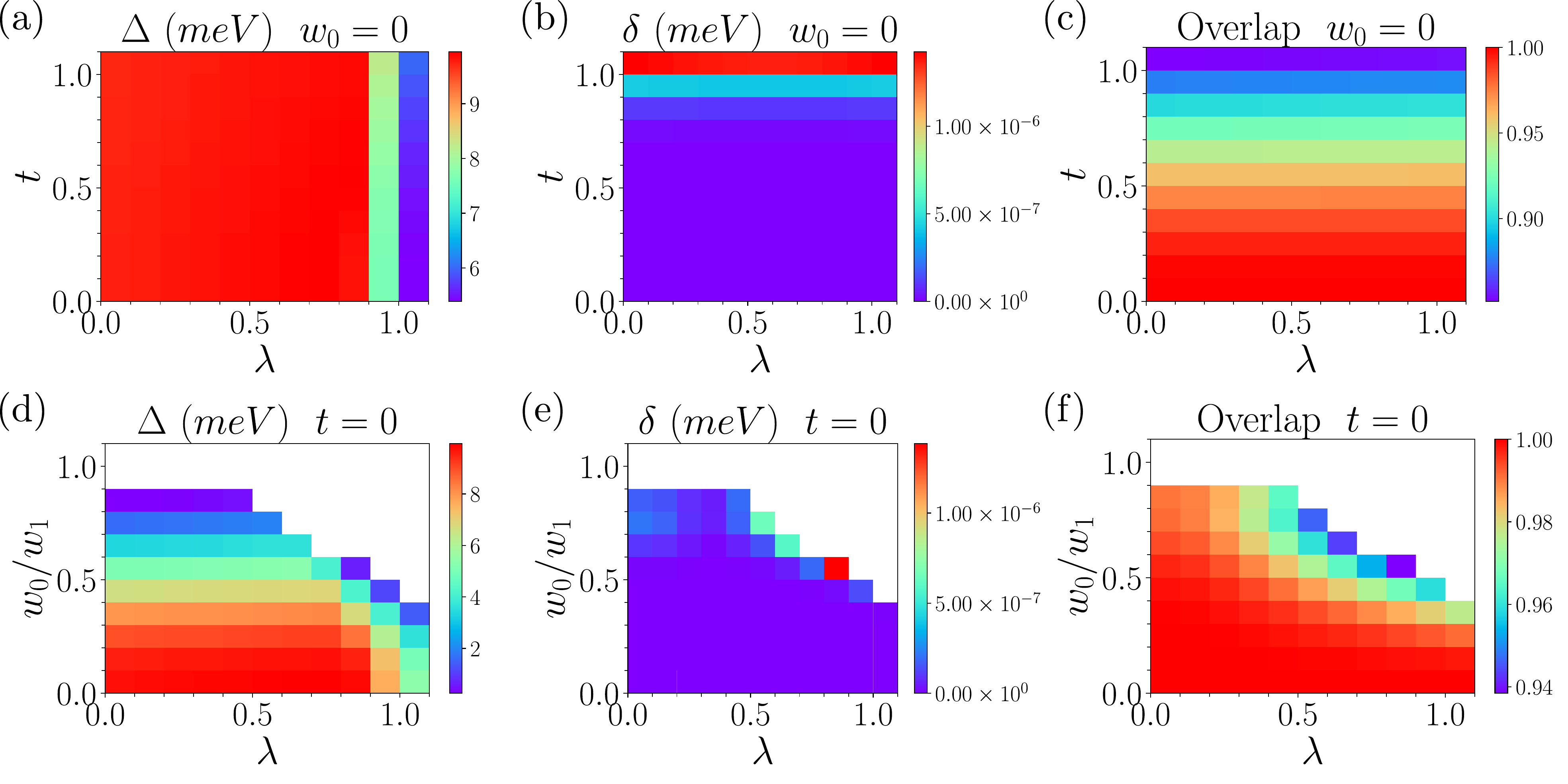}
    \caption{Phase diagrams at $\nu=-3$ on $4\times 3$ lattice in the chiral-nonflat limit (a, b and c) and nonchiral-flat limit (c, d and e) and in the spin and valley polarized sector. Here we show the finite size gap $\Delta$, the spread $\delta$, and the overlap between the two lowest states and the Chern insulator state wavefunctions. The quantities shown here are defined in Sec. \ref{sec:fullypolsectorsnu3}.}
    \label{ed:fig:phase_diagram_lambda_t_lambda_w0_app_4x3}
\end{figure}

\begin{figure}
    \centering
    \includegraphics[width=0.9\linewidth]{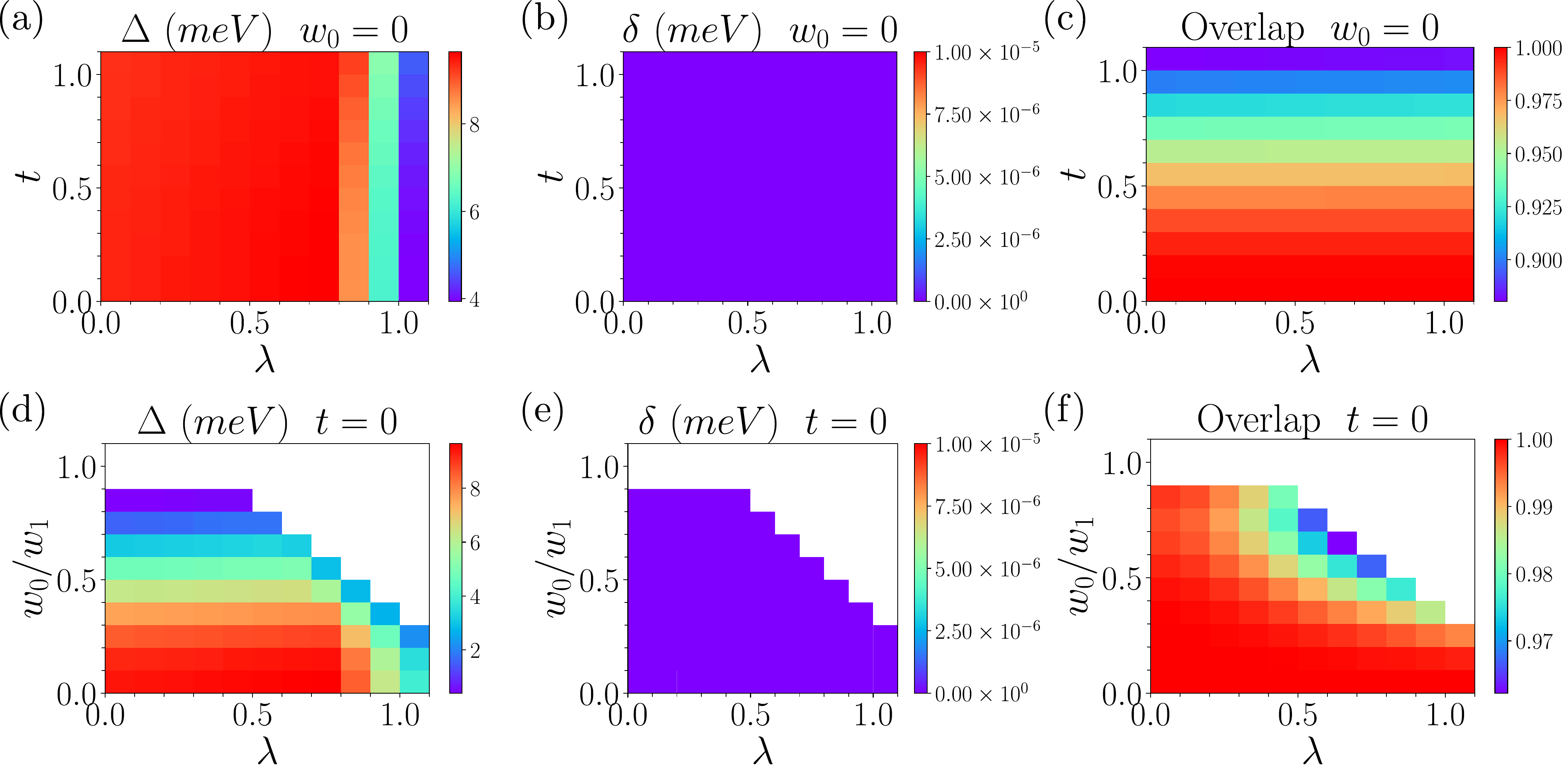}
    \caption{Phase diagrams at $\nu=-3$ on $3\times 3$ lattice in the chiral-nonflat limit (a, b and c) and nonchiral-flat limit (c, d and e) and in the spin and valley polarized sector. Similar to Fig.~\ref{ed:fig:phase_diagram_lambda_t_lambda_w0_app_4x3}, we show the finite size gap $\Delta$, the spread $\delta$ and the wavefunction overlap.}
    \label{ed:fig:phase_diagram_lambda_t_lambda_w0_app_3x3}
\end{figure}

By focusing on the spin and valley polarized sectors, we are able to study the phase diagrams on system sizes larger than $3\times 2$. Fig.~\ref{ed:fig:phase_diagram_lambda_t_lambda_w0_app_4x3} and \ref{ed:fig:phase_diagram_lambda_t_lambda_w0_app_3x3} are the phase diagrams calculated on $4\times 3$ and $3\times 3$ lattices, respectively. (Note that the corresponding $5\times 3$  phase diagrams have been omitted due to their similarities with the $4\times 3$ ones.) On both the system sizes, we find that the transition point of $w_0/w_1$ reduces from $\simeq 0.9$ to $\simeq0.3$ or $0.4$ when $\lambda > 0.5$. This shift of phase boundary due to finite $\lambda$ is similar to that on the $3\times2$ lattice shown Figs.~\ref{fig:phase_diagram_3x2_full_sector_lambda_sweep}c and d. This also hints that the area of Chern insulator phase in the phase diagram is not strongly affected by the system size. We also observe the spread between the two lowest energy states on $3\times 3$ lattice is indeed small for $\lambda$ values between 0 and 1. 
This is due to the complex $C_{3z}$ eigenvalues of the two ground states, which are only well defined on the $3\times3$ lattice but not the $4\times3$ and $3\times2$ lattices, and the time reversal symmetry makes the two states degenerate. 
Now we show that the two ground states must have complex $C_{3z}$ eigenvalues.
Starting from the $w_0=0,t=0$ ground states, which are Slater states $|\Psi_{-3}^{1,0}\rangle$, $|\Psi_{-3}^{0,1}\rangle$ defined in Eqs.~(\ref{ed:eq:nu3chern1}) and~(\ref{ed:eq:nu3chernm1}), the $C_{3z}$ eigenvalues of the ground states will not change with $w_0$ and $t$ before they reach the phase boundaries.
Thus the $C_3$ eigenvalues within the phase boundary are same as those of $|\Psi_{-3}^{1,0}\rangle$, $|\Psi_{-3}^{0,1}\rangle$. 
Due to the relation between Chern number $\nu_C$ and the $C_3$ eigenvalue $\xi$ for a Slater state \cite{fang_bulk_2012}, i.e., $\xi= e^{-i\frac{2\pi}3 \nu_C}$, we know the $C_{3z}$ eigenvalues of $|\Psi_{-3}^{1,0}\rangle$ and $|\Psi_{-3}^{0,1}\rangle$ must be complex, since they have $\nu_C=1$ and $\nu_C=-1$, respectively.

\section{Numerical Results for \texorpdfstring{$\nu=-2$}{nu=-2}}\label{ed:appsec:numerical-othernu}

\subsection{Spin polarized sector}\label{ed:appsec:nu2spin}

\begin{figure}
    \centering
    \includegraphics[width=0.5\linewidth]{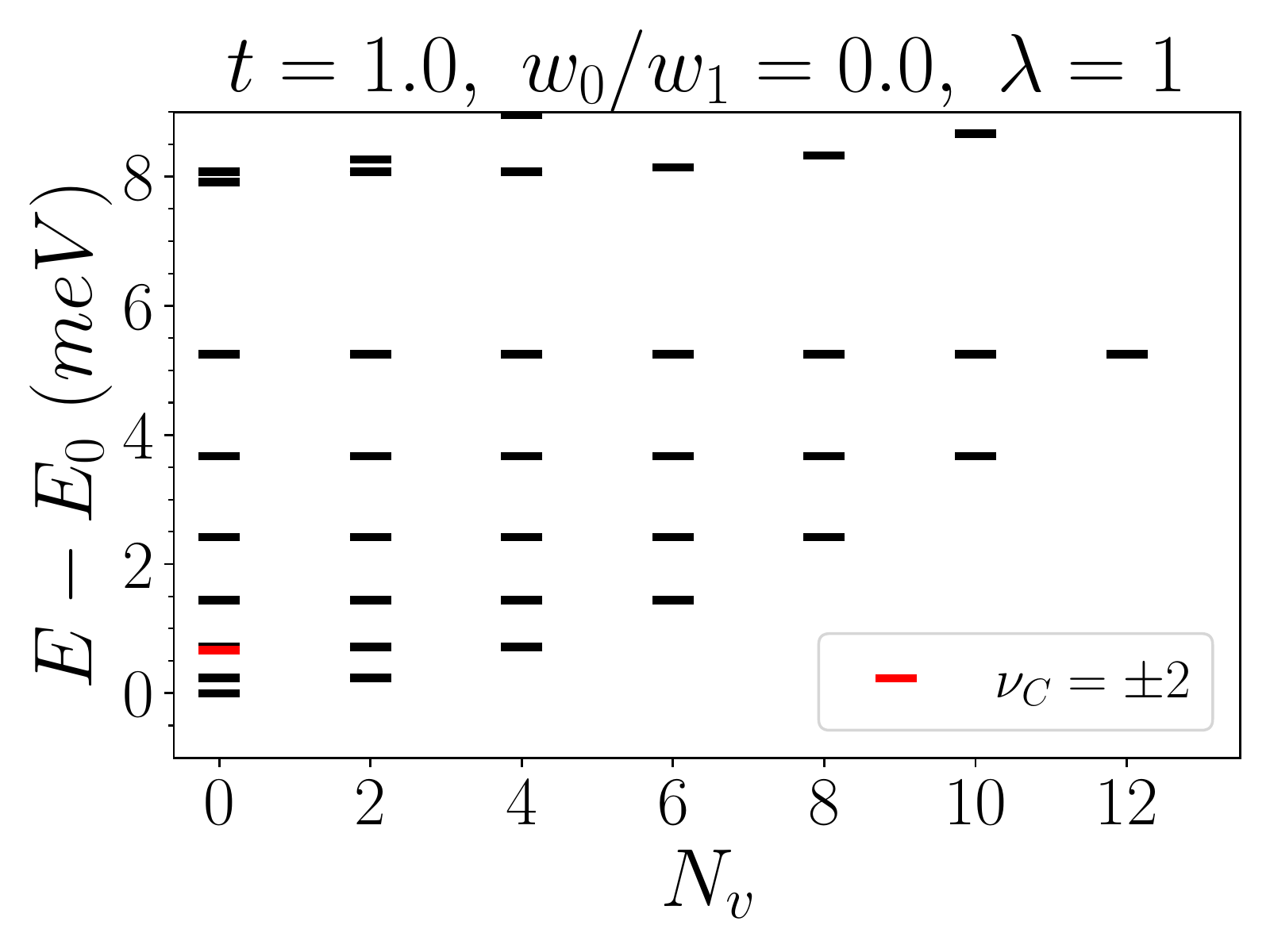}
    \caption{The low energy spectrum on $3\times 2$ lattice with $t=1, w_0 = 0$ and $\lambda=1$ in the spin polarized sector. The low energy states form $U(2)$ multiplets with $L=0, 1, \cdots, N_M$ and the ground state is a $U(2)$ singlet state. The two states with Chern number $\nu=\pm2$ are also shown in the figure represented by a red symbol.}
    \label{fig:spectrum_sp_chiral}
\end{figure}

\begin{figure}
    \centering
    \includegraphics[width=0.7\linewidth]{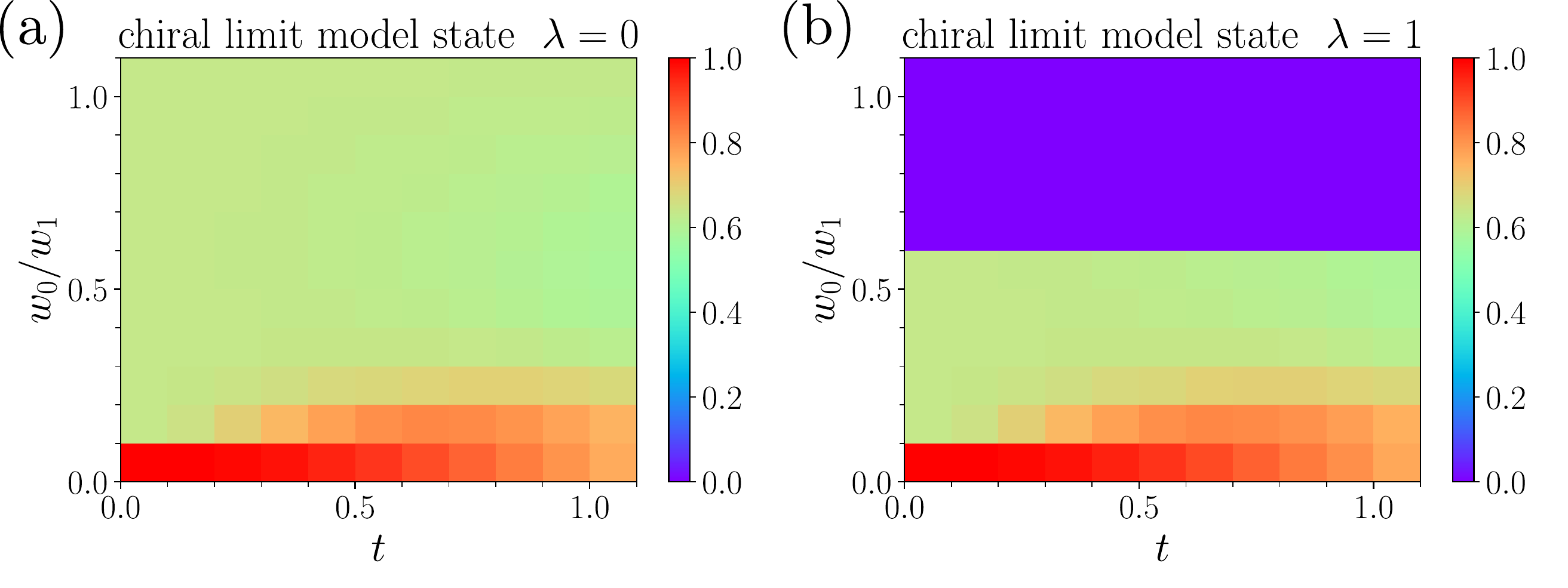}
    \caption{The phase diagrams at $\nu=-2$ filling on $3\times 2$ lattice in the spin polarized sector with FMC (a) and without FMC (b). The color code represents the overlap defined in Eq.~(\ref{eqn:chiral_model_state_overlap}). It can be shown that the overlap is close to 1 only when $w_0$ is close to zero, where the system in the spin polarized sector has a valley $SU(2)$ symmetry.}
    \label{fig:chiral_model_state_phase_diagram}
\end{figure}

In Sec.~\ref{sec:phasedigramnu-2spin} we discussed the valley coherent ground state in the non-chiral nonflat case in the spin polarized sector. We notice that the overlap between the inter valley coherent model state wavefunction and the ED ground state wavefunction in the chiral nonflat limit is smaller than the overlap when $w_0 \neq 0$. Because of the larger symmetry in the chiral limit, we need to consider another model state. As discussed in Ref.~\cite{ourpaper4} (see also Ref.~\cite{bultinck_ground_2020}), the perturbation theory shows that the following state is preferred in the chiral limit:
\begin{equation}
    |\Psi^{\rm chiral}_{\nu=-2}(\varphi, \gamma)\rangle = \prod_{\mathbf{k}}\left(e^{-\frac{i\gamma}{2}} d^\dagger_{\vk ,+1, +, \uparrow}\cos\frac{\varphi}{2} + e^{\frac{i\gamma}{2}}d^\dagger_{\vk, +1, -, \uparrow}\sin\frac{\varphi}{2}\right)\left(e^{-\frac{i\gamma}{2}} d^\dagger_{\vk ,-1, +, \uparrow}\sin\frac{\varphi}{2} + e^{\frac{i\gamma}{2}}d^\dagger_{\vk, -1, -, \uparrow}\cos\frac{\varphi}{2}\right)|0\rangle\,.
\end{equation}
However, this state is a symmetry breaking state, which depends on two angle parameters $\varphi$ and $\gamma$, and is not the eigenstate of the Cartan subalgebras. Thus it cannot be obtained by exact diagonalization. The Hamiltonian in the spin polarized sector in the chiral limit has a $U(2)$ symmetry in valley space. Thus the low energy spectrum exhibits $SU(2)$ multiplets, as observed in Fig.~\ref{fig:spectrum_sp_chiral} . Among these multiplets, the lowest one is an $SU(2)$ singlet. Therefore, we expand the model state wavefunction $|\Psi^{\rm chiral}_{\nu=-2}(\varphi, \gamma)\rangle$ on spherical harmonics $Y_L^m(\varphi, \gamma)$:
\begin{align}
    |\Psi^{\rm chiral}_{\nu=-2}(\varphi, \gamma)\rangle &= \sum_{L=0}^{N_M}\sum_{N_v/2=-L}^{L}\mathcal{N}_{L,N_v}Y_L^{N_v/2}(\varphi, \gamma) | \psi_{\rm chiral}(L, N_v)\rangle\\
    |\psi_{\rm chiral}(L, N_v)\rangle &\propto \int_0^{2\pi}d\gamma\int_0^{\pi} d\varphi~ \sin\varphi \left[Y^{\frac{N_v}{2}}_L(\varphi, \gamma)\right]^*|\Psi^{\rm chiral}_{\nu=-2}(\varphi, \gamma)\rangle
\end{align}
in which $\mathcal{N}_{L, N_v}$ are normalization factors and the components $|\psi_{\rm chiral}(L, N_v)\rangle$ with quantum numbers $L$ and $N_v$ are normalized wavefunctions. Both these normalization factors and components are independent of the two angles $\varphi$ and $\gamma$. $L$ is thus the valley $SU(2)$ ``angular momentum". Similar to the inter valley coherent model states, we define the overlap between the numerical ground state and the valley $SU(2)$ singlet model state ($L=0,N_v=0$) as
\begin{equation}\label{eqn:chiral_model_state_overlap}
    \mathrm{Overlap} = |\langle \psi_{\rm chiral}(L=0, N_v=0) |\psi_{ED} \rangle |^2\,.
\end{equation}
We provide the overlap for each value of $t$ and $w_0$ in Fig.~\ref{fig:chiral_model_state_phase_diagram}. It can be seen clearly that this model state agrees well with the ED ground state when $w_0 \approx 0$.

\subsection{Valley polarized sector}\label{ed:appsec:nu2valley}

In Sec.~\ref{sec:phasedigramnu-2valley} we have explored the phase diagram in the nonchiral-nonflat limit at filling factor $\nu=-2$ in the valley polarized sector. Those diagrams were obtained for the $\lambda=0$ (the FMC model) and $\lambda=1$ (the full TBG model). In this appendix, we will provide the valley polarized phase diagrams for the $\lambda$ interpolation in either nonchiral-flat limit or the chiral-nonflat limit.

\begin{figure*}[t]
    \centering
    \includegraphics[width=\linewidth]{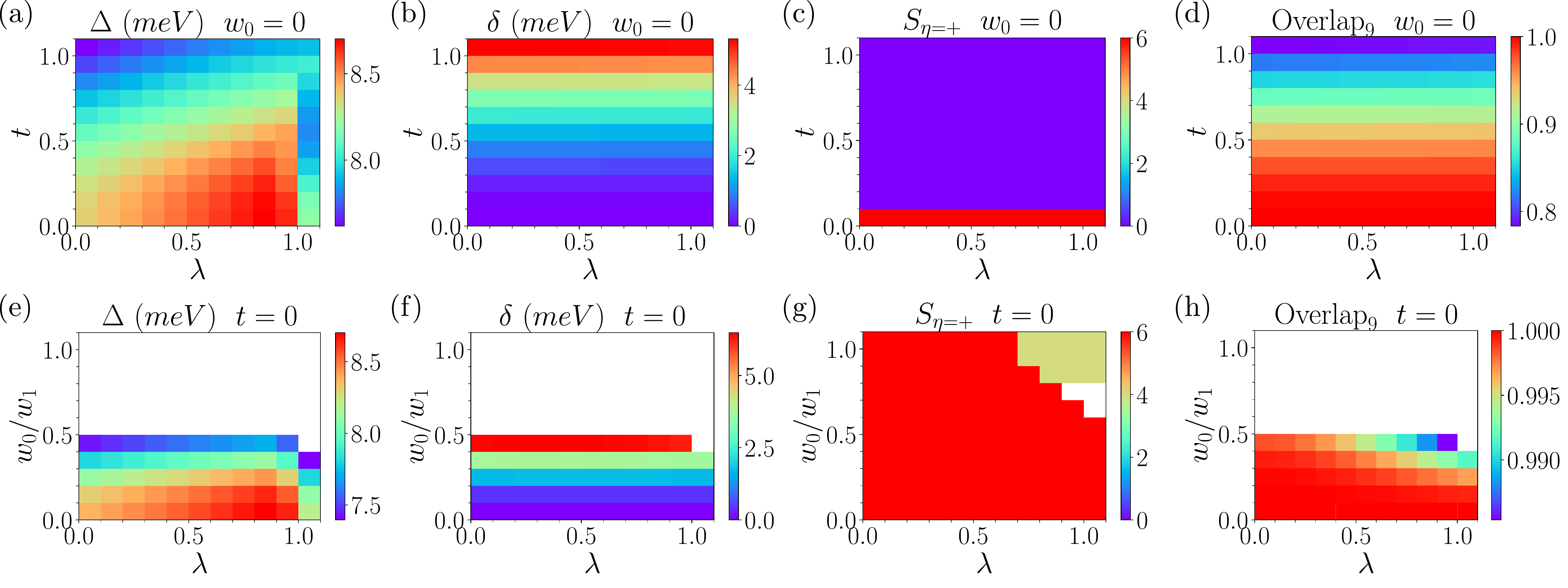}
    \caption{Phase diagrams at $\nu=-2$ on $3\times 2$ lattice in the chiral-nonflat limit (a, b, c and d) and the nonchiral-flat limit (e, f, g and h) and valley polarized sectors. These phase diagrams are function of the interpolating parameter $\lambda$. In each limit, we provide the finite size gap $\Delta$ (a and e), the spread $\delta$ (b and f), the total spin $S_{\eta=+}$ in valley $\eta=+$ and the overlap. The definitions of these quantities can be found in Sec.~\ref{sec:phasedigramnu-2}.}
    \label{ed:fig:phase_diagram_lambda_t_lambda_w0_3x2_-2}
\end{figure*}

These phase diagrams are shown in Fig.~\ref{ed:fig:phase_diagram_lambda_t_lambda_w0_3x2_-2}. As we have seen in Sec.~\ref{sec:phasedigramnu-2} and Fig.~\ref{ed:fig:phase_diagram_t_w0_3x2_-2}, the kinetic energy controlled by $t$ barely affects the Chern insulator. Unsurprisingly, we find the same feature in the chiral-nonflat limit (see Figs.~\ref{ed:fig:phase_diagram_lambda_t_lambda_w0_3x2_-2}a to d). The difference between the transition value of $w_0/w_1$ for $\lambda = 0$ and $\lambda = 1$ is small ($w_0/w_1 \simeq 0.5$ and $w_0/w_1 \simeq 0.4$ respectively). 
In these phase diagrams in the nonchiral-flat limit (Figs.~\ref{ed:fig:phase_diagram_lambda_t_lambda_w0_3x2_-2}e to h), the phase boundary barely depends on $\lambda$ as expected.

More interestingly, Figs.~\ref{ed:fig:phase_diagram_lambda_t_lambda_w0_3x2_-2}c and~g provide the total spin quantum number $S_{\eta=+}$ for the valley $\eta=+$. They validate again the analytical results (exact/perturbative) about the magnetic order in Ref.~\cite{ourpaper4}: when valley fully polarized, the system favors the ferromagnetic phase (where a single spin-valley flavor has two bands fully occupied) in the nonchiral-flat limit, and favors the spin-singlet (where each of the two spins in the occupied valley is half-occupied) in the chiral-nonflat limit. 

\section{Effect of normal ordering and particle-hole symmetry}\label{ed:appsec:phs}

In this appendix, we will compare the Hamiltonian Eq.~\ref{ed:eq:HIprojected} exhibiting particle-hole symmetric around the CNP and its normal ordered counterpart. The relation between these two Hamiltonians was discussed in Ref.~\cite{ourpaper3}. Here we will briefly analytically review this relation, followed by a more detailed numerical comparison. The two body interacting Hamiltonian in Eq.~\ref{ed:eq:HIprojected} can be also written as the following form:
\begin{equation}
H_{I}=H_I^{\text{norm}}+\Delta H^{(1)}+ \Delta H^{(2)}+\text{const.}\ .\label{eq:HInormal}
\end{equation}
where $H_I^{\text{norm}}$ is a normal-ordered term and $\Delta H^{(1)}_I$ and $\Delta H^{(2)}_I$ are quadratic terms of fermion operators. The total quadratic term 
\begin{equation}
\Delta H_I=\Delta H^{(1)}_I+\Delta H^{(2)}_I,\label{ed:eq:deltaHI}
\end{equation}
was proved to be, in Ref.~\cite{ourpaper3}, equal to the Hartree-Fock term from the filled bands below the flat bands ($\Delta H^{(1)}_I$ being the Hartree potential and $\Delta H^{(2)}_I$ the Fock potential). With the term $\Delta H_I$, the projected many-body Hamiltonian preserves the charge-conjugation symmetry around the CNP, a symmetry that is present for the unprojected Hamiltonian irrespective of the normal ordering. The situation is similar to the fractional quantum Hall effect and its lattice cousin, the fractional Chern insulator. For the former, using or not the normal ordering only differs by a chemical potential, preserving the particle-hole symmetry for the Hamiltonian projected in a Landau level. For the later, the difference between the normal ordered and the non-normal ordered Hamiltonian is a momentum dependent one-body term akin to a dispersion relation, spoiling the particle-hole symmetry for the band projected Hamiltonian \cite{Bernevig-PhysRevB.85.075128,Zhao-PhysRevLett.109.186805}.

\begin{figure}
    \centering
    \includegraphics[width=0.9\linewidth]{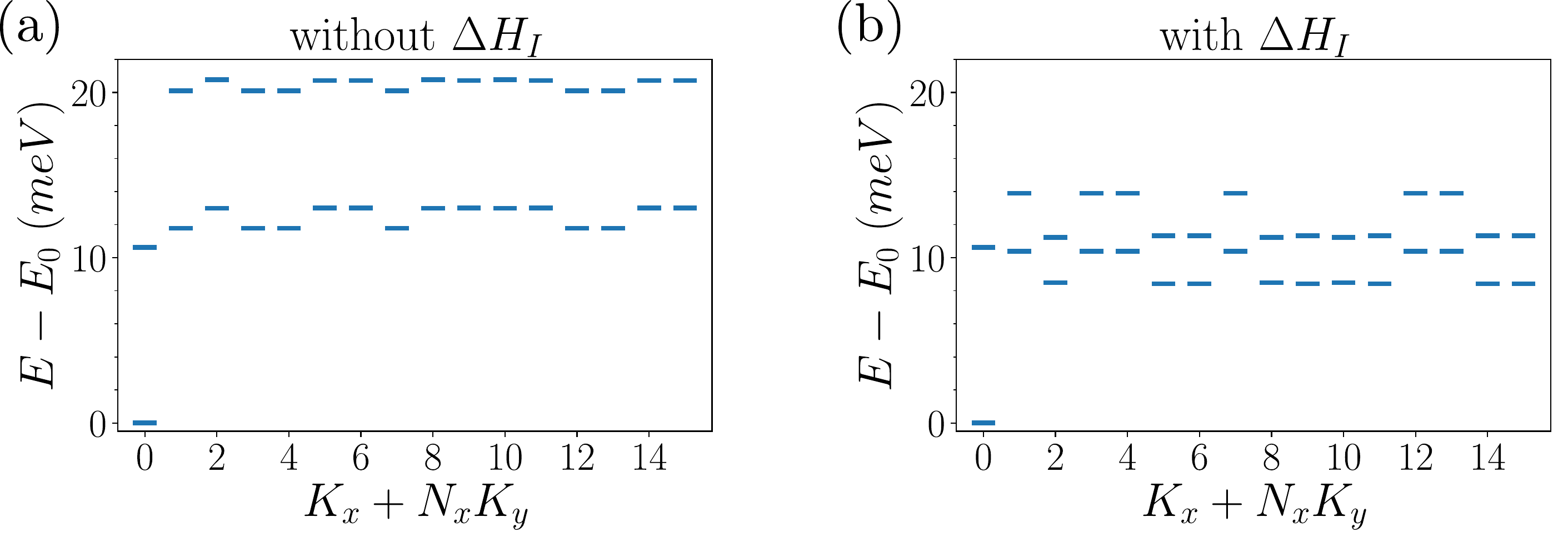}
    \caption{The spectrum for the full TBG model in the (first) chiral-flat limit at $\nu=-3$ filling on a $4\times 4$ lattice at twisting angle $\theta = 1.1014^\circ$ in the spin and valley fully polarized sectors. We use the normal-order Hamiltonian, which does not have Hartree-Fock terms, in subfigure a), and the particle-hole symmetric Hamiltonian, which has the Hartree-Fock terms, in subfigure b). It can be shown that the charge neutral gap tends to be smaller when we take the Hartree-Fock terms into consideration.}
    \label{ed:fig:4x4_HF}
\end{figure}

\begin{figure}
    \centering
    \includegraphics[width=0.9\linewidth]{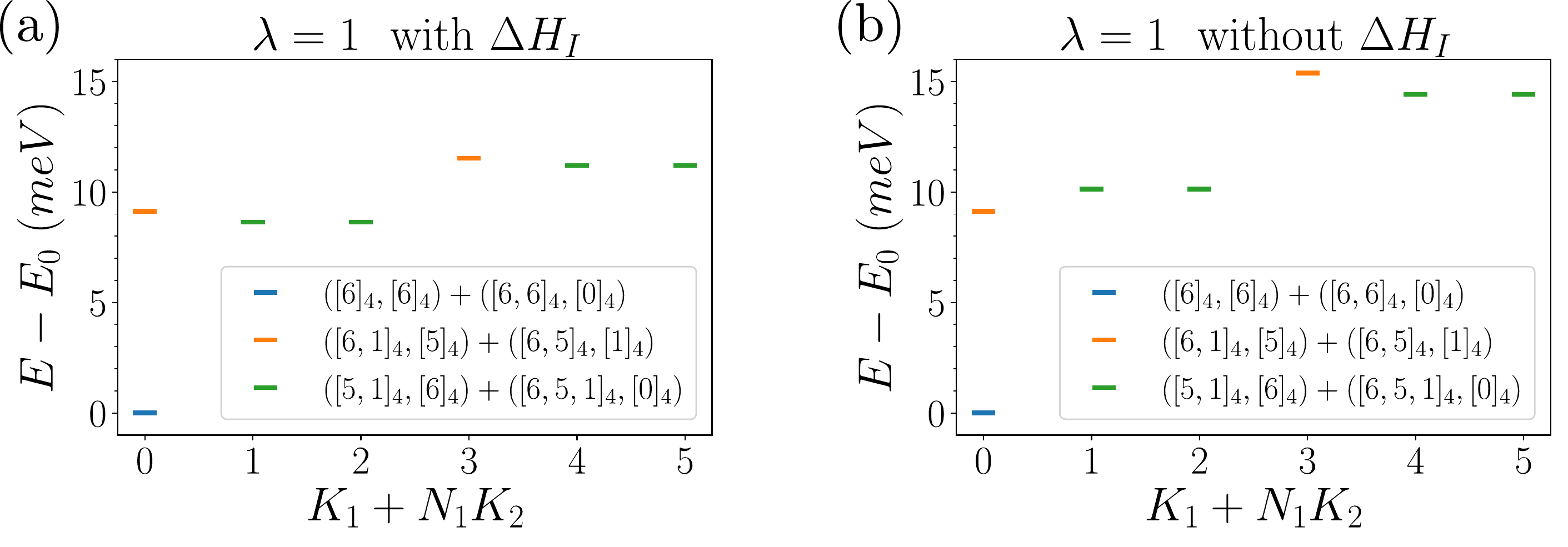}
    \caption{The energy spectra and irreps of low energy states on $3\times 2$ lattice at $\nu=-2$ filling with $\Delta H_I$ with (a) and without (b) $\Delta H_I$. Here we consider the full TBG model in the chiral-flat limit and the twisting angle is $\theta=1.1014^\circ$ (Note that (a) is just Fig.~\ref{ed:fig:irrep_3x2_lambda=1_N=12}.) In this example, we see that the $\Delta H_I$ term has no effect on the low energy irreps, and has almost no effect on the energies themselves.}
    \label{ed:fig:irrep_3x2_lambda=1_N=12_withoutHF}
\end{figure}

We now present the numerical results which show the effect of $\Delta H_I$ and how it affects the energy spectrum. We start with the spin and valley polarized sectors on a $4\times 4$ lattice in the chiral-flat limit at filling factor $\nu=-3$. The spectrum of the normal-ordered Hamiltonian and the full Hamiltonian at the (first) chiral-flat limit are shown in Fig.~\ref{ed:fig:4x4_HF}a and b, respectively. Although the ground state is identical in both cases, the low energy spectrum is globally compressed with $\Delta H_I$. We observe the same trend at filling factor $\nu=-2$. By performing ED in all symmetry sectors this time, we obtain the low energy states with their irreps. They are shown in Fig.~\ref{ed:fig:irrep_3x2_lambda=1_N=12_withoutHF} with and without $\Delta H_I$ (Note that Fig.~\ref{ed:fig:irrep_3x2_lambda=1_N=12_withoutHF}a is the same as in Fig.~\ref{ed:fig:irrep_3x2_lambda=1_N=12} and is just here for convenience). Once again we observe that the spectrum is not strongly affected. The irreps of the low-lying states are not changed, while the Goldstone branch finite momentum energy is slightly larger without $\Delta H_I$. 

\begin{figure}
    \centering
    \includegraphics[width=\linewidth]{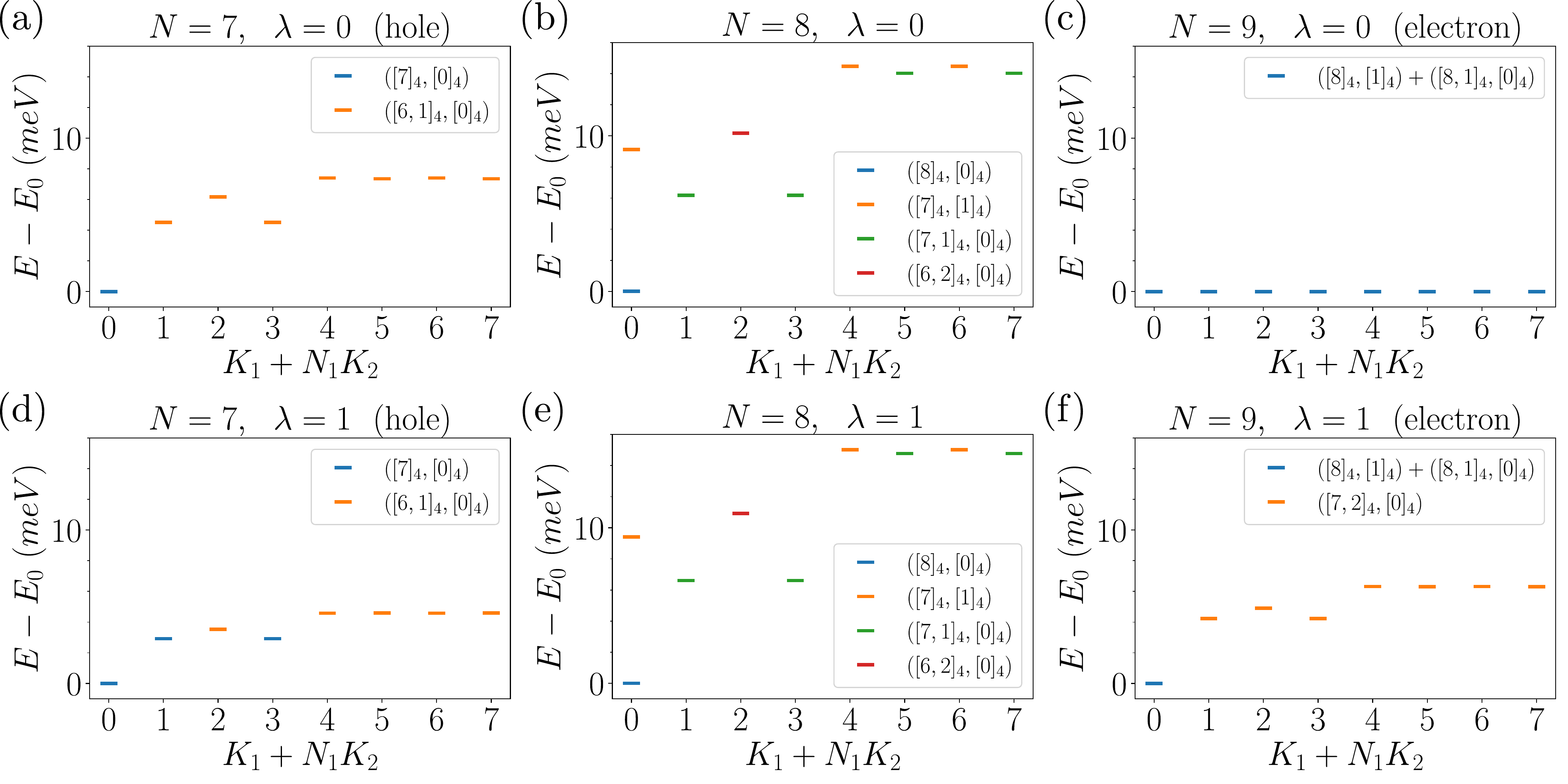}
    \caption{The energy spectra and irreps of the FMC model (a,b,c) and full TBG  model (d,e,f) on a $4\times 2$ lattice with total electron number $N = 7, 8$ and $9$ corresponding to the charge $-1$, $0$ and $+1$ excitations, respectively. Contrary to Fig.~\ref{ed:fig:charge_excitation_4x2}, the quadratic term $\Delta H_I$ has been discarded (other parameters are identical).  We use the notation "+" between irreps when they always appear with an exact degeneracy.}
    \label{ed:fig:charge_excitation_4x2_noHF}
\end{figure}

More interestingly, we also calculated the spectrum of the low energy states and their irreps of the FMC model and full TBG Hamiltonian at chiral-flat limit, on $4\times 2$ lattice with electron number $N = 7, 8, 9$, \emph{without} $\Delta H_I$. The results are shown in Fig.~\ref{ed:fig:charge_excitation_4x2_noHF} and should be compared to the results in Fig.~\ref{ed:fig:charge_excitation_4x2} where $\Delta H_I$ was included. In the normal ordered calculations, the charge $+1$ excitations and charge $-1$ excitations are no longer symmetric even when $\lambda = 0$. The irreps of these excitations are also no longer the same. Interestingly, we also notice that the dispersion of the charge $+1$ excitation is flat when $\lambda=0$ in the absence of $\Delta H_I$.

\begin{figure}
    \centering
    \includegraphics[width=10cm]{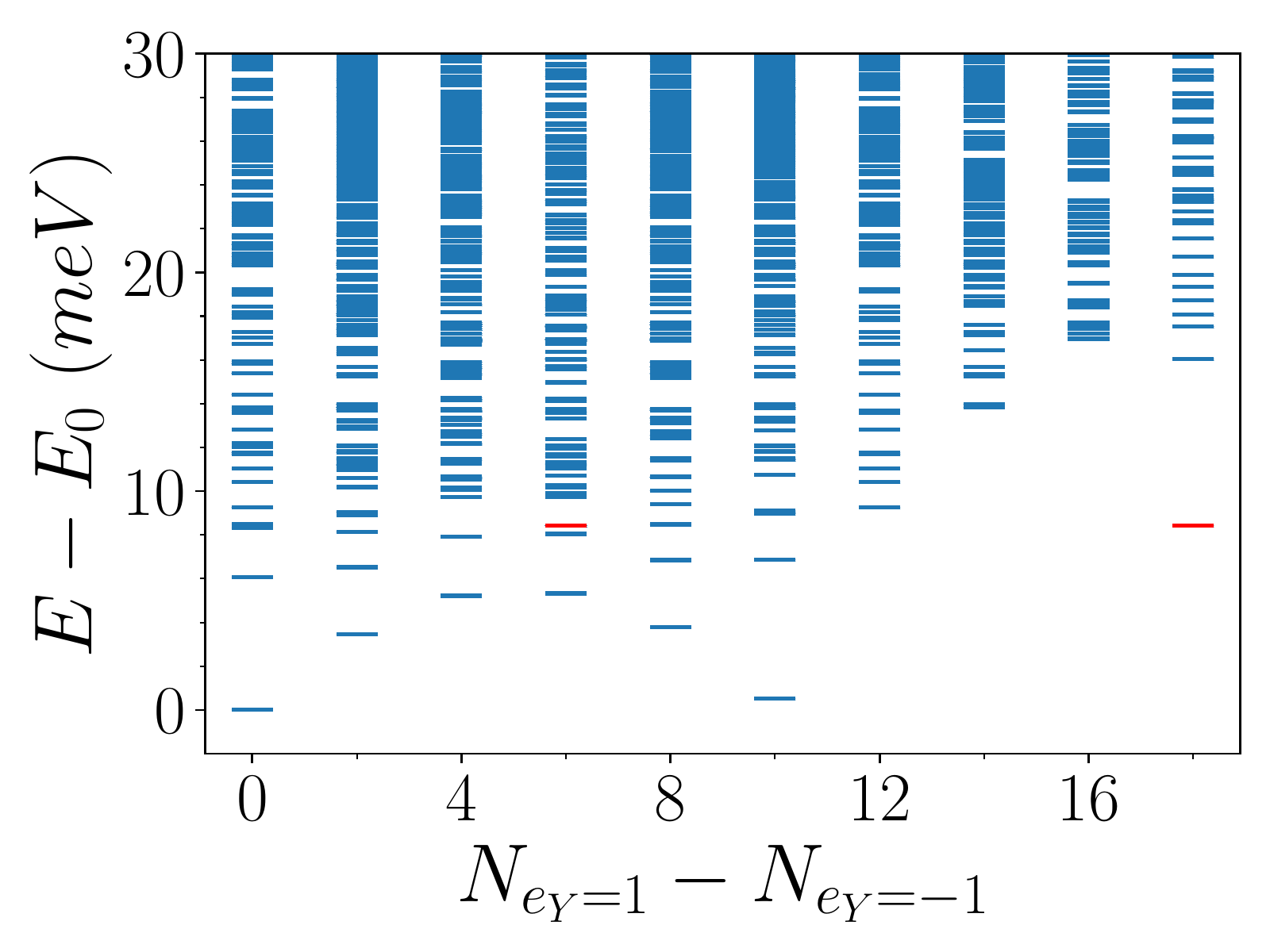}
    \caption{The low energy spectrum of symmetry sectors with a dimension smaller than $10^6$ at filling factor $\nu=-1$ ($N=18$) for the full TBG model on a $N_M = 3\times 2$ lattice in the chiral-flat limit. The spectrum is plotted versus the polarization of the Chern bands $N_{e_Y=+1} - N_{e_Y=-1}$ ($N_{e_Y}$ is the number of particles in bands with Chern number $e_Y$), irrespective of the other quantum numbers. Due to the $C_{2z}T$ symmetry, we only consider $N_{e_Y=+1} - N_{e_Y=-1} \ge 0$. The quadratic term $\Delta H_I$ is not considered. Red dashes are the Slater determinants which carry Chern numbers $\nu_C=3$ (at $N_{e_Y=+1} - N_{e_Y=-1}=3N_M$) and $\nu_C=1$ (at $N_{e_Y=+1} - N_{e_Y=-1}=N_M$). It should be compared to Fig.~\ref{ed:fig:spectrum3x2_-1_0}. In particular, the Chern state $\nu_C=1$ is no longer the lowest energy state in its own Chern band polarization.}
    \label{ed:fig:spectrum_3x2_-1_noHF}
\end{figure}

\begin{figure}
    \centering
    \includegraphics[width=0.9\linewidth]{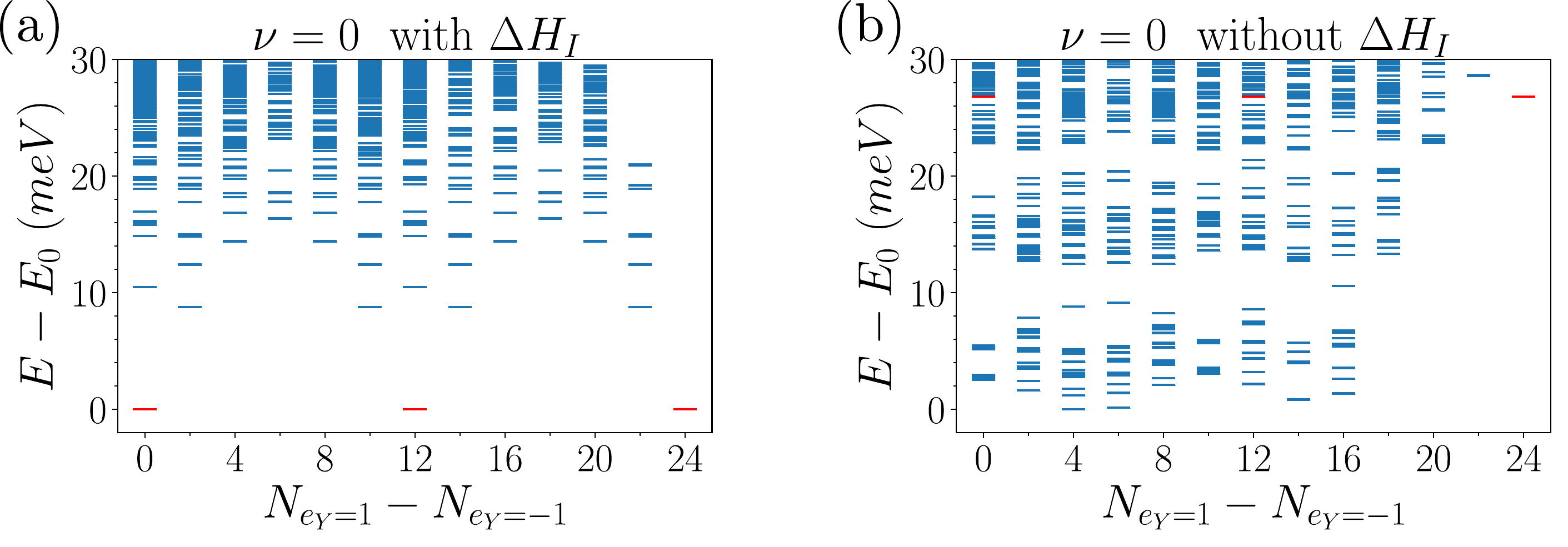}
    \caption{The low energy spectra of symmetry sectors with a dimension smaller than $10^6$ at filling factor $\nu=0$ ($N=24$) for the full TBG model on a $N_M=3\times 2$ lattice in chiral flat band limit. In subfigure (a) we considered the term $\Delta H_I$,  and in subfigure (b) $\Delta H_I$ is discarded. Similar to Figs.~\ref{ed:fig:spectrum3x2_-1_0} and~\ref{ed:fig:spectrum_3x2_-1_noHF}, the spectra are plotted versus the polarization of the Chern bands, and only $N_{e_Y=1}-N_{e_Y=-1}\ge 0$ are shown due to $C_{2z}T$ symmetry. Red dashes are the Chern insulator states with Chern number $\nu_C = 0$ (at $N_{e_Y=1}-N_{e_Y=-1} = 0$), $\nu_C=2$ (at $N_{e_Y=1}-N_{e_Y=-1} = 2N_M$) and $\nu_C=4$ (at $N_{e_Y=1}-N_{e_Y=-1} = 4N_M$). At this filling factor and in the absence of $\Delta H_I$, the Chern insulator states are not the ground states. In particular, the Chern states $\nu_C=0$ and $\nu_C=2$ are no longer the lowest energy states in their own Chern band polarization.}
    \label{ed:fig:nu_0_HF}
\end{figure}

We have also considered the filling factor $\nu=-1$ without $\Delta H_I$. A spectrum summarizing the symmetry sectors whose dimensions are smaller than $10^6$ is shown in Fig.~\ref{ed:fig:spectrum_3x2_-1_noHF}. It should be compared to Fig.~\ref{ed:fig:spectrum3x2_-1_0}. Like in Sec. \ref{sec:nu1}, we use red dashes to label these Chern insulator states. If $\Delta H_I$ is absent, we find that there exist states with energy lower than the Chern insulator states $\nu_C=\pm3$ and $\nu_C=\pm1$. Although the calculation does not consider all the possible symmetry sectors, we can already claim that the ground state irrep cannot be $([N_M, N_M, N_M]_4,[0]_4)$ or $([N_M, N_M ]_4, [N_M]_4)$. Indeed, these irreps can only be built from the Chern insulator states, which have already been shown to be excited states. Performing a similar calculation at $\nu=0$ in Fig.~\ref{ed:fig:nu_0_HF}, we reach exactly to the same conclusion, i.e., the Chern insulator states $\nu_C=\pm4$, $\nu_C=\pm2$ and $\nu_C=0$ are no longer the ground states once $\Delta H_I$ is discarded. These two examples illustrate the potential dramatic impact of $\Delta H_I$ on the low energy properties. 

\end{document}